\documentclass{jfm}
\usepackage{graphicx}
\usepackage{epstopdf, epsfig}
\usepackage{psfrag}

\usepackage{siunitx}
\usepackage{qtree}

\usepackage{subfigure}

\usepackage{upgreek}
\usepackage{revsymb}

\newcommand{\Tp}[1]{\Theta_{p{#1}}}

\newcommand{\dpp}[1]{d_{p#1}}
\newcommand{\Ii}{\mathsfbi{I}}
\newcommand{\bfit}[1]{\textbf{\textit{#1}}}

\newcommand{\kv}{\bfit{k}}

\newcommand{\kdotc}{\bfit{k}\cdot\bfit{w}}

\newcommand{\Uv}[1]{\bfit{U}_{p#1}}
\newcommand{\cv}[1]{\bfit{c}_{p#1}}
\newcommand{\Ev}{\bfit{E}}

\usepackage{amsmath}
\usepackage[dvipsnames]{xcolor}
\usepackage{comment}
\usepackage{bm}
\usepackage[colorinlistoftodos]{todonotes}
\usepackage[normalem]{ulem}

\usepackage{cleveref}
\crefformat{section}{\S#2#1#3} 
\crefformat{subsection}{\S#2#1#3}
\crefformat{subsubsection}{\S#2#1#3}

\shorttitle{Charge Transport Equation for Bidisperse Granular Flows}
\shortauthor{L. Ceresiat, J. Kolehmainen, A. Ozel}

\title{Charge Transport Equation for Bidisperse Collisional Granular Flows with Non-equipartitioned Fluctuating Kinetic Energy}

\author{Lise Ceresiat\aff{1}, Jari Kolehmainen\aff{2} \and Ali Ozel\aff{1}\corresp{\email{a.ozel@hw.ac.uk}}}
\affiliation{\aff{1}School of Engineering and Physical Sciences, Heriot-Watt University, Edinburgh EH14 4AS, UK
\aff{2} Department of Chemical and Biological Engineering, Princeton University, Princeton, NJ 08542, USA}

\begin{document}

\maketitle


\begin{abstract}
Starting from the Boltzmann-Enskog kinetic equations, the charge transport equation for bidisperse granular flows with contact electrification is derived with separate mean velocities, total kinetic energies, charges and charge variances for each solid phase. To close locally-averaged transport equations, a Maxwellian distribution is presumed for both particle velocity and charge. The hydrodynamic equations for bidisperse solid mixtures are first revisited and the resulting model consisting of the transport equations of mass, momentum, total kinetic energy, which is the sum of the granular temperature and the trace of fluctuating kinetic tensor, and charge is then presented. The charge transfer between phases and the charge build-up within a phase are modelled with local charge and effective work function differences between phases and the local electric field. The revisited hydrodynamic equations and the derived charge transport equation with constitutive relations are assessed through hard-sphere simulations of three-dimensional spatially homogeneous, quasi-one-dimensional spatially inhomogeneous bidisperse granular gases and a three-dimensional segregating bidisperse granular flow with conducting walls. 
\end{abstract}

\section{Introduction}
Granular materials acquire electrostatic charges after coming into frictional contact with themselves or with other materials. 
This process is called ``contact electrification" or short ``tribocharging". Tribocharging is naturally observed in Earth and Martian sandstorms~\citep{stow1969dust,melnik1998electrostatic}, and ash plumes of volcanic eruptions ~\citep{mendez2016effects,mendez2020microphysical}. It also is observed in industrial processes such as silo storage~\citep{gu_electrification_2017}, pneumatic conveying~\citep{yao2004electrostatics}, pharmaceutical blending and mixing~\citep{naik2016experimental}, electrostatic precipitation~\citep{mizuno2000electrostatic}, powder coating~\citep{barletta2006electrostatic}. Tribocharging also causes industrial implications; wall-sheeting in polyethylene fluidised bed reactors \citep{ciborowski1962electrostatic,hendrickson_electrostatics_2006}, particle segregation and mixing inefficiencies~\citep{forward2009charge}, potential hazard in packing containers~\citep{glor2005electrostatic}.

The governing physics of charge transfer are still under debate~\citep{williams2011triboelectric,lacks2011contact}. There are three main mechanisms suggested in the literature: (i) electron transfer~\citep{harper1967contact}, (ii) ion transfer~\citep{mccarty2008electrostatic}, and (iii) bulk material transfer~\citep{williams2012triboelectric}. All three have experimental evidence supporting them~\citep{matsusaka2010triboelectric}. In the electron transfer model, the driving force for electron transfer between contacting materials is the difference between the work functions of the materials. The electron transfer model probes the charge transfer between the conducting materials very well, but it is not applicable for insulators which have low charge mobility~\citep{duke_contact_1978, bailey_charging_2001}. The ion transfer mechanism proposes that insulators mainly exchange ions located on their surfaces during contact~\citep{mccarty2008electrostatic}. The ions are not necessarily part of the material, but can be tied to the environment properties (e.g. humidity) and during a mechanical contact between two surfaces, some of the ions may transfer from surface-to-surface that leads to different overall charges on the surfaces~\citep{wiles2004effects,mccarty2008electrostatic,waitukaitis2014size,schella2017influence}. When particles come into contact, they may also exchange material with one another. The material exchanged can have a non-zero charge difference that leads a charge transfer through a mechanism referred to as the bulk material transfer. While this possible mechanism has been known for some time~\citep{salaneck1976double}, its predictability and reproducibility is questionable~\citep{lowell1980contact}.  

One can ask whether tribocharging via electron or ion transfer mechanisms can be captured through a simple modelling framework, which is suitable for easy integration to large-scale granular flows. Recently, we developed a Computational Fluid Dynamics-Discrete Element Method (CFD-DEM) approach for gas-solid flows that accurately predicts the effects of tribocharging on flow hydrodynamics~\citep{kolehmainen2016hybrid,kolehmainen_triboelectric_2017}. In this approach, charge transfer between particles and charge build-up in the overall system are accounted for short-range electrostatic forces using the Coulomb force with neighbouring particles and long-range electrostatic forces via Poisson's equation~\citep{kolehmainen2016hybrid}. The charge accumulation on particles is modelled by an effective-work-function based model~\citep{laurentie_discrete_2013}. 
The effective work function is a lumped parameter that can be used to quantify charging rates and extents observed in specific experimental studies~\citep{laurentie_discrete_2013,naik2015combined,naik2016experimental, kolehmainen_effect_2017,sippola_experimental_2018} or quantum calculations~\citep{naik2015combined}. Similar CFD-DEM approaches were also developed by~\cite{pei2016cfd} and~\cite{grosshans2017accuracy}. We validated the computational framework against experimental measurements of charge on monodisperse particles in vibrated and fluidised beds~\citep{kolehmainen_triboelectric_2017, sippola_experimental_2018}. The studies show that the total charge in the system is well-predicted with the developed models. CFD-DEM simulations, however, are limited to relatively small systems in cm-range and not affordable for large industrial-scale systems due to the highly demanding computational effort. To achieve simulations of gas-solid flows with charged particles in larger systems, the kinetic-theory based Eulerian-Eulerian models (also called two-fluid) with tribocharging have been recently developed for monodisperse particles~\citep{kolehmainen_eulerian_2018,ray_euler-euler_2019}(the readers are referred to a rich literature on the two-fluid model without charge transfer, e.g.~\cite{savage_stress_1981,jenkins_theory_1983,lun_kinetic_1984,garzo1999dense}). \cite{singh2019electrification} has also developed hydrodynamic equations to study homogeneous and quasi-monodisperse aggregation of charged granular gases. In two-fluid models with tribocharging, the mean charge transport equation was derived from the Boltzmann equation with an assumption of Maxwellian distributions for particle velocities and charges, and coupled with the two-fluid hydrodynamic equations. These models allow for conduction of mean charge through collisions in the presence of electric field and the boundary condition capturing charge generation at the solid boundary. \cite{ray_euler-euler_2019} and \cite{montilla2020modelling} further proposed a model for the velocity-charge covariance that accounts for the self-diffusion of charge. \cite{kolehmainen_eulerian_2018} validated the constitutive equations for mean charge transfer through hard-sphere simulation results whereas \cite{ray_euler-euler_2019} validated the developed models through gas-solid fluidised bed experimental data~\citep{sowinski2012effect}. 

The recent charge transport models are only applicable for particles with a uniform size distribution. The gas-solid systems and granular flows containing particles with a variety of sizes and masses (polydisperse particles) experience specific clustering, deposition dynamics due to tribocharging that are not well understood. Furthermore, there is no consensus on the charge distribution based on particle size. As an example, ~\cite{salama2013investigation} and~\cite{schella2017influence} studied tribocharging of particles with bidisperse size distribution and concluded that larger particles tended to obtain a more negative charge than smaller particles. In contrast,~\cite{forward2009charge,zhao2003bipolar,lee2018collisional} and ~\cite{liu2020effect} observed the opposite behaviour. Very recently, ~\cite{ray_eulerian_2020} extended their monodisperse charge model for bidisperse particles to study steady-state solution of a bipolar charging of the particles with different sizes but the same material. The charge transport closures were derived by following the kinetic theory of~\cite{jenkins_balance_1987} for bidisperse granular flows with assuming that deviations of phase granular temperature from the equipartitioned granular temperature are small. However, several studies showed the importance of non-equipartition of granular temperature for bidisperse segregated granular flows~\citep{alam2003rheology,alam2005energy,galvin2005role,liu2007couette,serero2008classical}. The non-equipartition of granular temperature was also shown by~\cite{wildman2002coexistence} and~\cite{feitosa2002breakdown} experiments where binary mixtures of solid particles were agitated in vibrating fluidised beds. It was discussed that the non-equipartition of granular temperature further increased the driving forces associated to size segregation with the gradient terms of phase granular temperatures. The extensions of kinetic theory of granular flows with bidisperse particles and non-equipartitioned granular temperatures were proposed by ~\cite{garzo1999homogeneous,huilin2001kinetic,iddir_modeling_2005,garzo_enskog_2007,garzo_enskog_2007part2} for dilute and dense granular flows. In this study, we develop the charge transport equation for collisional bidisperse granular flows with separate mean velocities, charges, charge variances and fluctuating kinetic energies for each phase without accounting for the interstitial fluid effect. The developed model predictions are assessed through a set of hard-sphere simulations of bidisperse granular flows with various particle sizes, particle mass ratios and mixture solid volume fractions in a range from 0.2 to 0.4.

The structure of the paper is as follows; we revisit mass, momentum and granular temperature transport equations for bidisperse granular flows in $\cref{section:math}$. In the latter part of this section, we present the charge transport equation with constitutive relations for binary solid mixtures. Hard-sphere simulations of spatially homogeneous and inhomogeneous elastic granular flows are introduced in \cref{section:Validation} and their results are compared with the developed model predictions. In \cref{section:workfunction}, we discuss how the work function difference within a binary mixture generates charge in inhomogeneous flow. In \cref{Section:segflow}, we present the hard-sphere simulation results with the proposed model predictions for a segregating bidisperse granular flow bounded with conducting walls. In \cref{section:conclusion}, we summarise our findings and discuss further developments to the proposed model.

\section{Theoretical Derivation}\label{section:math}
\subsection{Boltzmann equation for charged particles with bidisperse size distribution}
Starting from the Boltzmann equation with the probability density function, we can describe the statistical behaviour of a binary mixture of particles in the dense regime. We denote the probability density function of particles by $f_{pi}( \boldsymbol{x}, {\boldsymbol{c}_{pi}}, q_{pi}, t)$  at position $\boldsymbol{x}$ with velocity $\boldsymbol{c}_{pi}$ and charge $q_{pi}$ on particles for the discrete phase $i$. The number of particles in the phase $i$ with velocity between $\boldsymbol{c}_{pi}$ and $\boldsymbol{c}_{pi}+d\boldsymbol{c}_{ip}$; and charge between $q_{pi}$ and $q_{pi} + dq_{pi}$ at position $\boldsymbol{x}$ and time $t$ is then given by $f_{pi} d \boldsymbol{c}_{pi} dq_{pi}$. The evolution of the probability density function follows the Boltzmann equation:
\begin{equation}
    \frac{\p f_{pi}}{\p t} + \nabla\cdot\Big(\bfit{c}_{pi}f_{pi}\Big)   + \frac{\p }{\p \bfit{c}_{pi}}\Big(\Big\langle\frac{d\bfit{c}_{pi}}{dt}\Big\rangle\Big|_{\bfit{c}_{pi}, q_{pi}}f_{pi}\Big) + \frac{\p }{\p q_{pi}}\Big(\Big\langle \frac{dq_{pi}}{dt}\Big\rangle\Big|_{\bfit{c}_{pi}, q_{pi}}f_{pi}\Big) = \Big(\frac{\p f_{pi}}{\p t}\Big)_{coll}.\label{Eq:BoltzmannEquation}
\end{equation}
The time derivative terms in $\langle.\rangle$ describe the rate of change of particle velocity and charge in the Lagrangian frame. The term on the right-hand-side is the rate of change of the probability density function with particle-particle collisions.   
\subsection{Discrete Particle Equations}
The rate of change of particle velocity is defined by the equation of motion as:
\begin{equation}
	m_{pi}\frac{d\bfit{c}_{pi}}{dt}= \bfit{F}_{ei}.
	\label{momparts}
\end{equation}
Here, the external forces acting on the discrete phase $i$ such as gravitational and fluid-solid interactions forces (e.g. drag and Archimedes forces) are not accounted and only the electrostatic force acting on a particle is accounted for as follows:
\begin{equation} 
	\bfit{F}_{ei} = q_{pi}\,\bfit{E}, \label{coloumbforce}
\end{equation} where $\bfit{E}$ is the resolved electric field (higher order terms due to polarization~\citep{kolehmainen2018effects} and magnetic forces~\citep{genc2014synthesis} were neglected). The resolved electric field is computed by solving a Poisson's equation
\begin{equation} 
	\nabla^2 \phi = - \frac{\rho_q}{\epsilon}, \label{poisson}
\end{equation}
for the electrical potential, $\phi$, where $\rho_q$ is the charge density; and $\epsilon$ is the electrical permittivity. Then, the resolved electric field is obtained by taking the gradient of the electrical potential: 
\begin{equation}
	\bfit{E}=-\nabla \phi. \label{Ei} 
\end{equation}
The charge transfer occurs only by collision, therefore, 
\begin{equation}
    \frac{dq_{pi}}{dt} = 0.
\end{equation}
\subsection{Moment equations}
Any macroscopic property of the discrete phase $i$ is defined using the probability density function and averaging properties over a range of velocity and charge is given as follows:
\begin{eqnarray}
    \langle \psi_{pi} \rangle & = & \frac{1}{n_{pi}}\int_{\mathbb{R}} \int_{\mathbb{R}^3} \psi_{pi}f_{pi}d\bfit{c}_{pi}dq_{pi},
\end{eqnarray}
where $n_{pi}$ is the number density of the phase $i$ particles. For each phase, mean velocity, $\bfit{U}_{pi}$, granular temperature, $\Tp{i}$, mean charge, $Q_{pi}$, and charge variance, $\mathcal{Q}_{pi}$, are then defined as:
\begin{eqnarray}
    \bfit{U}_{pi} & = & \frac{1}{n_{pi}}\int_{\mathbb{R}} \int_{\mathbb{R}^3} \bfit{c}_{pi}f_{pi}d\bfit{c}_{pi}dq_{pi},\\
    \Theta_{pi} & = & \frac{m_{pi}}{3n_{pi}}\int_{\mathbb{R}} \int_{\mathbb{R}^3} (\bfit{c}_{pi}'\cdot\bfit{c}_{pi}')f_{pi}d\bfit{c}_{pi}dq_{pi}, \\
    Q_{pi} & = & \frac{1}{n_{pi}}\int_{\mathbb{R}} \int_{\mathbb{R}^3} q_{pi}f_{pi}d\bfit{c}_{pi}dq_{pi},\\
    \mathcal{Q}_{pi} & = & \frac{m_{pi}}{n_{pi}}\int_{\mathbb{R}} \int_{\mathbb{R}^3} q_{pi}'q_{pi}'f_{pi}d\bfit{c}_{pi}dq_{pi},
\end{eqnarray}
where $\bfit{c}_{pi}'$ is the fluctuating phase velocity and $q_{pi}'$ is the fluctuating phase charge. The fluctuating phase velocity is defined as the difference between phase and mean velocities; $\bfit{c}_{pi}' = \bfit{c}_{pi} - \Uv{i}$. Averaging the Boltzmann Equation \eqref{Eq:BoltzmannEquation} over a range of velocities and charges and using the relation $n_{pi}m_{pi} = \alpha_{pi}\rho_{pi}$, the Enskog equation is obtained:
\begin{equation}\label{enskog}
    \frac{\p}{\p t}\Big(\alpha_{pi}\rho_{pi}\langle\psi_{pi}\rangle\Big) + \nabla\cdot\Big(\alpha_{pi}\rho_{pi}\langle\bfit{c}_{pi}\psi_{pi}\rangle\Big)  = \mathcal{C}(m_{pi}\psi_{pi}) + \alpha_{pi}\rho_{pi}\Big\langle\frac{d\bfit{c}_{pi}}{dt}\frac{\p \psi_{pi}}{\p \bfit{c}_{pi}}\Big\rangle,
\end{equation}
where $\alpha_{pi}$ is the solid volume fraction, $\rho_{pi}$ is the density and $m_{pi}$ is the mass of a particle in the discrete phase $i$. The two terms on the left-hand-side represent the transport of a quantity $\psi_{pi}$, the first term on the right-hand-side represents the rate of change of the quantity averaging over collisions and the last term represents the external force (herein, it is the electrostatic force) acting on the particles. To close the system, the collisional operator, $\mathcal{C}(m_{pi}\psi_{pi})$, needs to be modelled. For a binary mixture of particles, the rate of change of a property due to collisions can be decomposed into the flux and source terms by following \cite{jenkins_balance_1987}:
\begin{eqnarray}
    \mathcal{C}(m_{pi}\psi_{pi}) & = & \sum_{h = i,j}d_{pih}^2\int_{\bfit{k}\cdot\bfit{w} > 0}  m_{pi}(\psi_{pi}^+ - \psi_{pi})|\bfit{k}\cdot\bfit{w}|\nonumber\\
    & & \times f^*_{pih}(\bfit{c}_{pi}, \bfit{x}, \bfit{c}_{ph}, \bfit{x}+d_{pih}\bfit{k})d\bfit{k}\,d\bfit{c}_{pi}\,d\bfit{c}_{ph}\,dq_{pi}\,dq_{ph}\\
    & = & \sum_{h = i,j}\Big(- \nabla\cdot\bm{\theta}_{ih}(m_{pi}\psi_{pi}) + \chi_{ih}(m_{pi}\psi_{pi})\Big).\label{eq:collOperator}
\end{eqnarray}
The flux term, $\bm{\theta}_{ih}$, represents the redistribution of a quantity within and between phases while the source term, $\chi_{ih}$, represents the dissipation of the quantity $\psi_{pi}$ between the phases $i$ and $h$. These terms are derived by using the following integrals:
\begin{eqnarray}
\bm{\theta}_{ih}(m_{pi}\psi_{pi}) & = & -\frac{d_{pih}^3}{2}\int_{\bfit{k}\cdot\bfit{w} > 0} m_{pi}(\psi_{pi}^+ - \psi_{pi})|\bfit{k}\cdot\bfit{w}|\bfit{k}\nonumber\\
& & \times f^*_{pih}(\bfit{c}_{pi}, \bfit{x} - \frac{1}{2}d_{pih}\bfit{k}, \bfit{c}_{ph}, \bfit{x}+\frac{1}{2}d_{pih}\bfit{k})d\bfit{k}\,d\bfit{c}_{pi}\,d\bfit{c}_{ph}\,dq_{pi}\,dq_{ph},\label{Eq:Fluxdefinition}\\
\chi_{ih}(m_{pi}\psi_{pi}) & = & d_{pih}^2\int_{\bfit{k}\cdot\bfit{w} > 0}  m_{pi}(\psi_{pi}^+ - \psi_{pi})|\bfit{k}\cdot\bfit{w}| \nonumber\\
& & \times f^*_{pih}(\bfit{c}_{pi}, \bfit{x} - \frac{1}{2}d_{pih}\bfit{k}, \bfit{c}_{ph}, \bfit{x}+\frac{1}{2}d_{pih}\bfit{k})d\bfit{k}\,d\bfit{c}_{pi}\,d\bfit{c}_{ph}\,dq_{pi}\,dq_{ph}.\label{Eq:Sourcedefinition}
\end{eqnarray}
In these integrals, $\psi_{pi}^+$ refers a quantity after collision, \bfit{k} is the unit vector from the centre of two particles at contact, \bfit{w} is the relative velocity between two particles and $d_{pih}$ is the mean diameter defined as $(d_{pi} + d_{ph})/2$ where $d_{pi}$ and $d_{ph}$ are diameters of phase $i$ and $h$, respectively. The symbol $f^*_{pih}$ refers to the joint pair distribution function for phases $i$ and $h$ at contact point. With an assumption of random motion of particles, it is approximated with a Taylor's expansion at contact point as:
\begin{equation}
f^{*}_{pih} = g_0f_{pi}f_{ph}\bigg(1 + \frac{d_{pih}}{2}\bfit{k}\cdot\nabla\ln\Big( \frac{f_{ph}}{f_{pi}}\Big)\bigg),\label{Eq:JointDensityFunction}
\end{equation}
where $g_0$ is the radial distribution. To compute integrals, we presume that both charge and velocity distributions follow a Maxwellian distribution. The probability density function for the discrete phase $i$ is then defined as
\begin{eqnarray}
f_{pi}(\bfit{c}_{pi},q_{pi},\bfit{x},t) = n_{pi}& & \underbrace{\Big(\frac{m_{pi}}{2\upi\Tp{i}}\Big)^{3/2} \exp\Big(-\frac{m_{pi}}{2\Tp{i}}(\bfit{c}_{pi} - \bfit{U}_{pi})\cdot(\bfit{c}_{pi} - \bfit{U}_{pi})\Big)}_{f_{pi,c}}\nonumber\\
& & \underbrace{\Big(\frac{m_{pi}}{2\upi\mathcal{Q}_{pi}}\Big)^{1/2} \exp\Big(-\frac{m_{pi}}{2\mathcal{Q}_{pi}}(q_{pi} - Q_{pi})^2\Big)}_{f_{pi,q}}.\label{eq:densityFunction}
\end{eqnarray}
We aim to develop the constitutive closures and the charge transport equation for collisional granular flows with bidisperse charged particles where charge transfer occurs mainly via particle-particle collisions. Therefore the correlation between charge and velocity has been neglected in the probability density function. For dilute regime, this assumption is invalid and an additional term arising from self-diffusion of charge should be modelled. Readers are referred to a rigorous theoretical development achieved by a recent study of \cite{montilla2020modelling} on the modelling of the charge-velocity correlation for monosize particles.

\subsection{Revisiting hydrodynamic equations for bidisperse granular flows}
Before presenting the charge transport equation, we revisit the hydrodynamic equations for the granular flows with bidisperse size distribution. The transport equations for mass ($\psi_{pi}$ = 1), momentum ($\psi_{pi}$ = \bfit{c}$_{pi}$), granular temperature ($\psi_{pi}$ = $\frac{1}{2}\bfit{c}_{pi}'\cdot\bfit{c}_{pi}'$ = $\frac{3}{2}\Theta_{pi}/m_{pi}$) are derived from the Enskog equation \eqref{enskog}. The closure relations for the collision terms are derived by following \cite{iddir_modeling_2005}. However, there are slight differences in the derived constitutive equations discussed below.
If there is no exchange of mass or breaking of particles during collisions, the mass balance for the phase $i$ is written as: 
\begin{equation}
    \frac{\p}{\p t}\Big(\alpha_{pi}\rho_{pi}\Big) + \nabla\cdot\Big(\alpha_{pi}\rho_{pi}\bfit{U}_{pi}\Big)  = 0.
\end{equation}
The momentum balance for the phase $i$ is written as:
\begin{eqnarray}
    \alpha_{pi}\rho_{pi}\Big[\frac{\p}{\p t} + \bfit{U}_{pi}\cdot\nabla\Big]\bfit{U}_{pi}  
    & = & \sum_{h = i,j}\Big(- \nabla\cdot\bm{\uptheta}_{ih} + \bm{\chi}_{ih}\Big) - \nabla\cdot\Big(\alpha_{pi}\rho_{pi}\langle\bfit{c}_{pi}'\bfit{c}_{pi}'\rangle\Big) \nonumber\\
    & & + \frac{\alpha_{pi}\rho_{pi}}{m_{pi}}Q_{pi}\bfit{E}. 
    \label{refmompi}
\end{eqnarray}
Here, the electric field, \bfit{E}, is computed with \eqref{poisson} and \eqref{coloumbforce}. The first two terms on the right-hand-side represent the rate of change of momentum due to collisions and redistribution due to the random velocity fluctuations, respectively. The flux term for the collisional operator is defined as:
\begin{eqnarray}\label{fluxMom}
    \bm{\uptheta}_{ih} & = & n_{pi}n_{ph}\frac{m_{pi}m_{ph}}{(m_{pi} + m_{ph})}\Big(\frac{m_{pi}m_{ph}}{\Tp{i}\Tp{h}}\Big)^{3/2}(1+e_c)g_0\frac{\dpp{ih}^3}{48}\bigg[\upi M_1\Ii - \frac{2\dpp{ih}}{5}\sqrt{\upi}M_2\nonumber\\
    & \times & \frac{m_{pi}m_{ph}}{(m_{pi} + m_{ph})} \sum_{l=i,h}\bigg(\frac{1}{\Tp{l}}\Big[(\nabla\bfit{U}_{pl})^s + \frac{5}{6} \nabla\cdot\bfit{U}_{pl}\Ii\Big] \bigg)\bigg],
\end{eqnarray}
with 
\begin{equation}
    (\nabla\Uv{l})^s = \frac{1}{2}\Big((\nabla\Uv{l}) + (\nabla\Uv{l})^T\Big) -\frac{1}{3} \nabla\cdot\Uv{l}\Ii.
\end{equation}
The source term is given by:
\begin{eqnarray}\label{sourceMom}
    & & \bm{\chi}_{ih} = - n_{pi}n_{ph}\frac{m_{pi}m_{ph}}{(m_{pi} + m_{ph})}\Big(\frac{m_{pi}m_{ph}}{\Tp{i}\Tp{h}}\Big)^{3/2}(1+e_c)g_0\frac{\dpp{ih}^2}{6}\Bigg[\sqrt{\upi}(\bfit{U}_{pi} - \bfit{U}_{ph})M_3 \nonumber\\
    & & + \dpp{ih} \frac{\upi}{8}\bigg[M_1\Big(\nabla\ln\frac{n_{ph}}{n_{pi}} - \frac{3}{2}\nabla\ln\frac{\Tp{h}}{\Tp{i}}\Big) + \frac{1}{4} \bigg(3M_4\Big(\frac{m_{ph}\nabla\Tp{h}}{\Tp{h}^2} - \frac{m_{pi}\nabla\Tp{i}}{\Tp{i}^2}\Big)\nonumber\\
    & & + 5M_5\frac{m_{pi}m_{ph}}{(m_{pi} + m_{ph})^2}\Big(\frac{m_{pi}\nabla\Tp{h}}{\Tp{h}^2} - \frac{m_{ph}\nabla\Tp{i}}{\Tp{i}^2}\Big)+ \frac{10}{3}B M_6\frac{m_{pi}m_{ph}}{(m_{pi} + m_{ph})}\Big(\frac{\nabla\Tp{h}}{\Tp{h}^2} + \frac{\nabla\Tp{i}}{\Tp{i}^2}\Big)\bigg) \bigg]  \Bigg],\nonumber\\
\end{eqnarray}
where $e_c$ is the restitution coefficient with $e_c$ = 1  for fully elastic collisions. The coefficients $M_k$ ($k=1..6$) and $B$ are given in table \ref{tab:coefficients}. The derivation of these terms are not given here but the reader is referred to \cite{iddir_modeling_2005} for further details. 

The transport equation for granular temperature of the solid phase $i$ is given by
\begin{equation}\label{granularbi}
    \frac{3}{2} \frac{\p}{\p t}\Big(\alpha_{pi}\rho_{pi}\frac{\Tp{i}}{m_{pi}}\Big) + \frac{3}{2} \nabla\cdot\Big(\alpha_{pi}\rho_{pi}\bfit{U}_{pi}\frac{\Tp{i}}{m_{pi}}\Big) =  \sum_{h = i,j}\Big( -\nabla\cdot \bfit{q}_{ih} + \gamma_{ih}\Big)
\end{equation}
with the flux, $\bfit{q}_{ih}$, and source, $\gamma_{ih}$, terms defined as
\begin{eqnarray}\label{fluxGran}
    & &\bfit{q}_{ih} = - n_{pi}n_{ph}\frac{m_{pi}m_{ph}}{(m_{pi} + m_{ph})}\Big(\frac{m_{pi}m_{ph}}{\Tp{i}\Tp{h}}\Big)^{3/2}(1+e_c)g_0d_{pih}^3\Bigg( \frac{d_{pih}}{48}\sqrt{\upi}\bigg[\Big(\nabla\ln\Big[ \frac{n_{ph}}{n_{pi}}\Big]\nonumber\\
    & & + \frac{3}{2}\nabla\ln\Big[ \frac{\Tp{i}}{\Tp{h}}\Big]\Big) BM_7 + \frac{5}{4}\Big(m_{ph}\frac{\nabla\Tp{h}}{\Tp{h}^2} - m_{pi}\frac{\nabla\Tp{i}}{\Tp{i}^2}\Big)BM_8 + \frac{3m_{pi}m_{ph}}{2(m_{pi} + m_{ph})^2} BM_9\nonumber\\
    & & \times \Big(m_{pi}\frac{\nabla\Tp{h}}{\Tp{h}^2} - m_{ph}\frac{\nabla\Tp{i}}{\Tp{i}^2}\Big) + \frac{m_{pi}m_{ph}}{2(m_{pi} + m_{ph})}\Big(\frac{\nabla\Tp{h}}{\Tp{h}^2} + \frac{\nabla\Tp{i}}{\Tp{i}^2}\Big) M_{10}\bigg] \nonumber\\
    & & + \frac{(1-e_c)}{64}\frac{m_{ph}}{(m_{pi} + m_{ph})} \bigg[\frac{2\upi}{3}(\bfit{U}_{pi} - \bfit{U}_{ph})M_1 + \sqrt{\upi}d_{pih}\Big[\frac{2}{3}\Big(\nabla\ln\Big[ \frac{n_{ph}}{n_{pi}}\Big] + \frac{3}{2}\nabla\ln\Big[ \frac{\Tp{i}}{\Tp{h}}\Big]\Big)M_2\nonumber\\
    & & + \frac{1}{2}\Big(m_{ph}\frac{\nabla\Tp{h}}{\Tp{h}^2} - m_{pi}\frac{\nabla\Tp{i}}{\Tp{i}^2}\Big)M_{11} + \frac{m_{pi}m_{ph}}{(m_{pi} + m_{ph})^2}\Big(m_{pi}\frac{\nabla\Tp{h}}{\Tp{h}^2} - m_{ph}\frac{\nabla\Tp{i}}{\Tp{i}^2}\Big)M_{12}\nonumber\\
    & & + 2B \frac{m_{pi}m_{ph}}{(m_{pi} + m_{ph})}\Big(\frac{\nabla\Tp{h}}{\Tp{h}^2} + \frac{\nabla\Tp{i}}{\Tp{i}^2}\Big)M_{13}\Big]\bigg]\Bigg),
\end{eqnarray}
and
\begin{eqnarray}\label{sourceGran}
    & & \gamma_{ih} = n_{pi}n_{ph}\frac{m_{pi}m_{ph}}{(m_{pi} + m_{ph})}\Big(\frac{m_{pi}m_{ph}}{\Tp{i}\Tp{h}}\Big)^{3/2}(1+e_c)g_0d_{pih}^2\Bigg[\frac{\sqrt{\upi}}{4}BM_7 -  \frac{\upi d_{pih}}{160} \nonumber\\
    & & \times\bigg[\Big(m_{ph}\frac{\nabla\cdot\bfit{U}_{ph}}{\Tp{h}} - m_{pi}\frac{\nabla\cdot\bfit{U}_{pi}}{\Tp{i}}\Big)M_{14} +  5B\frac{m_{pi}m_{ph}}{(m_{pi} + m_{ph})}\Big(\frac{\nabla\cdot\bfit{U}_{ph}}{\Tp{h}} + \frac{\nabla\cdot\bfit{U}_{pi}}{\Tp{i}}\Big)M_6\bigg]\nonumber\\
    & &- \frac{1}{8}\frac{m_{ph}}{(m_{pi} + m_{ph})}(1-e_c)\bigg[\sqrt{\upi}M_2 - \frac{\upi d_{pih}}{8}\Big(m_{ph}\frac{\nabla\cdot\bfit{U}_{ph}}{\Tp{h}} - m_{pi}\frac{\nabla\cdot\bfit{U}_{pi}}{\Tp{i}}\Big)BM_6\nonumber\\
    & & - \frac{\upi d_{pih}}{8} \frac{m_{pi}m_{ph}}{(m_{pi} + m_{ph})}\Big(\frac{\nabla\cdot\bfit{U}_{ph}}{\Tp{h}} + \frac{\nabla\cdot\bfit{U}_{pi}}{\Tp{i}}\Big)M_5\bigg]\Bigg].   
\end{eqnarray}
The derived models are very similar to the ones proposed by \cite{iddir_modeling_2005} but there are differences in the high-order terms of the model coefficients (see table \ref{tab:coefficients}). The differences might result from the Taylor expansion of the relative velocity and the centre of mass velocity multiplication (see \eqref{Eq:TaylorExpansionff}) for integration of the collision operator. Our approximation is detailed in Appendix \ref{appA}. Unfortunately, \cite{iddir_modeling_2005} did not give an explicit explanation about how they treated this term. The assessment benchmark of these revisited constitutive equations through hard-sphere simulation results is given in \cref{section:Validation}. 
\subsection{Transport equation for phase mean charge }
In this section, we present the transport equation for mean charge for each phase. 
Assuming that charge and velocity distributions are uncorrelated and using \eqref{enskog}, the  transport equation for the mean charge is given by:
\begin{equation}
    \frac{\p}{\p t}\Big(\frac{\alpha_{pi}\rho_{pi}}{m_{pi}}Q_{pi}\Big) + \nabla\cdot\Big(\frac{\alpha_{pi}\rho_{pi}}{m_{pi}}Q_{pi}\bfit{U}_{pi}\Big) = \mathcal{C}(q_{pi}).\label{Eq:TransportCharge} 
\end{equation}
The charge transfer during collision between particles is based on the model proposed by \citet{laurentie_discrete_2013} with the phase effective work function, $\varphi_{ph=i,j}$, and the electric field, $\bfit{E}$. The transfer charge between particles $l$ and $m$ within the same phase (i.e. the phase $i$) is given by
\begin{eqnarray}
    q_{pi}^{(l)+} & = & q_{pi}^{(l)} + dq = q_{pi}^{(l)} + \mathcal{A}_{max}(\bfit{k}\cdot\bfit{w}) \epsilon_0\bigg(- \bfit{E}\cdot\bfit{k} + \frac{q_{pi}^{(m)} - q_{pi}^{(l)}}{\upi\epsilon_0d_{pi}^2}\bigg),\\
    q_{pi}^{(m)+} & = & q_{pi}^{(m)} - dq = q_{pi}^{(m)} - \mathcal{A}_{max}(\bfit{k}\cdot\bfit{w})\epsilon_0\bigg(- \bfit{E}\cdot\bfit{k} + \frac{q_{pi}^{(m)} - q_{pi}^{(l)}}{\upi\epsilon_0d_{pi}^2}\bigg).
\end{eqnarray}
For between different phases;
\begin{eqnarray}
    q_{pi}^{(l)+} & = & q_{pi}^{(l)} + dq = q_{pi}^{(l)} + \mathcal{A}_{max}(\bfit{k}\cdot\bfit{w}) \epsilon_0\bigg(\frac{\varphi_{pi} - \varphi_{pj}}{\delta_c e} - \bfit{E}\cdot\bfit{k} + \frac{1}{\upi\epsilon_0}\Big(\frac{q_{pj}^{(m)}}{d_{pj}^2} - \frac{q_{pi}^{(l)}}{d_{pi}^2}\Big)\bigg),\label{Eq:qiPlus}\\
    q_{pj}^{(m)+} & = & q_{pj}^{(m)} - dq = q_{pj}^{(m)} - \mathcal{A}_{max}(\bfit{k}\cdot\bfit{w})\epsilon_0\bigg(\frac{\varphi_{pi} - \varphi_{pj}}{\delta_c e} - \bfit{E}\cdot\bfit{k} + \frac{1}{\upi\epsilon_0}\Big(\frac{q_{pj}^{(m)}}{d_{pj}^2} - \frac{q_{pi}^{(l)}}{d_{pi}^2}\Big)\bigg).\label{Eq:qjPlus}\hspace{0.8cm}
\end{eqnarray}
In \eqref{Eq:qiPlus} and \eqref{Eq:qjPlus}, $\delta_c$ is the cutoff distance of electron transfer, $e$ is the elementary charge and $\epsilon_0$ is the electrical permittivity in a vacuum. $\mathcal{A}_{max}$ is the maximum overlapping area computed with help of the contact Hertz theory as \citep{kolehmainen_triboelectric_2017}:

\begin{equation}
    \mathcal{A}_{max} = 2\upi r_p^*\Big(\frac{15m_p^*}{16Y_p^*\sqrt{r_p^*}}\Big)^{2/5}|\bfit{k}\cdot\bfit{w}|^{4/5} =  \mathcal{A}^*|\bfit{k}\cdot\bfit{w}|^{4/5}.\label{Eq:maxArea}
\end{equation}

The maximum overlapping area is approximated by a collision of two elastic particles that follows a conversion of the particle kinetic energy into the potential energy of a Hertzian spring (it is assumed that the electric potential energy is negligible as compared to the potential energy in the spring). In \eqref{Eq:maxArea}, $\mathcal{A}^*$ is the effective area which is only a function of particle physical properties such as the effective Young modulus, $Y_p^*$, the effective mass, $m_p^*$, and the effective radius, $r_p^*$, that are defined as:
\begin{eqnarray}
\frac{1}{Y_p^*} = \frac{1 - \nu_{pi}^2}{Y_{pi}} + \frac{1 - \nu_{pj}^2}{Y_{pj}},\quad
\frac{1}{r_p^*} = \frac{1}{r_{pi}} + \frac{1}{r_{pj}},\quad
\frac{1}{m_p^*} = \frac{1}{m_{pi}} + \frac{1}{m_{pj}}.
\end{eqnarray}
The maximum contact area could be more accurately computed for granular material with viscoelastic particles by following \cite{schwager_coefficient_2008} or \cite{brilliantov2010kinetic}. However, due to the complex nature of contact electrification, an agreement on the charge transfer model in granular material has not been yet reached as discussed by \cite{matsusaka_triboelectric_2010}. The contact electrification also depends on many other particle properties as shape, roughness, surface functionalisation etc. Therefore, we preferred to use the elastic spheres in the phenomenological charge transfer model.

The closure of the collisional operator in \eqref{Eq:TransportCharge} is defined as follows: 
\begin{equation}
    \mathcal{C}(q_{pi}) = \sum_{h = i,j}\Big(- \nabla\cdot\bm{\theta}_{ih}^q(q_{pi}) + \chi_{ih}^q(q_{pi})\Big),\label{Eq:CollChargeDef}
\end{equation}
where the flux term, $\bm{\theta}_{ih}^q(q_{pi})$, represents the spatial redistribution of charge and the source term, $\chi_{ih}^q(q_{pi})$, represents the charge transfer between phases. The derivations of flux and source terms for the phase charge transport equation are discussed in Appendix-\ref{appA} and their final forms are given in \eqref{Eq:FluxCharge} and \eqref{Eq:SourceCharge}, respectively. Here, we present these equations in a compact form by following~\cite{kolehmainen_eulerian_2018}. The flux term, $\bm{\theta}_{ih}^q$, is then written as:
\begin{equation}\label{Eq:thetaFluxq}
   \bm{\theta}_{ih}^q = - \bm{\upsigma}^{\theta}_{ih}\cdot\bfit{E} - \kappa^{\theta}_{ih}\Big( \frac{\nabla Q_{ph}}{d_{ph}^2} + \frac{\nabla Q_{pi}}{d_{pi}^2} \Big) - \bfit{D}^{\theta}_{ih}\bigg(\frac{\varphi_{pi} - \varphi_{ph}}{\delta_c e} + \frac{1}{\upi\epsilon_0}\Big(\frac{Q_{ph}}{d_{ph}^2} - \frac{Q_{pi}}{d_{pi}^2}\Big)\bigg),
\end{equation}
with the triboelectric conductivity tensor, $\bm{\upsigma}^{\theta}_{ih}$, the triboelectric diffusivity, $\kappa^{\theta}_{ih}$, and the triboelectric phase coupling coefficient, $\bfit{D}^{\theta}_{ih}$. These coefficients are defined below. The first term on the-right-hand-side in \eqref{Eq:thetaFluxq} represents a current density due the electric field resulting from charge on particles. The second term results from the dispersion of charge while the third term arises due to charge difference between particles during a collision. The triboelectric phase coupling coefficient appears due to the non-equipartition of granular temperature (see \eqref{fluxTermNew3}). These terms are defined in explicit forms as:
\begin{eqnarray}
\bm{\upsigma}^{\theta}_{ih} & = & n_{pi}n_{ph}\Big(\frac{m_{pi}m_{ph}}{\Tp{i}\Tp{h}}\Big)^{3/2}\mathcal{A}^*\epsilon_0g_0\frac{d_{pih}^3}{8}\sqrt{\upi}\Bigg[ -\frac{5}{21}N_1\Ii + \frac{3}{1102}d_{pih}\frac{m_{pi}m_{ph}}{(m_{pi} + m_{ph})}N_5 \label{fluxTermNew1}\nonumber\\
& & \times\bigg[\sum_{l = i,j}\frac{1}{\Tp{l}}\bigg( (\nabla\bfit{U}_{pl}) + (\nabla\bfit{U}_{pl})^T + \nabla\cdot\bfit{U}_{pl}\Ii\bigg)\bigg]\Bigg],\\
\kappa^{\theta}_{ih} & = & n_{pi}n_{ph}\Big(\frac{m_{pi}m_{ph}}{\Tp{i}\Tp{h}}\Big)^{3/2}\mathcal{A}^*g_0\frac{d_{pih}^4}{\sqrt{\upi}}\frac{5}{336} N_1,\label{fluxTermNew}\\
\bfit{D}^{\theta}_{ih} & = & n_{pi}n_{ph}\Big(\frac{m_{pi}m_{ph}}{\Tp{i}\Tp{h}}\Big)^{3/2}\mathcal{A}^*\epsilon_0g_0\frac{5}{112}d_{pih}^4\sqrt{\upi}\bigg[\frac{1}{3}\bigg(\nabla\ln\Big( \frac{n_{ph}}{n_{pi}}\Big) + \frac{3}{2}\nabla\ln\Big( \frac{\Tp{i}}{\Tp{h}}\Big)\bigg) N_1\nonumber\\
& &  + \frac{1}{8}\Big(m_{ph}\frac{\nabla\Tp{h}}{\Tp{h}^2} - m_{pi}\frac{\nabla\Tp{i}}{\Tp{i}^2}\Big)N_2 + \frac{1}{6}\frac{m_{pi}m_{ph}}{(m_{pi} + m_{ph})^2} \Big(m_{pi}\frac{\nabla\Tp{h}}{\Tp{h}^2} - m_{ph}\frac{\nabla\Tp{i}}{\Tp{i}^2}\Big)N_3 \nonumber\\
& & + \frac{1}{3}B\frac{m_{pi}m_{ph}}{(m_{pi} + m_{ph})}\Big(\frac{\nabla\Tp{h}}{\Tp{h}^2} + \frac{\nabla\Tp{i}}{\Tp{i}^2}\Big)N_4\bigg]\label{fluxTermNew3}.
\end{eqnarray}
The source term, $\chi^q_{ih}$, is written as:
\begin{equation}\label{Eq:sourceCharge}
   \chi^q_{ih} = - \bm{\sigma}^{\chi}_{ih}\cdot\bfit{E} + D^{\chi}_{ih} \bigg(\frac{\varphi_{pi} - \varphi_{ph}}{\delta_c e} + \frac{1}{\upi\epsilon_0}\Big(\frac{Q_{ph}}{d_{ph}^2} - \frac{Q_{pi}}{d_{pi}^2}\Big)\bigg) ,
\end{equation}
with the triboelectric source conductivity vector, $\bm{\sigma}^{\chi}_{ih}$, and the triboelectric source phase coupling coefficient, $D^{\chi}_{ih}$ that are defined as:

\begin{eqnarray}
D^{\chi}_{ih} & = & n_{pi}n_{ph}\Big(\frac{m_{pi}m_{ph}}{\Tp{i}\Tp{h}}\Big)^{3/2}\mathcal{A}^*\epsilon_0g_0d_{pih}^2\frac{5\sqrt{\upi}}{28}\Bigg[ N_1 - \frac{7d_{pih}}{57}\frac{m_{pi}m_{ph}}{(m_{pi} + m_{ph})}\Big(\frac{\nabla\cdot\bfit{U}_{ph}}{\Tp{h}} + \frac{\nabla\cdot\Uv{i}}{\Tp{i}}\Big)N_5\Bigg], \nonumber \label{sourceTermNew1}\\
& & \\
\bm{\sigma}^{\chi}_{ih} & = & n_{pi}n_{ph}\Big(\frac{m_{pi}m_{ph}}{\Tp{i}\Tp{h}}\Big)^{3/2}\mathcal{A}^*\epsilon_0g_0d_{pih}^3\frac{5\sqrt{\upi}}{168}\Bigg[ \Big[\nabla\ln\Big( \frac{n_{ph}}{n_{pi}}\Big) + \frac{3}{2}\nabla\ln\Big( \frac{\Tp{i}}{\Tp{h}}\Big)\Big]N_1 \nonumber\\
    & & + \frac{3}{4}\Big(m_{ph}\frac{\nabla\Tp{h}}{\Tp{h}^2} - m_{pi}\frac{\nabla\Tp{i}}{\Tp{i}^2}\Big) N_2 + \frac{m_{pi}m_{ph}}{2(m_{pi} + m_{ph})^2}\Big(m_{pi}\frac{\nabla\Tp{h}}{\Tp{h}^2} - m_{ph}\frac{\nabla\Tp{i}}{\Tp{i}^2}\Big) N_3 \nonumber\\
    & & + \frac{m_{pi}m_{ph}}{(m_{pi} + m_{ph})}\Big(\frac{\nabla\Tp{h}}{\Tp{h}^2} + \frac{\nabla\Tp{i}}{\Tp{i}^2}\Big)B N_4\bigg]\Bigg] \label{sourceTermNew2}.
\end{eqnarray}
The coefficients $N_k$ ($k=1..5$) and $B$ in the flux and source terms are listed in table \ref{tab:coefficients}.

\begin{table}
  \begin{center}
\def~{\hphantom{0}}
  \begin{tabular}{lll}
    $A$ & = & {\large$\frac{m_{pi}\Tp{j} + m_{pj}\Tp{i}}{2\Tp{i}\Tp{j}}$}, ~ $D$ ~ = ~ {\large$\frac{m_{pi}m_{pj}(m_{pj}\Tp{j} + m_{pi}\Tp{i})}{2(m_{pi} + m_{pj})^2\Tp{i}\Tp{j}}$}, ~ $B$ ~  = ~ {\large$\frac{m_{pi}m_{pj}(\Tp{j} - \Tp{i})}{2(m_{pi} + m_{pj})\Tp{i}\Tp{j}}$}\\ [1.5ex]
    $M_1$ & = & {\large$\frac{1}{A^{3/2}D^{5/2}} + \frac{5 B^2}{A^{5/2}2D^{7/2}} + \frac{35 B^4}{8A^{7/2}D^{9/2}}$ + ...}\\[1.5ex]
    $M_2$ & = & {\large $\frac{1}{A^{3/2}D^3} + \frac{3B^2}{A^{5/2}D^4} + \frac{6B^4}{A^{7/2}D^5}$ + ...}\\[1.5ex]
    $M_3$ & = & {\large$\frac{1}{A^{3/2}D^2} + \frac{2B^2}{A^{5/2}D^3} + \frac{3B^4}{A^{7/2}D^4} + ...$}\\[1.5ex]
    $M_4$ & = &  {\large$\frac{1}{A^{5/2}D^{5/2}} + \frac{25 B^2}{6 A^{7/2}D^{7/2}} + \frac{245 B^4}{24 A^{9/2}D^{9/2}} + ...$}\\[1.5ex]
    $M_5$ & = & {\large$\frac{1}{A^{3/2}D^{7/2}} + \frac{7 B^2}{2A^{5/2}D^{9/2}} + \frac{63 B^4}{8 A^{7/2}D^{11/2}} + ...$}\\[1.5ex]
    $M_6$ & = & {\large$\frac{1}{A^{5/2}D^{7/2}} + \frac{7B^2}{2A^{7/2}D^{9/2}} + ...$}\\[1.5ex]
    $M_7$ & = & {\large$\frac{1}{A^{5/2}D^3} + \frac{3B^2}{A^{7/2}D^4} +  ...$}\\[1.5ex]
    $M_8$ & = & {\large$\frac{1}{A^{7/2}D^3} + \frac{21B^2}{5A^{9/2}D^4} + ...$}\\[1.5ex]
    $M_9$ & = & {\large$\frac{1}{A^{5/2}D^4} + \frac{4B^2}{A^{7/2}D^5} + ...$}\\[1.5ex]
    $M_{10}$ & = & {\large$\frac{1}{A^{5/2}D^3}  + \frac{9B^2}{A^{7/2}D^4}  + \frac{30B^4}{A^{9/2}D^5} + ...$}\\[1.5ex]
    $M_{11}$ & = & {\large$\frac{1}{A^{5/2}D^3} + \frac{5B^2}{A^{7/2}D^4} + \frac{14B^4}{A^{9/2}D^5} + ...$}\\[1.5ex]
    $M_{12}$ & = & {\large$\frac{1}{A^{3/2}D^4} + \frac{4B^2}{A^{5/2}D^5} + \frac{10B^4}{A^{7/2}D^6} + ...$}\\[1.5ex]
    $M_{13}$ & = & {\large$\frac{1}{A^{5/2}D^4} + \frac{4B^2}{A^{7/2}D^5} + ...$}\\[1.5ex]
    $M_{14}$ & = & {\large$\frac{1}{A^{5/2}D^{5/2}} + \frac{15B^2}{2A^{7/2}D^{7/2}} + \frac{175B^4}{8A^{9/2}D^{9/2}} + ...$}\\[1.5ex]
    $N_1$ & = & {\large$\frac{1}{A^{3/2}D^{12/5}}\Gamma(\frac{12}{5}) + \frac{B^2}{A^{5/2}D^{17/5}}\Gamma(\frac{17}{5}) + \frac{B^4}{2A^{7/2}D^{22/5}}\Gamma(\frac{22}{5}) + ...$}\\[1.5ex]
    $N_2$ & = & {\large$\frac{1}{A^{5/2}D^{12/5}}\Gamma(\frac{12}{5}) + \frac{5B^2}{3A^{7/2}D^{17/5}}\Gamma(\frac{17}{5}) + \frac{7B^4}{6A^{9/2}D^{22/5}}\Gamma(\frac{22}{5}) + ...$ }\\[1.5ex]
    $N_3$ & = & {\large$\frac{1}{A^{3/2}D^{17/5}}\Gamma(\frac{17}{5}) + \frac{B^2}{A^{5/2}D^{22/5}}\Gamma(\frac{22}{5}) + \frac{B^4}{2A^{7/2}D^{27/5}}\Gamma(\frac{27}{5}) + ...$}\\[1.5ex]
   $N_4$ & = & {\large$\frac{1}{A^{5/2}D^{17/5}}\Gamma(\frac{17}{5}) + \frac{B^2}{A^{7/2}D^{22/5}}\Gamma(\frac{22}{5}) + ...$}\\[1.5ex]
   $N_5$ & = & {\large$\frac{1}{A^{3/2}D^{29/10}}\Gamma(\frac{29}{10}) + \frac{B^2}{A^{5/2}D^{39/10}}\Gamma(\frac{39}{10}) + \frac{B^4}{2A^{7/2}D^{49/10}}\Gamma(\frac{49}{10}) + ...$}\\[1.5ex]
  \end{tabular}
  \caption{Model coefficients of flux and source terms for phase momentum, granular temperature and charge transport equations. The model coefficients, $M_k$ ($k=1..6$) and $B$, are used in \eqref{fluxMom} and \eqref{sourceMom}. The model coefficients, $M_k$ ($k=1..14$) and $B$ are used in \eqref{fluxGran} and \eqref{sourceGran}. The model coefficients, $N_k$ ($k=1..5$) and $B$, are used in \eqref{fluxTermNew1}-\eqref{fluxTermNew3}, \eqref{sourceTermNew1} and \eqref{sourceTermNew2}. The symbol $\Gamma(.)$ refers to the Gamma function.}
  \label{tab:coefficients}
  \end{center}
\end{table}

\subsection{Constitutive relations with linear departures from mixture properties}
The model derived above has been given under a general form assuming the probability function to be Maxwellian distribution with distinct granular temperature and mean velocity for each solid phase. These assumptions about the probability function are valid as long as the flow is nearly elastic and close to equilibrium. To extend the range of application for the model, it should include a non-Maxwellian distribution (i.e. \cite{galvin2005role, garzo_enskog_2007}) to take into account the inelastic collisions and their consequences on the collisional terms as well as the deviation from equilibrium. It is important to underline that our charge model assumed there is no correlation between the charge and velocity probability function, therefore analytical derivations of charge collision integrals given in Appendix \ref{appA} are independent of the hydrodynamic model for either nearly elastic or inelastic collisions.

In this study, the assessment for the model including the charge transport equation for a binary mixture is done assuming that the particles are fully elastic and the system deviated slightly from equilibrium. To be consistent with this underlying assumption, the model could be simplified with the granular temperature and mean velocity of each solid phase undergoing a linear variation from the mixture value by following \cite{jenkins_balance_1987}. The granular temperature and mean velocity for phase $h$ ($h = i,j$) can be defined as:
\begin{equation}
    \Tp{h} = \Tp{m} + \delta_{\Tp{h}}, ~ \Uv{h} = \Uv{m} + \delta_{\Uv{h}}
\end{equation}
where $\delta_{\Tp{h}}$ and $\delta_{\Uv{h}}$ are the linear deviations for the granular temperature and mean velocity respectively and the subscript $m$ refers to the mixture property. 
As these deviations being small, the coefficients $M_k$ ($k = [1,14]$) and $N_l$ ($l = [1,5]$) listed in table \ref{tab:coefficients} can be simplified by accounting for the first term only; the higher-order terms being proportional to the difference between the granular temperature via the coefficient $B$ can be neglected ($B \propto (\delta_{\Tp{j}} - \delta_{\Tp{i}}) \approx 0$). Assuming that the gradient term multiplied by the linear deviations turns to zero ($\delta_{\Tp{h}}\nabla \approx 0$), all collisional terms given above can be reduced in new expressions as a function of the mixture properties. For the momentum transport equation, the flux and source expressions become respectively:
\begin{eqnarray}
    \bm{\uptheta}_{ih}^{\text{m}} & = & \upi n_{pi}n_{ph}(1+e_c)g_0\frac{d_{pih}^3}{3}\bigg[\Big(\Tp{m} + \frac{m_{pi}\delta_{\Tp{j}} + m_{ph}\delta_{\Tp{i}}}{(m_{pi} + m_{ph})}\Big)\Ii - \frac{2d_{pih}}{5}\Big( \frac{(m_{pi}m_{ph})}{\upi(m_{pi} + m_{ph})}\Tp{m}\Big)^{1/2} \nonumber\\
    & & \times\Big[\Big((\nabla\Uv{m}) + (\nabla\Uv{m})^T\Big) + \nabla\cdot\Uv{m}\Ii \Big]\bigg], \\ 
    \bm{\chi}_{ih}^{\text{m}} & = & \upi n_{pi}n_{ph}(1+e_c)g_0\frac{d_{pih}^3}{3}\Tp{m}\bigg[\frac{4}{d_{pih}}(\delta_{\Uv{j}} - \delta_{\Uv{i}})\Big(\frac{m_{pi}m_{ph}}{2\upi(m_{pi} + m_{ph})\Tp{m}}\Big)^{1/2} \nonumber \\
    & & + \nabla\ln\frac{n_{pi}}{n_{ph}} + \frac{(m_{ph} - m_{pi})}{(m_{pi} + m_{ph})}\nabla\ln\Tp{m} \bigg].
\end{eqnarray}
For the granular temperature transport equation, the flux and source terms are given by:
\begin{eqnarray}
    \bfit{q}_{ih}^{\text{m}} & = & \frac{n_{pi}n_{pj}}{(m_{pi} + m_{pj})}(1+e_c)g_0 d_{pij}^4\Tp{m}\bigg[ -\frac{2}{3}\Big(\frac{2\upi\Tp{m}m_{pi}m_{pj}}{(m_{pi}+m_{pj})}\Big)^{1/2}\nabla\ln\Tp{m} + m_{pj}(1-e_c) \nonumber\\
    & & \times\Big[\frac{\upi}{6d_{pij}}(\delta_{u,j} - \delta_{u,i}) + \frac{1}{4}\Big(\frac{2\upi(m_{pi} + m_{pj})\Tp{m}}{m_{pi}m_{pj}}\Big)^{1/2}\nonumber\\
    & & \times\Big(\frac{2}{3}\nabla\ln\Big(\frac{n_{pi}}{n_{pj}}\Big) + \frac{(m_{pj} - m_{pi})}{(m_{pi} + m_{pj})}\nabla\ln\Tp{m}\Big)\Big]\bigg], \\
    \gamma_{ih}^{\text{m}} & = & 2n_{pi}n_{ph}g_0d_{pih}^2(1+e_c)\Tp{m}\Bigg[2\frac{(\delta_{\Theta,h} - \delta_{\Theta,i})}{\Tp{m}(m_{pi}+m_{ph})}\Big(\frac{2\upi m_{pi}m_{ph}\Tp{m}}{(m_{pi}+m_{ph})}\Big)^{1/2} \nonumber\\
    & & -  \frac{\upi d_{pih}(m_{ph} - m_{pi})}{10(m_{pi}+m_{ph})}\nabla\cdot\Uv{m} - \frac{m_{ph}}{(m_{pi}+m_{ph})}(1-e_c)\bigg[\Big(\frac{2\upi(m_{pi} + m_{ph})\Tp{m}}{m_{pi}m_{ph}}\Big)^{1/2}\nonumber\\
    & & - \frac{\upi d_{pih}}{2} \nabla\cdot\Uv{m}\bigg] + \frac{\upi d_{pih}}{2(m_{pi}+m_{ph})}\bigg[ \frac{1}{5}\Big(m_{pi}\nabla\cdot\delta_{u,i} - m_{ph}\nabla\cdot\delta_{u,h}\Big)\nonumber\\
    & & + \frac{m_{ph}}{2}(1-e_c) \Big(\nabla\cdot\delta_{u,h} + \nabla\cdot\delta_{u,i}\Big)\bigg]\Bigg].
\end{eqnarray}

In addition in our generic model, the collisional terms in the mean charge transport equation can be also expressed in term of small deviations from the mixture parameter for the granular temperature and the mean velocity. Following \eqref{Eq:thetaFluxq}, the triboelectric conductivity tensor, diffusivity and phase coupling coefficient in the flux term for unlike particles collision expression can be shortened as:  
\begin{eqnarray}
    \bm{\upsigma}^{\theta \text{m}}_{ih}  & = & \sqrt{\upi}g_0d_{pih}^3n_{pi}n_{ph}\mathcal{A}^*\epsilon_0 \Big(2\Tp{m}\frac{(m_{pi} + m_{ph})}{m_{pi}m_{ph}}\Big)^{9/10}\Gamma(\frac{12}{5})\bigg[ -\frac{5}{21}\Ii \nonumber\\
    & & + d_{pih}\frac{6}{551}\Big(\frac{6m_{pi}m_{ph}}{5(m_{pi}+m_{ph})\Tp{m}}\Big)^{1/2}\Big[ (\nabla\Uv{m}) + (\nabla\Uv{m})^T + \nabla\cdot\Uv{m}\Ii\Big] \bigg],\\
    \kappa^{\theta \text{m} }_{ih} & = & \frac{5}{42}\sqrt{\upi}g_0d_{pih}^4n_{pi}n_{ph}\mathcal{A}^*\epsilon_0 \Big(2\Tp{m}\frac{(m_{pi} + m_{ph})}{m_{pi}m_{ph}}\Big)^{9/10}\Gamma(\frac{12}{5}),\\
    \bfit{D}^{\theta \text{m} }_{ih} & = & \frac{1}{14}\sqrt{\upi}g_0d_{pih}^4n_{pi}n_{ph}\mathcal{A}^*\epsilon_0 \Big(2\Tp{m}\frac{(m_{pi} + m_{ph})}{m_{pi}m_{ph}}\Big)^{9/10}\Gamma(\frac{12}{5})\bigg[\frac{5}{3}\nabla\ln\Big( \frac{n_{ph}}{n_{pi}}\Big) \nonumber\\
    & & -\frac{11}{4}\frac{(m_{ph} - m_{pi})}{(m_{pi} + m_{ph})}\nabla\ln\Tp{m} \bigg].
\end{eqnarray}
For similar particles collision ($h = i$), the triboelectric phase coupling coefficient turns to zero and the others coefficients become:  
\begin{eqnarray}
    \bm{\upsigma}^{\theta \text{m} }_{ii} & = & 2^{9/5}\sqrt{\upi}g_0d_{pi}^3n_{pi}^2\mathcal{A}^*\epsilon_0 \Big(\frac{\Tp{m}}{m_{pi}}\Big)^{9/10}\Gamma(\frac{12}{5})\bigg[ -\frac{5}{21}\Ii \nonumber\\
    & & + d_{pi}\frac{6}{551}\Big(\frac{3m_{pi}}{5\Tp{m}}\Big)^{1/2}\Big[ (\nabla\Uv{m}) + (\nabla\Uv{m})^T + \nabla\cdot\Uv{m}\Ii\Big] \bigg],\\
    \kappa^{\theta \text{m} }_{ii} & = & 2^{4/5}\frac{5}{21}\sqrt{\upi}g_0d_{pi}^4n_{pi}^2\mathcal{A}^*\epsilon_0 \Big(\frac{\Tp{m}}{m_{pi}}\Big)^{9/10}\Gamma(\frac{12}{5}).
\end{eqnarray}
For the source term, the coefficients in \eqref{Eq:sourceCharge} for unlike particle collision expression become:
\begin{eqnarray}
    \bm{\sigma}^{\chi \text{m} }_{ih} & = & g_0\mathcal{A}^*\epsilon_0n_{pi}n_{ph}d_{pih}^3\frac{5\sqrt{\upi}}{21}\Big(2\Tp{m}\frac{(m_{pi} + m_{ph})}{m_{pi}m_{ph}}\Big)^{9/10}\Gamma(\frac{12}{5})\bigg[\nabla\ln\Big( \frac{n_{ph}}{n_{pi}}\Big) \nonumber\\
    & & - \frac{9}{10}\frac{(m_{ph} - m_{pi})}{(m_{pi} + m_{ph})}\nabla\ln\Tp{m} \bigg], \\
    D^{\chi \text{m} }_{ih} & = & g_0\mathcal{A}^*\epsilon_0n_{pi}n_{ph}d_{pih}^2\frac{10\sqrt{\upi}}{7}\Big(2\Tp{m}\frac{(m_{pi} + m_{ph})}{m_{pi}m_{ph}}\Big)^{9/10}\Gamma(\frac{12}{5})\nonumber\\
    & & \times\bigg[ 1 - \frac{28d_{pih}}{57}\Big(\frac{6m_{pi}m_{ph}}{5(m_{pi}+m_{ph})\Tp{m}}\Big)^{1/2} \nabla\cdot\Uv{m} \bigg]. \nonumber\\
\end{eqnarray}
For similar particles collisions, the source term for the mean charge transport equation turns to zero. 

\section{Model assessment}\label{section:Validation}
We assess the generic models through Lagrangian hard-sphere simulations for three case studies: (i) spatially homogeneous elastic granular gases with a random initial distribution of monodisperse and bidisperse solids with a bimodal charge distribution, (ii) quasi-1D bidisperse elastic granular gases with spatial gradients, and (iii) a three-dimensional segregating inelastic bidisperse granular flow with conducting walls. 
For these simulations, we varied the particle diameter ratio, $R_d = d_{pj}/d_{pi}$, the solid volume fraction ratio, $R_{\alpha} = \alpha_{pj}/\alpha_{pi}$, the particle density ratio, $R_{\rho} = \rho_{pj}/\rho_{pi}$ of phases $i$ and $j$, and phase initial charges. Recalling that the proposed model is applicable only for moderately dense and dense flows where charge transfer is mainly driven by collisions, the model was assessed for granular flows with the mixture solid volume fraction in a range from 0.2 to 0.4. The readers are referred to Appendix \ref{appB} for details of the Lagrangian hard-sphere method. Briefly, the Lagrangian hard-sphere solver is based on the time-stepped algorithm where the predicted locations are computed for each particle to ensure no overlapping occurs. In case of overlapping between two particles, the positions of these particles are reversed to the previous time step and the velocities are corrected by following the collision rule. The new locations of particles are then updated using the corrected velocity. If the overlapping distance is more than 10\% percent of particle radius, the time-step is decreased. The hard-sphere simulations with a mixture volume fraction larger than 0.4 produced several collisions where the ratio of overlapping was greater than 5\% of the particle diameter even for very small time-step. This was deemed unacceptable, therefore these simulations were excluded from this study.
To compute the electric field at contact point, we first map particle charges into the Eulerian cells to compute charge densities and then solve the Poisson's equation with a spectral method for fully periodic simulations (Sections \ref{section:homogenousSimulation} and \ref{section:Quasi1DSimulation}). For bounded simulations with conducting walls, we use a finite difference method to discretize the Poisson equation with the electric potential set to zero at walls.

As we present hard-sphere simulation results and Eulerian model predictions, we use the following dimensionless quantities for the phase $h$ ($h = i,j$):
\begin{equation}\label{scalingTimeVelGran}
    t^* = \frac{t}{d_{pm}}\sqrt{\frac{\Tp{m}}{m_{p m}}},\quad U^*_{ph} =  \frac{U_{ph}}{\sqrt{\Tp{m}/m_{pm}}},\quad \Tp{h}^* = \frac{\Tp{h}}{\Tp{m}}.
\end{equation}
The subscript, $m$, refers to the mixture quantities defined as:
\begin{eqnarray}
    \Tp{m} & = & \frac{n_{pi}\Tp{i} + n_{pj}\Tp{j}}{n_{pi} + n_{pj}}, \label{scaledtheta}\\
    m_{pm} & = & \frac{n_{pi}m_{pi} + n_{pj}m_{pj}}{n_{pi} + n_{pj}},\\
    d_{pm} & = & \frac{d_{pi} + d_{pj}}{2},\\
    Q_{pm} & = & \frac{n_{pi}Q_{pi} + n_{pj}Q_{pj}}{n_{pi} + n_{pj}}.\label{Eq:meancharge}
\end{eqnarray}
For spatially homogeneous and quasi-1D granular gas simulations, the mean charge of the system is set to zero, $Q_{pm}$ = 0. We scale the mean phase charge as
\begin{equation}\label{scaledcharge}
    Q^*_{ph} =  \frac{Q_{ph}}{Q_{p}^0},
\end{equation}
with a reference charge, $Q_{p}^0$ = 1 fC.

For all simulations below, the Poisson's ratio and the Young's modulus were kept constant ($\nu_{ph}$ = 0.42 and $Y_{ph}$ = \SI{0.5}{\mega\pascal}). The radial distribution function, $g_0$, proposed by \citet{jenkins_balance_1987} 
\begin{equation}
    g_0 = \frac{1}{(1 - \mu)} + 3 \Big(\frac{d_{pi}d_{pj}}{d_{pi} + d_{pj}}\Big)\frac{\xi}{(1 - \mu)^2} + 2 \Big(\frac{d_{pi}d_{pj}}{d_{pi} + d_{pj}}\Big)^2\frac{\xi^2}{(1 - \mu)^3},
\end{equation}
with the coefficients $\mu$ and $\xi$
\begin{equation}
    \mu = \frac{\pi}{6}\Big(n_{pi}d_{pi}^3 + n_{pj}d_{pj}^3\Big), ~ \xi ~ = ~ \frac{\pi}{6}\Big(n_{pi}d_{pi}^2 + n_{pj}d_{pj}^2\Big),
\end{equation}
was used for the Eulerian predictions in the following sections. 
\subsection{Spatially homogeneous bidisperse granular gas simulations}\label{section:homogenousSimulation}
We performed hard-sphere simulations of elastic granular gases in a fully periodic cubic box with a dimension of $32\,d_{pj}\times32\,d_{pj}\times32\,d_{pj}$. Here, $d_{pj}$ refers to the larger particle with a diameter of \SI{300}{\micro\meter}. The particle density and diameter, the domain-averaged solid volume fraction, the granular temperature, the initial charge, the number of particles ($N_p$) and the number of collisions per particles ($n_c/N_p$) for each phase are listed for three simulation cases in table \ref{Table:InitConditionHomog}. For all simulations, the particles were randomly distributed in the domain  and velocities were initialised with a Maxwellian distribution with a zero mean velocity for each solid phase. The initial charges followed a bimodal distribution with imposing the mixture charge, $Q_{pm}$, equal to zero. The difference of work function was set to zero, as well. As the mixture charge was zero all at times, the macroscopic electric field turns to zero, therefore, there was no electrostatic force acting on particles. For the phase $i$, we set the initial mean charge equal to the reference charge, $Q_p^0$, while the initial mean charge for the phase $j$ was imposed by the ratio of particle number density as:
\begin{equation}
  \left\{
    \begin{array}{ll}
    Q_{pi}(x, 0) & = Q_p^0 \\[3pt]
    Q_{pj}(x, 0) & =  - \frac{n_{pi}}{n_{pj}}Q_p^0. 
  \end{array} \right. \label{eq:chargeConditionHomog}
\end{equation}

\begin{table}
    \centering
    \begin{tabular}{ccccccccc}
        Case & Phase &$d_{p}$ & $\rho_{p}$  & $\langle \alpha_{p} \rangle$ & $\langle \Theta_{p} \rangle{\small \times\SI{e-10}{}}$ & $\langle Q_{p} \rangle$ & $N_p$ & $n_c/N_p$
        \\
         &  &[\SI{}{\micro\meter}] & [\SI{}{\kilo\gram\per\meter^3}] & [-] & [\SI{}{\kilo\gram~\meter^2\per\second^2}] &  [\SI{}{\femto\coulomb}] &  &
        \\
        \hline
        A & $i$ & 300 & 1500 & 0.1 & 3.55 & -1 & \SI{6258}{} & $\approx$ 51\\
          & $j$ & 300 & 1500 & 0.1   & 3.55 & 1 & \SI{6258}{} &\\ 
        \hline
        B & $i$ & 60 & 1500 & 0.05  & 0.028 & -1 & \SI{391143}{} & $\approx$ 132\\
          & $j$ & 300 & 1500 & 0.15   & 3.55 & 42.3 & \SI{9250}{} & \\ 
        \hline
        C & $i$ & 100 & 150 & 0.05  & 0.013 & -1 & \SI{84493}{} &  $\approx$ 240\\
          & $j$ & 300 & 1500 & 0.15  & 3.55 & 9 & \SI{9351}{} &
    \end{tabular}
    \caption{Particle properties and flow parameters for three spatially homogeneous flow configurations. Case-A refers to a monodisperse case (two solid phase classes with the same particle properties but with opposite initial charges); Case-B refers to a bidisperse case with a particle diameter ratio of $R_d$ = 5 and the same particle density; Case-C refers to a bidisperse case with a particle diameter ratio of $R_d$ = 3 and a particle density ratio of $R_{\rho}$ = 10. For all cases, a bimodal distribution of charge is imposed by following \eqref{eq:chargeConditionHomog} with a total charge equal to zero, $Q_{pm}$ = 0 (electrically neutral condition) and the restitution coefficient is set to unity, $e_c$ = 1. 
    \label{Table:InitConditionHomog}}
\end{table}

\begin{figure}
    \centering
    \psfrag{y3}[][c][1]{$Q_{ph}^*\,[-]$}
    \psfrag{x1}[][c][1]{$t^{*}\,[-]$}
    \includegraphics[scale=0.225,angle=-90]{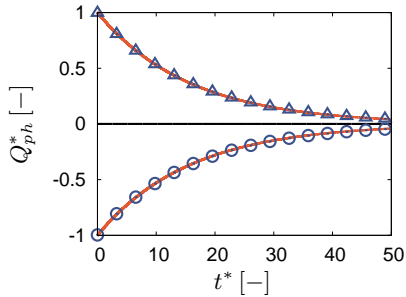}
    \caption{Evolution of scaled charge of the phase $h\,(h=i,j)$ for Case-A. Particle properties and flow parameters are given in table \ref{Table:InitConditionHomog}. The orange lines show the solutions of \eqref{dChargepidt} and \eqref{dChargepjdt}.  ${\color{CadetBlue} \odot}$: hard-sphere simulation results for the phase $i$ and ${\color{CadetBlue} \Delta}$: hard-sphere simulation results for the phase $j$. Charge was scaled by using \eqref{scaledcharge}. \label{fig:figureMono}}
\end{figure}

For spatially homogeneous flow with elastic collision ($e_c=1$) and without mean convection, the set of equations given in the previous sections will be simplified to ordinary differential equations for granular temperature and charge evolution for each phase as:
\begin{eqnarray}
    \frac{d\Tp{i}}{dt} & = &  n_{pj}\frac{m_{pi}m_{pj}}{m_{pi} + m_{pj}}\Big(\frac{m_{pi}m_{pj}}{\Tp{i}\Tp{j}}\Big)^{3/2}\frac{\sqrt{\upi}}{3}g_0d_{pij}^2BM_7, \label{dThetapidt}\\
    \frac{d\Tp{j}}{dt} & = &  - n_{pi}\frac{m_{pi}m_{pj}}{m_{pi} + m_{pj}}\Big(\frac{m_{pi}m_{pj}}{\Tp{i}\Tp{j}}\Big)^{3/2}\frac{\sqrt{\upi}}{3}g_0d_{pij}^2BM_7, \label{dThetapjdt}\\
    \frac{dQ_{pi}}{dt} & = & n_{pj}\Big(\frac{m_{pi}m_{pj}}{\Tp{i}\Tp{j}}\Big)^{3/2}\frac{\mathcal{A}^*}{\sqrt{\upi}}g_0\frac{5d_{pij}^2}{14}\Big(\frac{Q_{pj}}{d_{pj}^2} - \frac{Q_{pi}}{d_{pi}^2}\Big)N_1, \label{dChargepidt}\\
    \frac{dQ_{pj}}{dt} & = & - n_{pi}\Big(\frac{m_{pi}m_{pj}}{\Tp{i}\Tp{j}}\Big)^{3/2}\frac{\mathcal{A}^*}{\sqrt{\upi}}g_0\frac{5d_{pij}^2}{14}\Big(\frac{Q_{pj}}{d_{pj}^2} - \frac{Q_{pi}}{d_{pi}^2}\Big)N_1. \label{dChargepjdt}
\end{eqnarray}
We first started with the monodisperse flow configuration represented by Case-A in table \ref{Table:InitConditionHomog} where we had two solid classes with the same particle properties and the domain-averaged solid volume fraction but opposite initial charges. The granular temperature for each phase was also identical, therefore the charge only evolved due to the mean charge difference between phases. The total solid volume fraction was set to $\langle\alpha_p\rangle = 0.2$ and half of the particles were assigned initial charge of $Q_p$ = \SI{-1}{\femto\coulomb} while the other half were assigned opposite charge of $Q_p$ = \SI{1}{\femto\coulomb}. The scaled charge evolutions by simulation and model prediction for each identical phase are shown in figure \ref{fig:figureMono}. The orange line shows the solutions of \eqref{dChargepidt} and \eqref{dChargepjdt} while the symbols show hard-sphere simulation results for phases $i$ (${\color{CadetBlue} \odot}$) and $j$ (${\color{CadetBlue} \Delta}$). The mean charge of each phase follows an exponential trend in time until they reach the total charge which is equal to zero.

\begin{figure}
    \centering
    \psfrag{y3}[][c][1]{$Q_{ph}^*$ [-]}
    \psfrag{y2}[][c][1]{$\Tp{h}^*$ [-]}
    \psfrag{x1}[][c][1]{$t^{*}\,[-]$}
    \begin{tabular}{cc}
        (a) & (b) \vspace{-0.5cm}\\ 
        \includegraphics[scale=0.225,angle=-90]{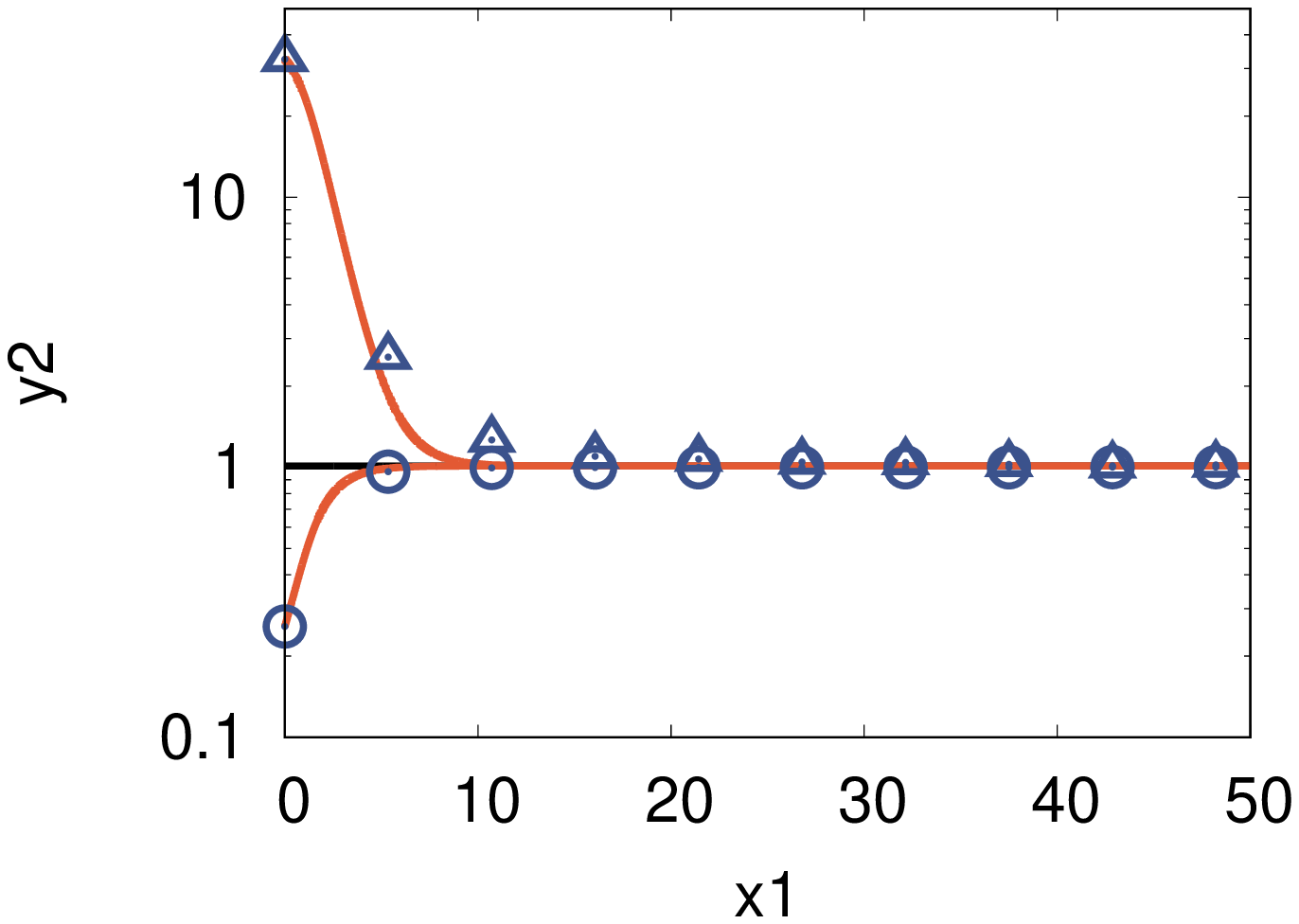} & \includegraphics[scale=0.225,angle=-90]{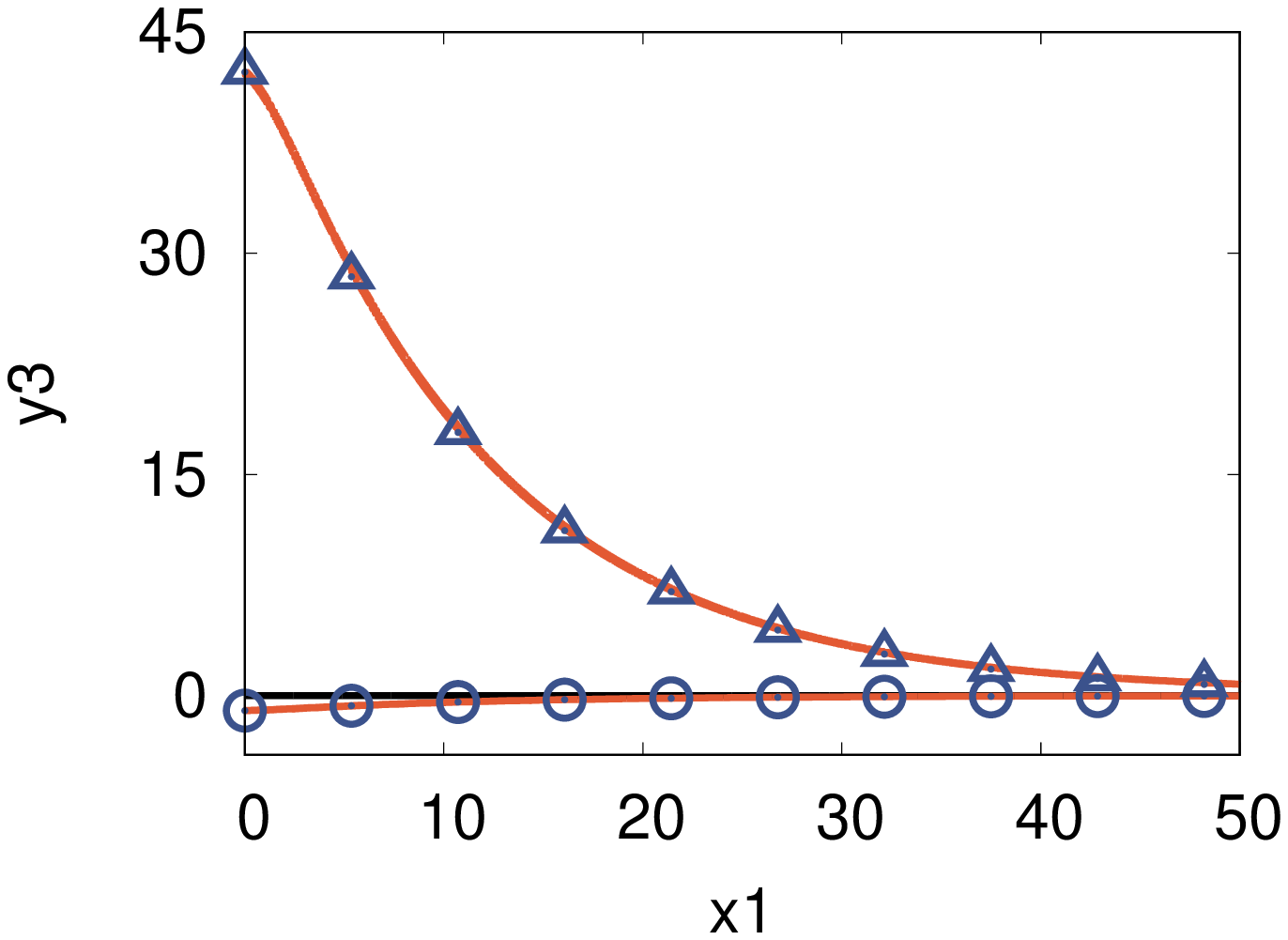} \\
    \end{tabular}
    \caption{Evolution of (a) scaled granular temperature and (b) scaled charge of the phase $h\,(h=i,j)$ for Case-B. Particle properties and flow parameters are given in table \ref{Table:InitConditionHomog}. The orange lines show the solutions of  \eqref{dThetapidt}, \eqref{dThetapjdt} in (a) and show the solutions of  \eqref{dChargepidt} and \eqref{dChargepjdt} in (b). ${\color{CadetBlue} \odot}$: hard-sphere simulation results for the phase $i$ and ${\color{CadetBlue} \Delta}$: hard-sphere simulation results for the phase $j$. Granular temperature and charge were scaled by using  \eqref{scalingTimeVelGran} and \eqref{scaledcharge}, respectively. \label{fig:caseB}}
\end{figure}
In Case-B, we performed a simulation of bidisperse solid mixtures with a particle diameter ratio of $R_d = 5$ and compared results with solutions of an equation set given in \eqref{dThetapidt}, \eqref{dThetapjdt}, \eqref{dChargepidt} and \eqref{dChargepjdt}. The total solid volume fraction was set to $\langle\alpha_p\rangle = 0.2$ with a large-to-small particle solid volume fraction ratio of $R_{\alpha}$ = 3. We also imposed different initial granular temperatures for each phase. The initial mean charges for the phases $i$ and $j$ were equal to $Q_{pi}$ = \SI{-1}{\femto\coulomb} and $Q_{pj}$ = \SI{42.3}{\femto\coulomb}, respectively. Granular temperature and charge equations were coupled with sequential solutions. Figure \ref{fig:caseB} shows the time evolution of the scaled granular temperature and the scaled charge for each phase. The granular temperature for each phase rapidly reaches the equilibrium state (dimensionless granular temperature which is equal to one) at $t^{*}=10$ (figure \ref{fig:caseB}(a)). In contrast, the mean charge goes to zero with a slower trend (figure \ref{fig:caseB}(b)). One can argue that the mean charge for the phase $i$ follows an exponential decay after the granular temperature reaches equilibrium, which is similar to the monodisperse case (Case-A). 

\begin{figure}
\centering
    \centering
    \psfrag{y3}[][c][1]{$Q_{ph}^*$ [-]}
    \psfrag{y2}[][c][1]{$\Tp{h}^*$ [-]}
    \psfrag{x1}[][c][1]{$t^{*}\,[-]$}
    \begin{tabular}{cc}
        (a) & (b) \vspace{-0.5cm}\\ 
        \includegraphics[scale=0.225,angle=-90]{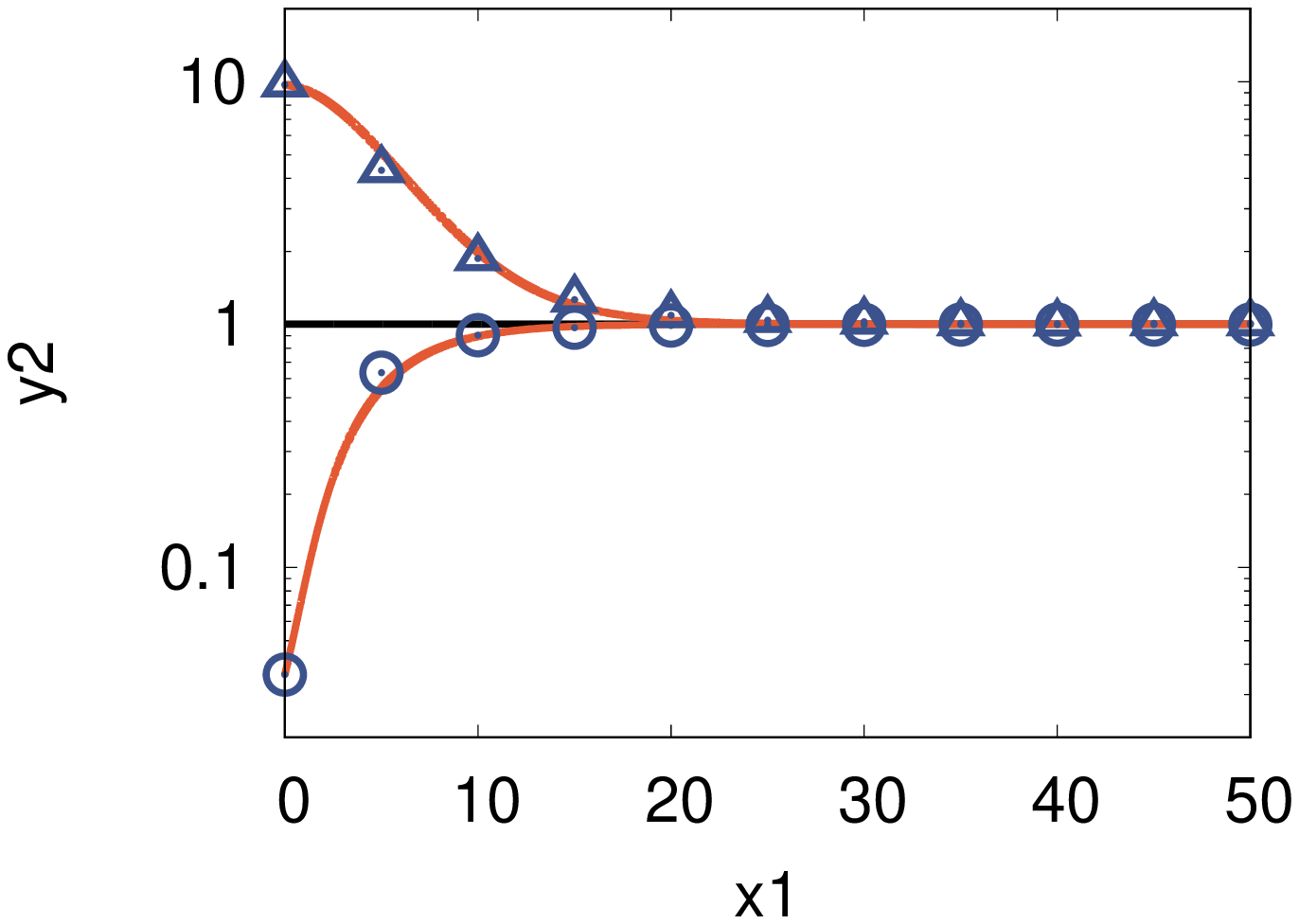} & \includegraphics[scale=0.225,angle=-90]{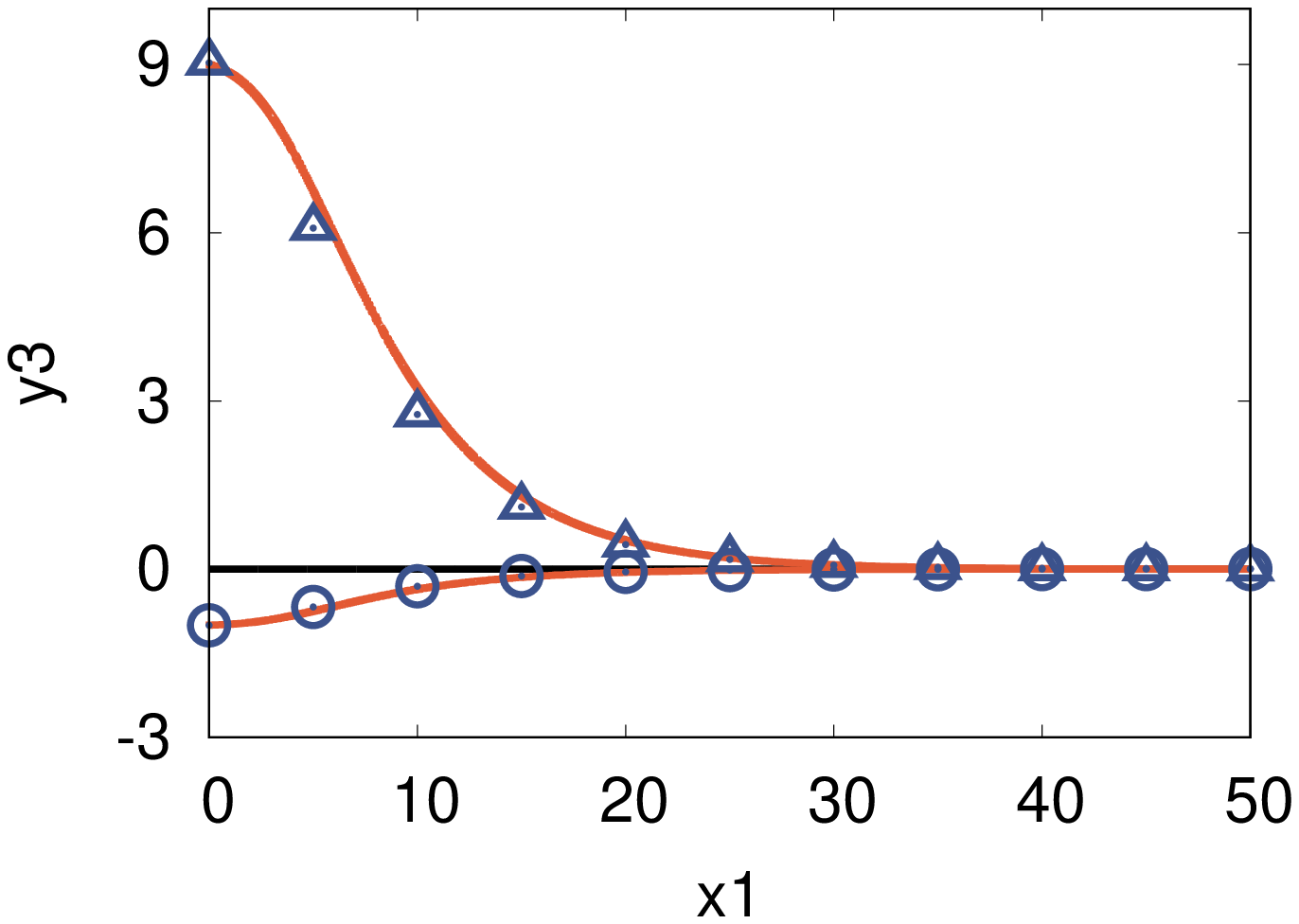} 
    \end{tabular}
    \caption{Evolution of (a) scaled granular temperature and (b) scaled charge of the phase $h\,(h=i,j)$ for Case-C. Particle properties and flow parameters are given in table \ref{Table:InitConditionHomog}. The orange lines show the solutions of \eqref{dThetapidt}, \eqref{dThetapjdt} in (a) and the solutions of \eqref{dChargepidt} and \eqref{dChargepjdt} in (b). ${\color{CadetBlue} \odot}$: hard-sphere simulation results for the phase $i$ and ${\color{CadetBlue} \Delta}$: hard-sphere simulation results for the phase $j$. Granular temperature and charge were scaled by using \eqref{scalingTimeVelGran} and \eqref{scaledcharge}, respectively. \label{fig:caseC}}
\end{figure}

In Case-C, we imposed the particle diameter-ratio of $R_d = 3$ and the particle density-ratio of $R_{\rho} = 10$ at the same time. The total solid fraction and the large-to-small particle solid fraction ratio were identical to of Case B. The initial mean charges were $Q_{pi}$ = \SI{-1}{\femto\coulomb} and $Q_{pj}$ = \SI{9}{\femto\coulomb}. The granular temperature and mean charge evolution are shown in figure \ref{fig:caseC}. A similar pattern as Case-B was observed with a quick evolution to the equilibrium temperature and slower evolution for the mean charge. It can be observed that before $t^*$ = 10, the mean charge evolution does not follow the exponential decrease due to the variation of the granular temperature. The Supplementary Material 1 and 2 show particle motions and charge evolution by the hard-sphere simulation for Case-C. For all spatially homogeneous granular gas cases, the model predictions are in excellent agreement with hard-sphere simulation results. 

These simulation cases also allowed us to probe the truncation terms in $M_7$ and $N_1$ (see table \ref{tab:coefficients}). We computed the mean errors between hard-sphere simulation results and model predictions for phase granular temperature and phase mean charge for different truncation orders and truncated the expansion if the mean error is lower than 5\%.

\subsection{Quasi-1D bidisperse granular gas simulations with spatial gradients}\label{section:Quasi1DSimulation}
As a second assessment benchmark, we performed hard-sphere simulations of bidisperse elastic granular gases in a fully periodic rectangular box with a dimension of $384\,d_{pj}\times 12\,d_{pj}\times 12\,d_{pj}$ ($d_{pj}$ is the diameter of the larger particle). For these simulations, we imposed the solid volume fraction for each phase with a step function through the domain and granular temperature difference between phases to validate gradient terms in the derived models. Similar to homogeneous granular gas cases, the velocities were initiated with a Maxwellian distribution with a mean velocity equal to zero for each phase. The total charge was equal to zero with an initial mean charge of each phase imposed as follows:
\begin{equation}
    \Big(n_{pi}^L + n_{pi}^R\Big)Q_{pi}(x, 0) + \Big(n_{pj}^L + n_{pj}^R\Big)Q_{pj}(x, 0)  = 0. \label{Eq:chargeInitBimodal}
\end{equation}
Here, $n_{ph}^L$ and $n_{ph}^R$ refer to particle number densities of the phase $h$ for left and right sides of the domain. If the solid phase is initially charged; the charge follows a Maxwellian distribution with a pre-defined non-zero mean value and varies with a step function in the domain. The phase charge variance is then equal to a small value; $\mathcal{Q}_p/m_{ph}$ = \SI{1e-32}{\coulomb^2}. 
The simulation campaign for quasi-1D bidisperse granular gases where we varied particle properties, domain-averaged solid volume fractions, granular temperatures and initial mean charges are listed in table \ref{Table:InitialSpatial}. For all these cases, the collisions were elastic ($e_c=1$) and the electrostatic force was not taken into account.
\begin{table}
    \centering
    {\small
    \begin{tabular}{ccccccccccc}
        Case & $d_p$  & $\rho_p$ &$\alpha_{p}^L$ & $\alpha_{p}^R$ & $\Theta_{p}^L {\small \times\SI{e-10}{}}$ & $\Theta_{p}^R{\small \times\SI{e-10}{}}$ & Q$_{p}^L$ & Q$_{p}^R$ & $N_p$ & $n_c/N_p$ \\
         & [\SI{}{\micro\meter}] & [\SI{}{\kilo\gram\per\meter^3}] 
         & [-] & [-] & [\SI{}{\kilo\gram~\meter^2\per\second^2}] & [\SI{}{\kilo\gram~\meter^2\per\second^2}] & [\SI{}{\femto\coulomb}] & [\SI{}{\femto\coulomb}] & [-] & [-] \\
    \hline
        D & 300 & 1500 & 0.1 & 0.2967 & 2.1 & 2.1 & -1 & 0.33764 & \SI{20945}{} & $\approx$ 57\\
    \hline
        E & 100 & 1500 & 0.02 & 0.06 & 0.0785 & 0.0785 & -3 & 1  & \SI{114065}{} & $\approx$ 206\\
          & 300 & 1500 & 0.2948 & 0.0992 & 2.1 & 2.1 & 0.0 & 0.0 & \SI{20785}{} & \\
    \hline
        F & 50  & 1500 & 0.005 & 0.015  & 0.00971 & 0.00971 & -1 & -1  & \SI{228099}{} & $\approx$ 83 \\
          & 300 & 1500 & 0.296 & 0.099 & 2.1 & 2.1 & 10.94 & 10.94 & \SI{20801}{} & \\
    \hline
        G & 100 & 150 & 0.02 & 0.06 & 0.00786 & 0.00786 & 1 & 1  & \SI{114065}{} & $\approx$ 206\\
          & 300 & 1500 & 0.2948 & 0.0992 & 2.1 & 2.1 & -5.51 & -5.51 & \SI{20752}{} & 
    \end{tabular}}
    \caption{Particle properties and flow parameters for quasi-1D granular gas simulations. Case-D refers to a monodisperse case; Case-E refers to a bidisperse case with a particle diameter ratio of $R_d$ = 3; Case-F refers to a bidisperse case with a particle diameter ratio of $R_d$ = 6; Case-G refers to a bidisperse case with a particle diameter ratio of $R_d$ = 3 and a particle density ratio of $R_{\rho}$ = 10. For all cases, a bimodal distribution of charge is imposed by following \eqref{eq:chargeConditionHomog} with a total charge equal to zero, $Q_{pm}$ = 0 (electrically neutral condition) and the restitution coefficient is set to unity, $e_c$ = 1. \label{Table:InitialSpatial}}
\end{table}

By following \cite{fox2014multiphase}, instead of solving the granular temperature balance equation for each phase, we solve the total kinetic energy, $E_{ph}$, for a solid phase $h$, which is a conserved quantity. The total kinetic energy tensor for a solid phase $h$ is defined as
\begin{equation}
    \mathsfbi{E}_{ph} = \frac{1}{2}\Big(\bm{\upsigma}_{ph} + \Uv{h}\otimes\Uv{h}\Big),
\end{equation}
with the fluctuating kinetic energy tensor, $\bm{\upsigma}_{ph}$. The granular temperature is given by the trace of $\bm{\upsigma}_{ph}$ as
\begin{equation}
    \frac{\Tp{h}}{m_{ph}} = \frac{1}{3}\mathrm{tr}{(\bm{\upsigma}_{ph})}.
\end{equation}
Hence, the total kinetic energy is given by
\begin{equation}\label{totalkineticenergy}
    E_{ph} = \frac{3}{2}\frac{\Tp{h}}{m_{ph}} + \frac{1}{2}\mathrm{tr}(\Uv{h}\otimes\Uv{h}).
\end{equation}
The complete set of mass, momentum, total kinetic energy and charge transport equation for a phase $h$ ($h = i,j$) is written in a conservative form as:

\begin{equation}
\resizebox{.9\hsize}{!}{$\left\{\begin{array}{ll}
    \frac{\partial}{\partial t} \Big[ \alpha_{ph}\Big] + \nabla\cdot \Big[\alpha_{ph} \Uv{h}\Big] & = 0 \\[2pt]
    \frac{\partial}{\partial t} \Big[ \alpha_{ph} \Uv{h}\Big]  + \nabla\cdot \Big[\alpha_{ph} (\Uv{h}\otimes\Uv{h}) + \frac{1}{\rho_{ph}}\Big(P_{ph}^{kin}\Ii + \sum\limits_{l = i,j}\bm{\uptheta}_{hl} \Big) \Big] & = \frac{1}{\rho_{ph}}\sum\limits_{l = i,j}\bm{\chi}_{hl} \\[2pt]
    \frac{\partial}{\partial t} \Big[ \alpha_{ph} E_{ph}\Big]  + \nabla\cdot \Big[\alpha_{ph}E_{ph}\Uv{h} + \frac{\Uv{h}}{\rho_{ph}}\cdot\Big(P_{ph}^{kin}\Ii + \sum\limits_{l = i,j}\bm{\uptheta}_{hl}\Big) + \frac{1}{\rho_{ph}}\sum\limits_{l = i,j}\bfit{q}_{hl}\Big] & = \frac{1}{\rho_{ph}}\sum\limits_{l = i,j}(\bm{\chi}_{hl}\cdot\Uv{h} + \gamma_{hl}) \\[2pt]
    \frac{\partial}{\partial t} \Big[ \alpha_{ph}Q_{ph}\Big] + \nabla\cdot \Big[\alpha_{ph}Q_{ph}\Uv{h} + \frac{m_{ph}}{\rho_{ph}}  \sum\limits_{l = i,j}\bm{\theta}_{hl}^{q}\Big] & = \frac{m_{ph}}{\rho_{ph}} \sum\limits_{l = i,j}\chi_{hl}^q 
    \end{array} \right. $}   
\end{equation} 

In the first block of the equation set, we have time derivative terms for the conserved quantities, $\alpha_{ph}, \alpha_{ph}\Uv{h}, \alpha_{ph} E_{ph}$ and $\alpha_{ph}Q_{ph}$. In the second block, the spatial fluxes with the collisional flux terms representing variable exchange within and between phases are given, as well as the kinetic granular pressure, $P_{ph}^{kin}$, which is defined as (e.g. \cite{gidaspow_multiphase_1994}) 
\begin{equation}\label{Eq:kineticPressure}
    P_{ph}^{kin}=\alpha_{ph}\,\rho_{ph}\,\frac{\Tp{h}}{m_{ph}}.
\end{equation}
On the right-hand-side, we have non-conservative source terms representing the dissipation of quantity between different phases. The phase solid volume fraction, $\alpha_{ph}$, the phase velocity, $\Uv{h}$, the phase total kinetic energy, $E_{ph}$, and the phase mean charge, $Q_{ph}$, are then found from the conserved quantities. The phase granular temperature, $\Tp{h}$, is computed by \eqref{totalkineticenergy}. The given conservative forms can be solved using any finite-volume method. Here, the time derivatives were discretized with the Euler method whereas the spatial fluxes were computed with a Lax-Friedrichs scheme with van Albada slope limiter. 
\begin{figure}
    \centering
    \psfrag{x1}[][][0.8]{$x/L$}
    \begin{tabular}{cccc}
        (a) & (b) & (c) & (d) \vspace{-0.25cm}\\
        \psfrag{y1}[][bl][0.9]{$\alpha_{p}$}
        \hspace{-0.3cm}\includegraphics[scale=0.15,angle=-90]{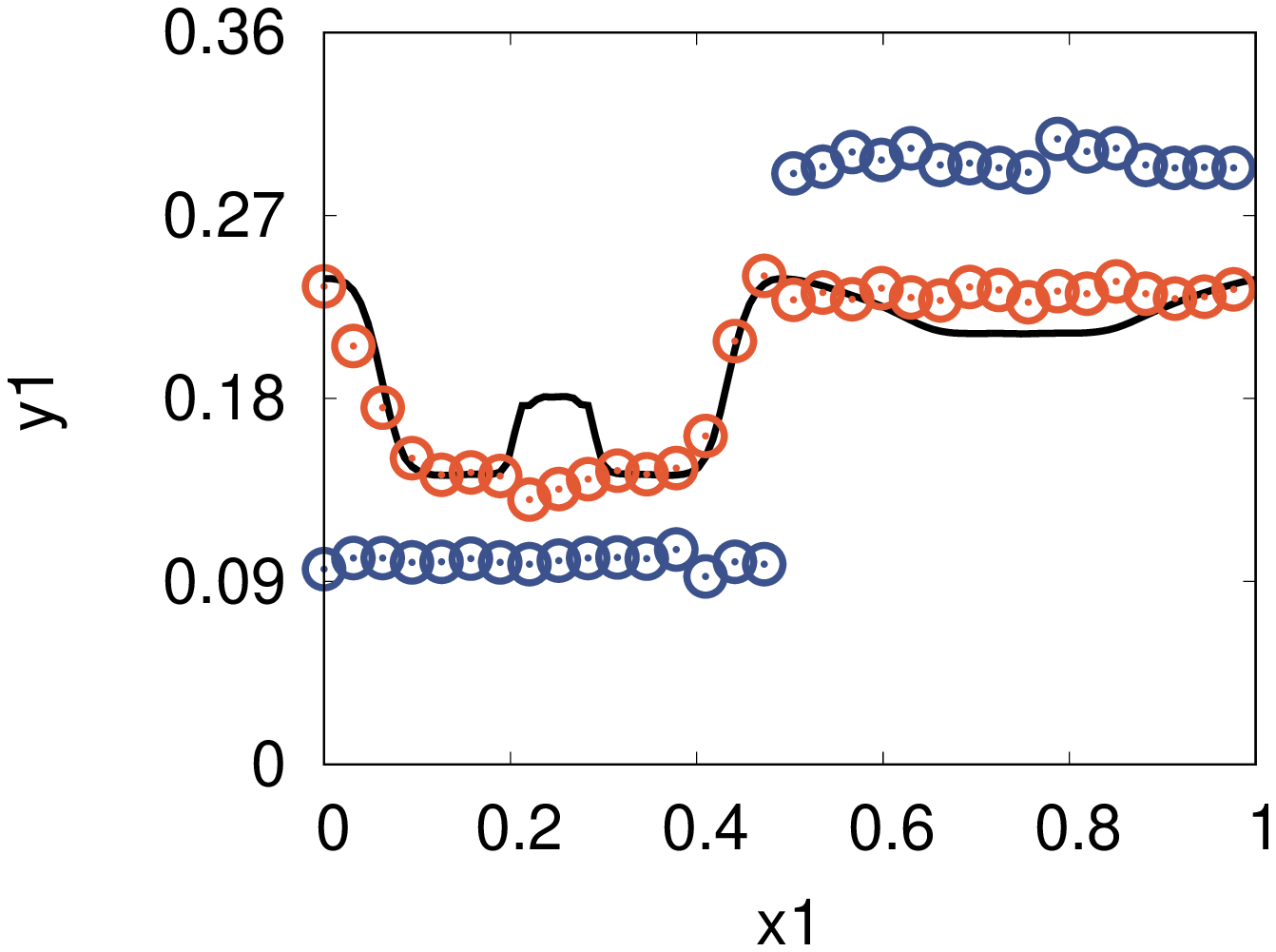} & \psfrag{y2}[][bl][0.9]{$U_{p}^*$}
        \hspace{-0.6cm}\includegraphics[scale=0.15,angle=-90]{Figures/CaseD/Fig4a1.eps} & \psfrag{y3}[][bl][0.9]{$\Tp{}^*$}
        \hspace{-0.6cm}\includegraphics[scale=0.15,angle=-90]{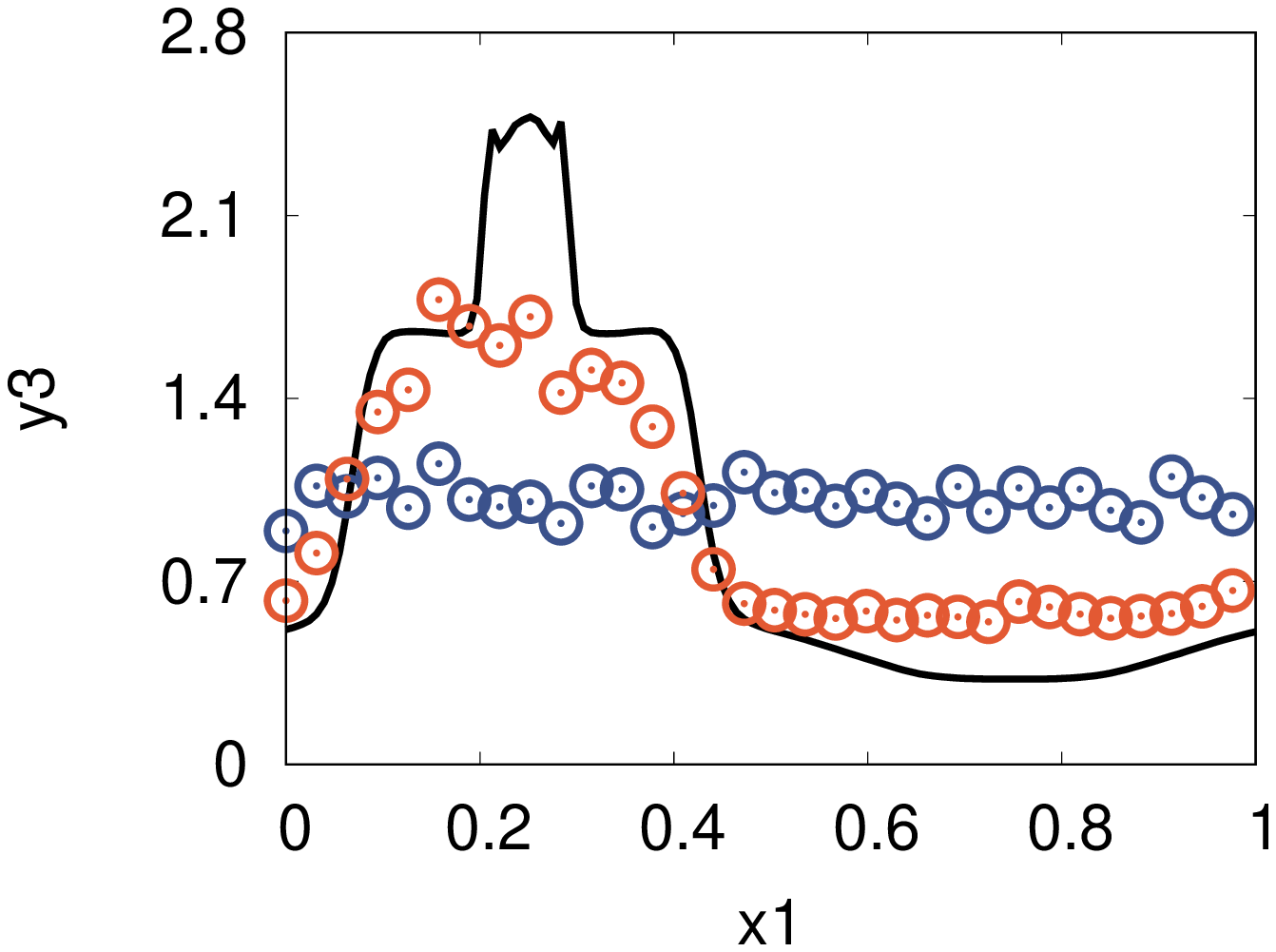} & \psfrag{y4}[][bl][0.9]{$Q_{p}^*$}
        \hspace{-0.6cm}\includegraphics[scale=0.15,angle=-90]{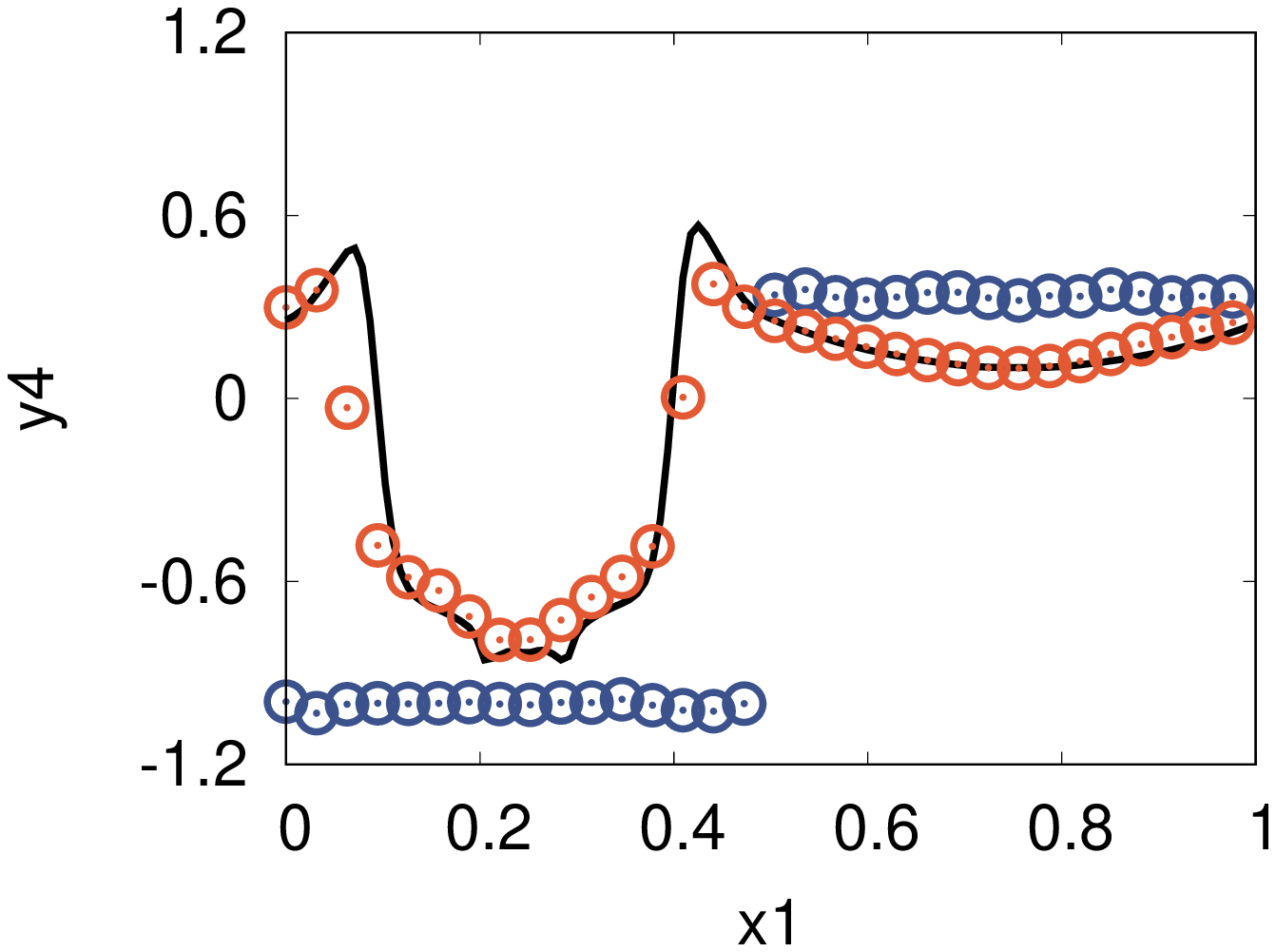}  \\
    \end{tabular}
    \begin{tabular}{cccc}
        \psfrag{y1}[][bl][0.9]{$\alpha_{p}$}
        \hspace{-0.3cm}\includegraphics[scale=0.15,angle=-90]{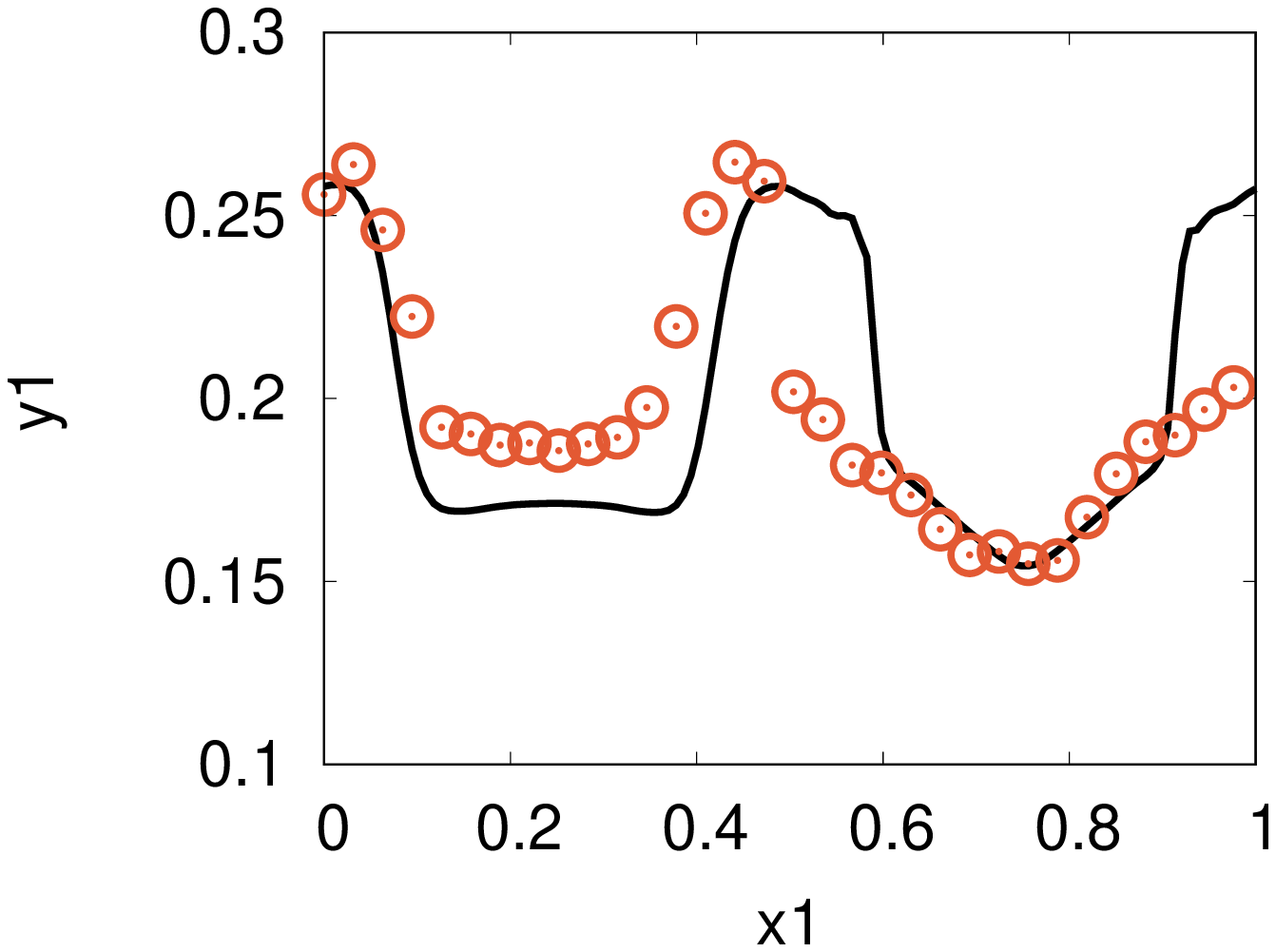} & \psfrag{y2}[][bl][0.9]{$U_{p}^*$}
        \hspace{-0.6cm}\includegraphics[scale=0.15,angle=-90]{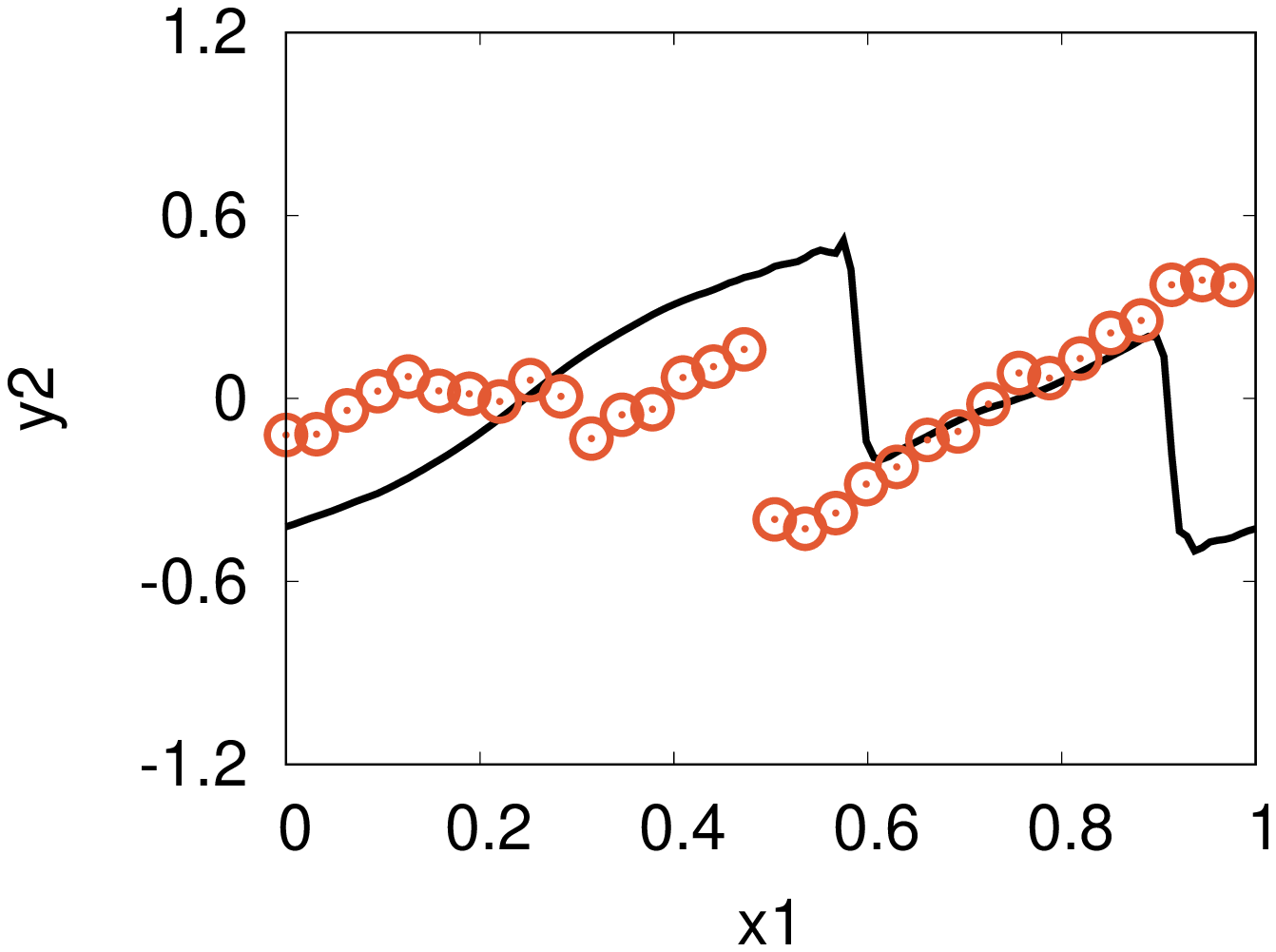} & \psfrag{y3}[][bl][0.9]{$\Tp{}^*$}
        \hspace{-0.6cm}\includegraphics[scale=0.15,angle=-90]{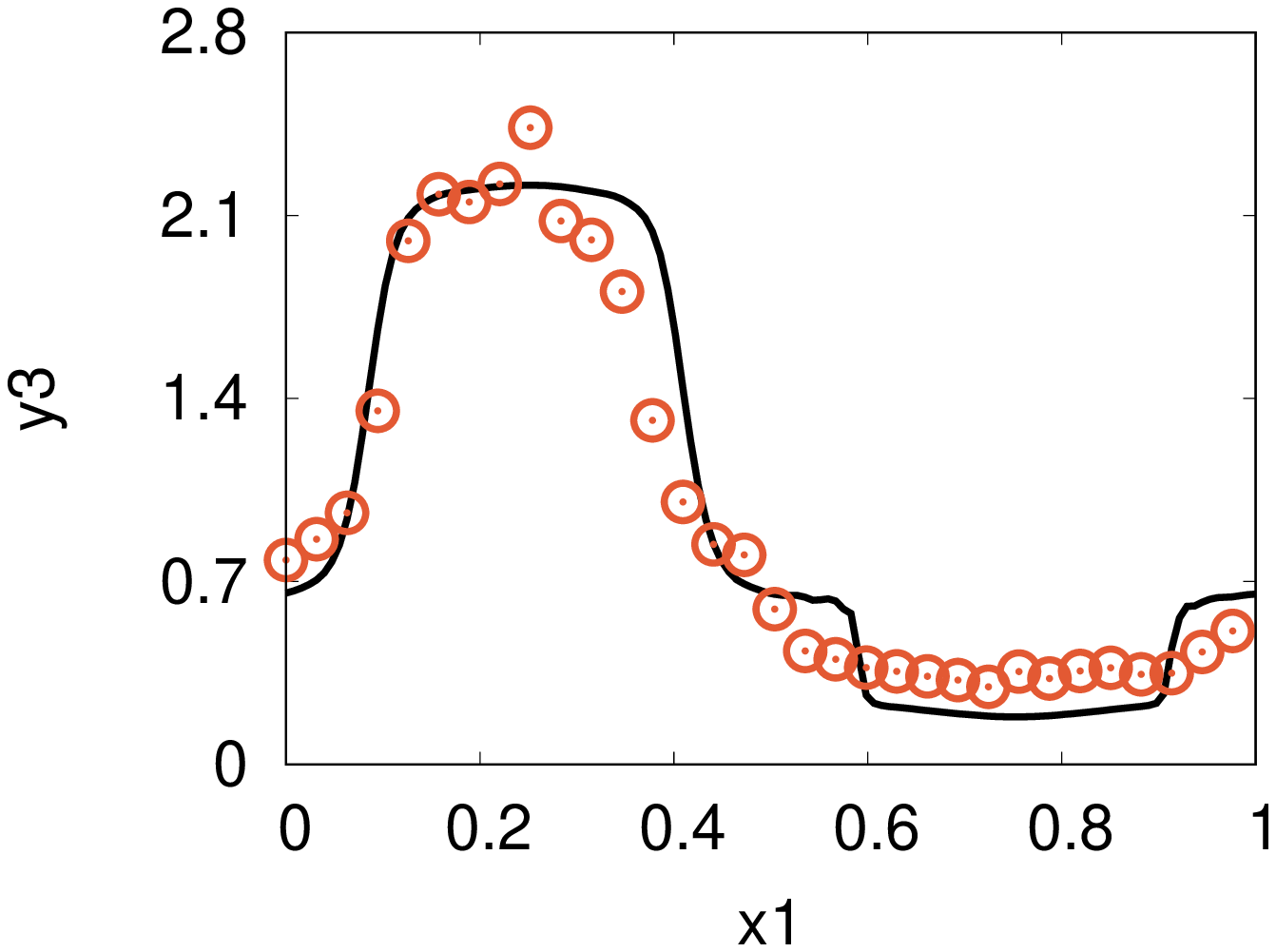} & \psfrag{y4}[][bl][0.9]{$Q_{p}^*$}
        \hspace{-0.6cm}\includegraphics[scale=0.15,angle=-90]{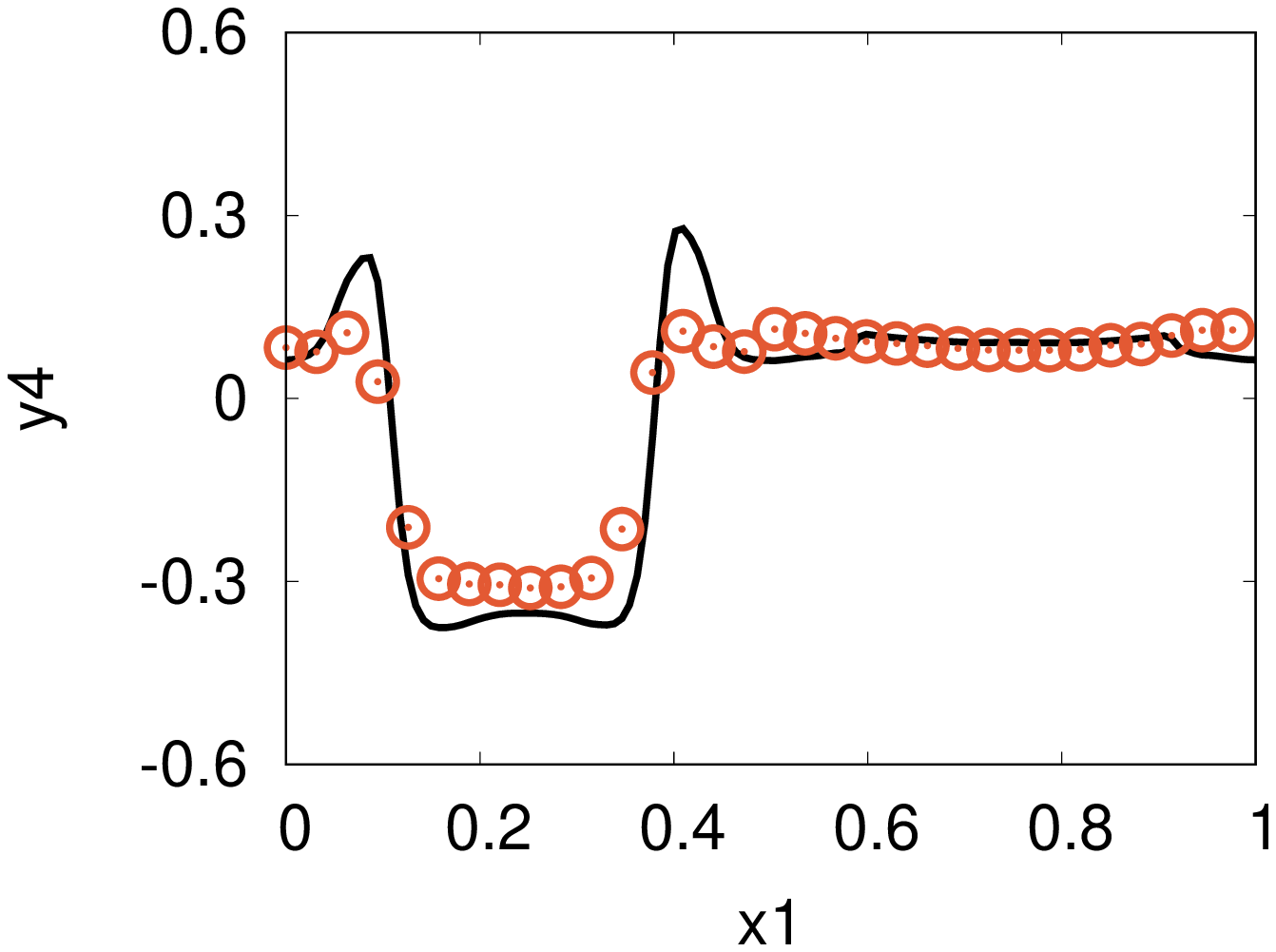}\\
    \end{tabular}
    \caption{Evolution of (a) phase solid volume fraction, (b) phase scaled velocity, (c) phase scaled granular temperature and (d) phase scaled mean charge for Case-D at two time instants. Top and bottom figures refer to the simulation results and the model predictions at $t^*=33.33$ and $t^*=66.66$, respectively. ${\color{CadetBlue}\boldsymbol{\odot}}$: initial conditions for the hard sphere simulation. ${\color{Black}\rule[0.5ex]{10pt}{0.75pt}}$: Eulerian model predictions and ${\color{RedOrange}\boldsymbol{\odot}}$: hard sphere simulation results. Variables were scaled by using \eqref{scalingTimeVelGran}, \eqref{scaledtheta} and \eqref{scaledcharge}.
    \label{Fig:ResultMonoSpatial}}
\end{figure}

We first compared hard-sphere simulation results with the developed model predictions for a quasi-1D granular gas simulation of monodisperse particles named Case-D in table \ref{Table:InitialSpatial}. The charge and solid volume fraction were imposed with a step function and the total charge inside the system was equal to zero. The initial conditions for hard-sphere simulation with blue dot-points (${\color{CadetBlue} \odot}$) are shown in figure \ref{Fig:ResultMonoSpatial}. This figure also shows the evolution of solid volume fraction, velocity, granular temperature and charge by the simulation and the model predictions (solid lines) at $t^* = 33.33$ (top) and $t^* = 66.6$ (bottom). The phase velocity rapidly evolves due to the solid volume fraction gradient and the wave-like behaviour is seen through the periodic domain in time (figure \ref{Fig:ResultMonoSpatial}(b)). The non-uniform granular temperature profile also develops as a wave-like shape (figure \ref{Fig:ResultMonoSpatial}(c)). Initially, the granular pressure induces solid flux that leads to spatial redistribution of the granular temperature due to stress production term in the fluctuating energy equation and ``thermal'' diffusion. Wave-like solid fluxes from left and right boundaries also lead to local increases and decreases in granular temperature. As expected, the charge distribution dissipates and goes to zero (figure \ref{Fig:ResultMonoSpatial}(d)). All these behaviours are very well captured by our model predictions. However, the Eulerian model overestimated the solid volume fraction and the granular temperature at $x/L=0.2$ and $t^* = 33.33$ (figures \ref{Fig:ResultMonoSpatial}(a) and c). Additionally, the location of velocity sharp gradient $t^* = 66.66$  was slightly mispredicted by the model. 

\begin{figure}
    \centering
    \psfrag{x1}[][][0.8]{$x/L$}
    \begin{tabular}{cccc}
        \psfrag{y1}[][bl][0.9]{$\alpha_{pi}$}
        \hspace{-0.3cm}\includegraphics[scale=0.15,angle=-90]{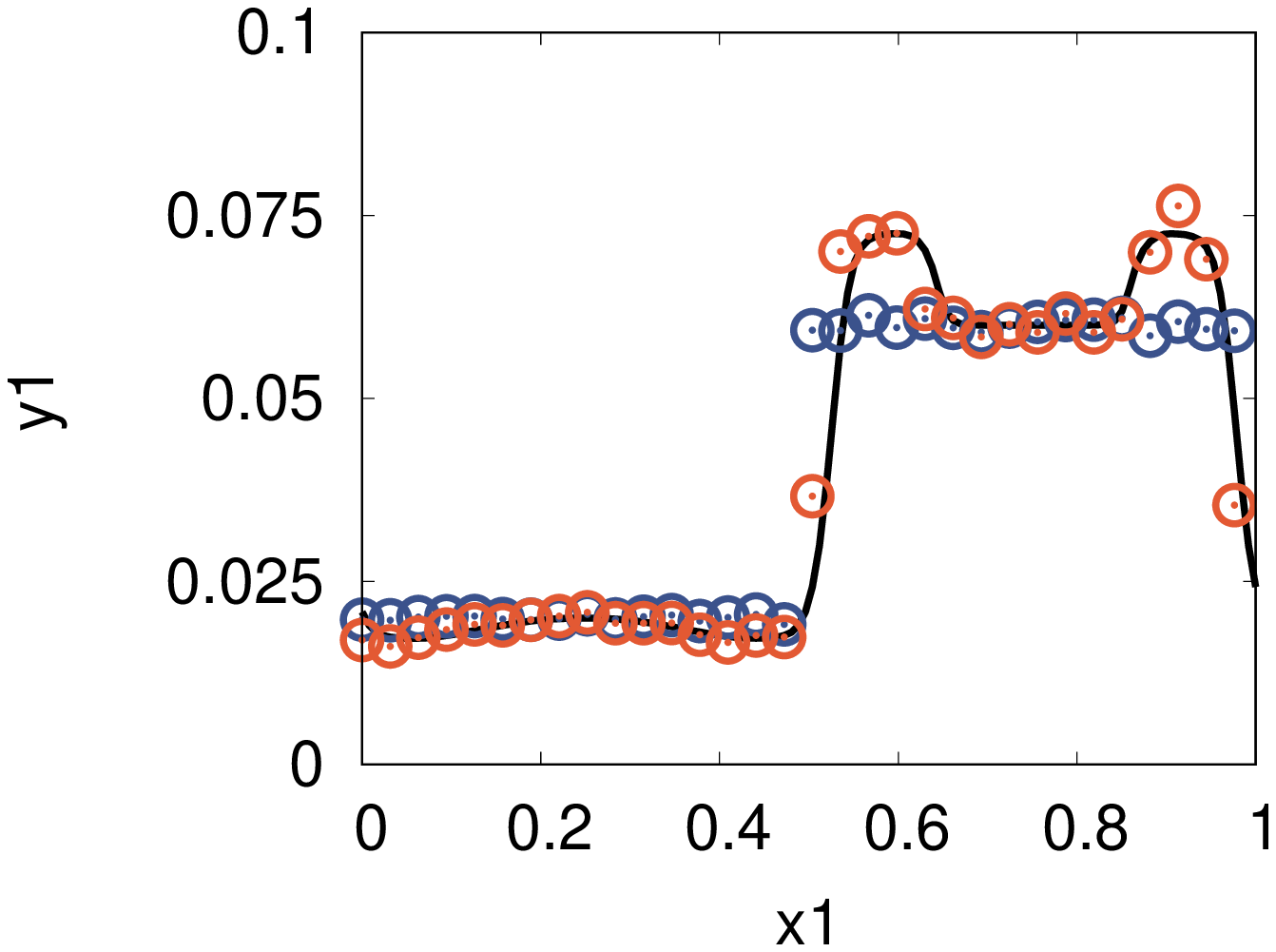} & \psfrag{y2}[][bl][0.9]{$U_{pi}^*$}
        \hspace{-0.6cm}\includegraphics[scale=0.15,angle=-90]{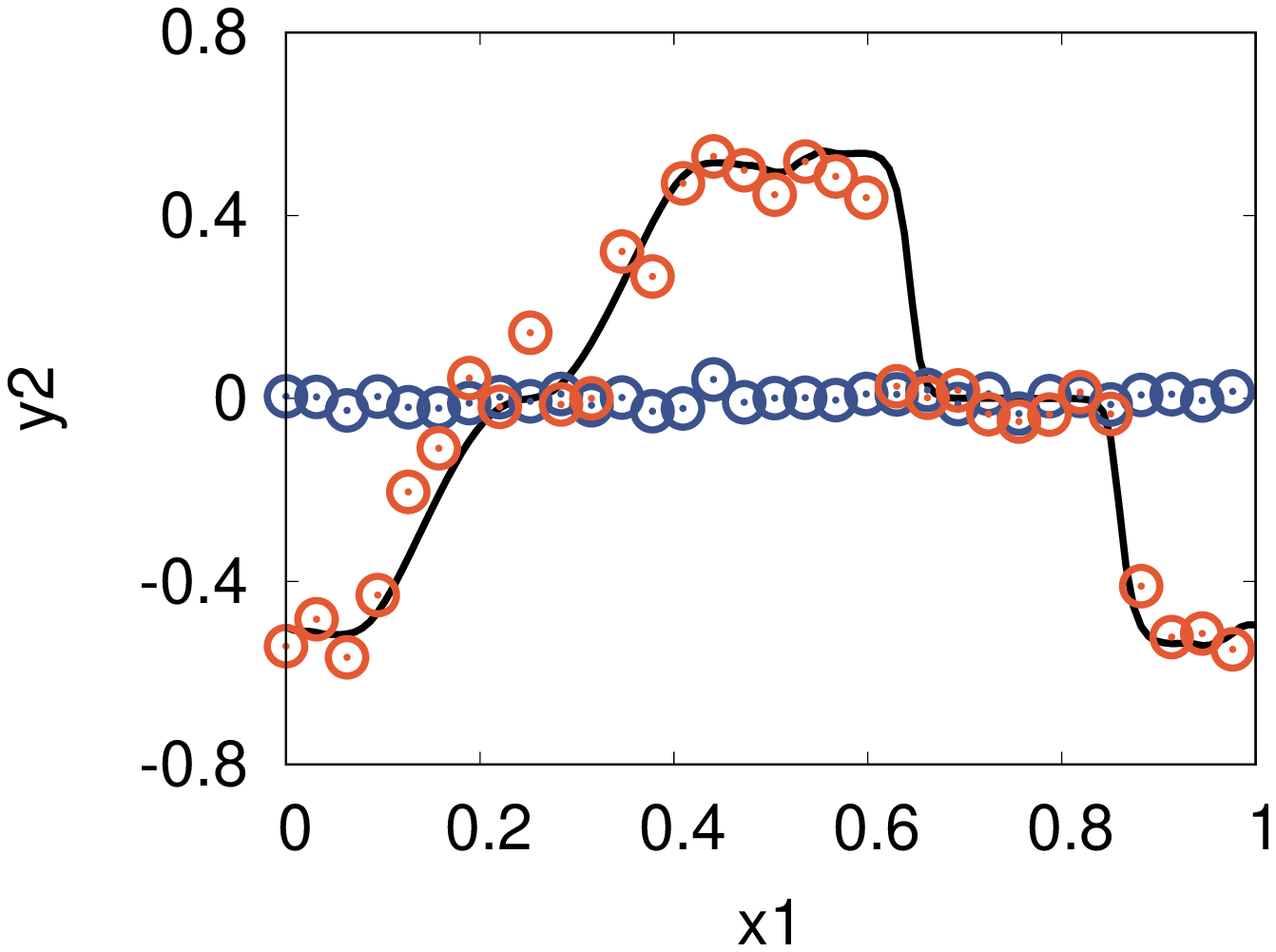} & \psfrag{y3}[][bl][0.9]{$\Tp{i}^*$}
        \hspace{-0.6cm}\includegraphics[scale=0.15,angle=-90]{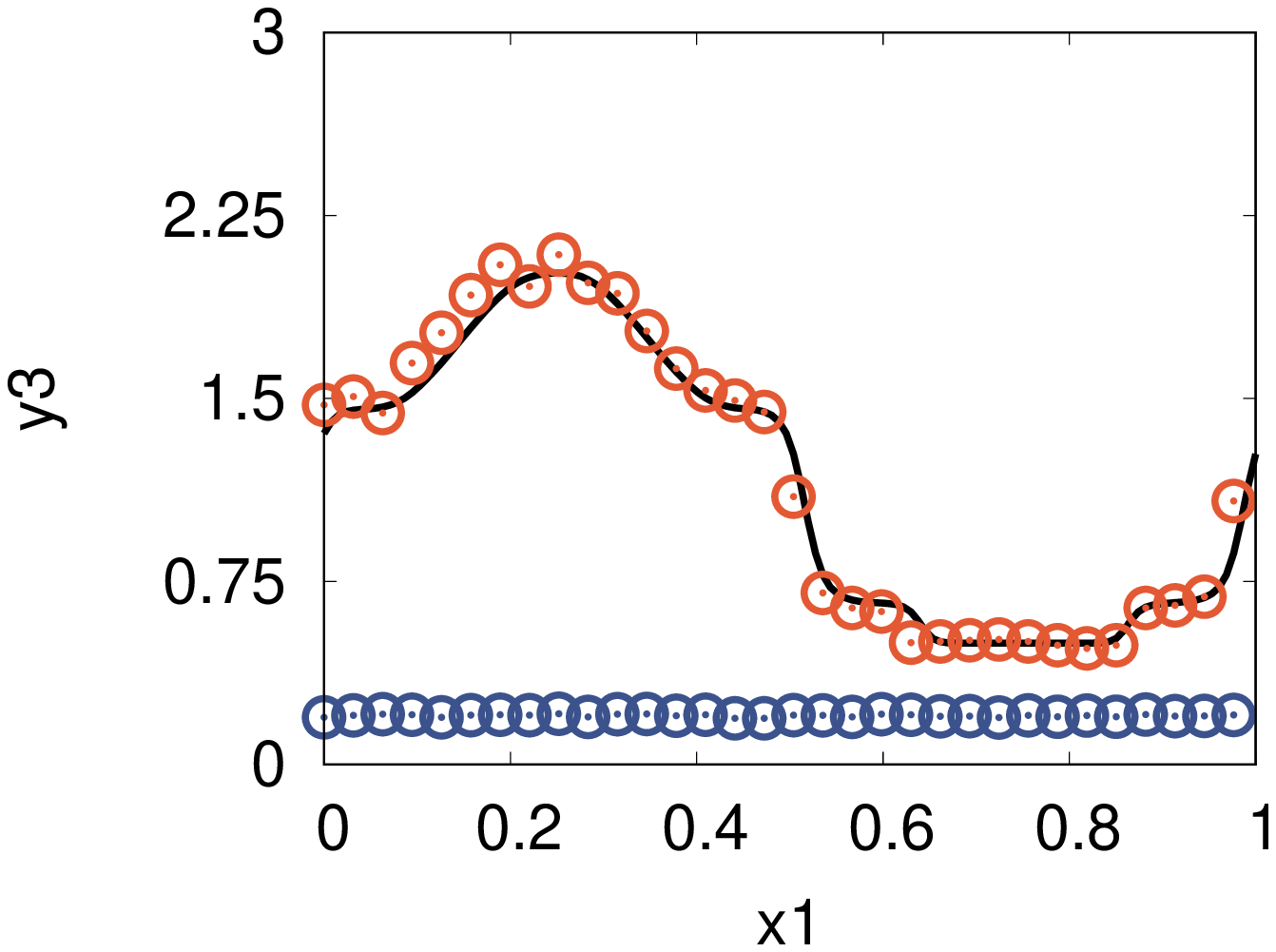} & \psfrag{y4}[][bl][0.9]{$Q_{pi}^*$}
        \hspace{-0.6cm}\includegraphics[scale=0.15,angle=-90]{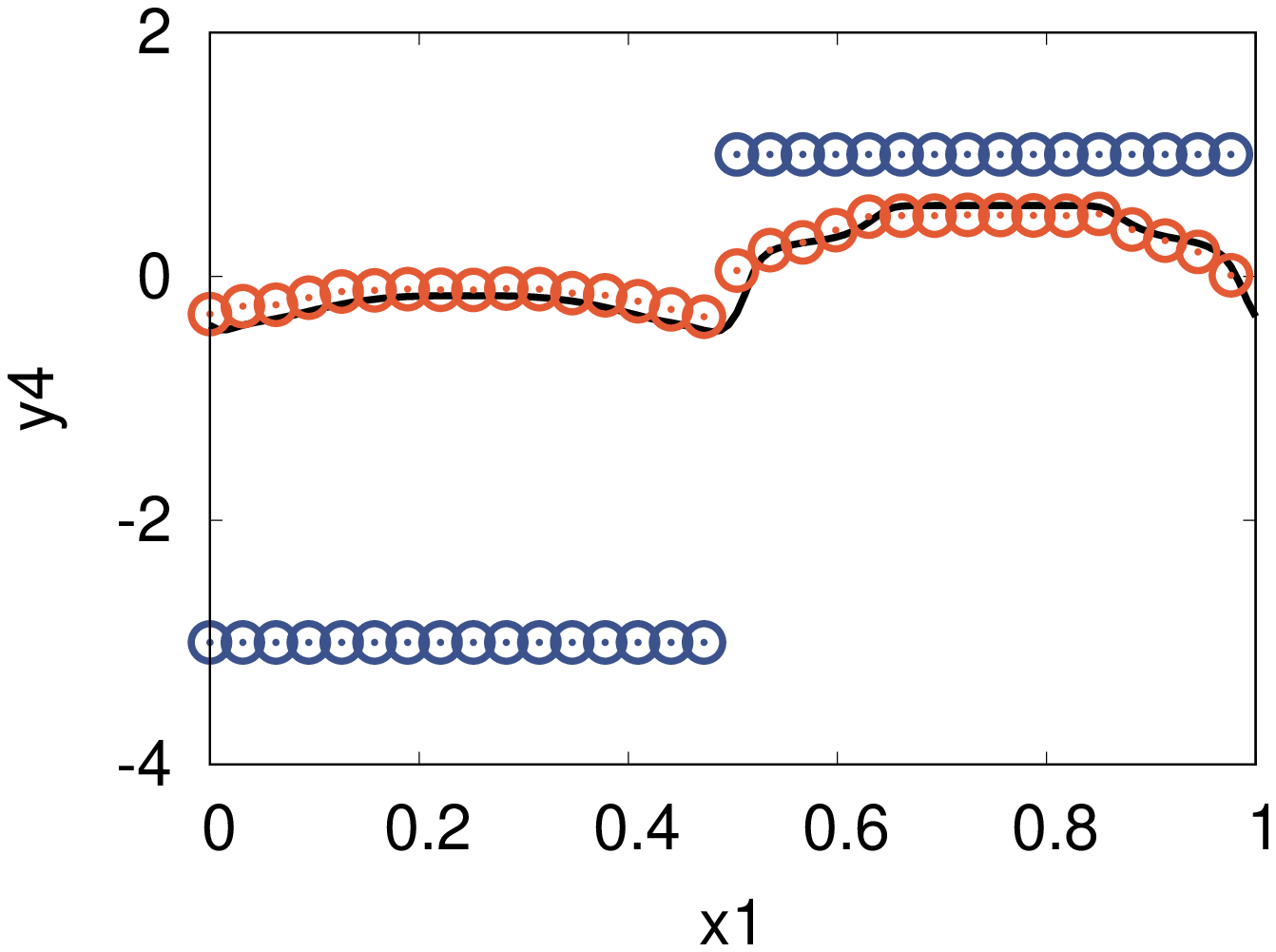} \\
        \psfrag{y1}[][bl][0.9]{$\alpha_{pj}$}
        \hspace{-0.3cm}\includegraphics[scale=0.15,angle=-90]{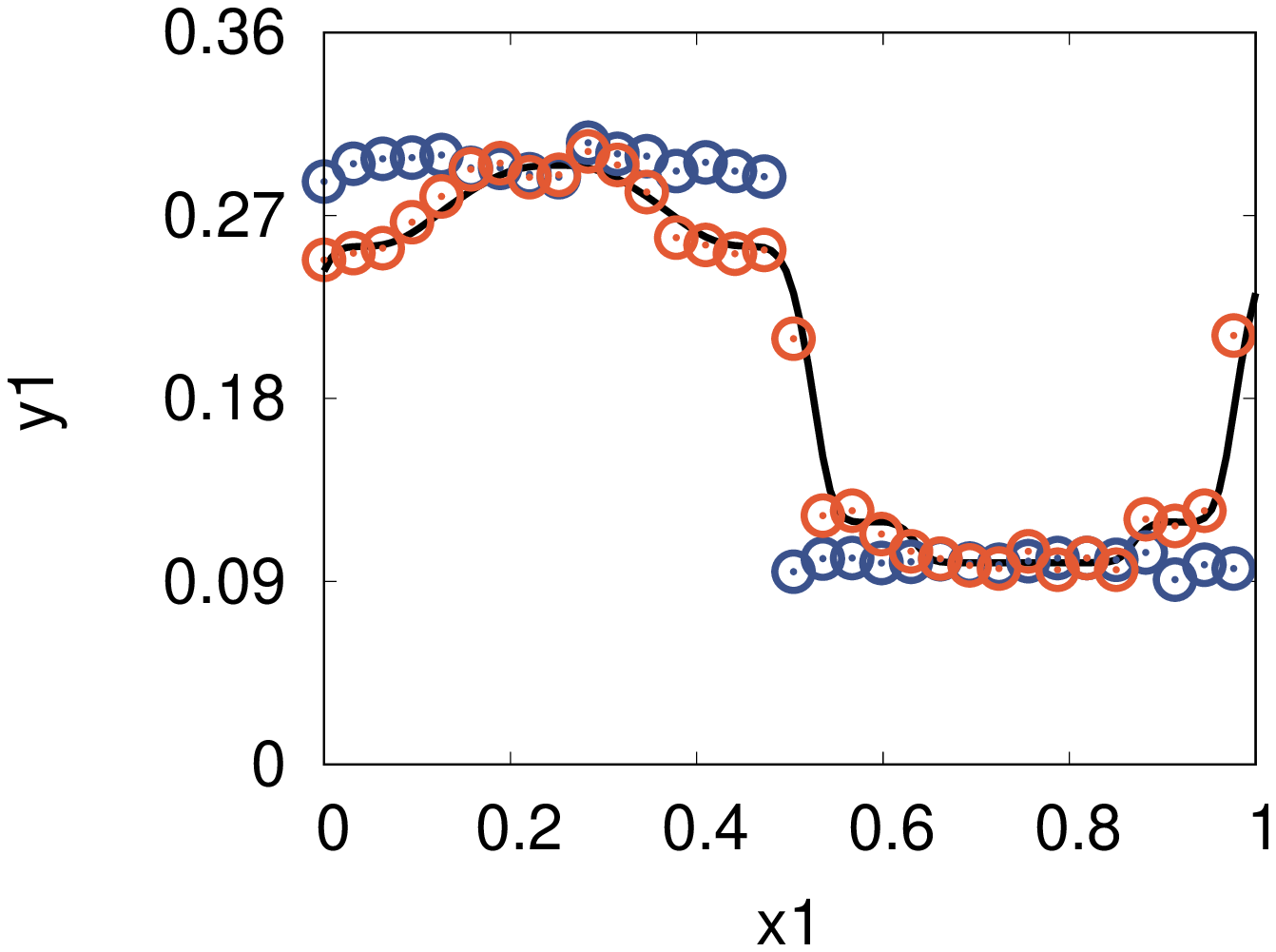} & \psfrag{y2}[][bl][0.9]{$U_{pj}^*$}
        \hspace{-0.6cm}\includegraphics[scale=0.15,angle=-90]{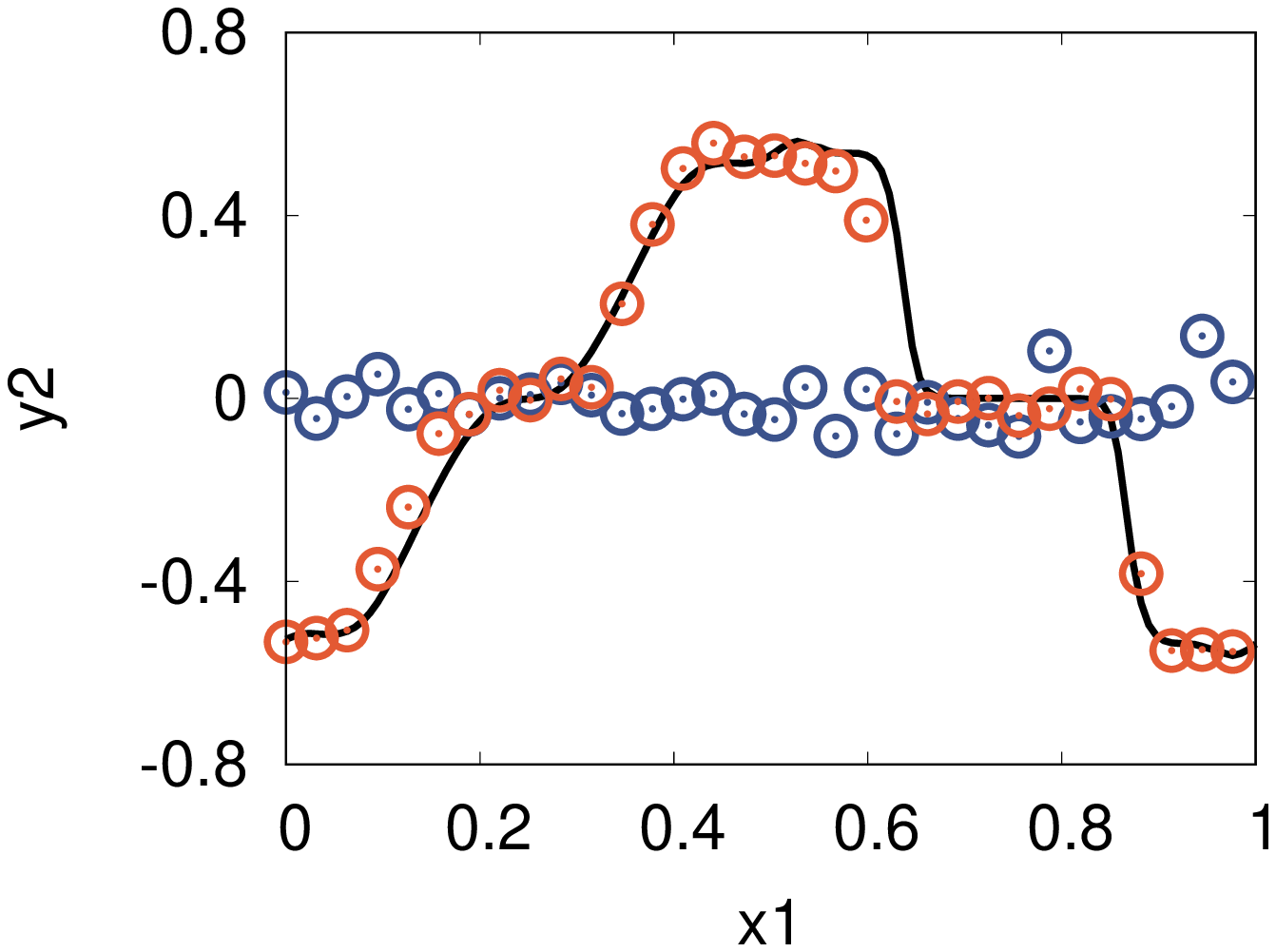} & \psfrag{y3}[][bl][0.9]{$\Tp{j}^*$}
        \hspace{-0.6cm}\includegraphics[scale=0.15,angle=-90]{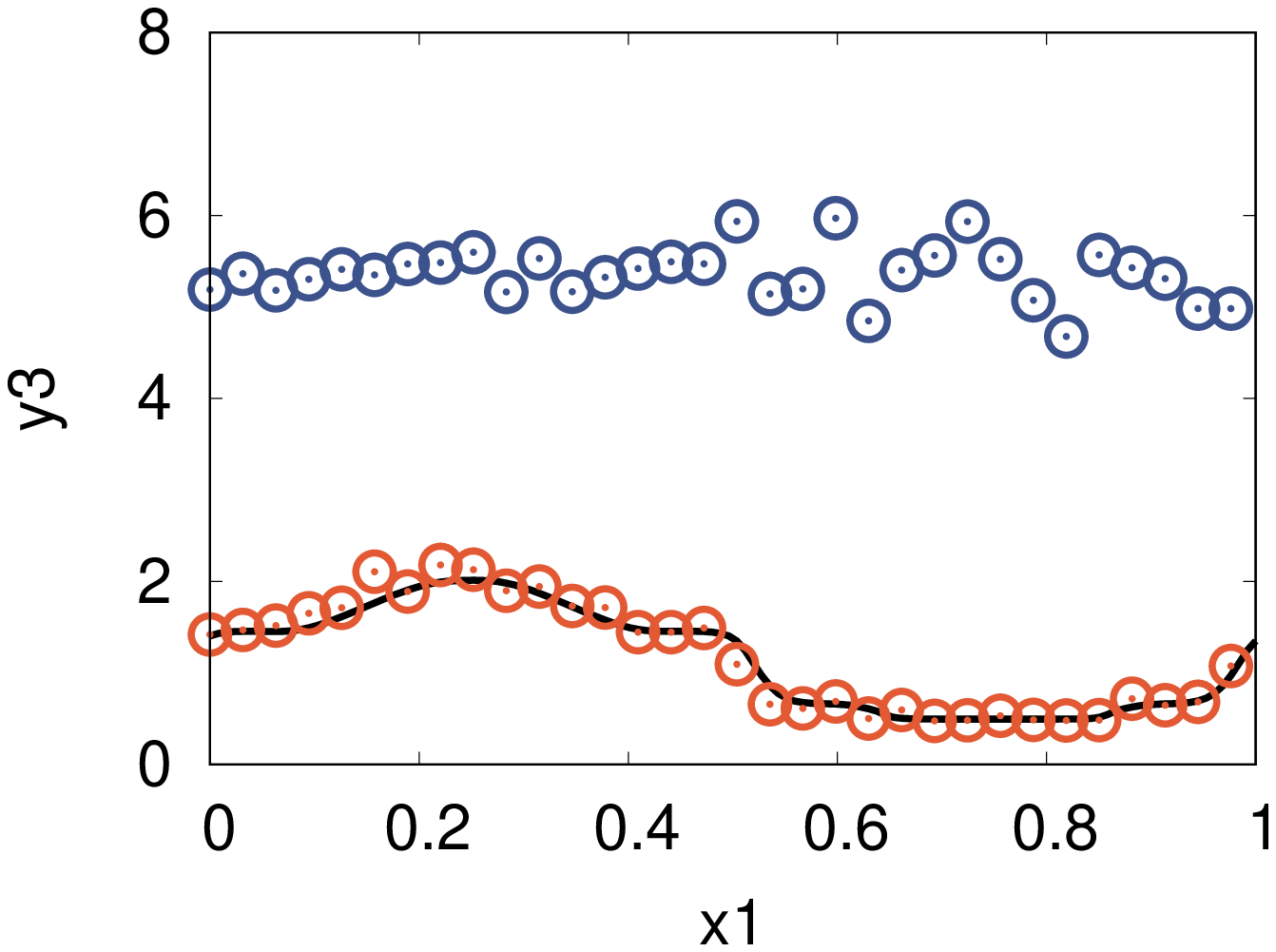} & \psfrag{y4}[][bl][0.9]{$Q_{pj}^*$}
        \hspace{-0.6cm}\includegraphics[scale=0.15,angle=-90]{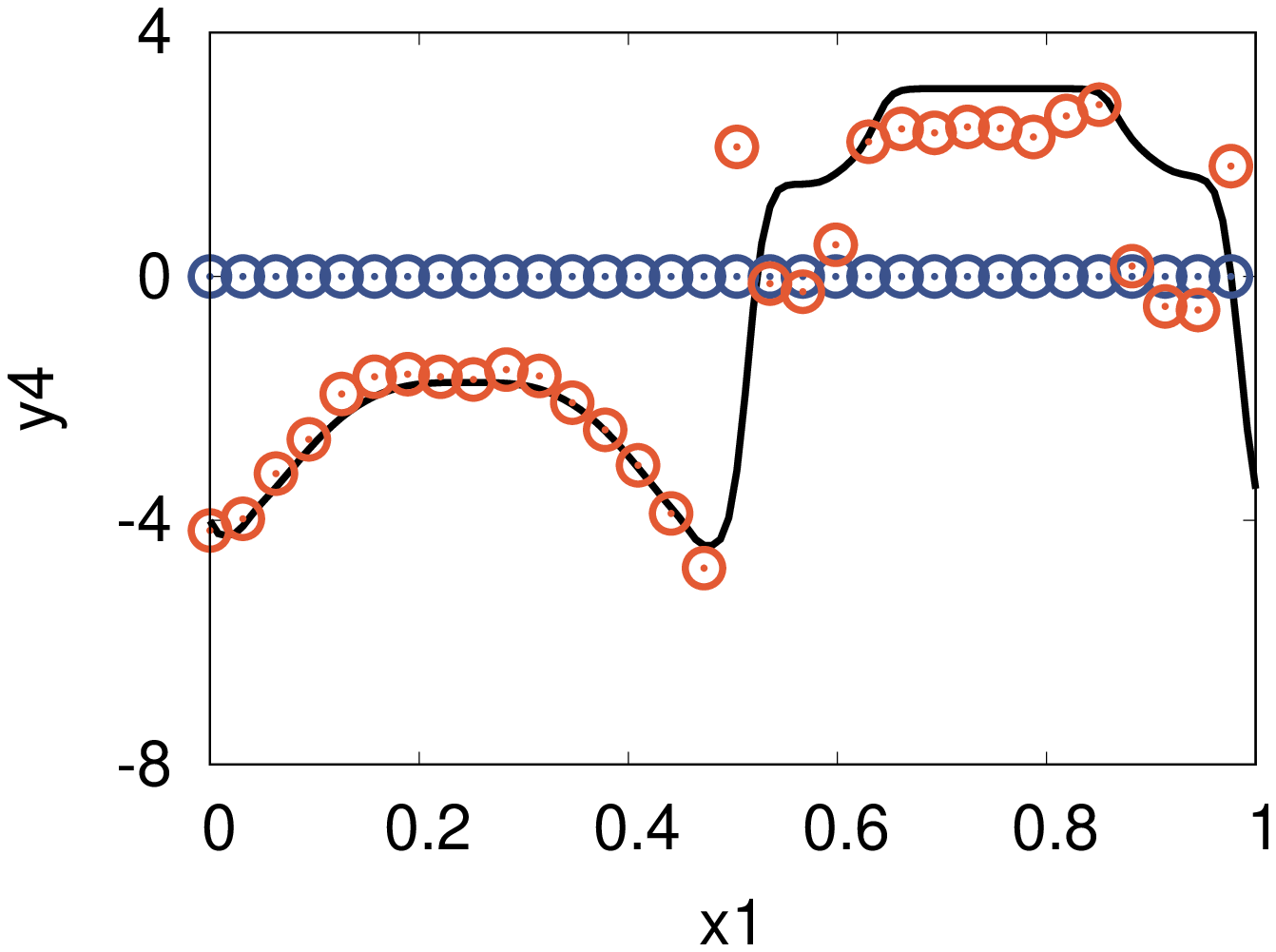} \\
    \end{tabular}
    \caption{Evolution of (a) phase solid volume fraction, (b) phase scaled velocity, (c) phase scaled granular temperature and (d) phase scaled mean charge of the phase $h\,(h=i,j)$ for Case-E at $t^*=25$. ${\color{CadetBlue}\boldsymbol{\odot}}$: initial conditions for the hard sphere simulation. ${\color{Black}\rule[0.5ex]{10pt}{0.75pt}}$: Eulerian model predictions and ${\color{RedOrange}\boldsymbol{\odot}}$: hard sphere simulation results. Variables were scaled by using \eqref{scalingTimeVelGran}, \eqref{scaledtheta} and \eqref{scaledcharge}.
    \label{Fig:ResultRd3Spatialt25}}
    \centering
    \psfrag{x1}[][][0.8]{$x/L$}
    \begin{tabular}{cccc}
        \psfrag{y1}[][bl][0.9]{$\alpha_{pi}$}
        \hspace{-0.3cm}\includegraphics[scale=0.15,angle=-90]{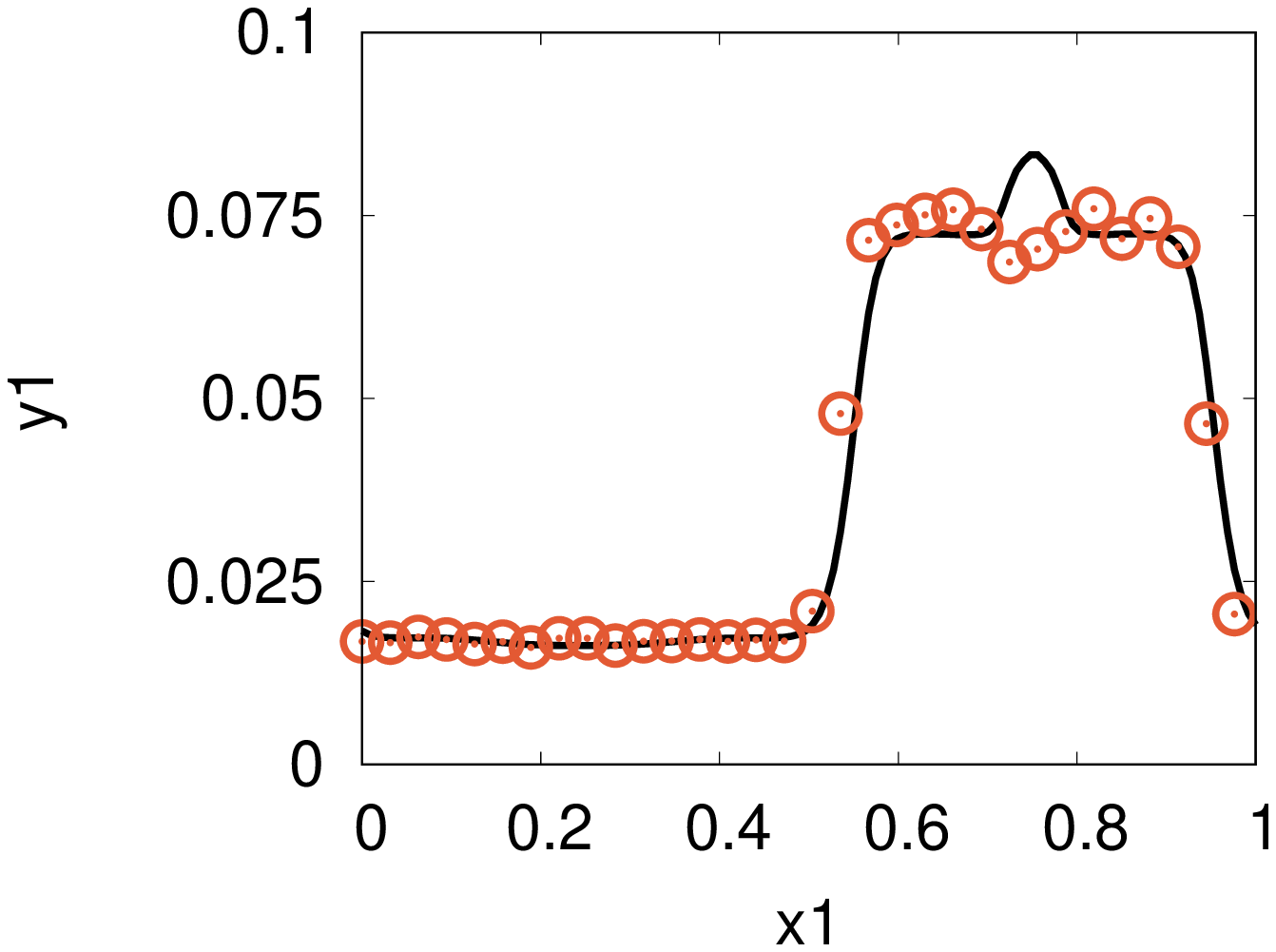} & \psfrag{y2}[][bl][0.9]{$U_{pi}^*$}
        \hspace{-0.6cm}\includegraphics[scale=0.15,angle=-90]{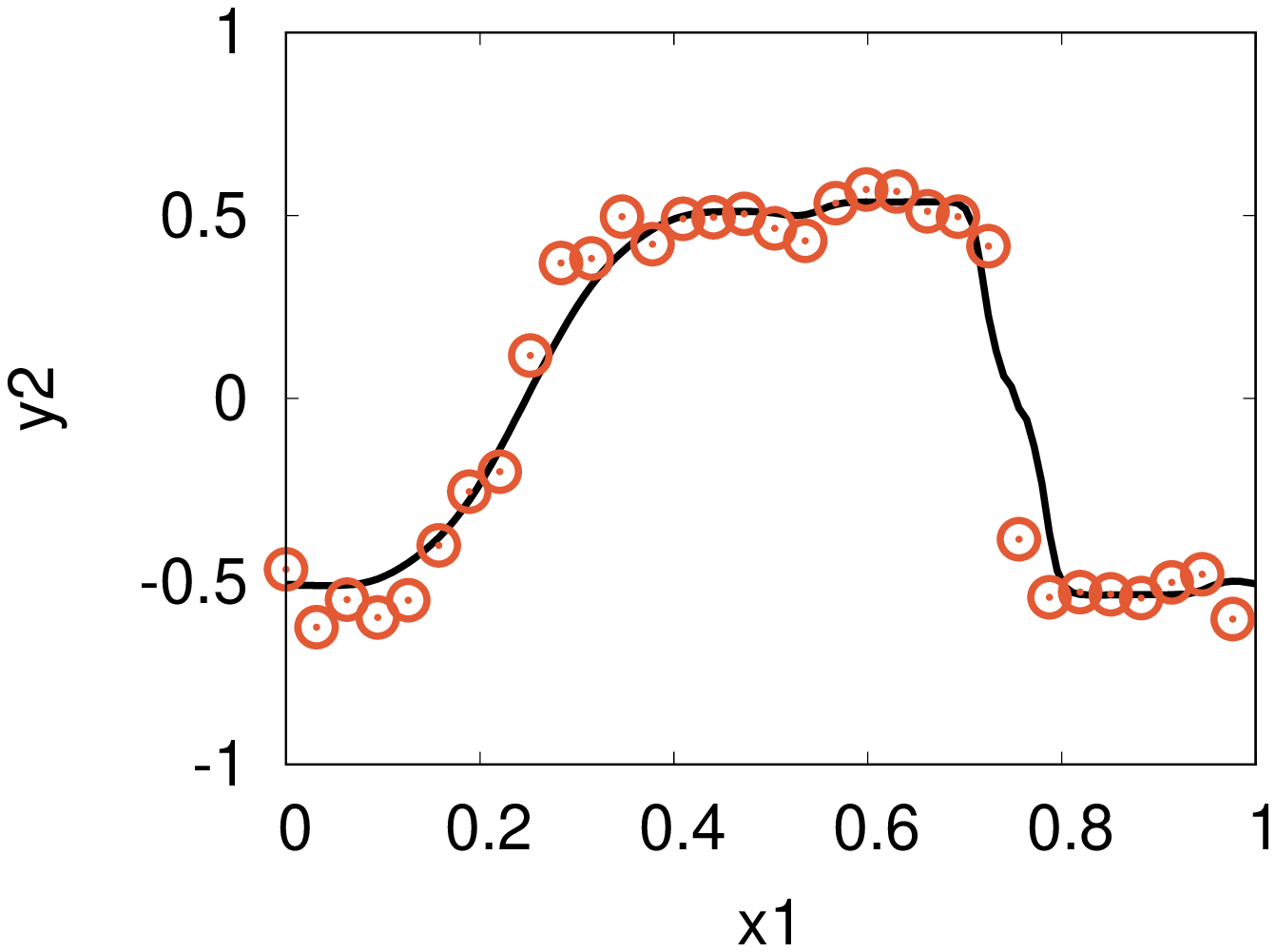} & \psfrag{y3}[][bl][0.9]{$\Tp{i}^*$}
        \hspace{-0.6cm}\includegraphics[scale=0.15,angle=-90]{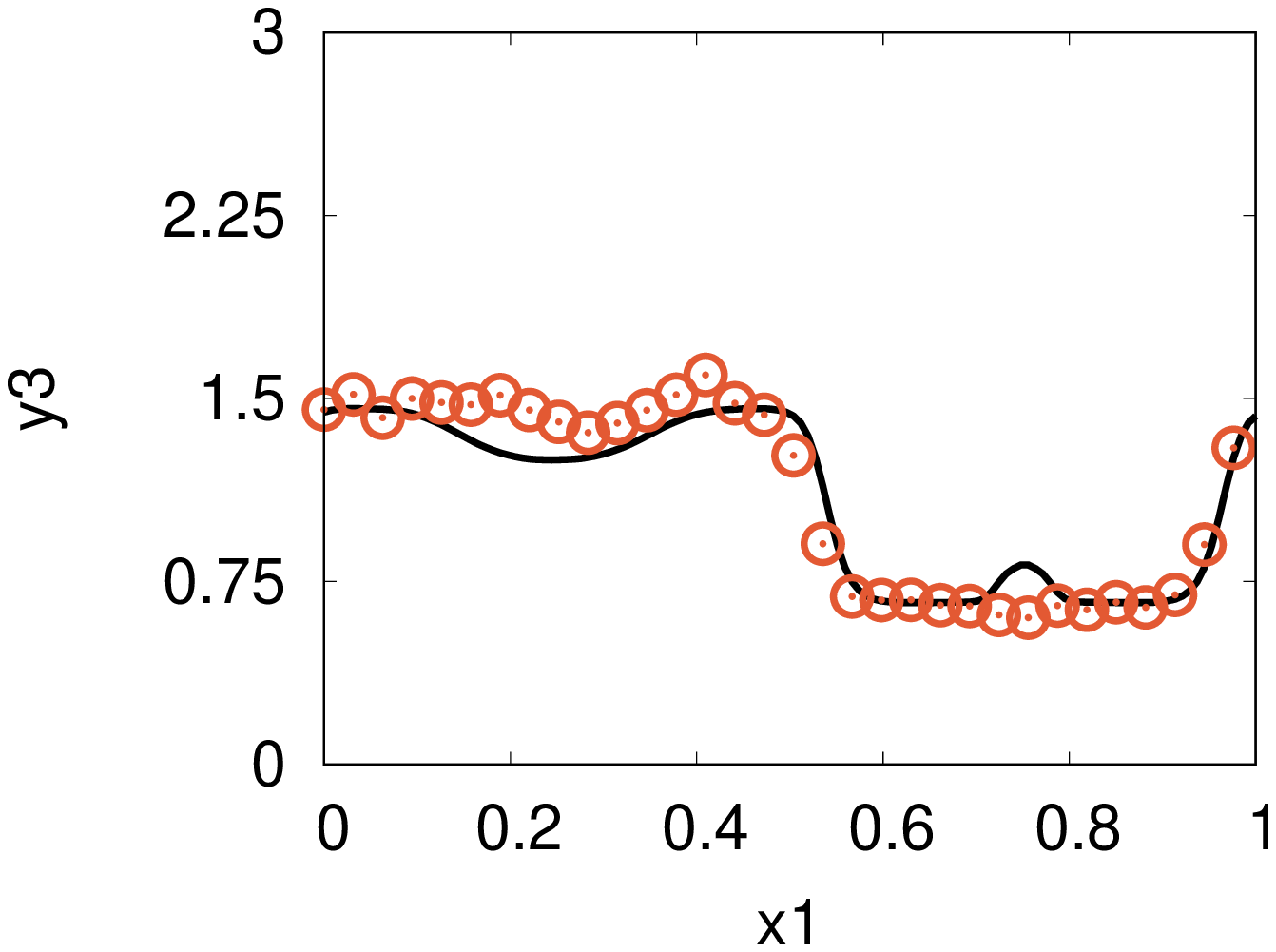} & \psfrag{y4}[][bl][0.9]{$Q_{pi}^*$}
        \hspace{-0.6cm}\includegraphics[scale=0.15,angle=-90]{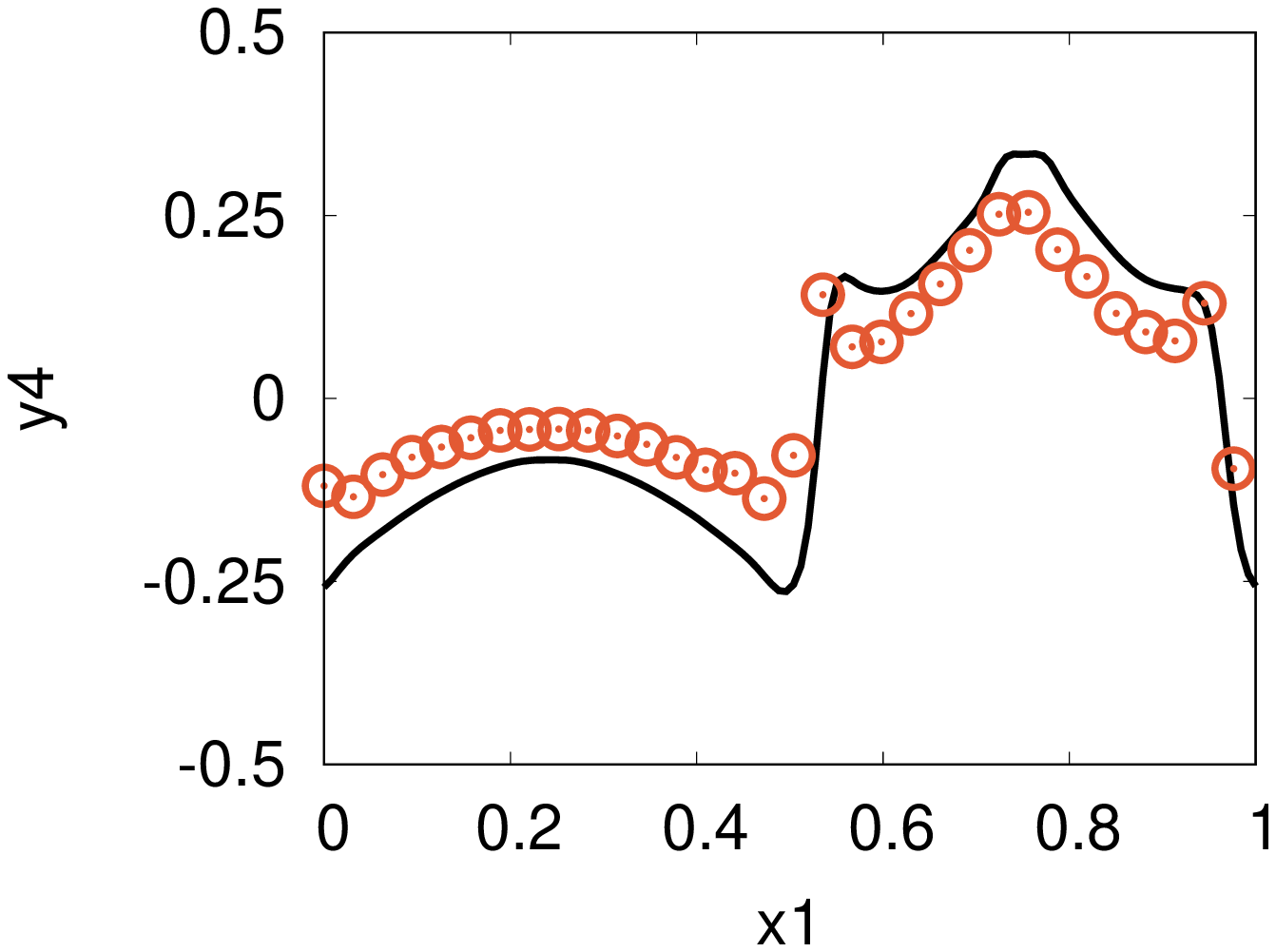} \\
        \psfrag{y1}[][bl][0.9]{$\alpha_{pj}$}
        \hspace{-0.3cm}\includegraphics[scale=0.15,angle=-90]{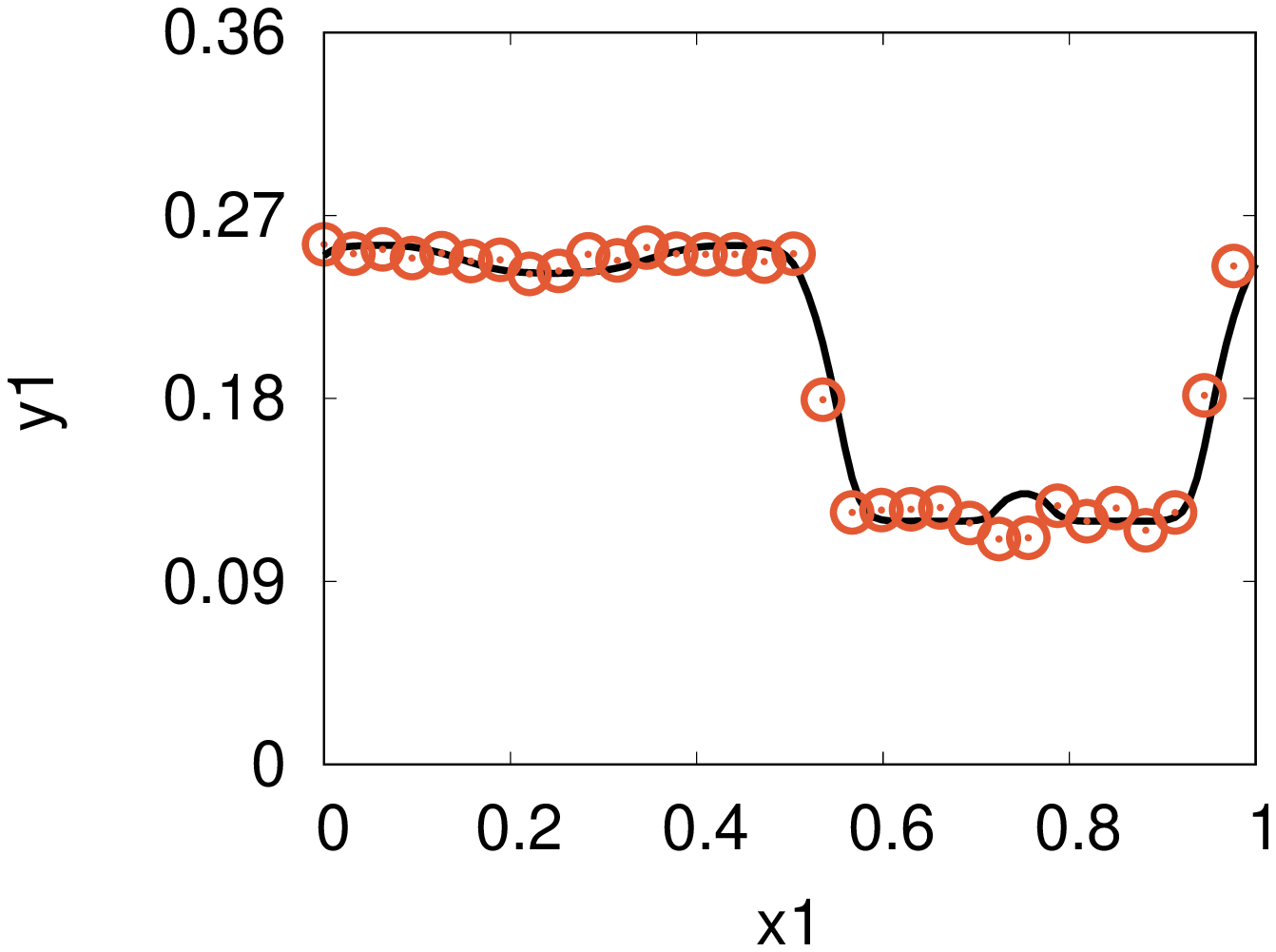} & \psfrag{y2}[][bl][0.9]{$U_{pj}^*$}
        \hspace{-0.6cm}\includegraphics[scale=0.15,angle=-90]{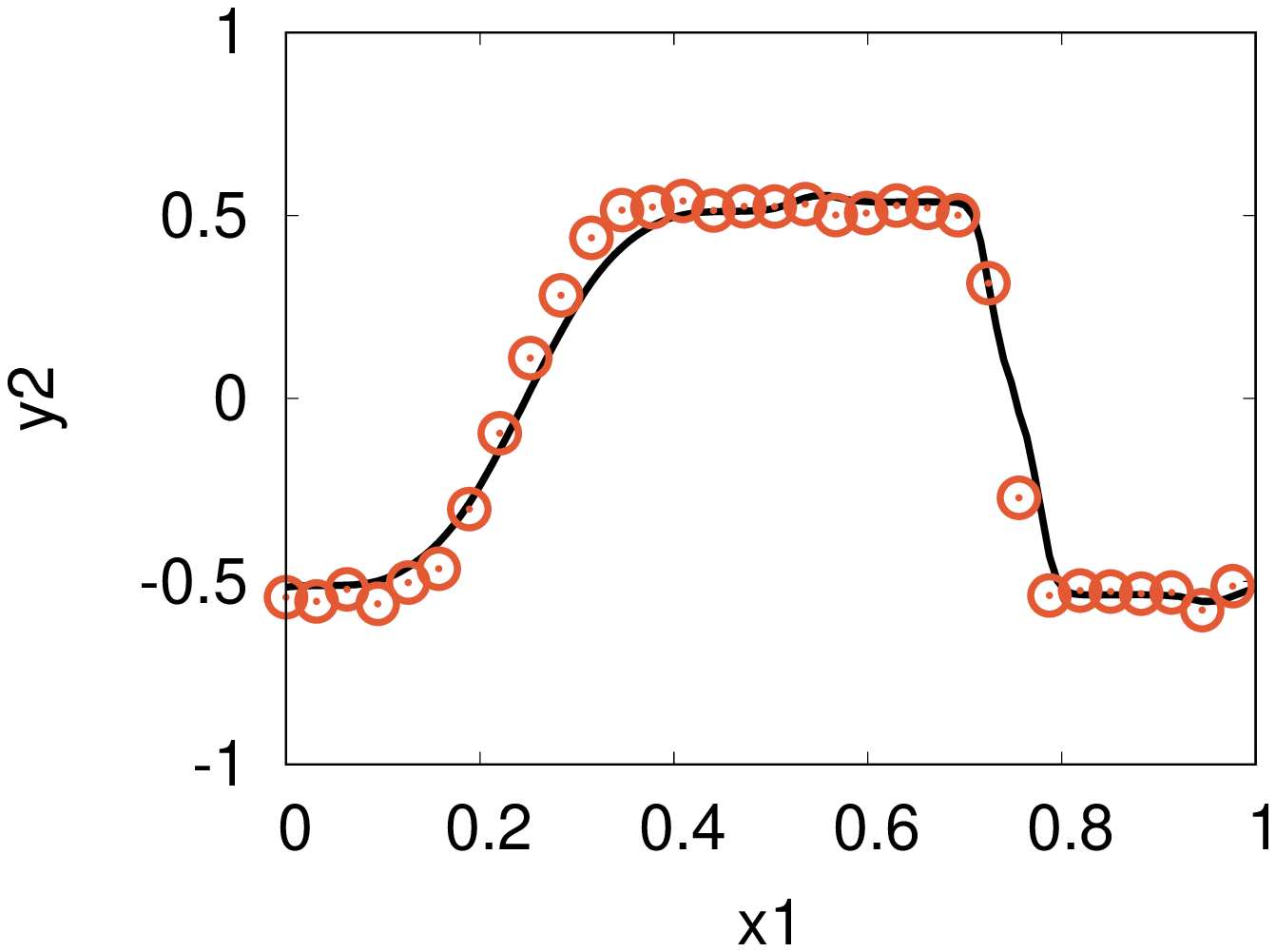} & \psfrag{y3}[][bl][0.9]{$\Tp{j}^*$}
        \hspace{-0.6cm}\includegraphics[scale=0.15,angle=-90]{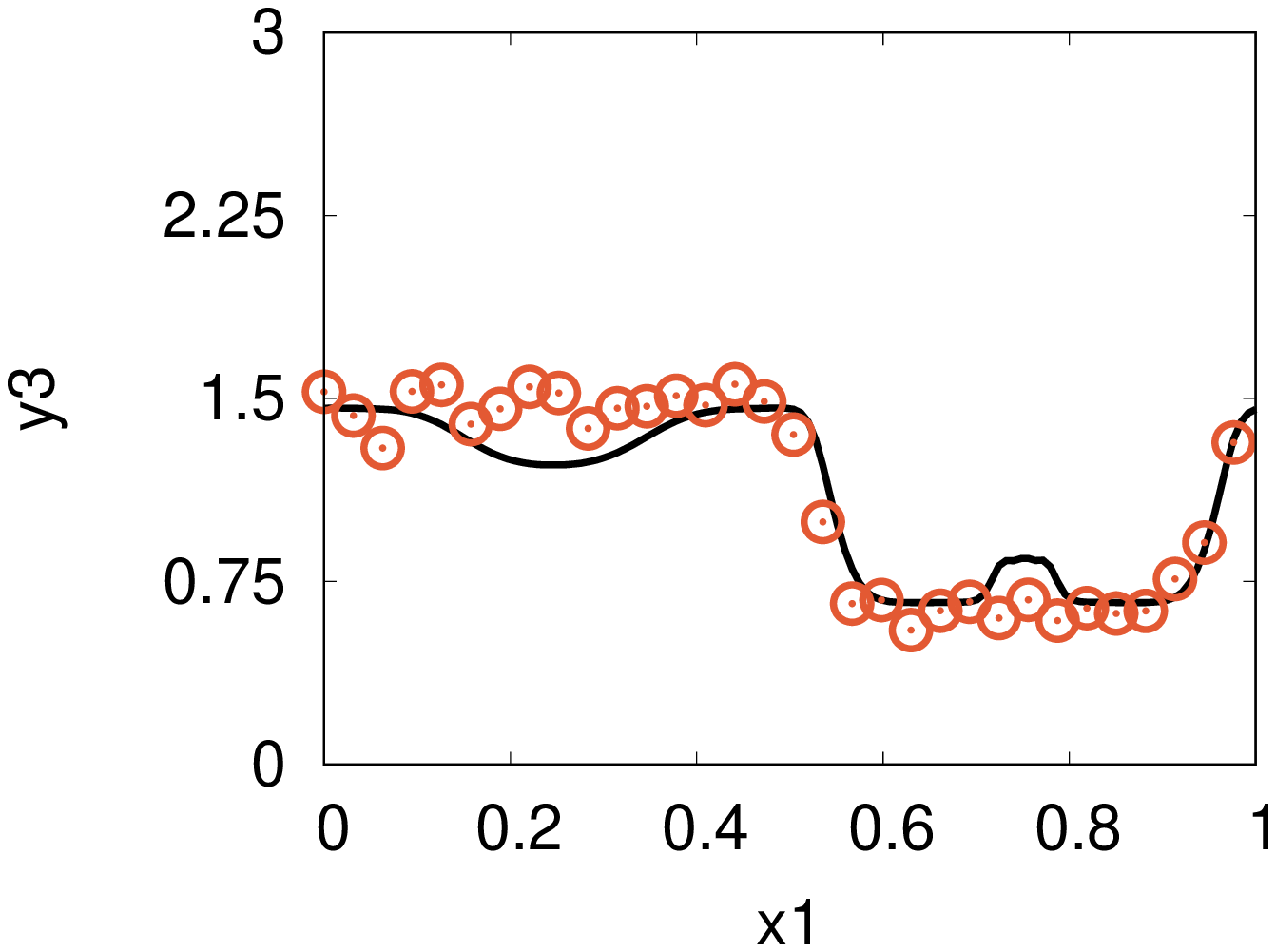} & \psfrag{y4}[][bl][0.9]{$Q_{pj}^*$}
        \hspace{-0.6cm}\includegraphics[scale=0.15,angle=-90]{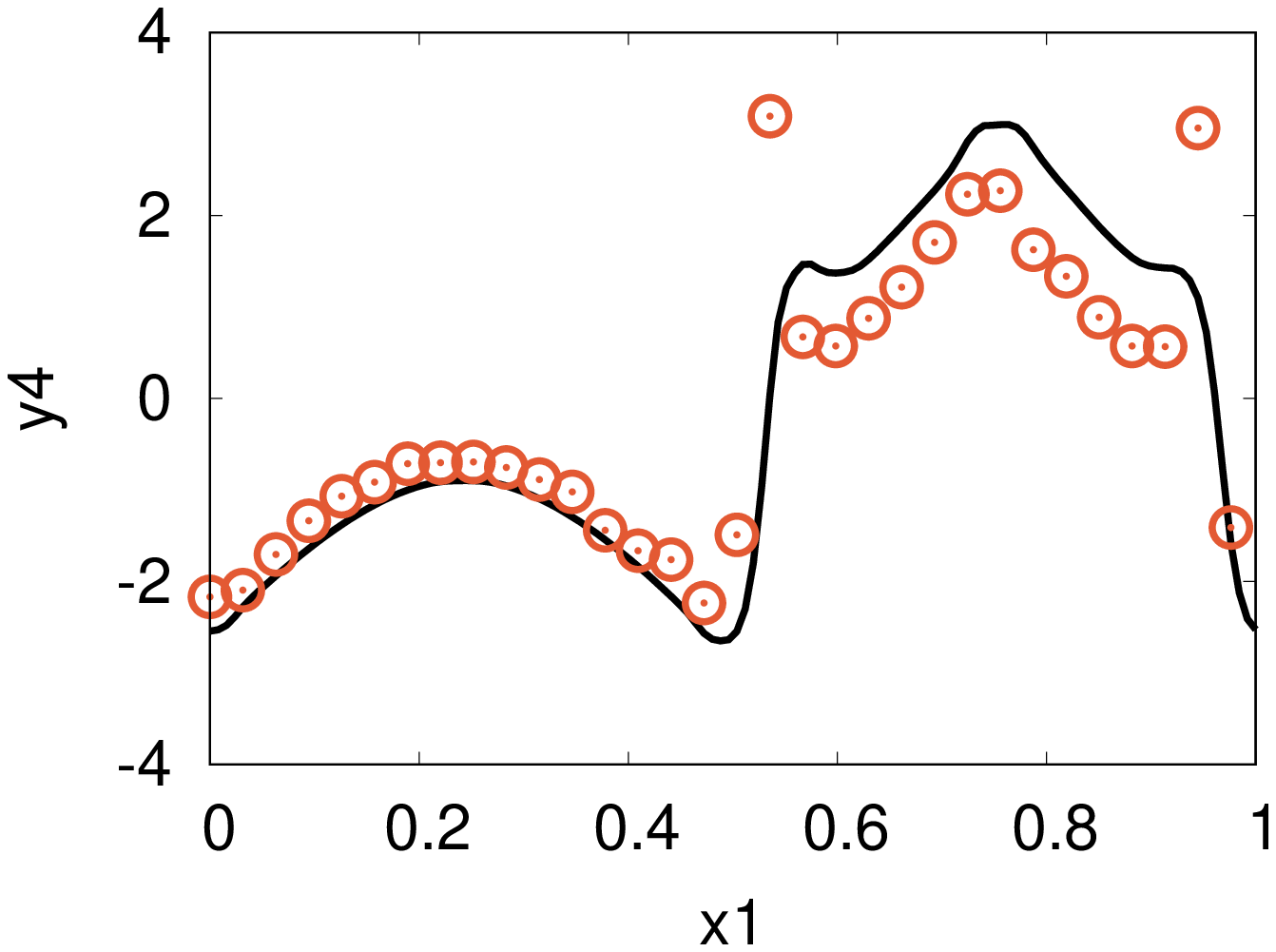} \\
    \end{tabular}
    \caption{Evolution of (a) phase solid volume fraction, (b) phase scaled velocity, (c) phase scaled granular temperature and (d) phase scaled mean charge of the phase $h\,(h=i,j)$ for Case-E at $t^*=50$. ${\color{CadetBlue}\boldsymbol{\odot}}$: initial conditions for the hard sphere simulation. ${\color{Black}\rule[0.5ex]{10pt}{0.75pt}}$: Eulerian model predictions and ${\color{RedOrange}\boldsymbol{\odot}}$: hard sphere simulation results. Variables were scaled by using \eqref{scalingTimeVelGran}, \eqref{scaledtheta} and \eqref{scaledcharge}.
    \label{Fig:ResultRd3Spatialt50}}
\end{figure}
Case-E presents a quasi-1D simulation case of a bidisperse solid mixture with a particle diameter ratio of $R_d=3$. The total solid volume fraction $\langle \alpha_p\rangle$ was equal to 0.237 and initially, only the phase $i$ particles were charged. The simulation results and the model predictions for the phases $i$ and $j$ at $t^* = 25$ and $t^* = 50$ are shown in figures \ref{Fig:ResultRd3Spatialt25} and \ref{Fig:ResultRd3Spatialt50}, respectively. One can see that both solid phases rapidly reach the mixture mean velocity and granular temperature. After phases reach the equilibrium state, the flow is mainly driven by the gradient of solid volume fraction which is a slow process for this specific case (figures \ref{Fig:ResultRd3Spatialt25}(a) and \ref{Fig:ResultRd3Spatialt50}(a)). The charge evolution shows the charge repartition between the solid phases and the phase $j$ picks up a large amount of charge before decreasing towards zero charge in time (figures \ref{Fig:ResultRd3Spatialt25}(d) and \ref{Fig:ResultRd3Spatialt50}(d)). This second stage is slower as the granular temperature of each phase has reached the mixture value; the contribution from the electric field in \eqref{Eq:thetaFluxq} tends to zero and the exchange of charge is driven mainly by the difference of charge between phase. To conclude, the hard-sphere simulation results are very well predicted by the Eulerian model. 

\begin{figure}
    \centering
    \psfrag{x1}[][][0.8]{$x/L$}
    \begin{tabular}{cccc}
        (a) & (b) & (c) & (d) \vspace{-0.25cm}\\
        \psfrag{y1}[][bl][0.9]{$\alpha_{pi}$}
        \hspace{-0.3cm}\includegraphics[scale=0.15,angle=-90]{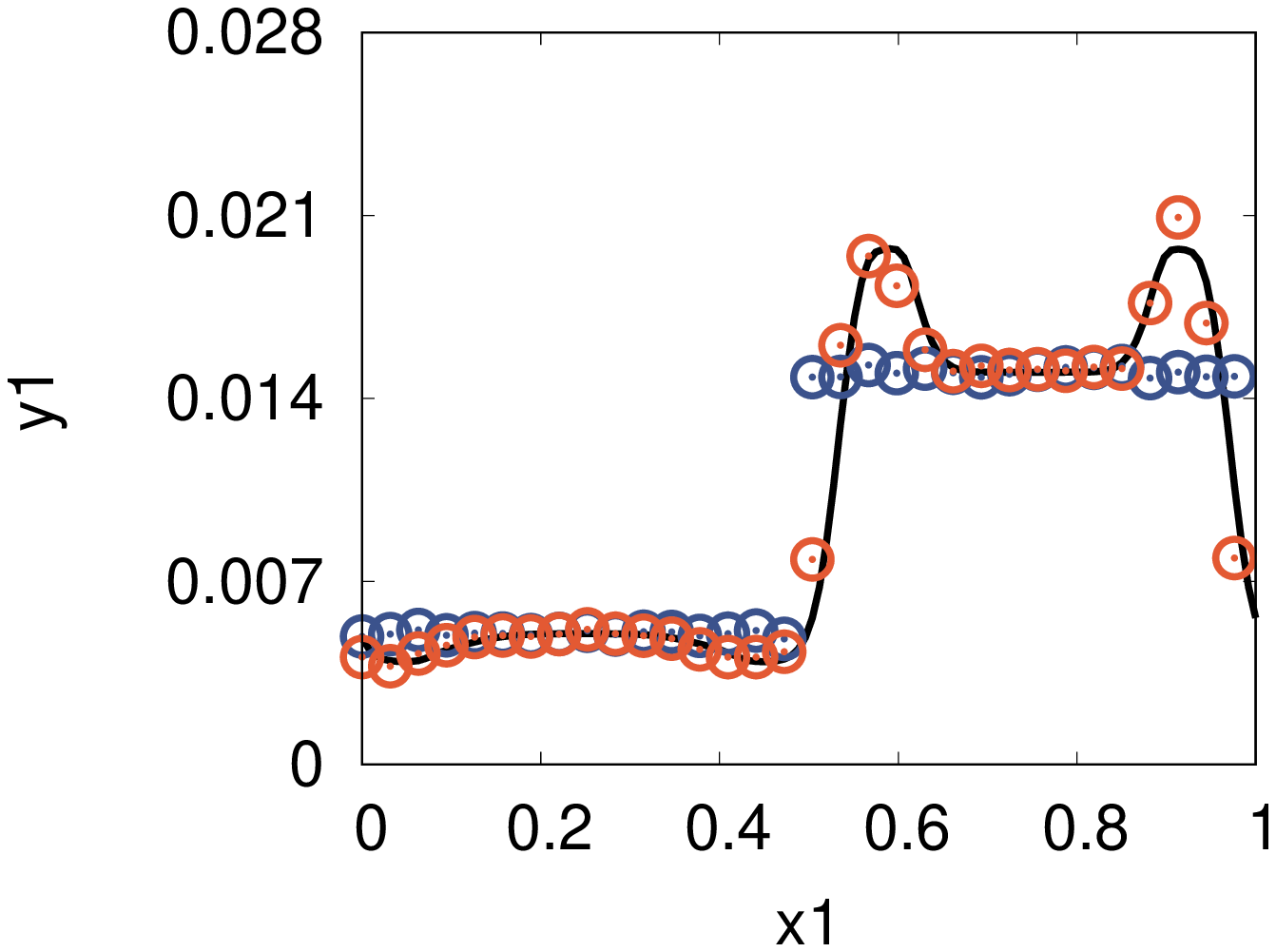} & \psfrag{y2}[][bl][0.9]{$U_{pi}^*$}
        \hspace{-0.6cm}\includegraphics[scale=0.15,angle=-90]{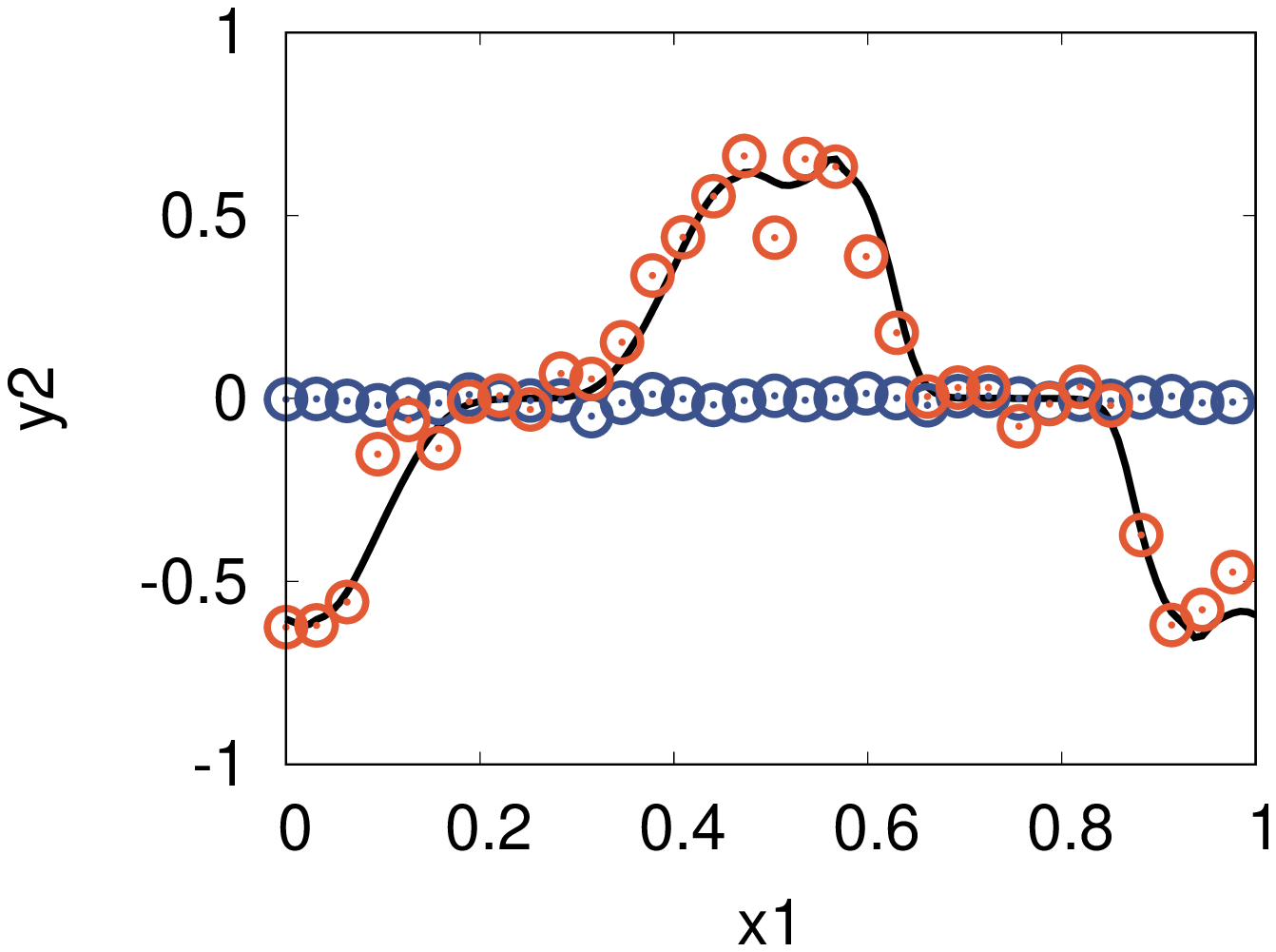} & \psfrag{y3}[][bl][0.9]{$\Tp{i}^*$}
        \hspace{-0.6cm}\includegraphics[scale=0.15,angle=-90]{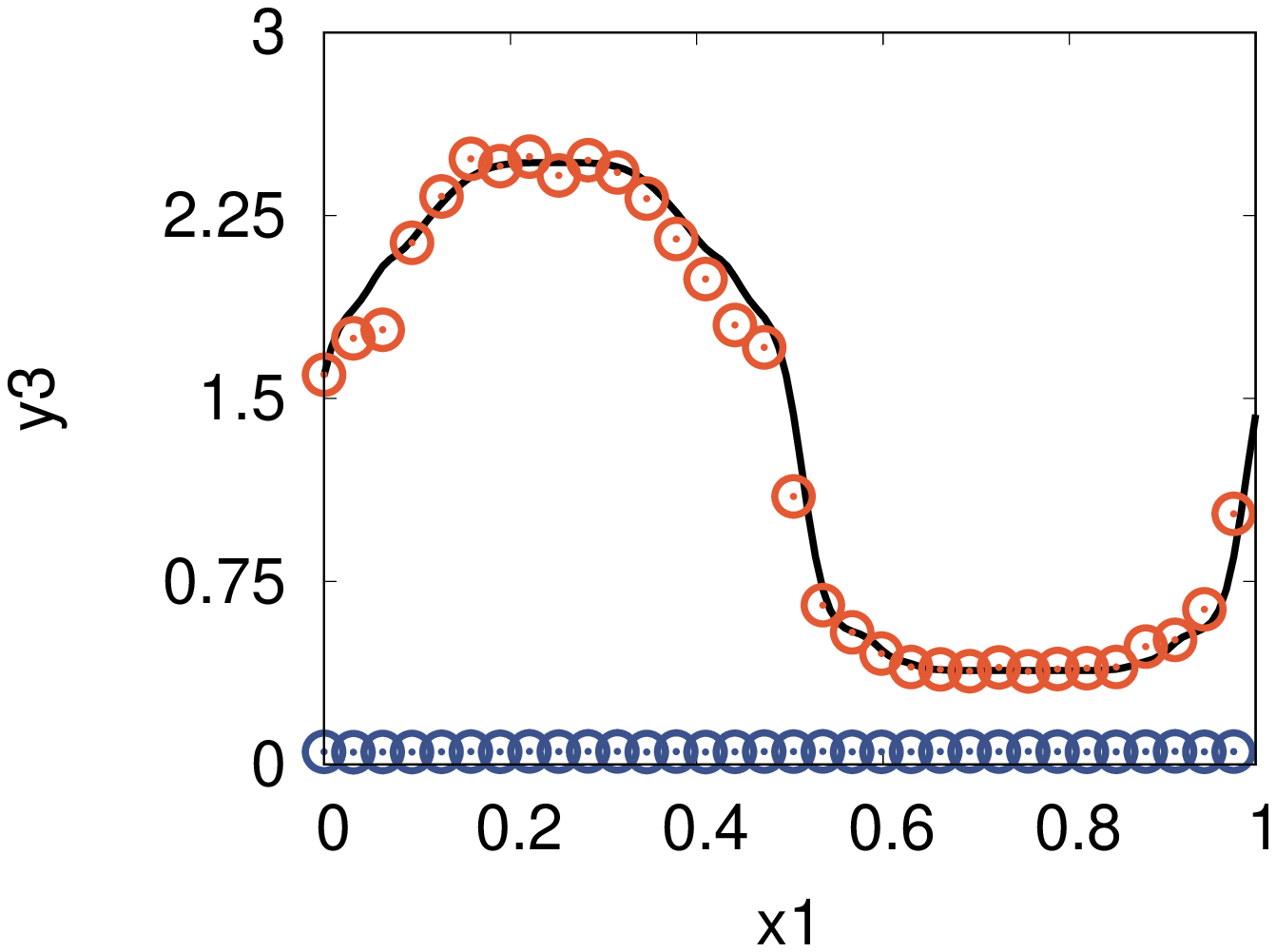} & \psfrag{y4}[][bl][0.9]{$Q_{pi}^*$}
        \hspace{-0.6cm}\includegraphics[scale=0.15,angle=-90]{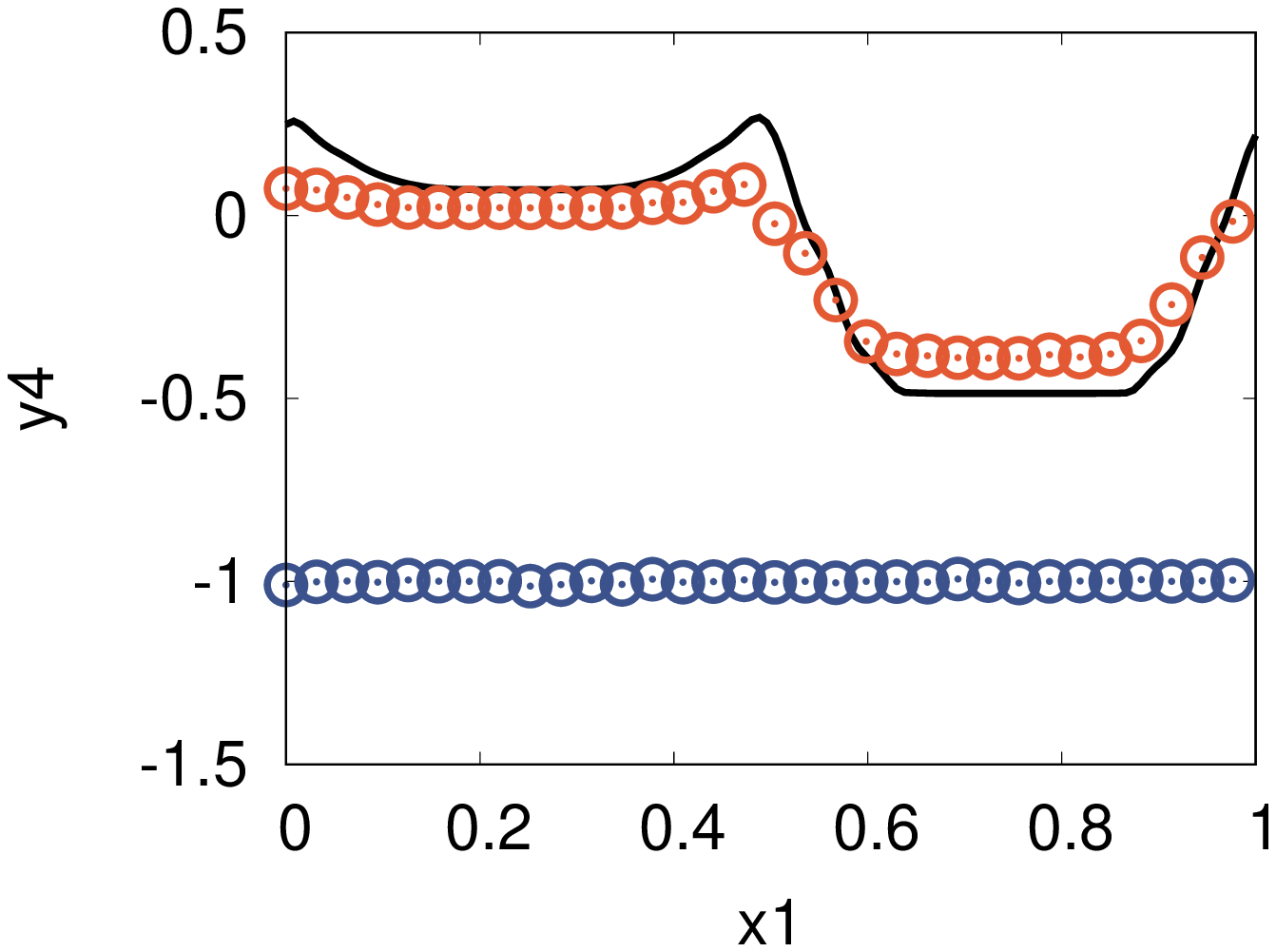} \\
        \psfrag{y1}[][bl][0.9]{$\alpha_{pj}$}
        \hspace{-0.3cm}\includegraphics[scale=0.15,angle=-90]{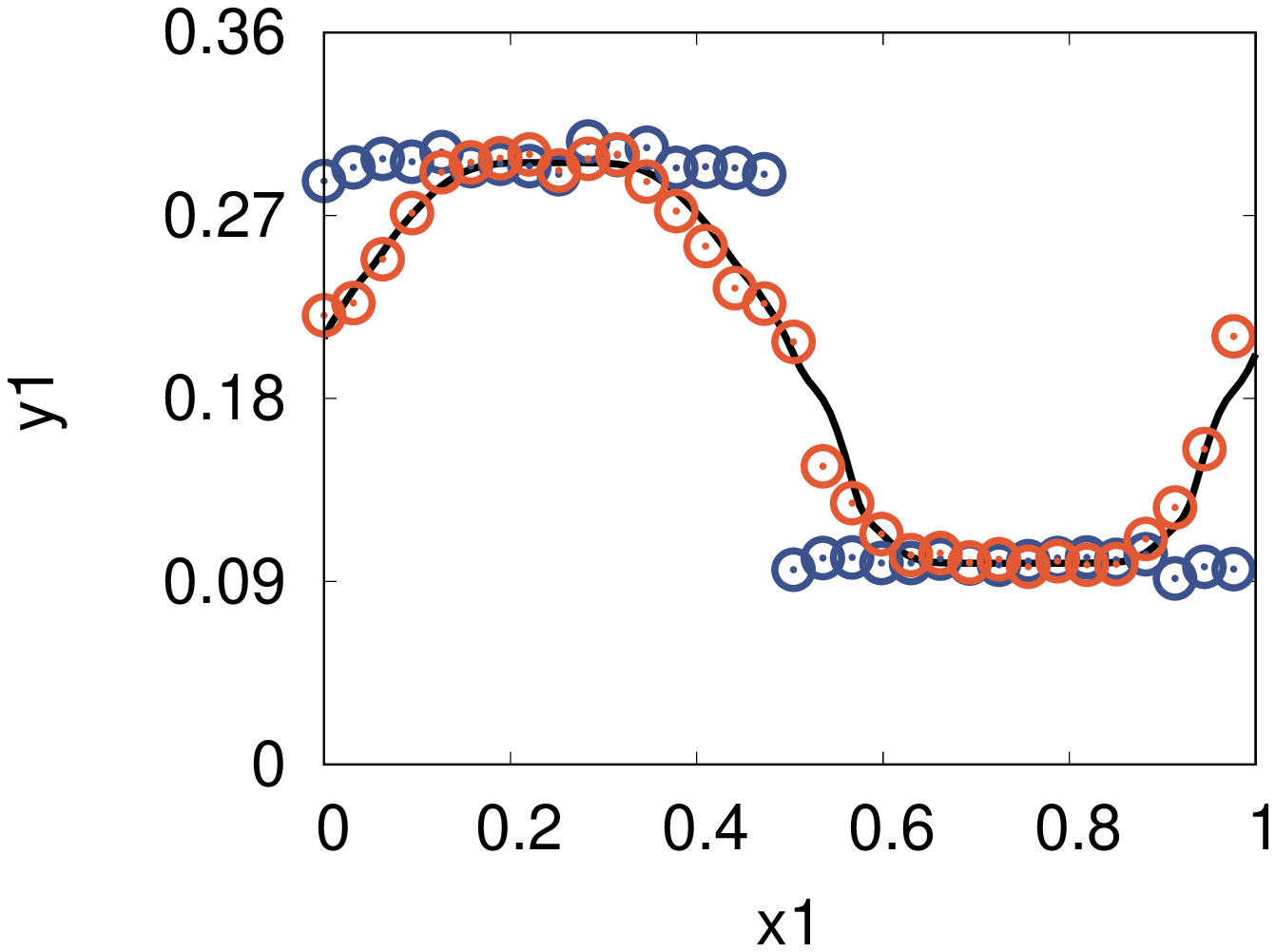} & \psfrag{y2}[][bl][0.9]{$U_{pj}^*$}
        \hspace{-0.6cm}\includegraphics[scale=0.15,angle=-90]{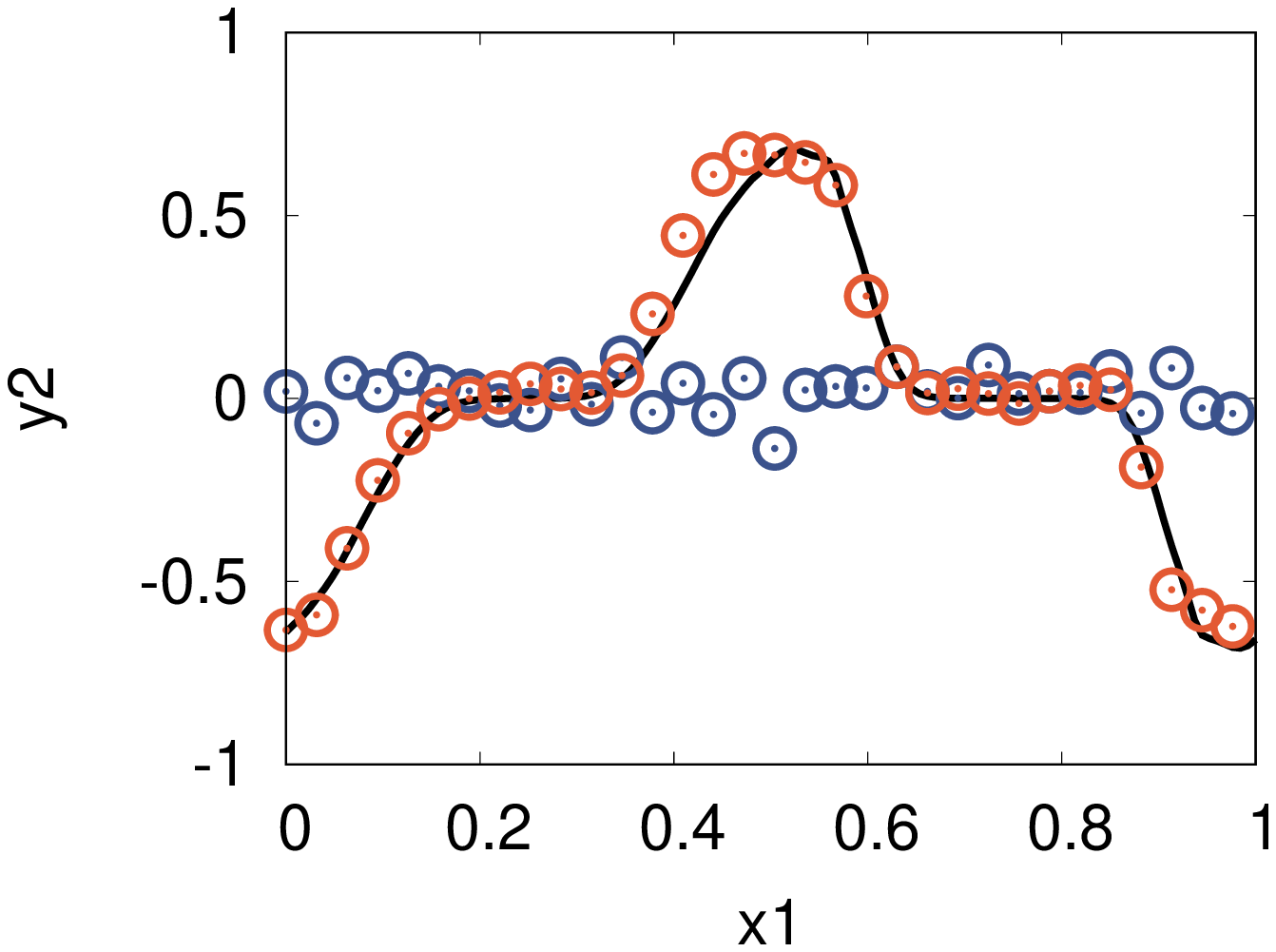} & \psfrag{y3}[][bl][0.9]{$\Tp{j}^*$}
        \hspace{-0.6cm}\includegraphics[scale=0.15,angle=-90]{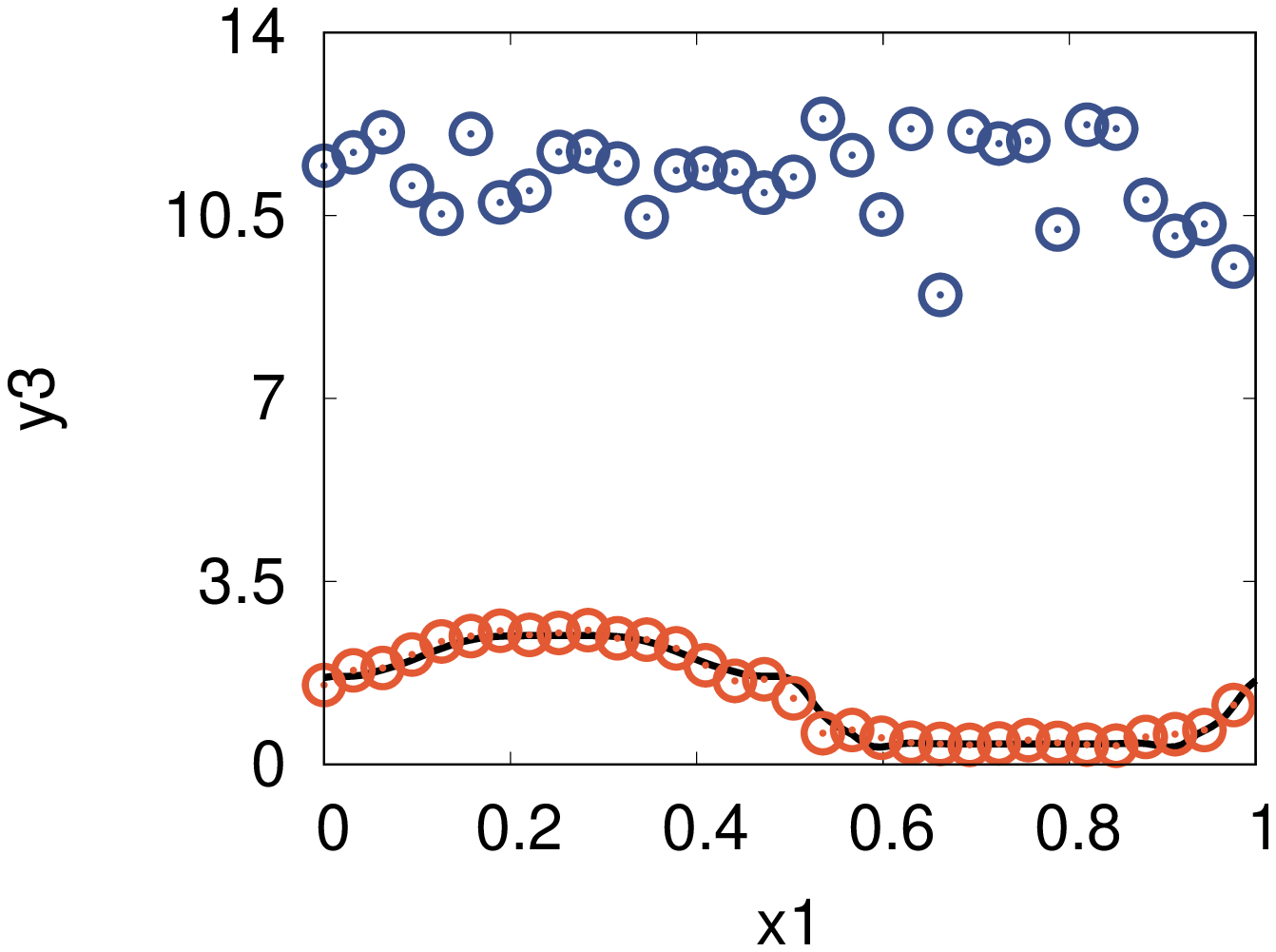} & \psfrag{y4}[][bl][0.9]{$Q_{pj}^*$}
        \hspace{-0.6cm}\includegraphics[scale=0.15,angle=-90]{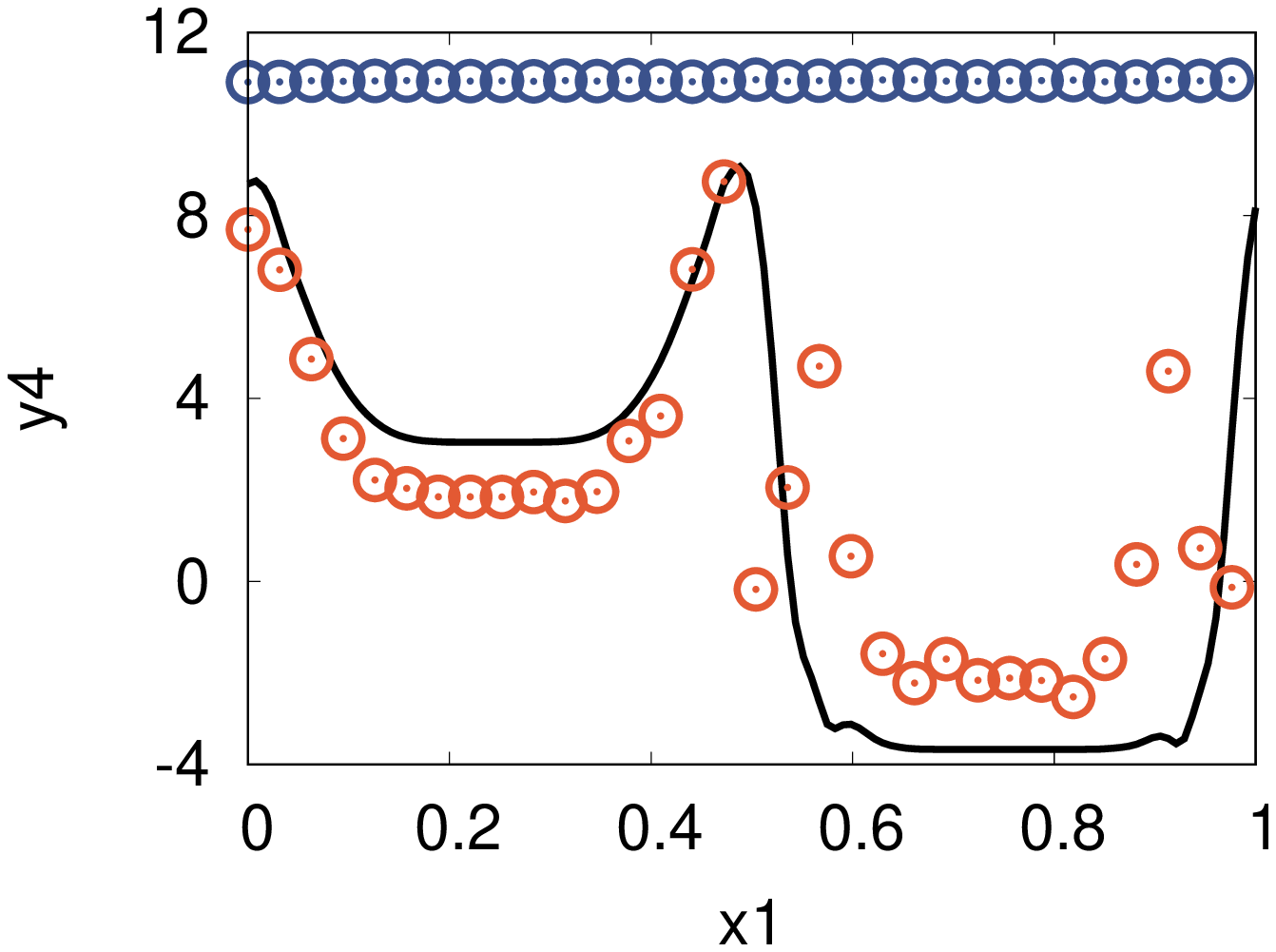} \\
    \end{tabular}
    \caption{Evolution of (a) phase solid volume fraction, (b) phase scaled velocity, (c) phase scaled granular temperature and (d) phase scaled mean charge of the phase $h\,(h=i,j)$ for Case-F at $t^*=28.5$. ${\color{CadetBlue}\boldsymbol{\odot}}$: initial conditions for the hard sphere simulation. ${\color{Black}\rule[0.5ex]{10pt}{0.75pt}}$: Eulerian model predictions and ${\color{RedOrange}\boldsymbol{\odot}}$: hard sphere simulation results. Variables were scaled by using \eqref{scalingTimeVelGran}, \eqref{scaledtheta} and \eqref{scaledcharge}.\label{Fig:ResultRd6Spatialt25}}
    \centering
    \psfrag{x1}[][][0.8]{$x/L$}
    \begin{tabular}{cccc}
        (a) & (b) & (c) & (d) \vspace{-0.25cm}\\
        \psfrag{y1}[][bl][0.9]{$\alpha_{pi}$}
        \hspace{-0.3cm}\includegraphics[scale=0.15,angle=-90]{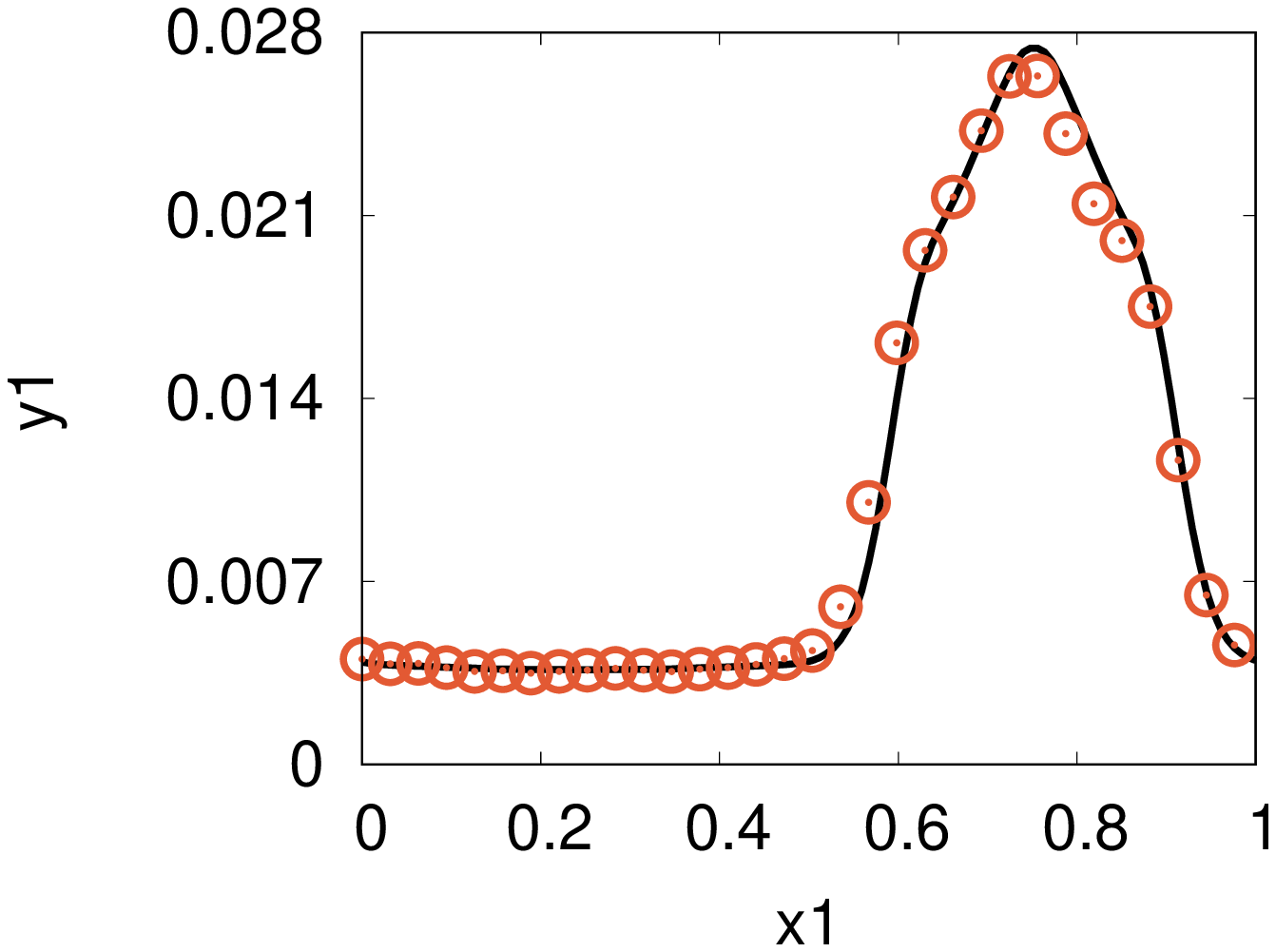} & \psfrag{y2}[][bl][0.9]{$U_{pi}^*$}
        \hspace{-0.6cm}\includegraphics[scale=0.15,angle=-90]{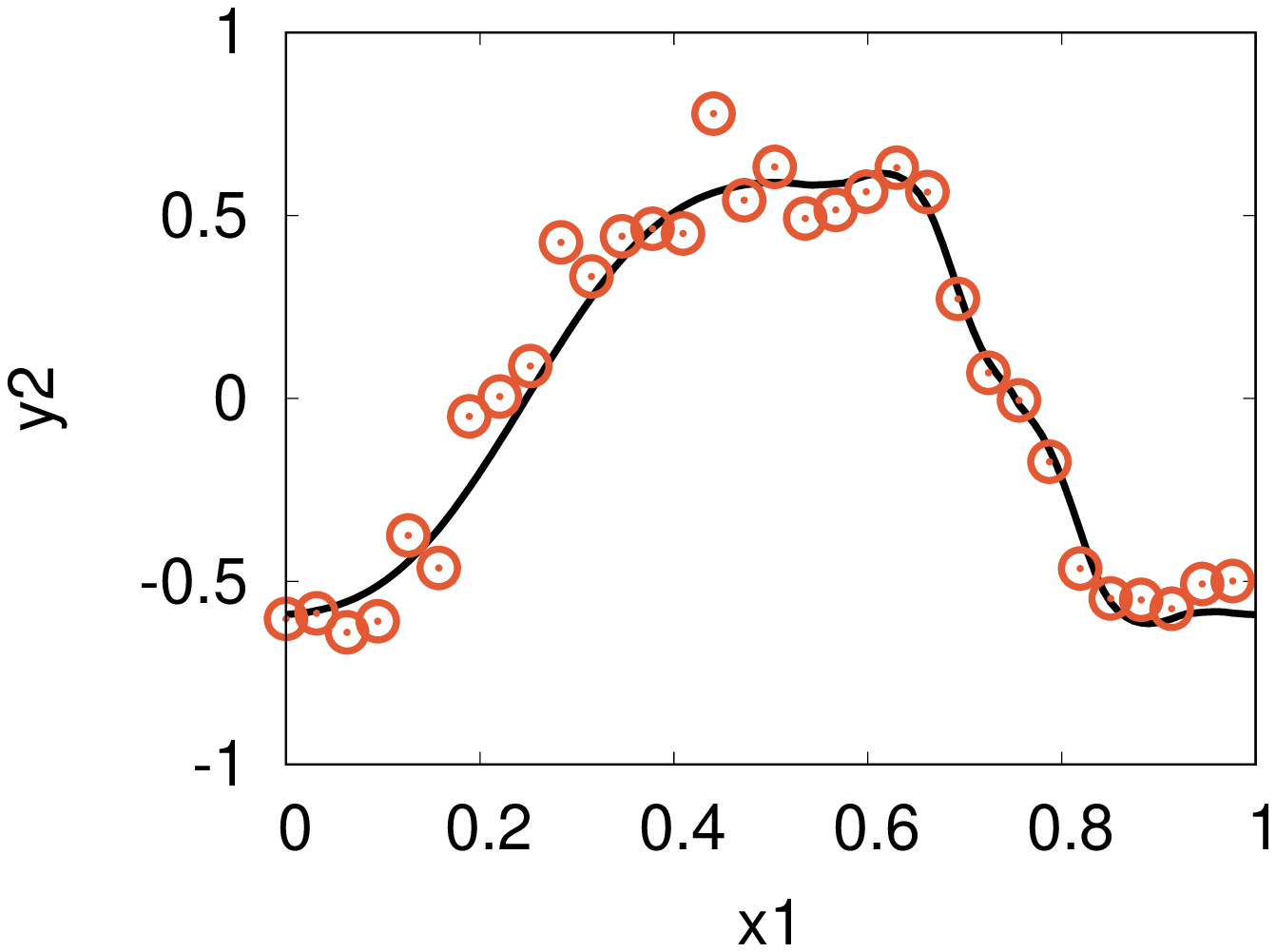} & \psfrag{y3}[][bl][0.9]{$\Tp{i}^*$}
        \hspace{-0.6cm}\includegraphics[scale=0.15,angle=-90]{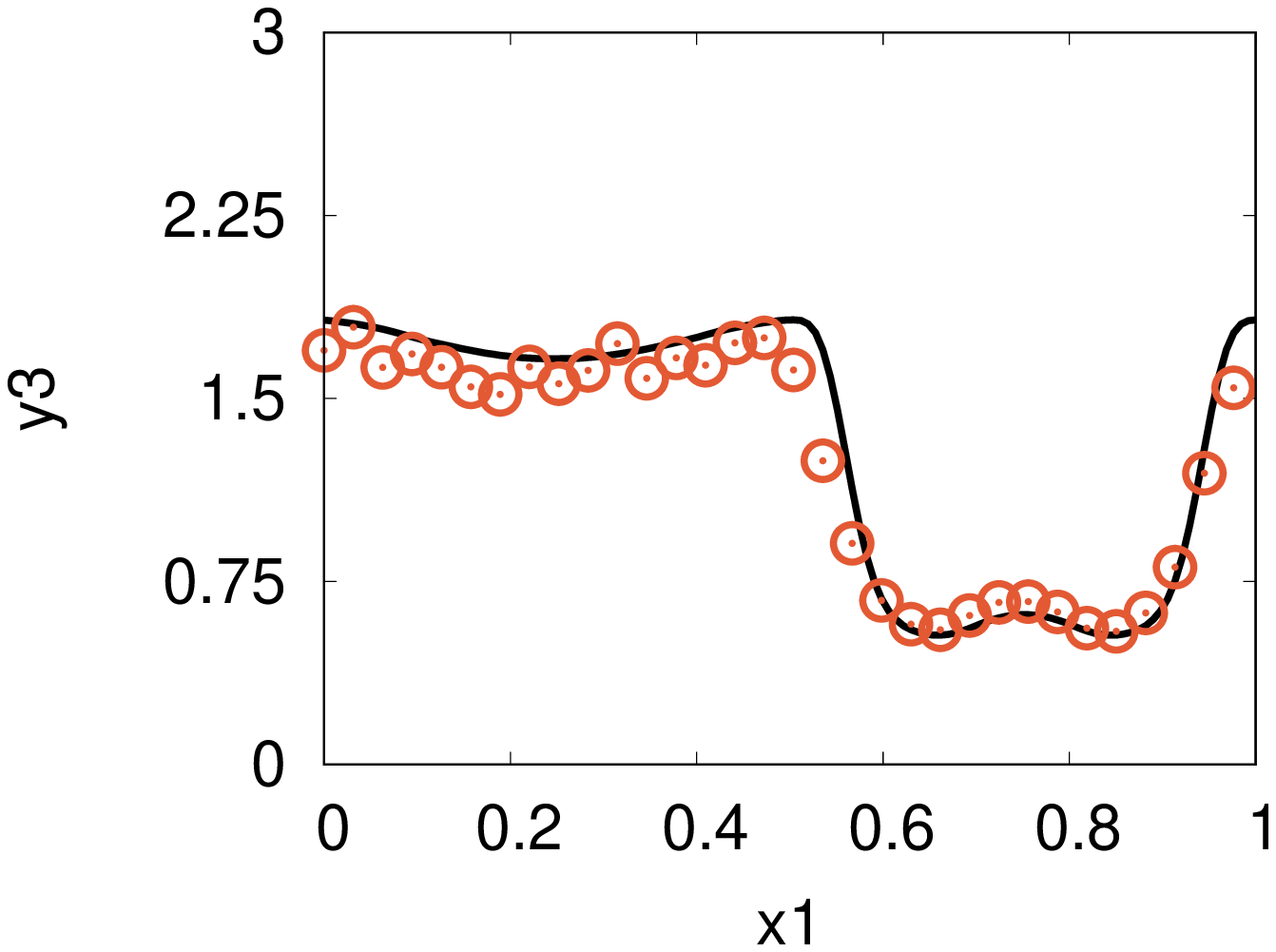} & \psfrag{y4}[][bl][0.9]{$Q_{pi}^*$}
        \hspace{-0.6cm}\includegraphics[scale=0.15,angle=-90]{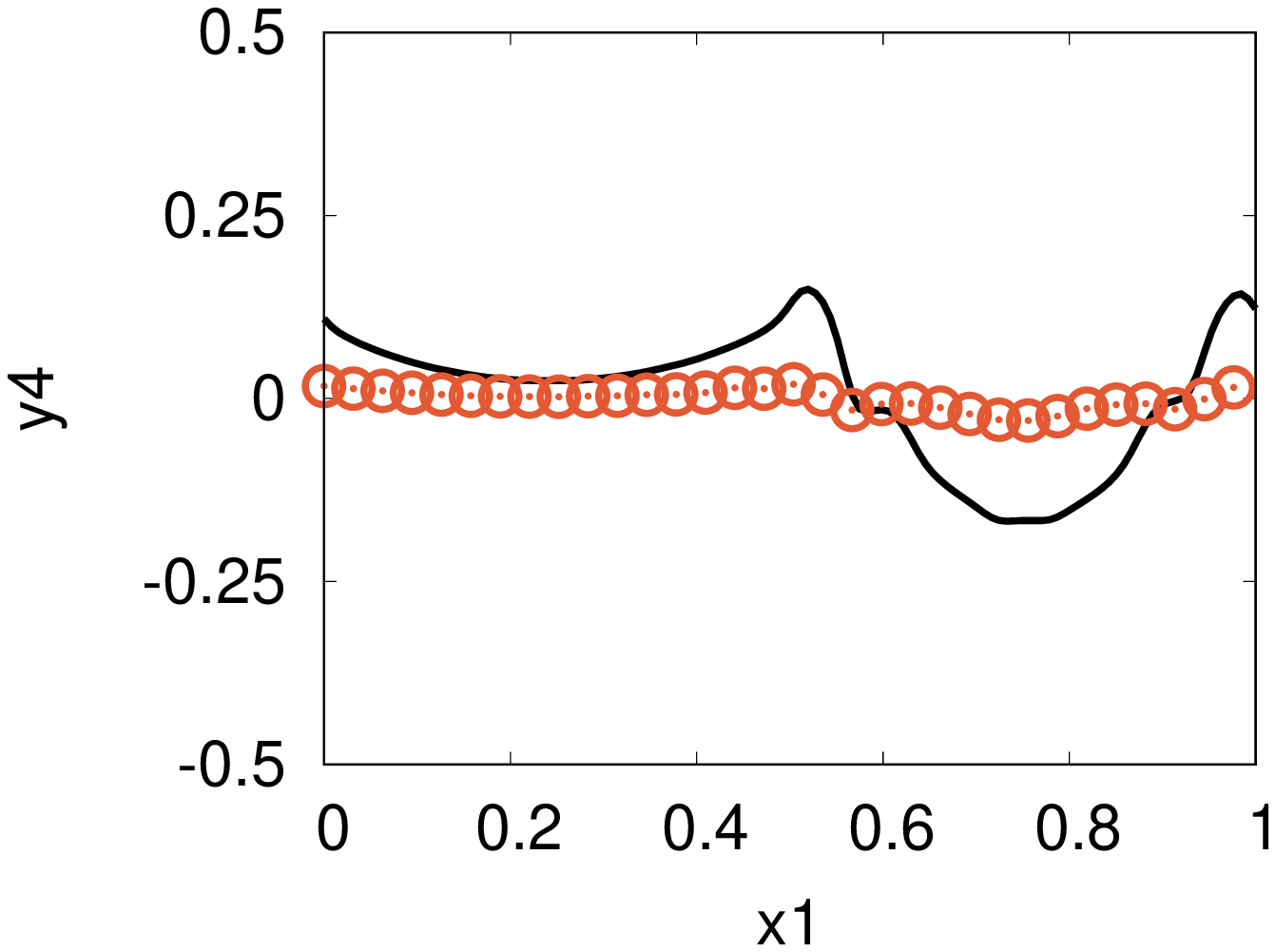} \\
        \psfrag{y1}[][bl][0.9]{$\alpha_{pj}$}
        \hspace{-0.3cm}\includegraphics[scale=0.15,angle=-90]{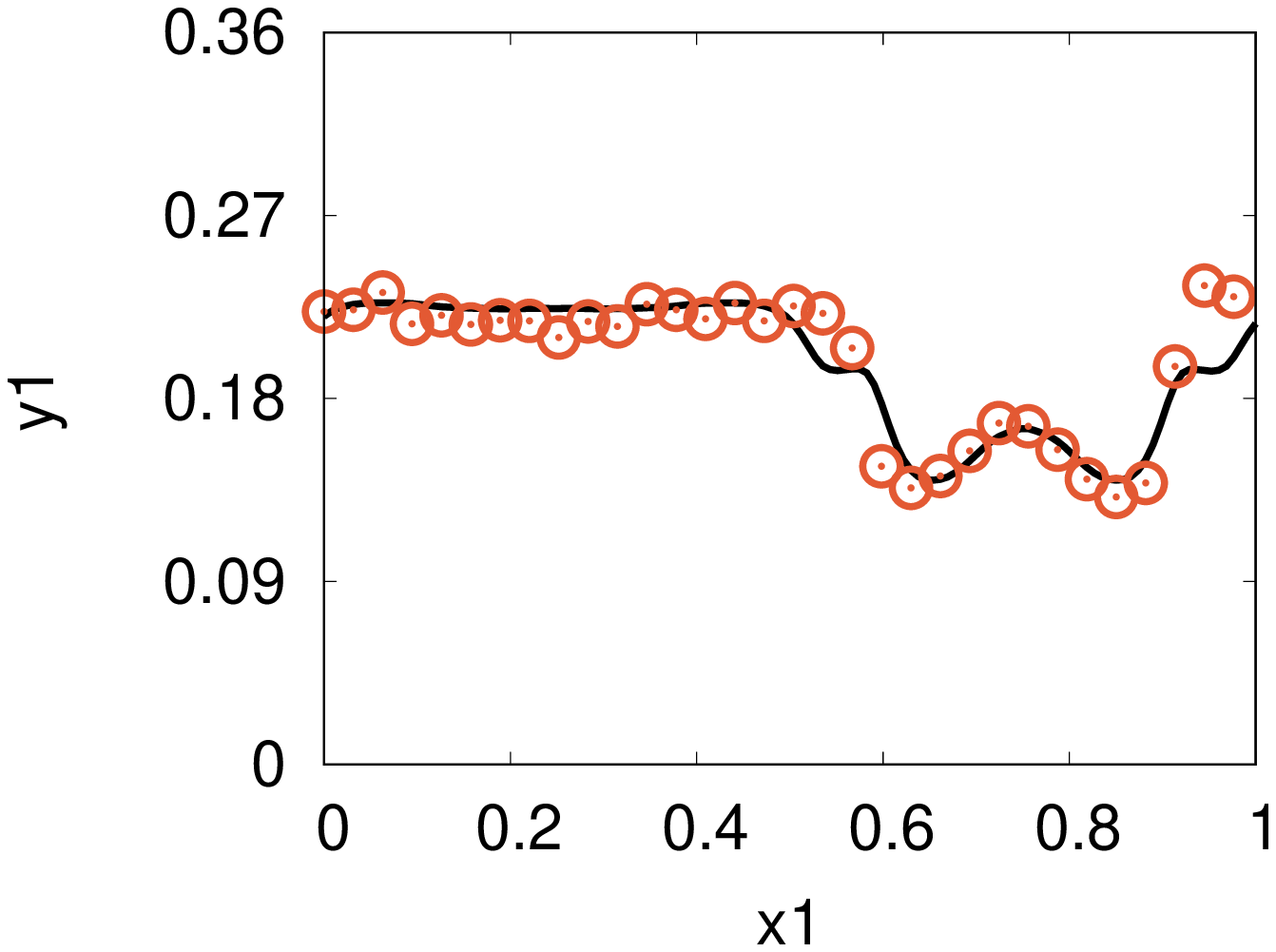} & \psfrag{y2}[][bl][0.9]{$U_{pj}^*$}
        \hspace{-0.6cm}\includegraphics[scale=0.15,angle=-90]{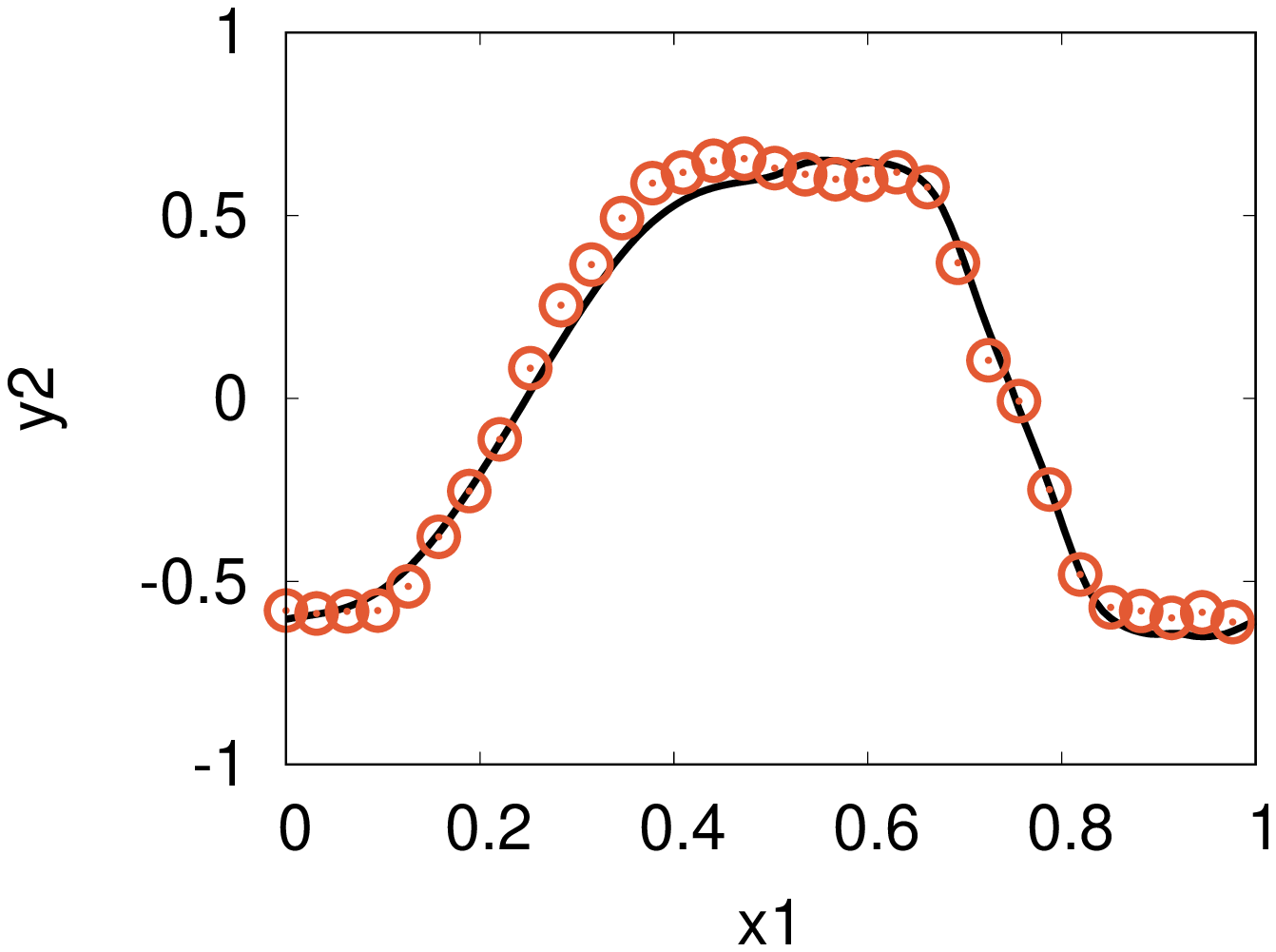} & \psfrag{y3}[][bl][0.9]{$\Tp{j}^*$}
        \hspace{-0.6cm}\includegraphics[scale=0.15,angle=-90]{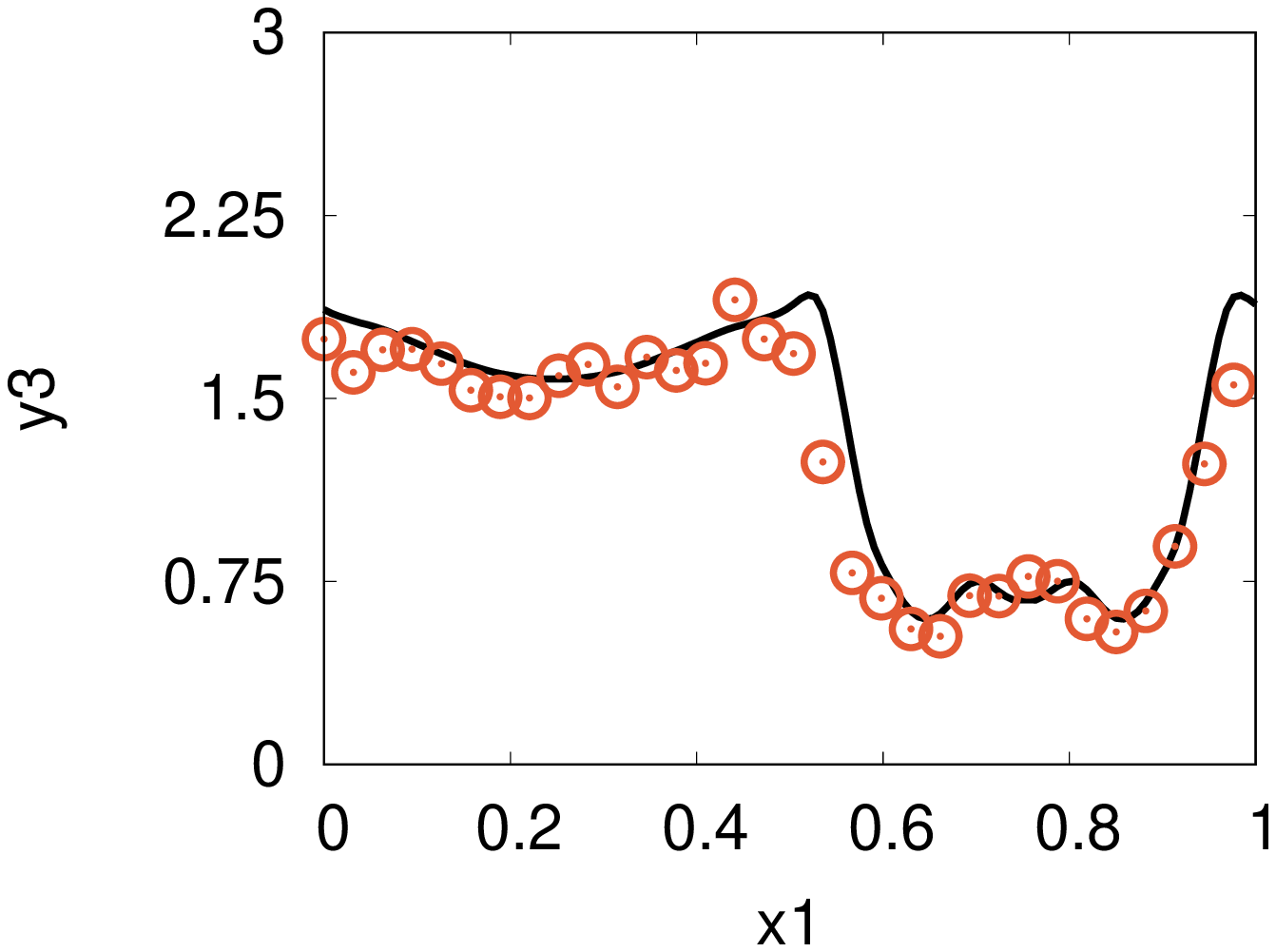} & \psfrag{y4}[][bl][0.9]{$Q_{pj}^*$}
        \hspace{-0.6cm}\includegraphics[scale=0.15,angle=-90]{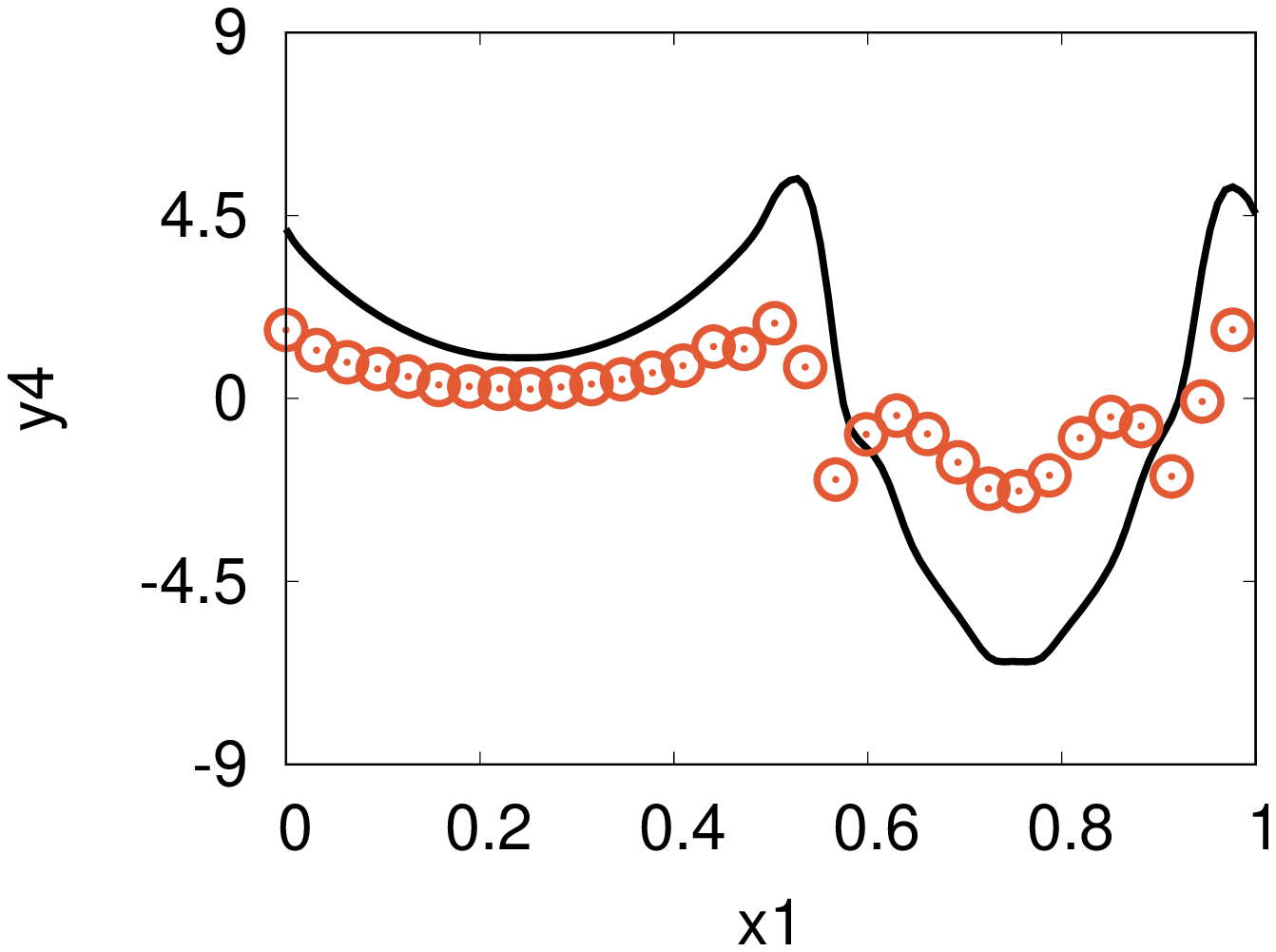} \\
    \end{tabular}
    \caption{Evolution of (a) phase solid volume fraction, (b) phase scaled velocity, (c) phase scaled granular temperature and (d) phase scaled mean charge of the phase $h\,(h=i,j)$ for Case-F at $t^*=85.5$. ${\color{CadetBlue}\boldsymbol{\odot}}$: initial conditions for the hard sphere simulation. ${\color{Black}\rule[0.5ex]{10pt}{0.75pt}}$: Eulerian model predictions and ${\color{RedOrange}\boldsymbol{\odot}}$: hard sphere simulation results. Variables were scaled by using \eqref{scalingTimeVelGran}, \eqref{scaledtheta} and \eqref{scaledcharge}.
    \label{Fig:ResultRd6Spatialt50}}
\end{figure}
We compared the simulation results and the model predictions for a larger particle diameter ratio and the charged particles for both phases in Case-F. The particle diameter ratio, $R_d$, was set to 6 and the average solid fraction was slightly less than of Case-E. The simulation results and the model predictions for the phases $i$ and $j$ at $t^{*}=28.5$ and $t^*=85.5$ are shown in figures \ref{Fig:ResultRd6Spatialt25} and \ref{Fig:ResultRd6Spatialt50}, respectively. Similar to Case-E, the phase granular temperatures and the phase velocities quickly saturate to the mixture values. The predictions of hydrodynamic variables evolution are in very good agreement with the simulation results. However, there is a discrepancy between the results for charge evolution and particularly, the phase charges are overestimated at $t^*=85.5$ by the model predictions. This difference might be explained by the self-diffusion of charge with velocity-charge correlation \citep{montilla2020modelling} in the dilute regime which is not modelled for this case.

\begin{figure}
    \centering
    \psfrag{x1}[][][0.8]{$x/L$}
    \begin{tabular}{cccc}
        (a) & (b) & (c) & (d) \vspace{-0.25cm}\\
        \psfrag{y1}[][bl][0.9]{$\alpha_{pi}$}
        \hspace{-0.3cm}\includegraphics[scale=0.15,angle=-90]{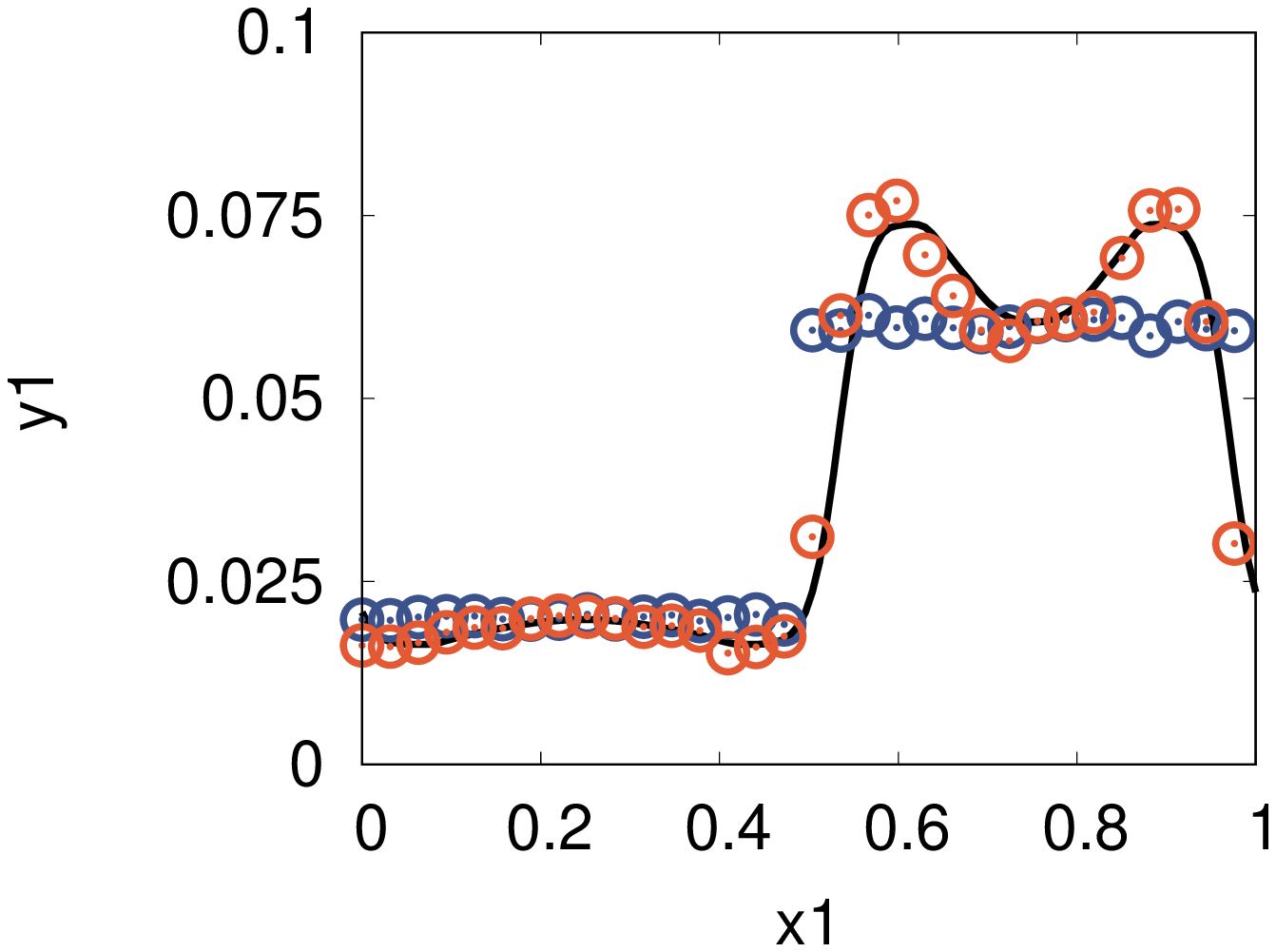} & \psfrag{y2}[][bl][0.9]{$U_{pi}^*$}
        \hspace{-0.6cm}\includegraphics[scale=0.15,angle=-90]{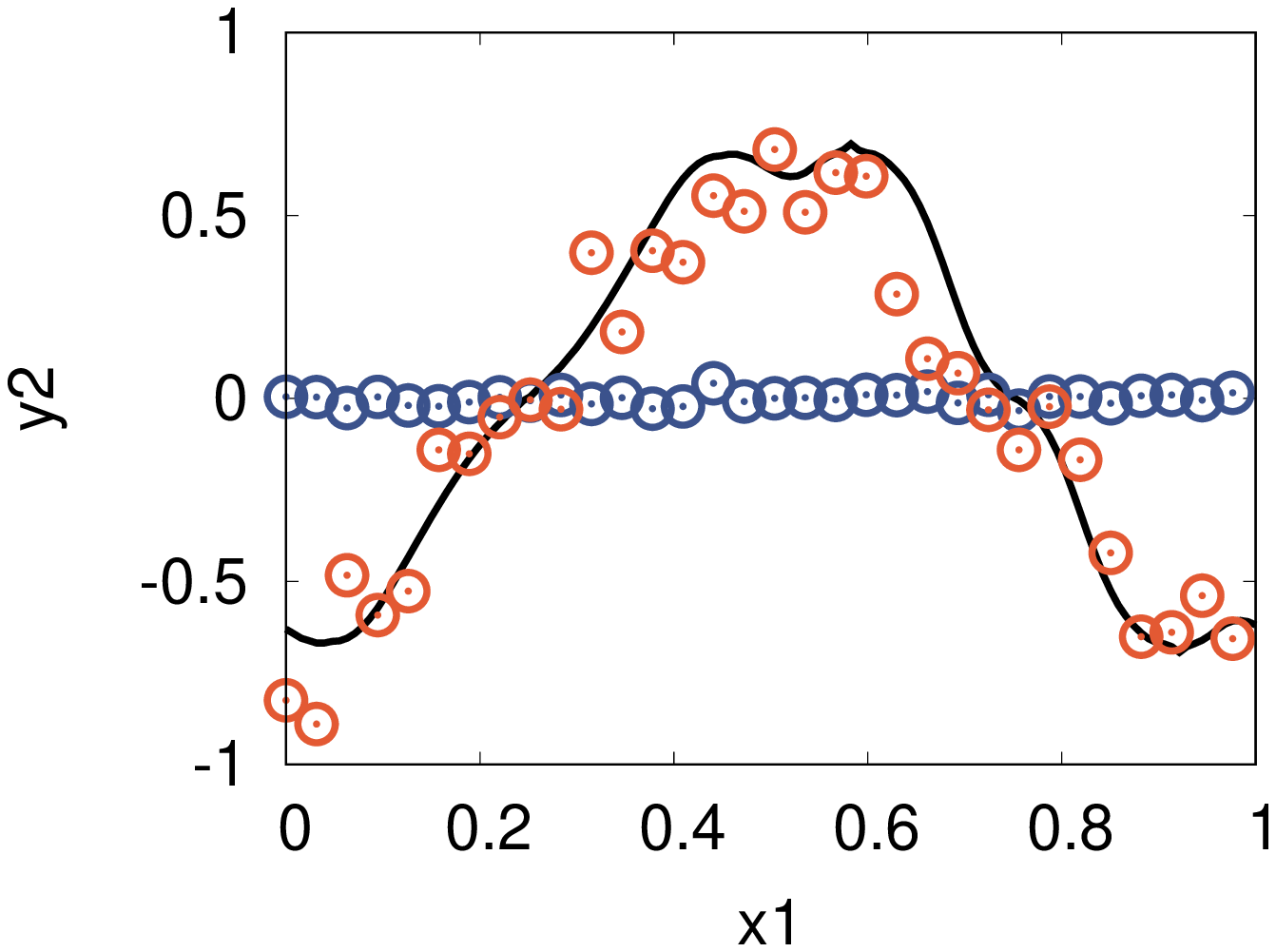} & \psfrag{y3}[][bl][0.9]{$\Tp{i}^*$}
        \hspace{-0.6cm}\includegraphics[scale=0.15,angle=-90]{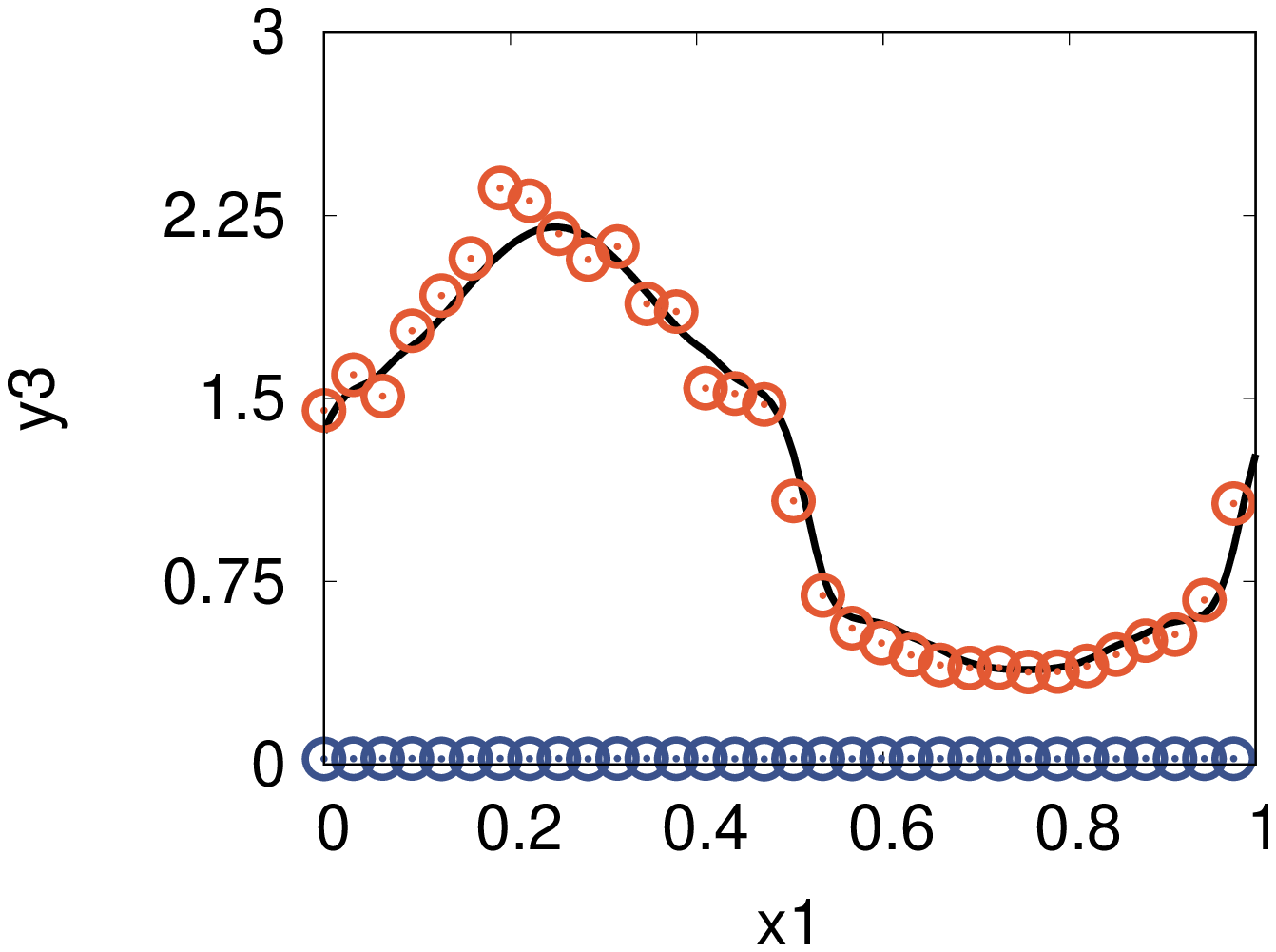} & \psfrag{y4}[][bl][0.9]{$Q_{pi}^*$}
        \hspace{-0.6cm}\includegraphics[scale=0.15,angle=-90]{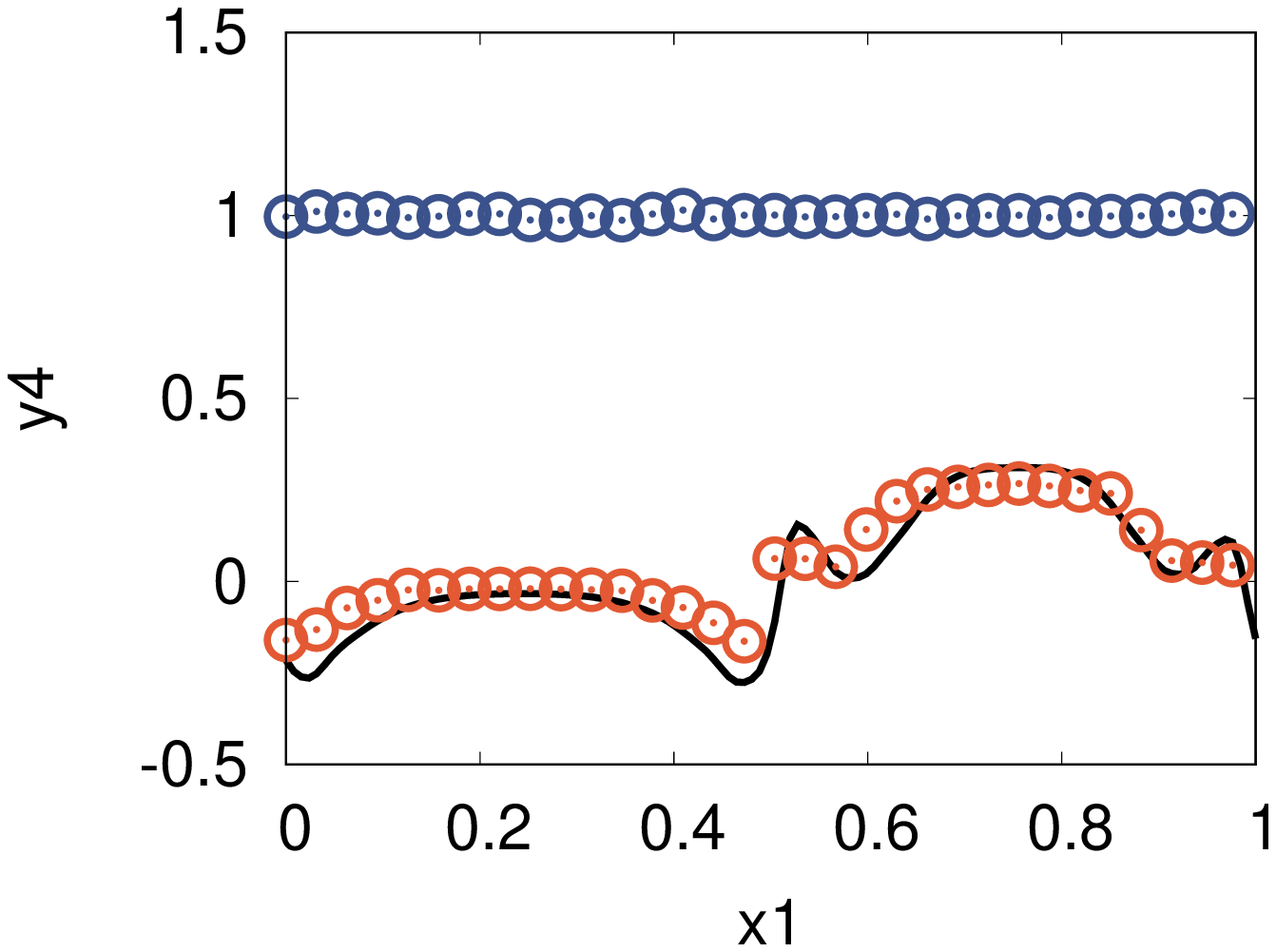} \\
        \psfrag{y1}[][bl][0.9]{$\alpha_{pj}$}
        \hspace{-0.3cm}\includegraphics[scale=0.15,angle=-90]{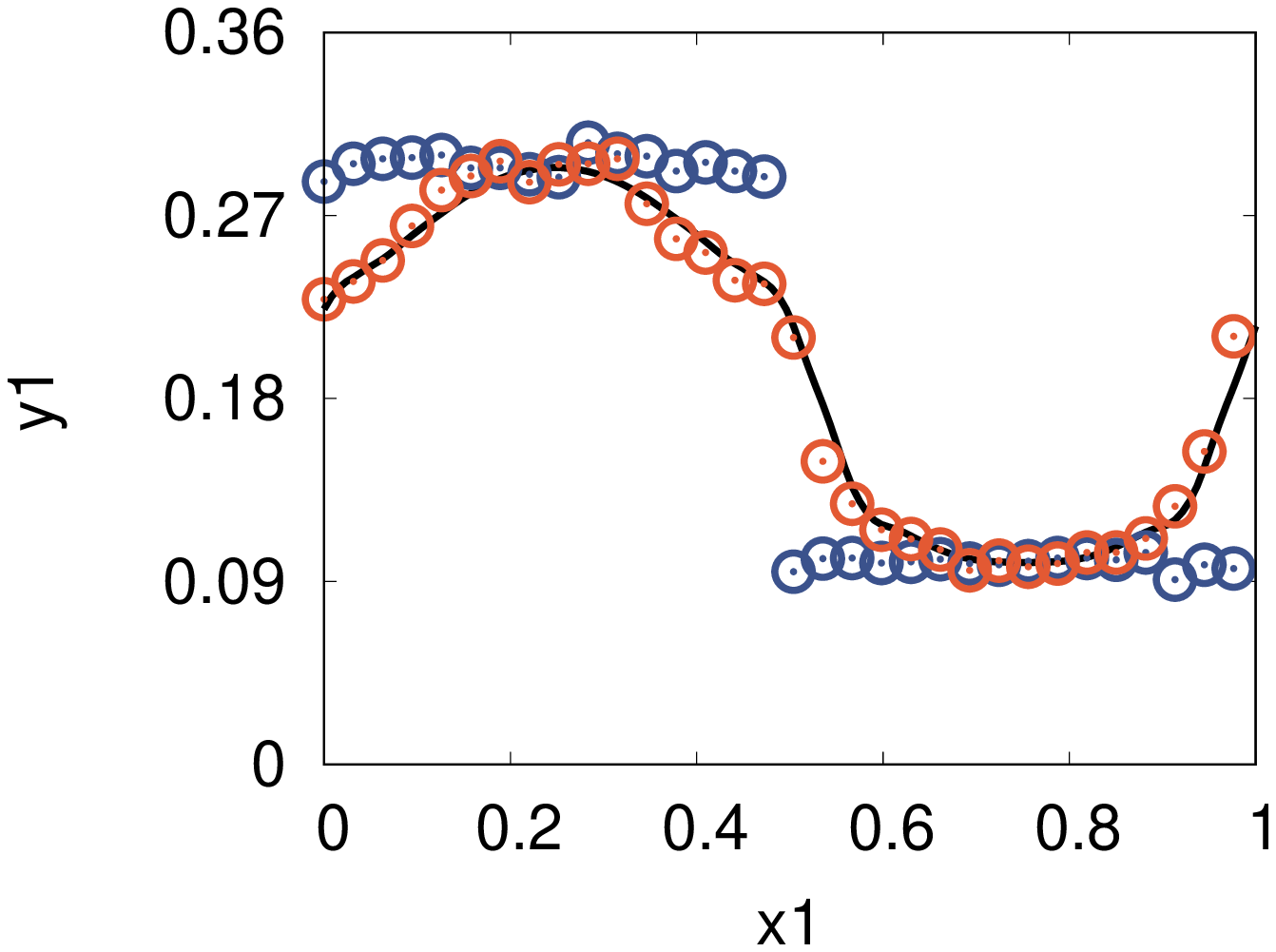} & \psfrag{y2}[][bl][0.9]{$U_{pj}^*$}
        \hspace{-0.6cm}\includegraphics[scale=0.15,angle=-90]{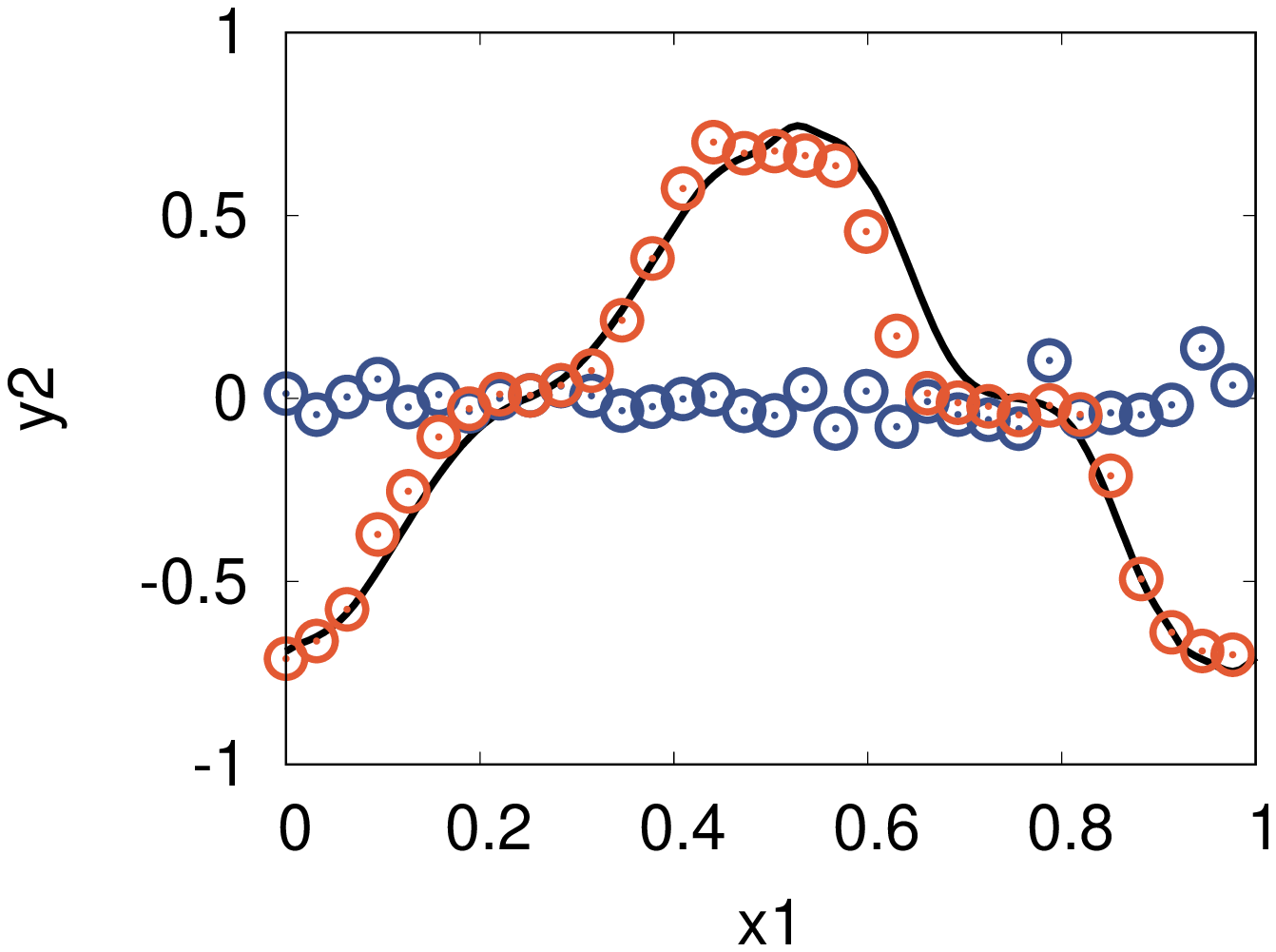} & \psfrag{y3}[][bl][0.9]{$\Tp{j}^*$}
        \hspace{-0.6cm}\includegraphics[scale=0.15,angle=-90]{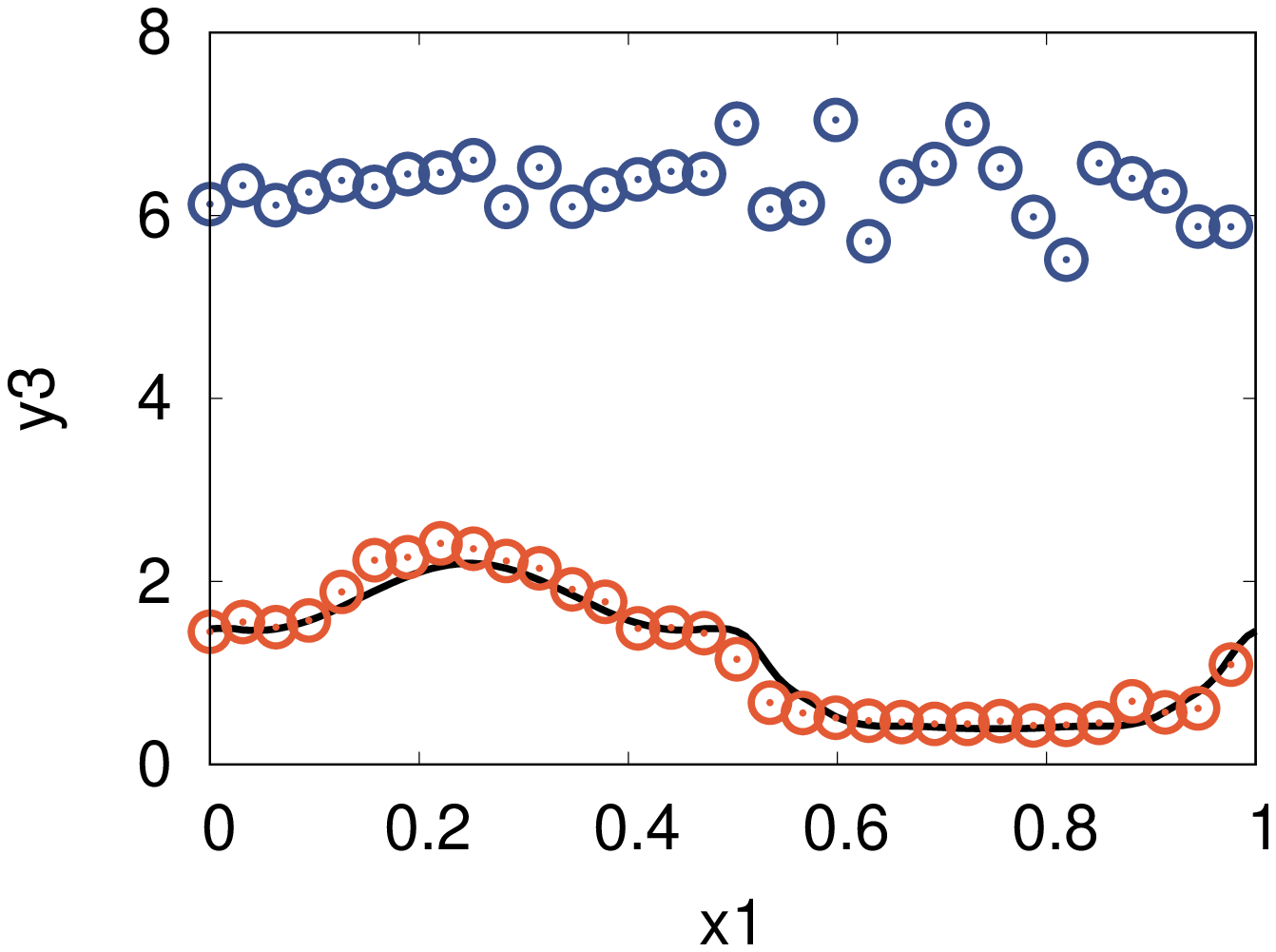} & \psfrag{y4}[][bl][0.9]{$Q_{pj}^*$}
        \hspace{-0.6cm}\includegraphics[scale=0.15,angle=-90]{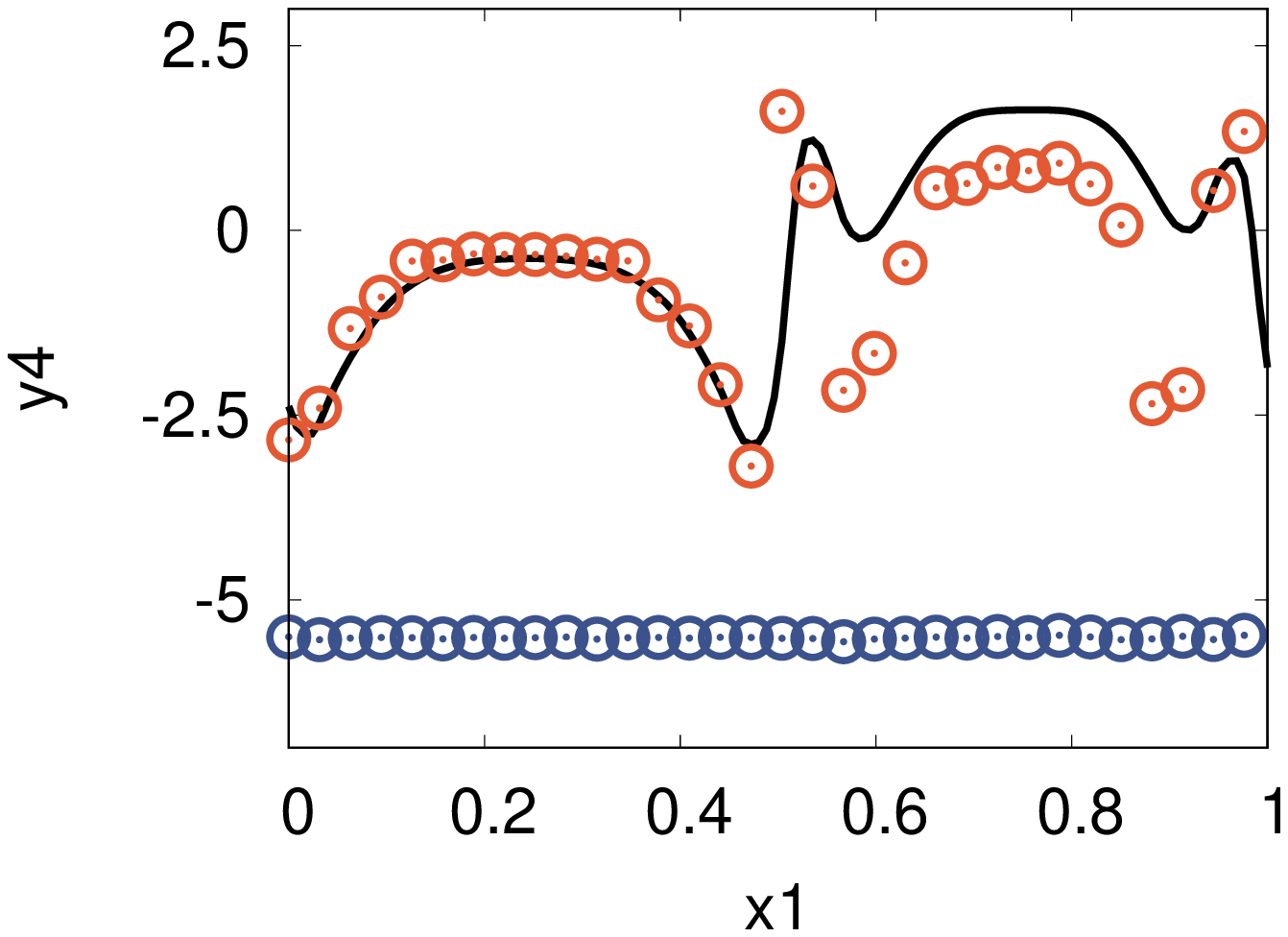} \\
    \end{tabular}    
    \caption{Evolution of (a) phase solid volume fraction, (b) phase scaled velocity, (c) phase scaled granular temperature and (d) phase scaled mean charge of the phase $h\,(h=i,j)$ for Case-G at $t^*=25$. ${\color{CadetBlue}\boldsymbol{\odot}}$: initial conditions for the hard sphere simulation. ${\color{Black}\rule[0.5ex]{10pt}{0.75pt}}$: Eulerian model predictions and ${\color{RedOrange}\boldsymbol{\odot}}$: hard sphere simulation results. Variables were scaled by using \eqref{scalingTimeVelGran}, \eqref{scaledtheta} and \eqref{scaledcharge}.\label{Fig:chargeInitBimodalt25}}
    \centering
    \psfrag{x1}[][][0.8]{$x/L$}
    \begin{tabular}{cccc}
        \psfrag{y1}[][bl][0.9]{$\alpha_{pi}$}
        \hspace{-0.3cm}\includegraphics[scale=0.15,angle=-90]{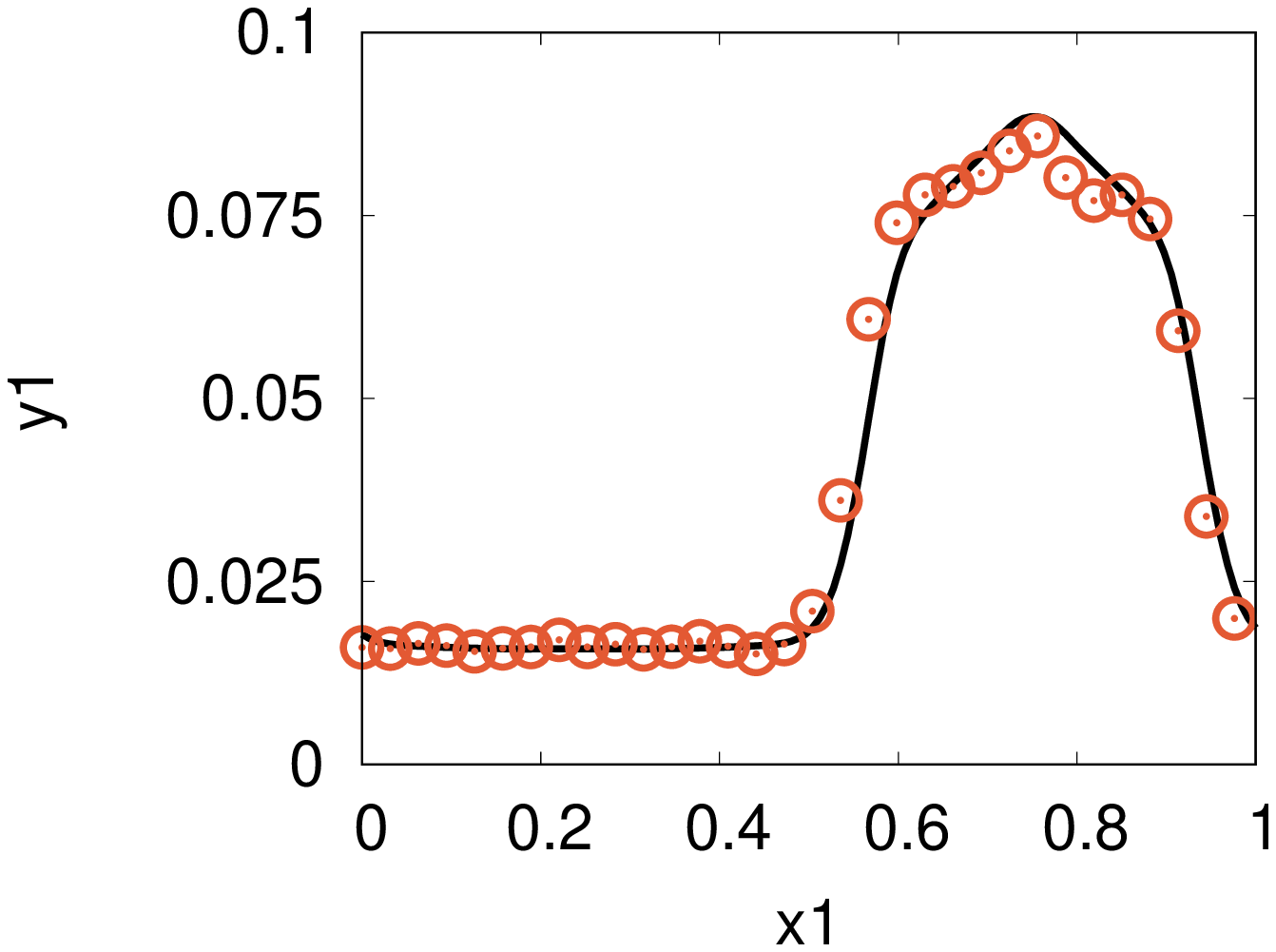} & \psfrag{y2}[][bl][0.9]{$U_{pi}^*$}
        \hspace{-0.6cm}\includegraphics[scale=0.15,angle=-90]{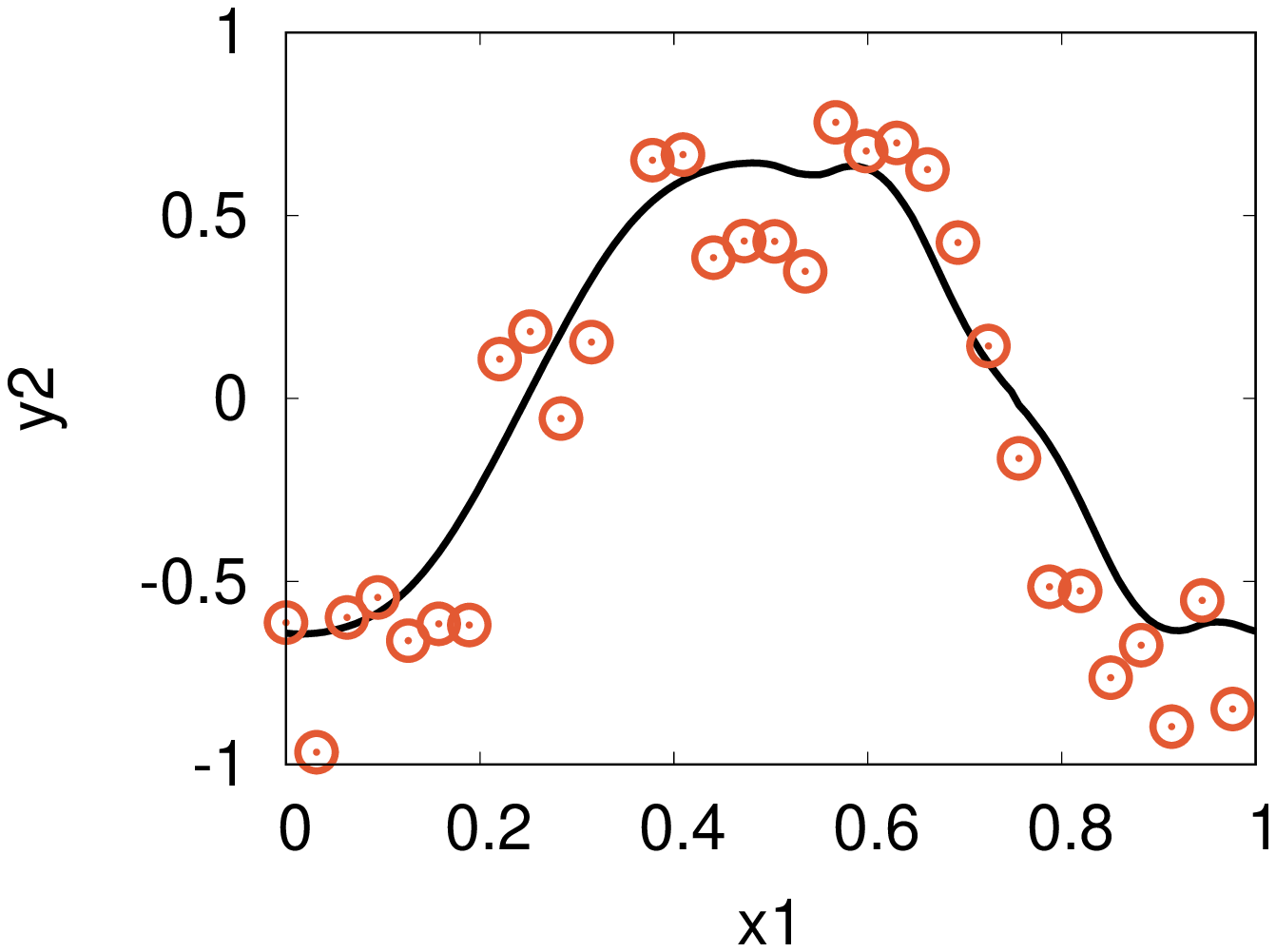} & \psfrag{y3}[][bl][0.9]{$\Tp{i}^*$}
        \hspace{-0.6cm}\includegraphics[scale=0.15,angle=-90]{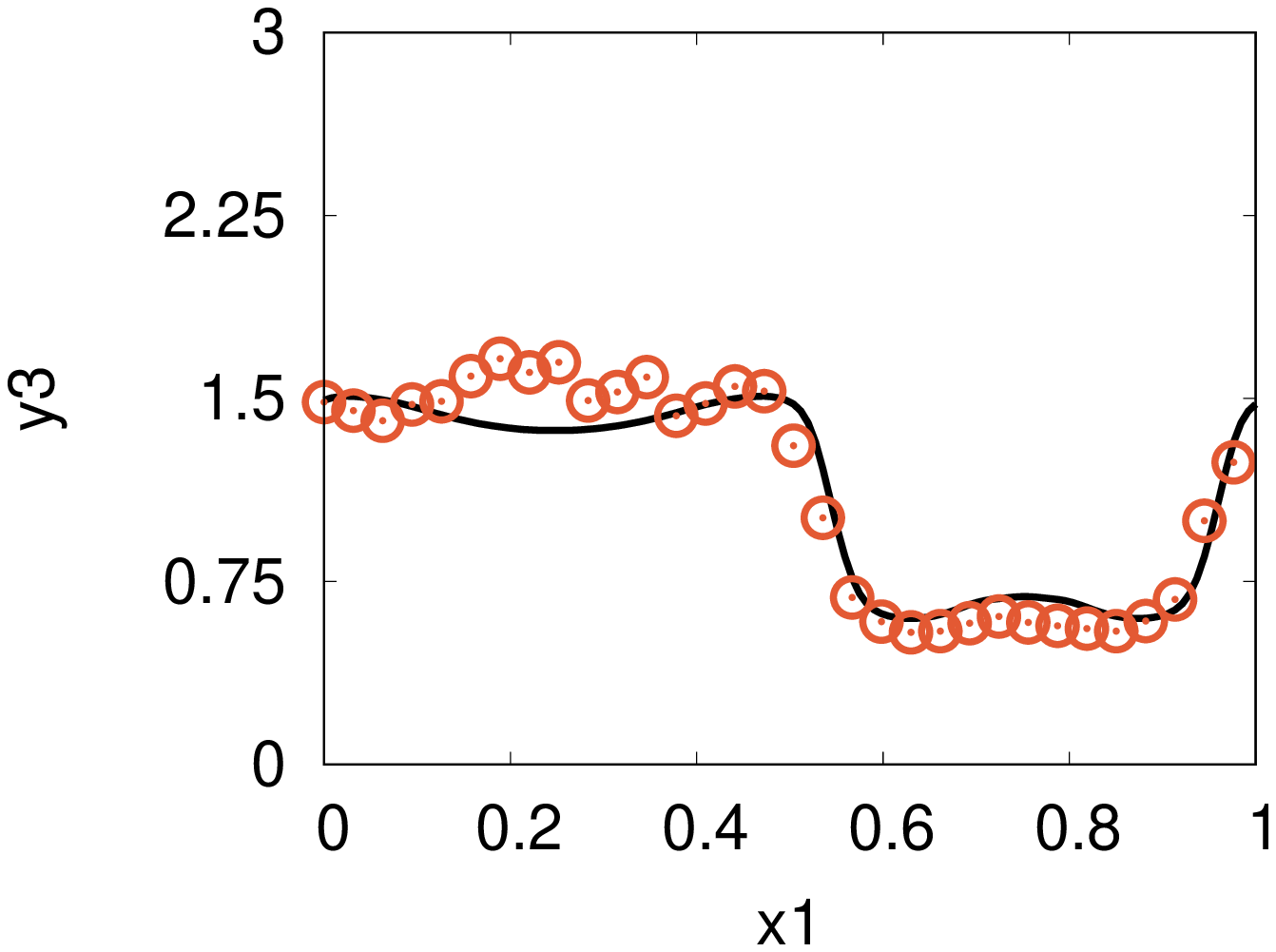} & \psfrag{y4}[][bl][0.9]{$Q_{pi}^*$}
        \hspace{-0.6cm}\includegraphics[scale=0.15,angle=-90]{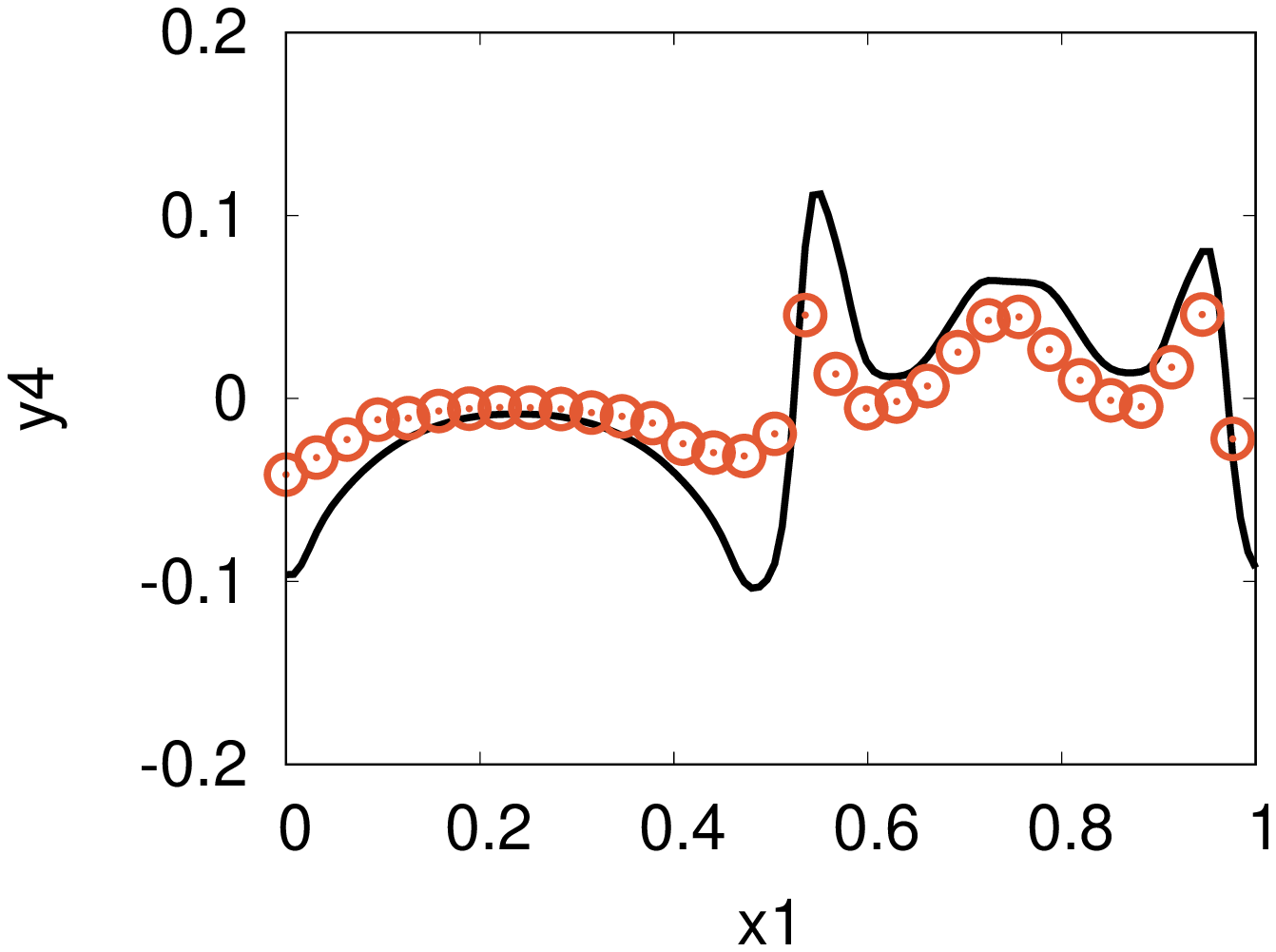} \\
        \psfrag{y1}[][bl][0.9]{$\alpha_{pj}$}
        \hspace{-0.3cm}\includegraphics[scale=0.15,angle=-90]{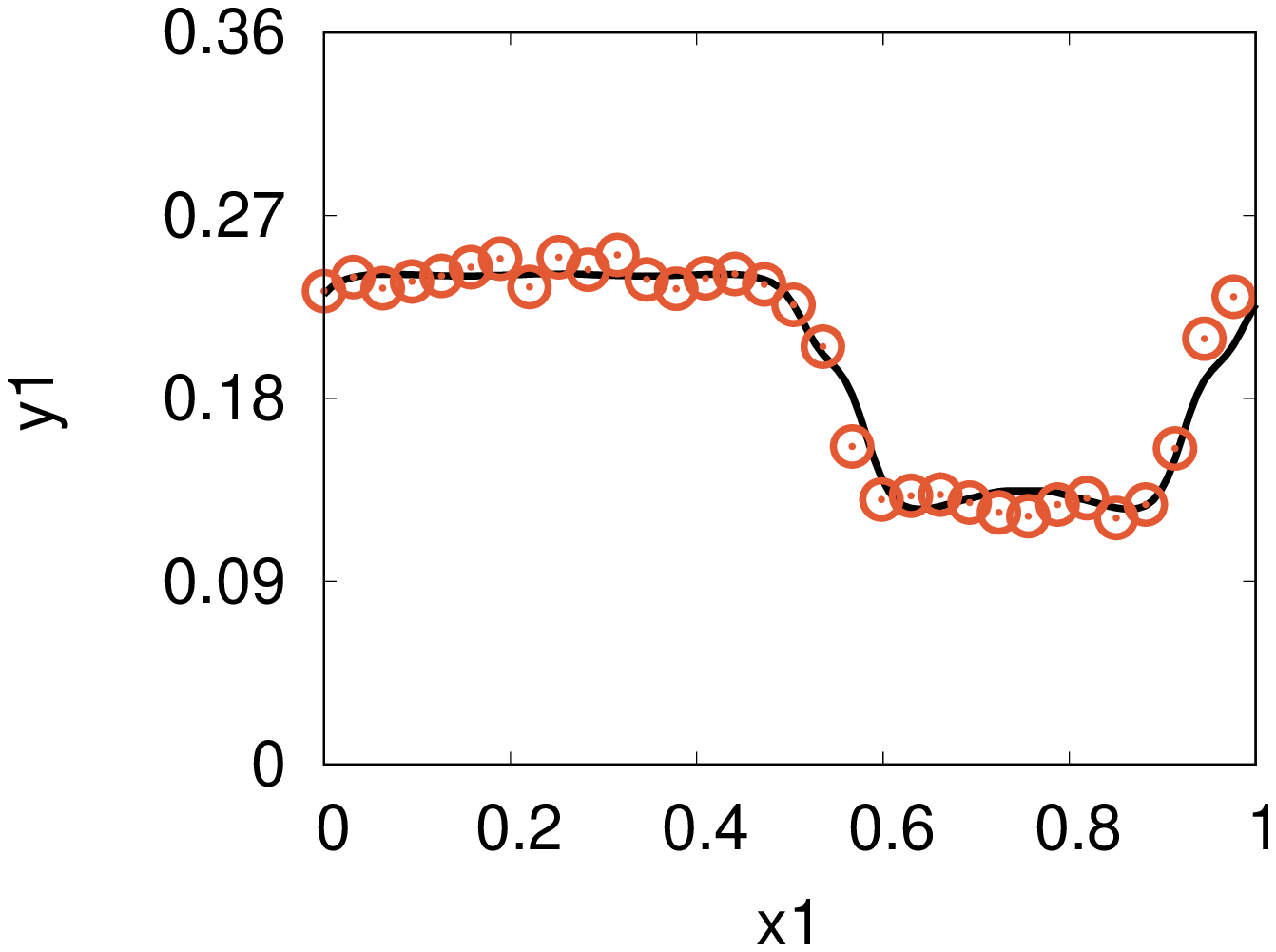} & \psfrag{y2}[][bl][0.9]{$U_{pj}^*$}
        \hspace{-0.6cm}\includegraphics[scale=0.15,angle=-90]{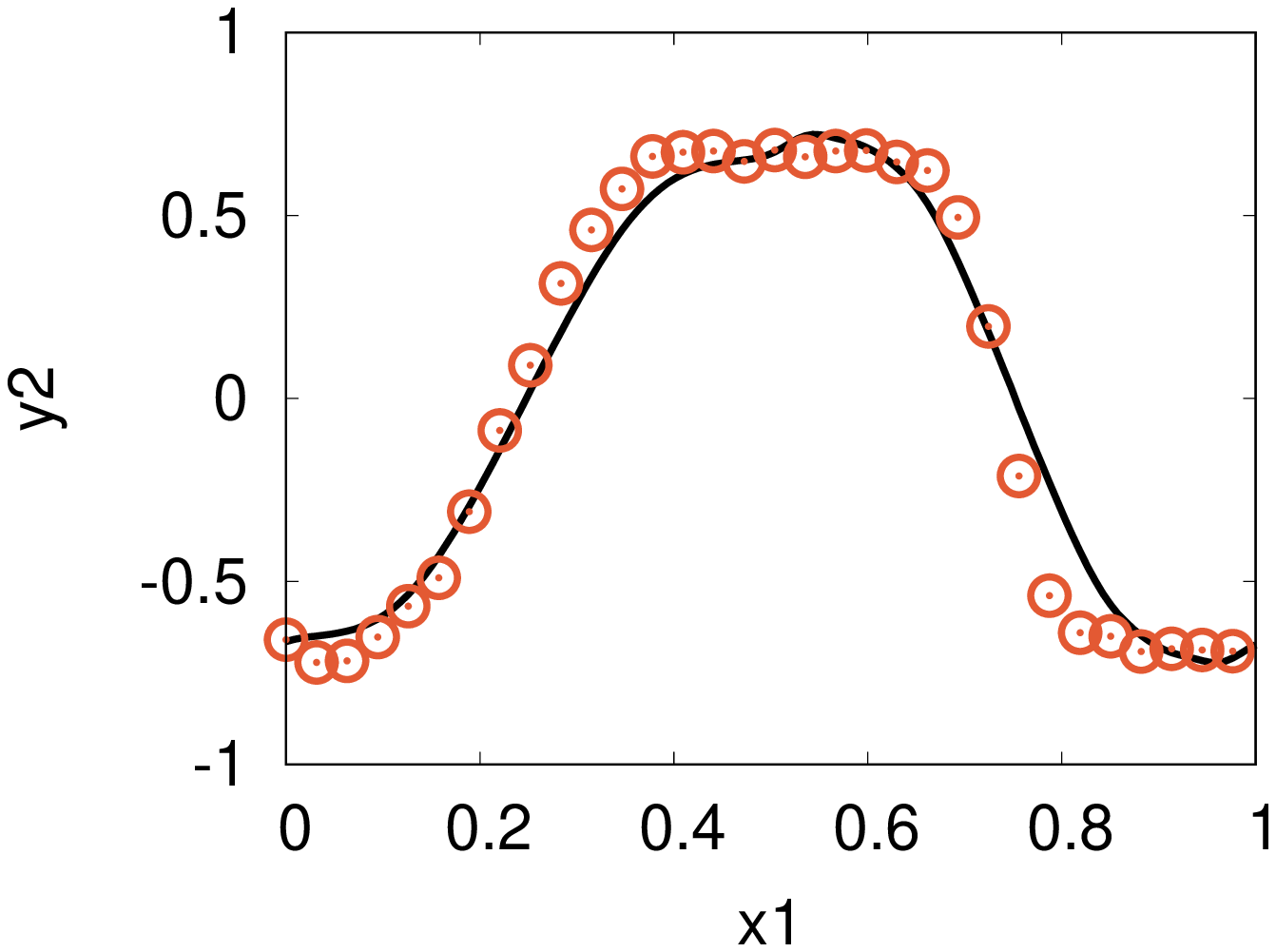} & \psfrag{y3}[][bl][0.9]{$\Tp{j}^*$}
        \hspace{-0.6cm}\includegraphics[scale=0.15,angle=-90]{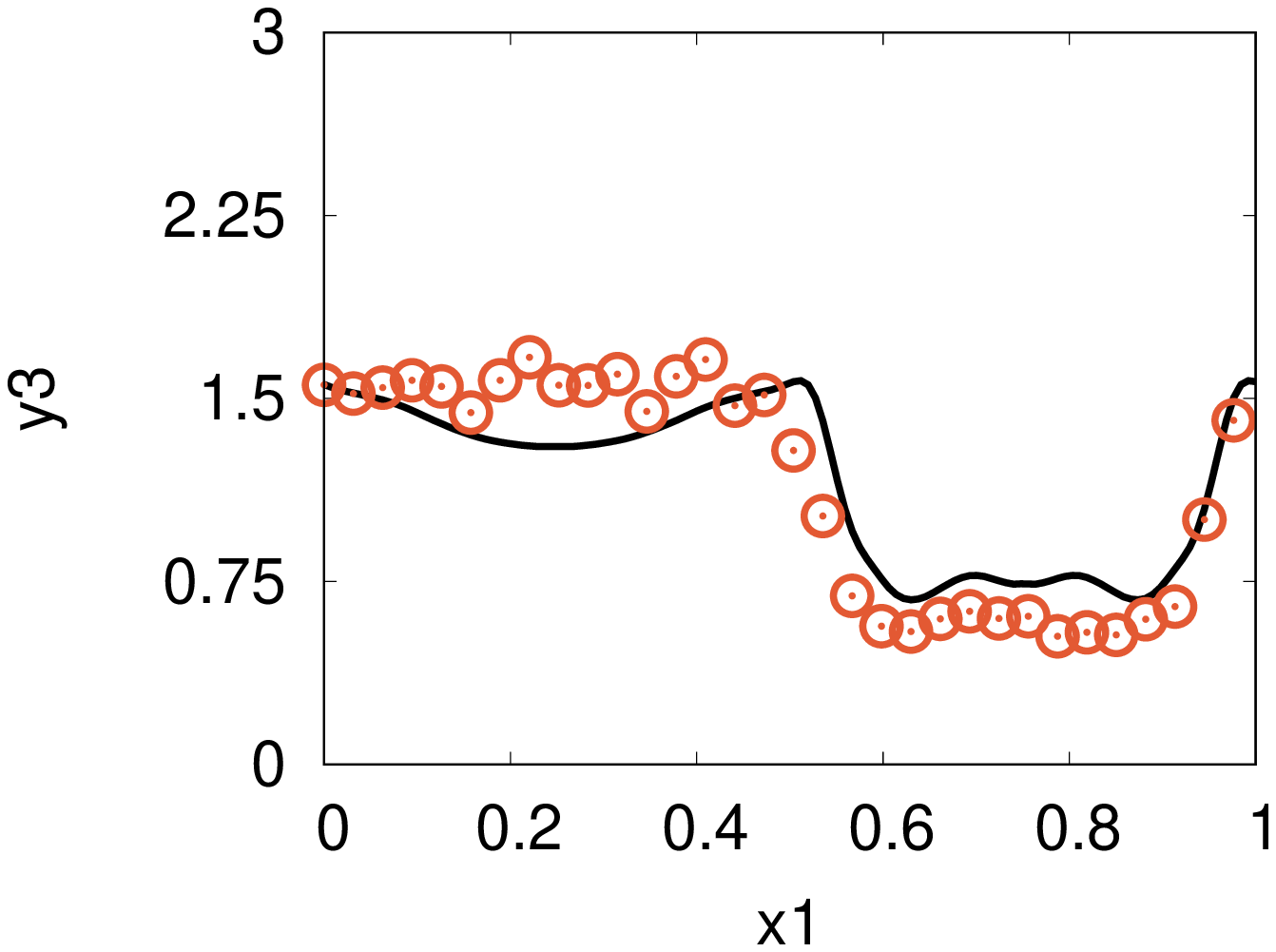} & \psfrag{y4}[][bl][0.9]{$Q_{pj}^*$}
        \hspace{-0.6cm}\includegraphics[scale=0.15,angle=-90]{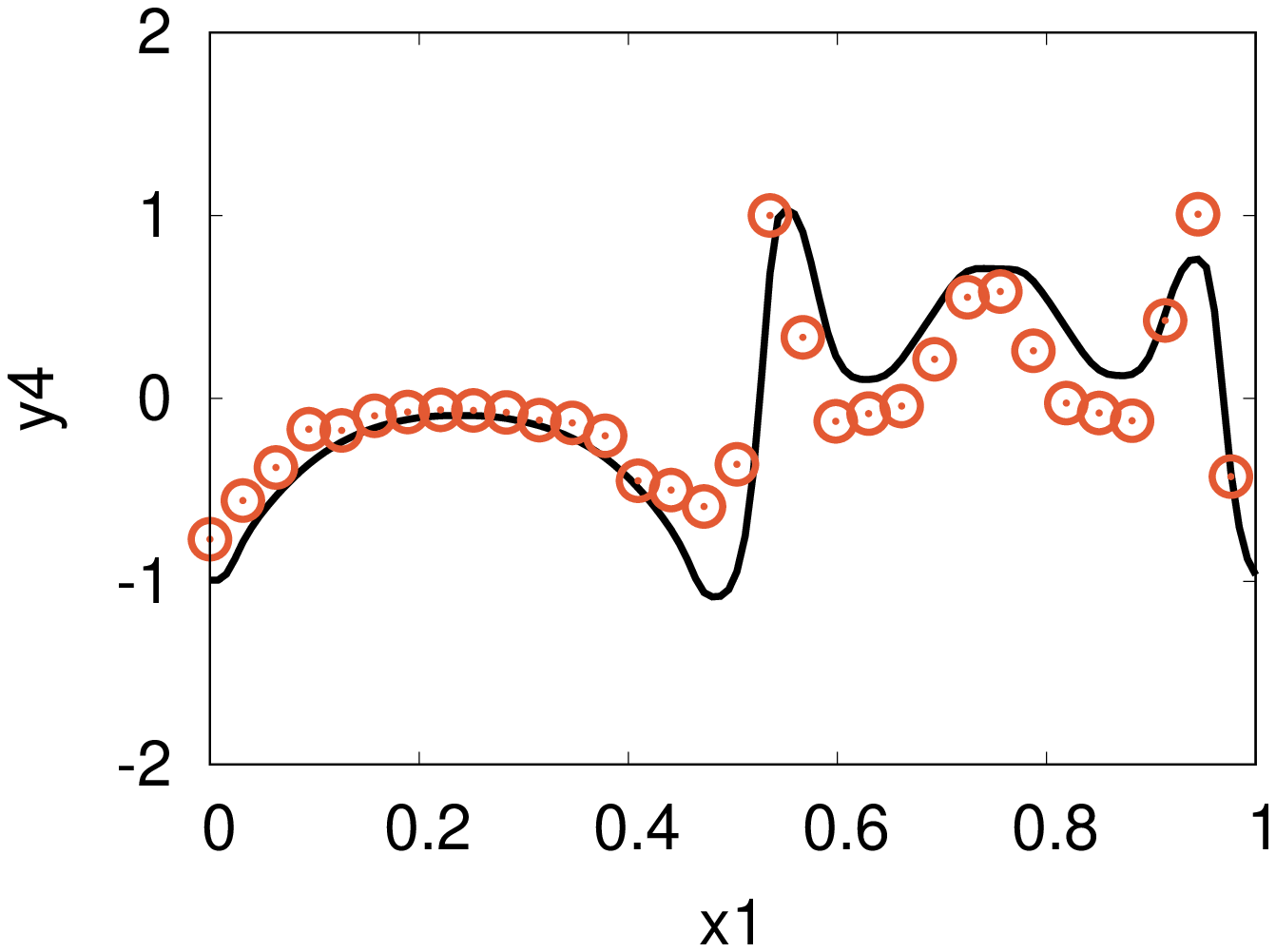} \\
    \end{tabular}
    \caption{Evolution of (a) phase solid volume fraction, (b) phase scaled velocity, (c) phase scaled granular temperature and (d) phase scaled mean charge of the phase $h\,(h=i,j)$ for Case-G at $t^*=25$. ${\color{CadetBlue}\boldsymbol{\odot}}$: initial conditions for the hard sphere simulation. ${\color{Black}\rule[0.5ex]{10pt}{0.75pt}}$: Eulerian model predictions and ${\color{RedOrange}\boldsymbol{\odot}}$: hard sphere simulation results. Variables were scaled by using \eqref{scalingTimeVelGran}, \eqref{scaledtheta} and \eqref{scaledcharge}.\label{Fig:chargeInitBimodalt50}}
\end{figure}
In Case-G, we set the particle diameter-ratio, $R_d$, to 3 and set the  density ratio, $R_{\rho}$, to 10. The average solid volume fraction and the number of particles are identical to Case-E. Each phase is initially charged following \eqref{Eq:chargeInitBimodal}. The simulation results and the model predictions for Case-G at $t^{*}=25$ and $t^*=50$ are shown in figures \ref{Fig:chargeInitBimodalt25} and \ref{Fig:chargeInitBimodalt50}, respectively. The charge for both phases shows a ``wavy'' pattern in the domain and these trends are very well captured with our model. The animation of charge evolution by the hard-sphere simulation for Case-G is given in Supplementary Material 3.

\subsubsection{Quasi-1D bidisperse granular gas simulation with a work function difference}\label{section:workfunction}
\begin{figure}
    \centering
    \psfrag{x1}[][][0.8]{$x/L$}
    \begin{tabular}{cccc}
        (a) & (b) & (c) & (d) \vspace{-0.25cm}\\
        \psfrag{y1}[][bl][0.9]{$\alpha_{pi}$}
        \hspace{-0.3cm}\includegraphics[scale=0.15,angle=-90]{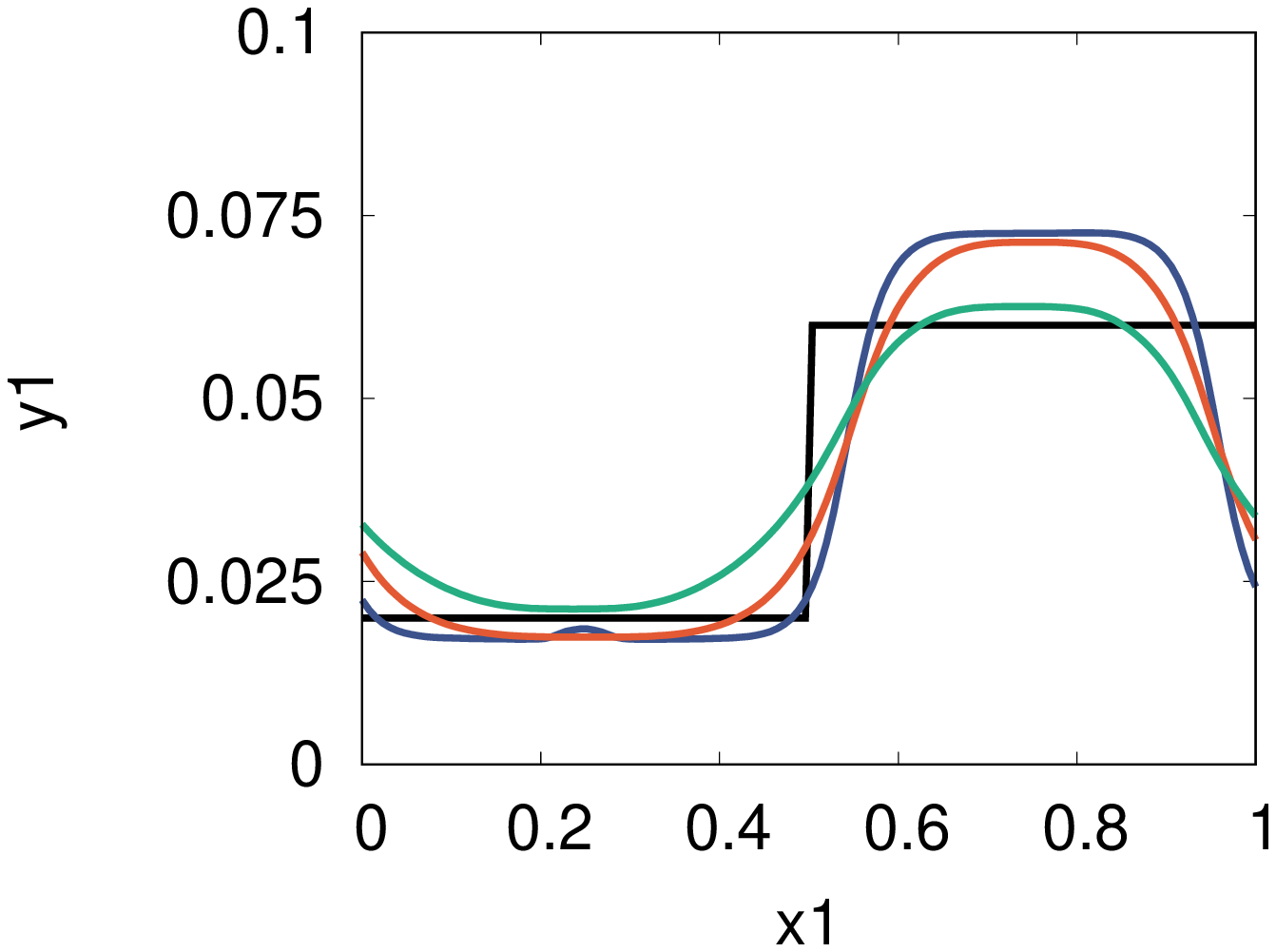} & \psfrag{y2}[][bl][0.9]{$U_{pi}^*$}
        \hspace{-0.6cm}\includegraphics[scale=0.15,angle=-90]{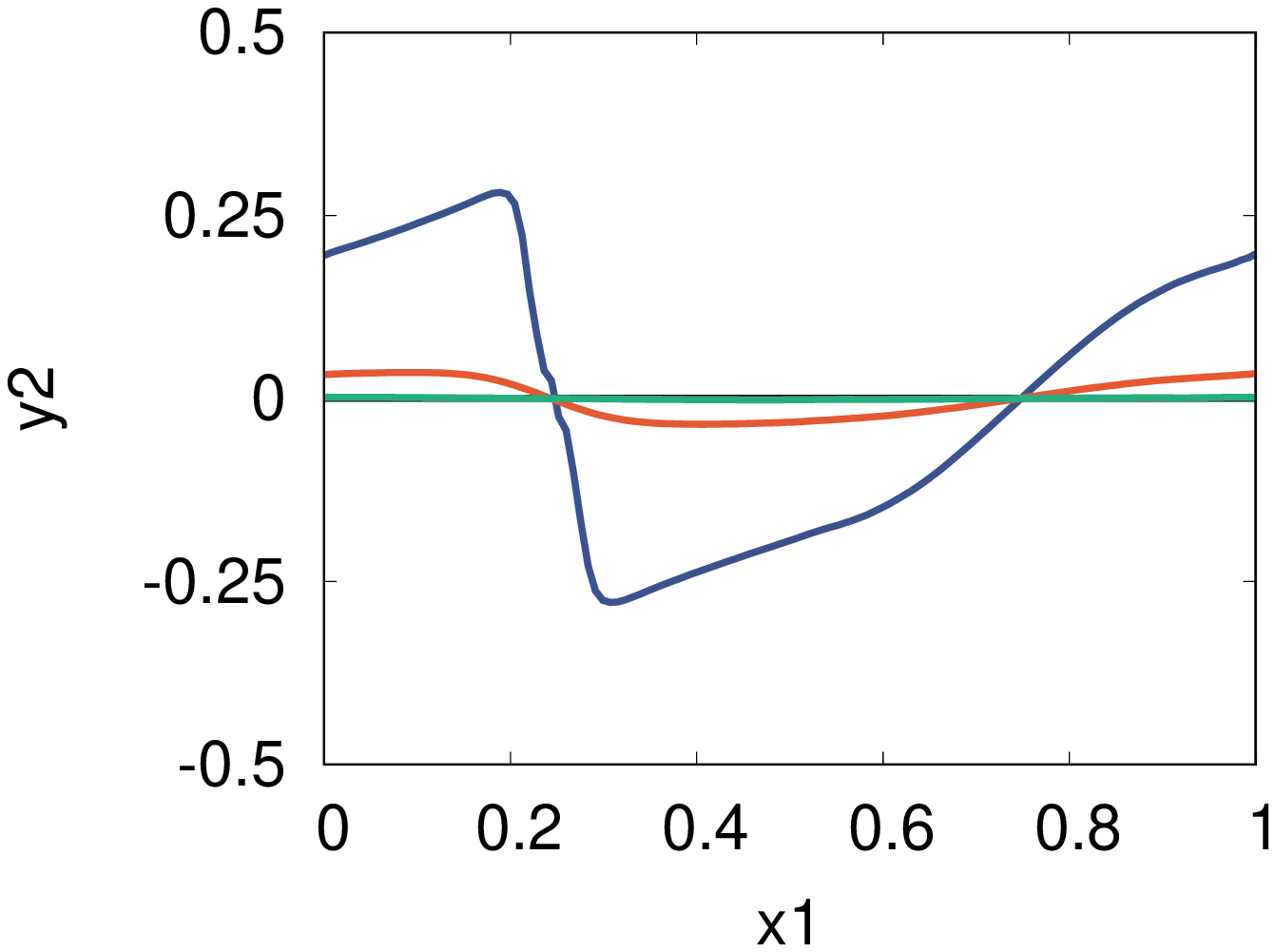} & \psfrag{y3}[][bl][0.9]{$\Theta_{pi}^*$}
        \hspace{-0.6cm}\includegraphics[scale=0.15,angle=-90]{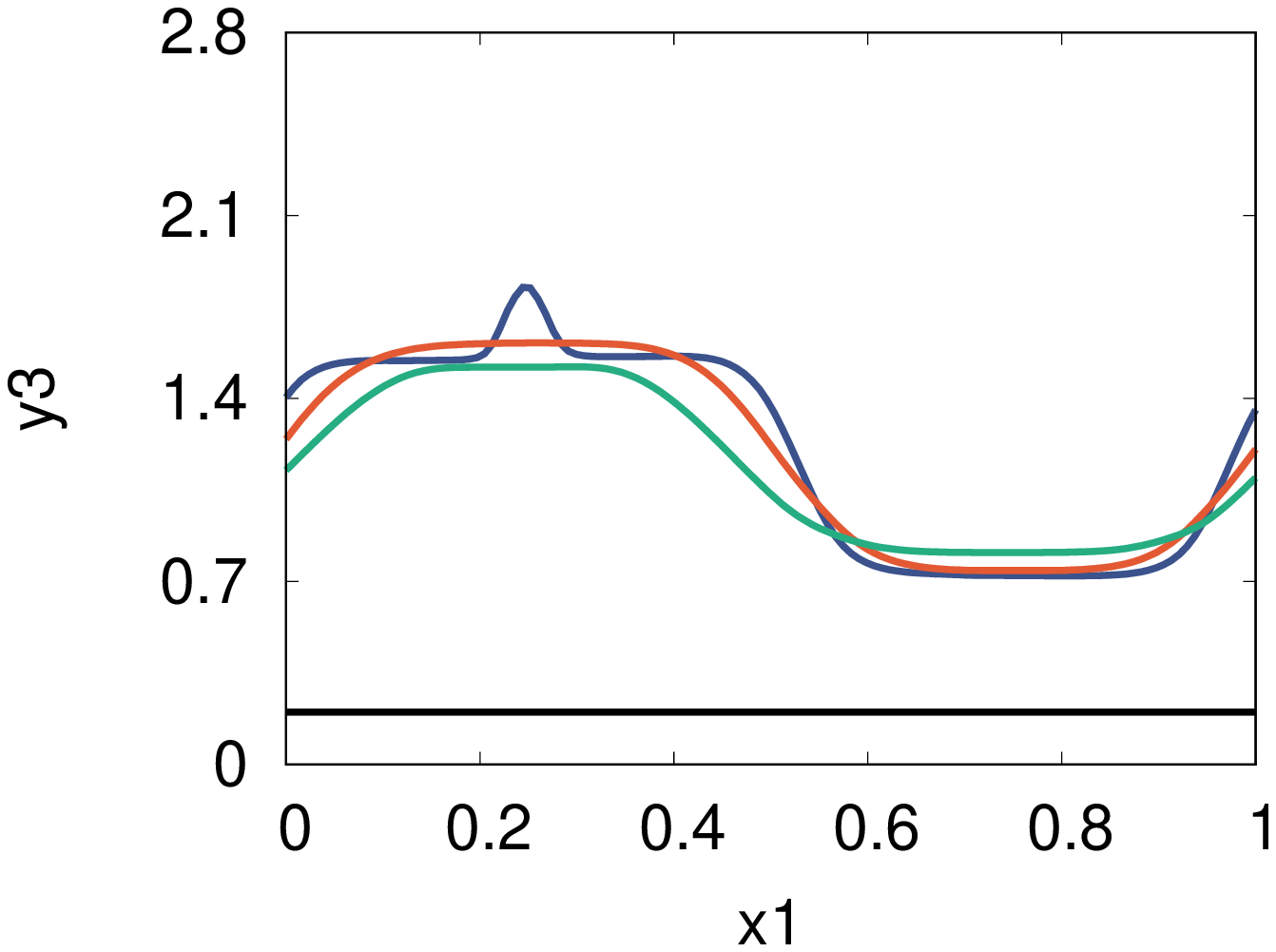} & \psfrag{y4}[][bl][0.9]{$Q_{pi}^*$}
        \hspace{-0.6cm}\includegraphics[scale=0.15,angle=-90]{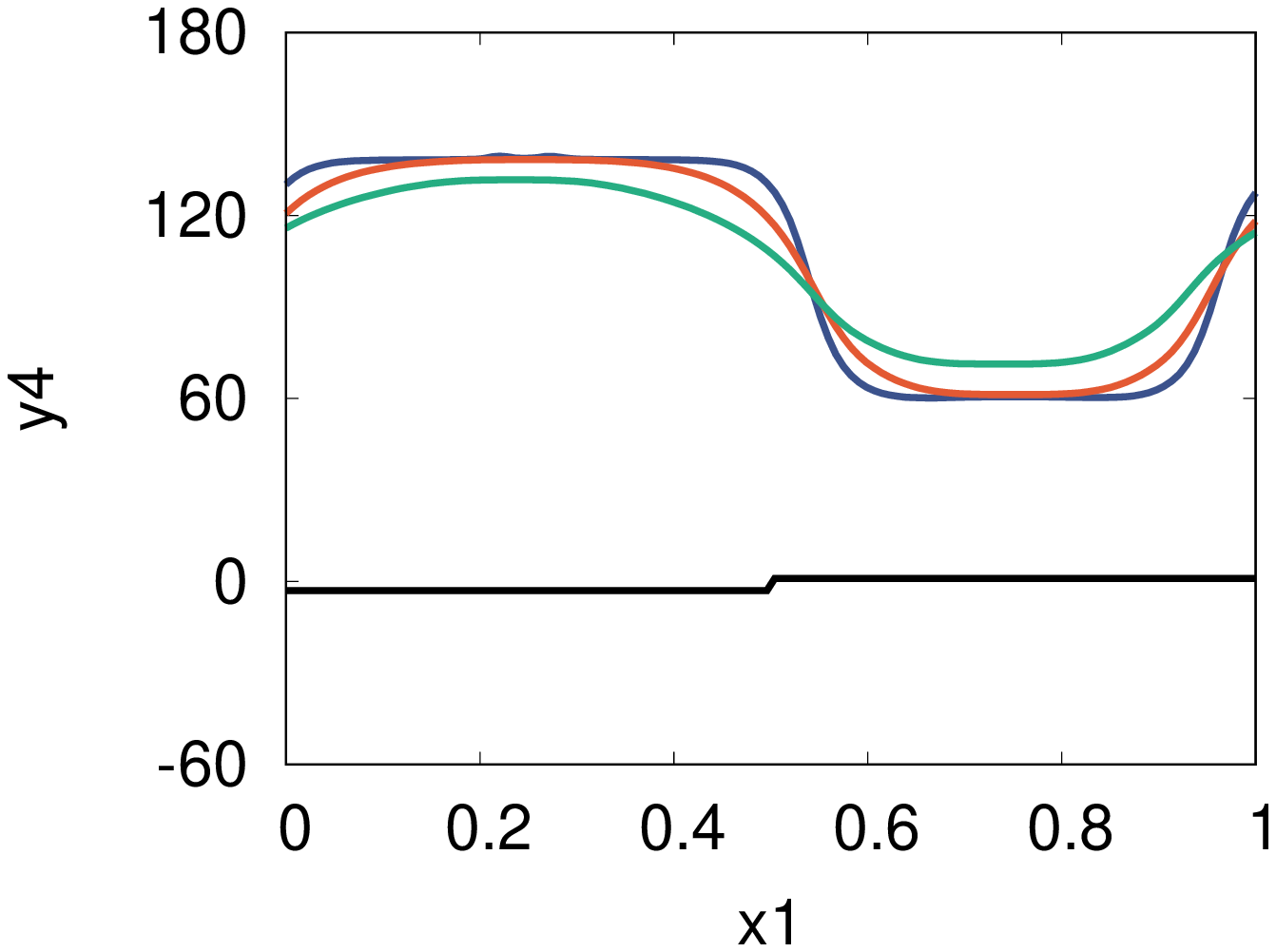} \\
        \psfrag{y1}[][bl][0.9]{$\alpha_{pj}$}
        \hspace{-0.3cm}\includegraphics[scale=0.15,angle=-90]{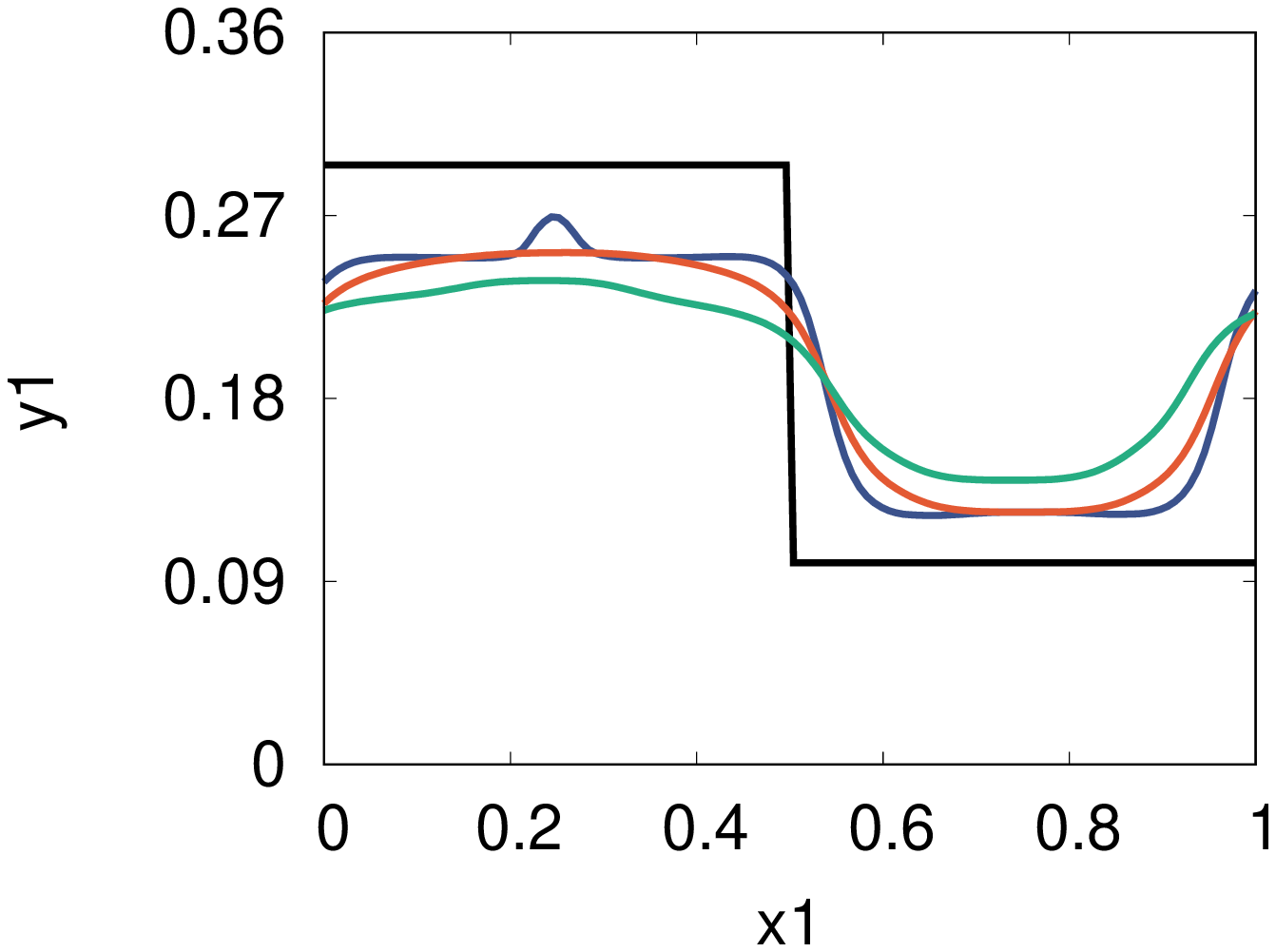} & \psfrag{y2}[][bl][0.9]{$U_{pj}^*$}
        \hspace{-0.6cm}\includegraphics[scale=0.15,angle=-90]{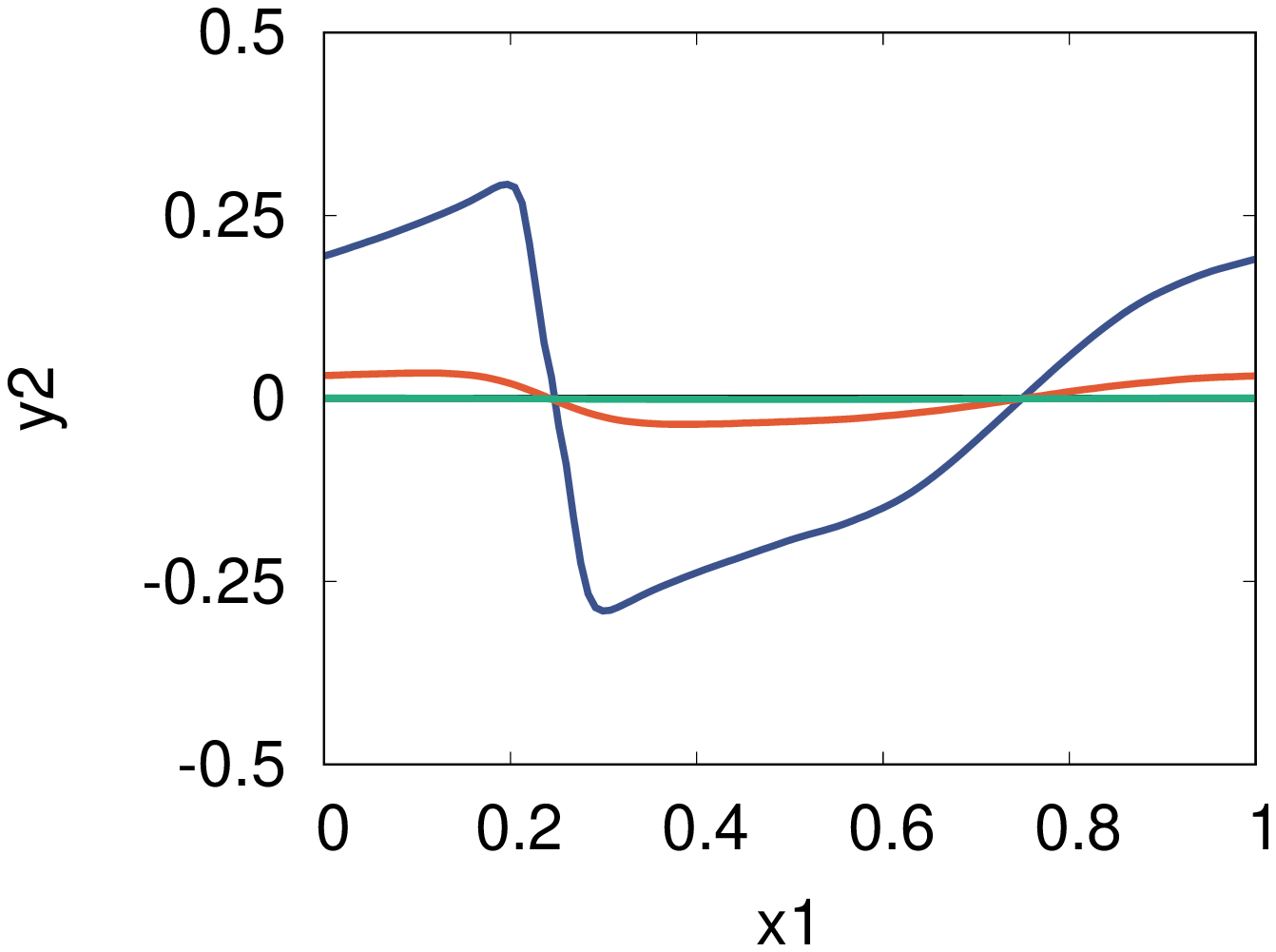} & \psfrag{y3}[][bl][0.9]{$\Theta_{pj}^*$}
        \hspace{-0.6cm}\includegraphics[scale=0.15,angle=-90]{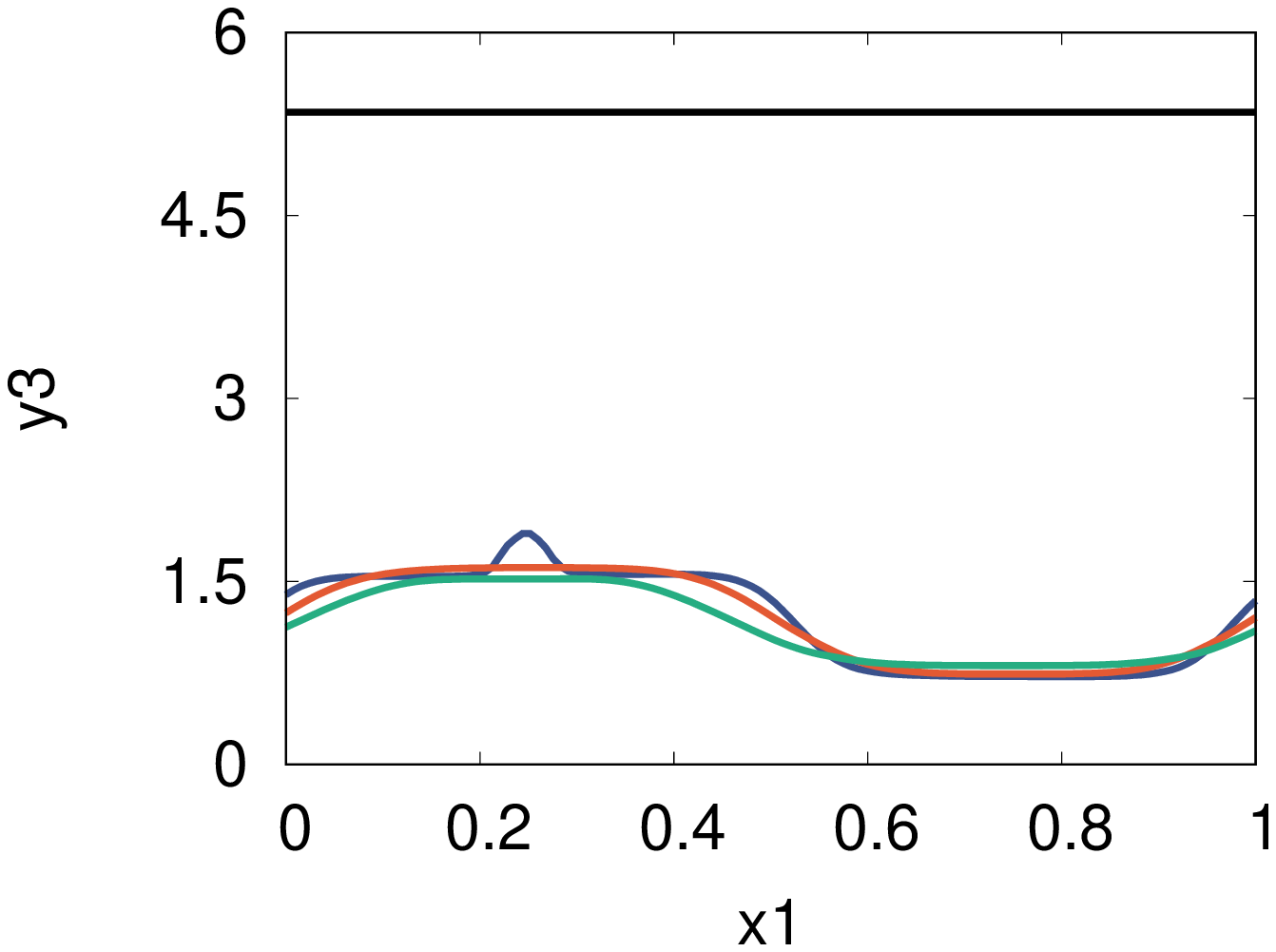} & \psfrag{y4}[][bl][0.9]{$Q_{pj}^*$}
        \hspace{-0.6cm}\includegraphics[scale=0.15,angle=-90]{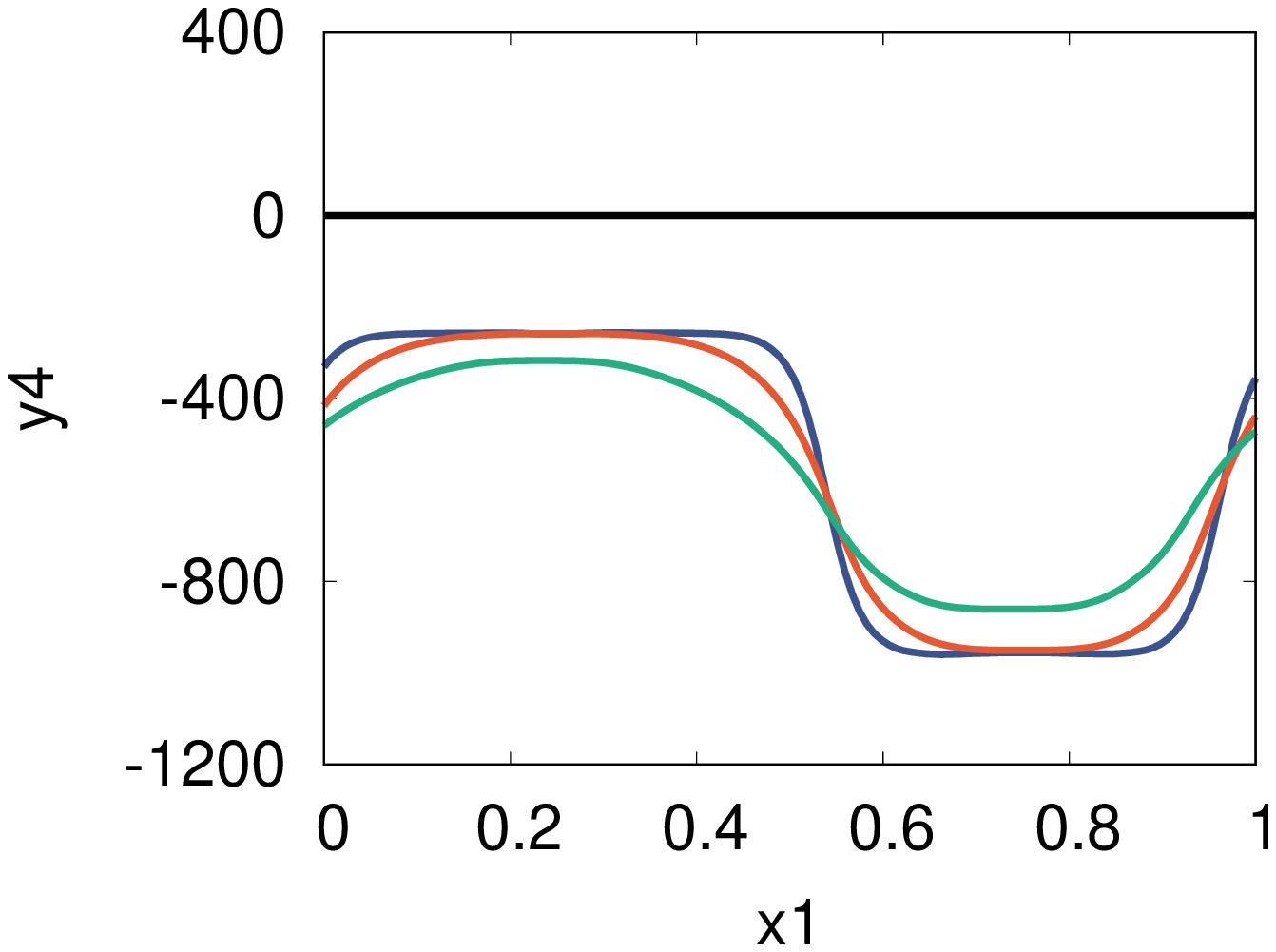} \\
    \end{tabular}
    \caption{Evolution of (a) phase solid volume fraction, (b) phase scaled velocity, (c) phase scaled granular temperature and (d) phase scaled mean charge of the phase $h\,(h=i,j)$ for Case-E with a negative work function difference between phases by the Eulerian predictions at various time instants. Work functions for phases $i$ and $j$ are 3.9 and 4.2 \SI{}{\electronvolt}, respectively. Black, blue, red and green lines refer the Eulerian predictions at $t^* = 0, 223.5, 2\,235, 13\,410$, respectively. Variables were scaled by using \eqref{scalingTimeVelGran}, \eqref{scaledtheta} and \eqref{scaledcharge}.\label{fig:WorkFunctionNeg}}
\end{figure}

In this short section, we show how the work function difference between phases generates charge with the Eulerian model. Starting from Case-E given in table \ref{Table:InitialSpatial}, the work functions of 3.9 and 4.2 \SI{}{\electronvolt} were imposed for the phases $i$ and $j$, respectively. The Eulerian simulation was performed with a negative work function difference $\Delta \varphi_p =\varphi_{pi}-\varphi_{pj}$ for duration of $t^*$ = 13\,410 (equal to \SI{60}{\second} in physical time). It is worth to note that the hard-sphere simulation is not computationally affordable for this duration but it is well established that the charge build-up is a slow process (e.g. it takes more than few minutes to reach the saturated charge in the vibrated beds shown by \cite{kolehmainen_effect_2017}). Therefore, the simulations with longer duration or steady-state solutions are necessary to have better understanding of effects of the tribocharging on the hydrodynamics. 

The evolution of hydrodynamic variables and charges for Case-E with a work function difference by the Eulerian predictions is shown in figure \ref{fig:WorkFunctionNeg}. The solid volume fraction for each phase slowly reaches a flat profile (figure \ref{fig:WorkFunctionNeg}-(a)) and the phase velocities dissipate quickly and become zero at $t^*>13\,410$ (figure \ref{fig:WorkFunctionNeg}-(b)). The granular temperatures reach a equilibrium value which is slightly lower than of Case-E (figure \ref{Fig:ResultRd3Spatialt50}-(c)). Due to the work function difference, a bipolar charge distribution occurs and each phase reaches an equilibrium value (figure \ref{fig:WorkFunctionNeg}-(d)). For this case, we also accounted for the electrostatic force in the phase momentum equations. After a short duration ($t^* > 223.5$), the electric field became very small in the domain, therefore, the electrostatic force has a limited effect on the momentum and energy evolution. We also performed the same case with a positive function difference (not shown here) and obtained a very similar evolution for hydrodynamic variables with an inverse bipolar charge distribution (the phase $i$ has a negative charge whereas the phase $j$ has a positive charge).    

\subsubsection{Knudsen number analysis}\label{section:knudsen}
We solved the hydrodynamic equations along $x$-direction for quasi-one dimensional simulations and here, the validity of these computations in the continuum regime was tested with the Knudsen number analysis. The Knudsen number, $Kn$, is defined as
\begin{equation}
    Kn = \frac{\lambda}{L} 
\end{equation}
where $\lambda$ is the mean free path and $L$ is a macroscopic length scale. In a dense granular fluid, the mean free path is given by \cite{garzo2005instabilities} as:
\begin{equation}
    \lambda= \frac{d_p}{6\sqrt{2}\alpha_p\,g_0(\alpha_p)}.
\end{equation}
\cite{fullmer2017clustering} and \cite{wang2019quantifying} chose $\alpha_p/|\nabla\alpha_p|$, $U_p/|\nabla U_p|$ and $\Tp{}/|\nabla \Tp{}|$ for the characteristic length scale, $L$. By following these studies, we define the three Knudsen numbers based on the gradients of solid volume fraction, velocity and granular temperature for the phase $i$ as follows:
\begin{eqnarray}
    {Kn}_{\alpha_{pi}} & = & \frac{5}{6\sqrt{2}}\frac{d_{pi}\,|\nabla \alpha_{pi}|}{\alpha_{pi}^2\,g_0(\alpha_{pi})}, \label{knalp} \\
    {Kn}_{U_{pi}} & = & \frac{5}{12}\frac{d_{pi}\,|\nabla U_{pi}|}{\alpha_{pi}\,g_0(\alpha_{pi})\sqrt{\Tp{i}/m_{pi}}},  \label{knvel} \\
    {Kn}_{\Tp{i}} & = & \frac{5}{6\sqrt{2}}\frac{d_{pi}\,|\nabla \Tp{i}/m_{pi}|}{\alpha_{pi}\,g_0(\alpha_{pi})\Tp{i}/m_{pi}}. \label{knthe}
\end{eqnarray}

\begin{figure}
    \centering
    \psfrag{x1}[][][0.8]{$x/L$}
    \begin{tabular}{ccc}
        (a) & (b) & (c) \vspace{-0.25cm}\\
        \psfrag{y1}[b][b][0.9]{$Kn_{\alpha_p}$}
        \hspace{-0.3cm}\includegraphics[scale=0.18,angle=-90]{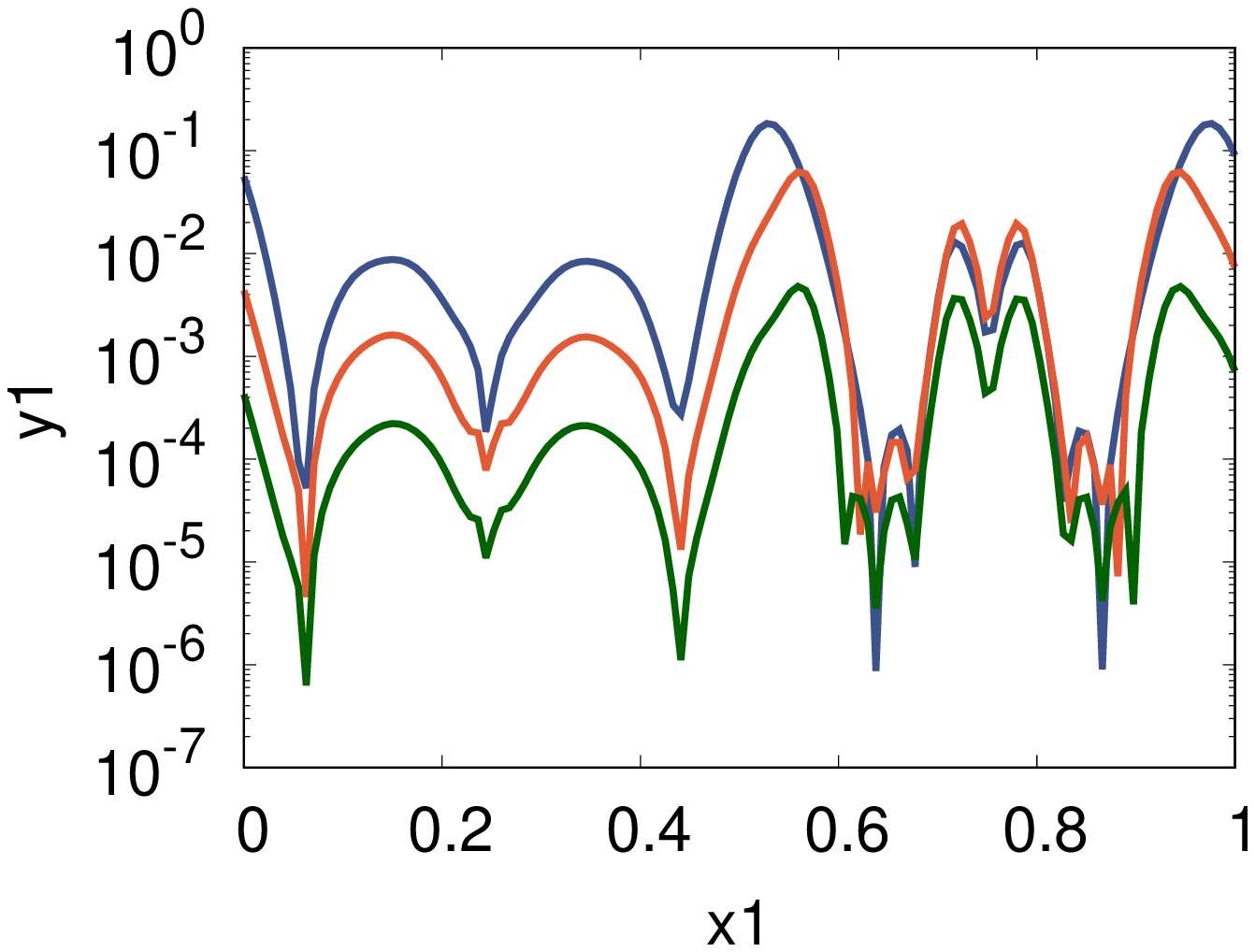} & \psfrag{y2}[b][b][0.9]{$Kn_{U_p}$}
        \hspace{-0.4cm}\includegraphics[scale=0.18,angle=-90]{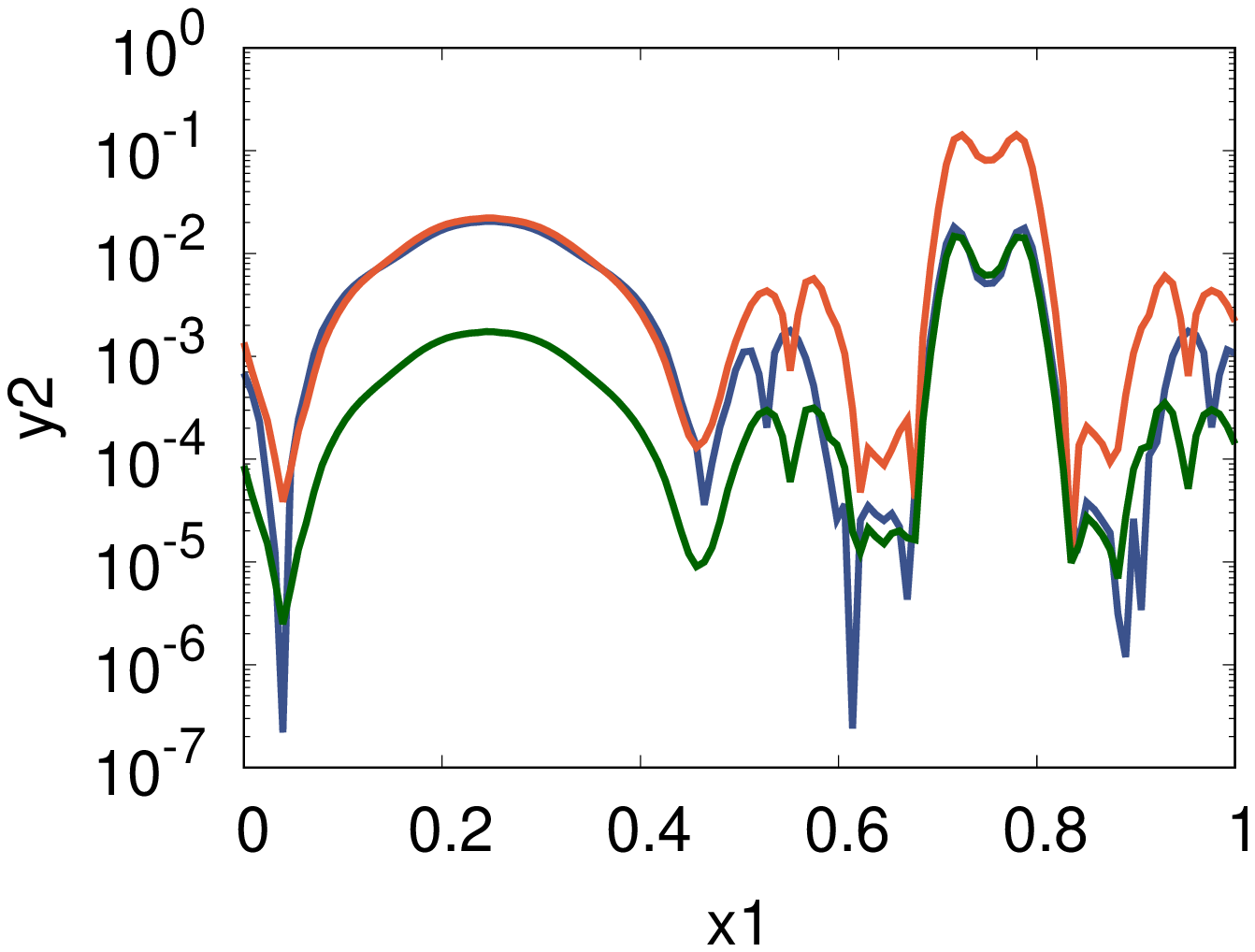} & \psfrag{y3}[b][b][0.9]{$Kn_{\Tp{}}$}
        \hspace{-0.4cm}\includegraphics[scale=0.18,angle=-90]{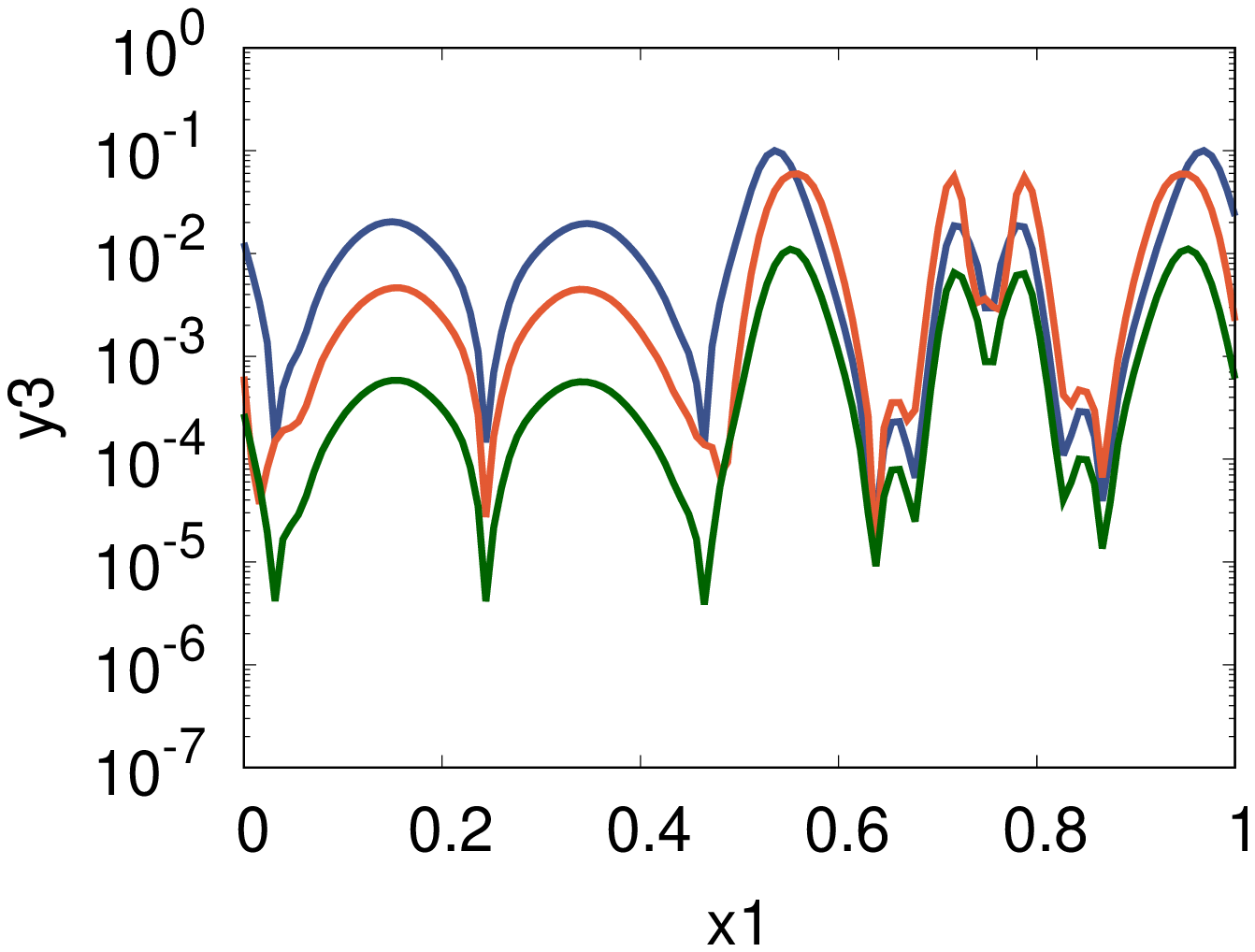}
        \\
        \psfrag{y1}[b][b][0.9]{$Kn_{\alpha_p}$}
        \hspace{-0.3cm}\includegraphics[scale=0.18,angle=-90]{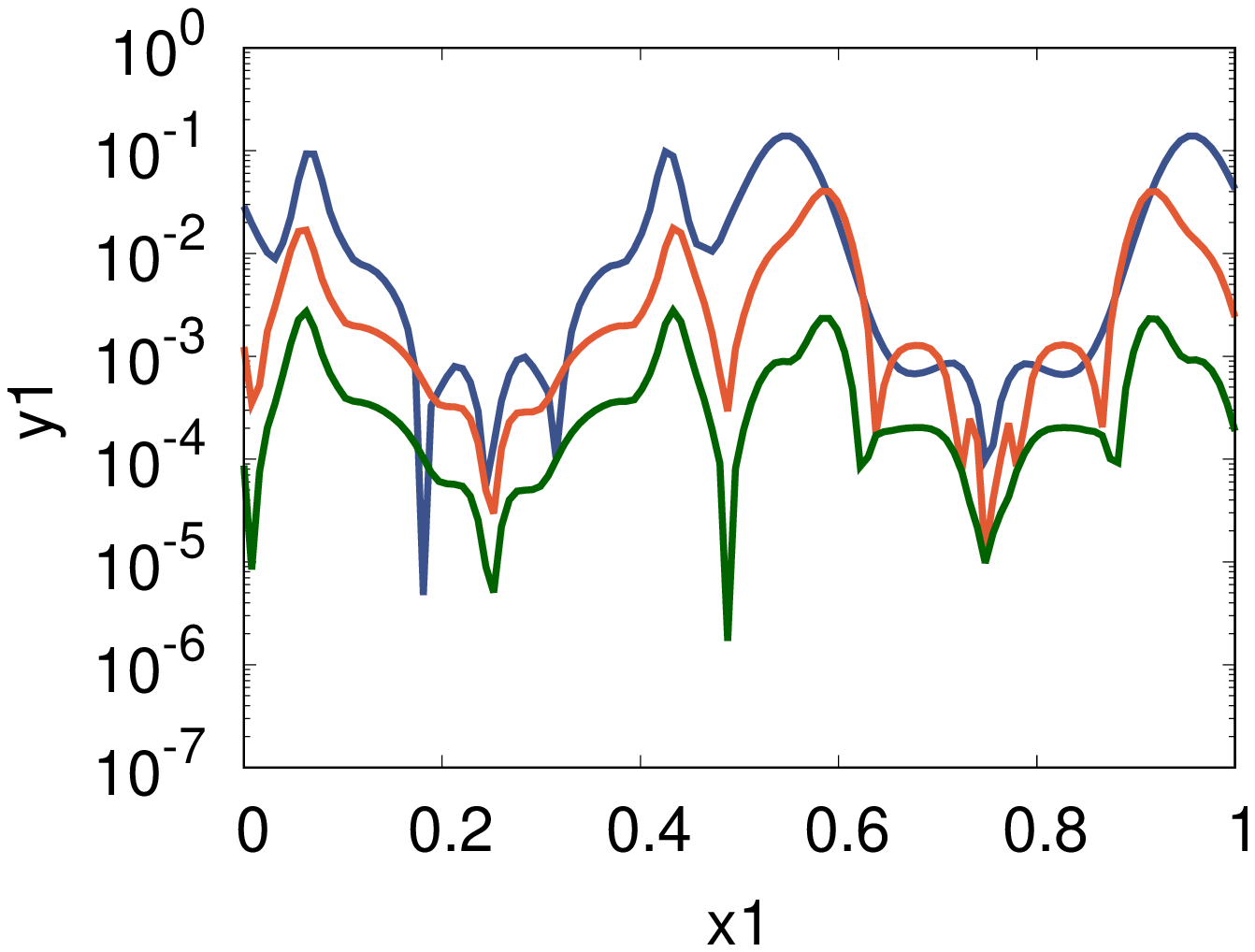} & \psfrag{y2}[b][b][0.9]{$Kn_{U_p}$}
        \hspace{-0.4cm}\includegraphics[scale=0.18,angle=-90]{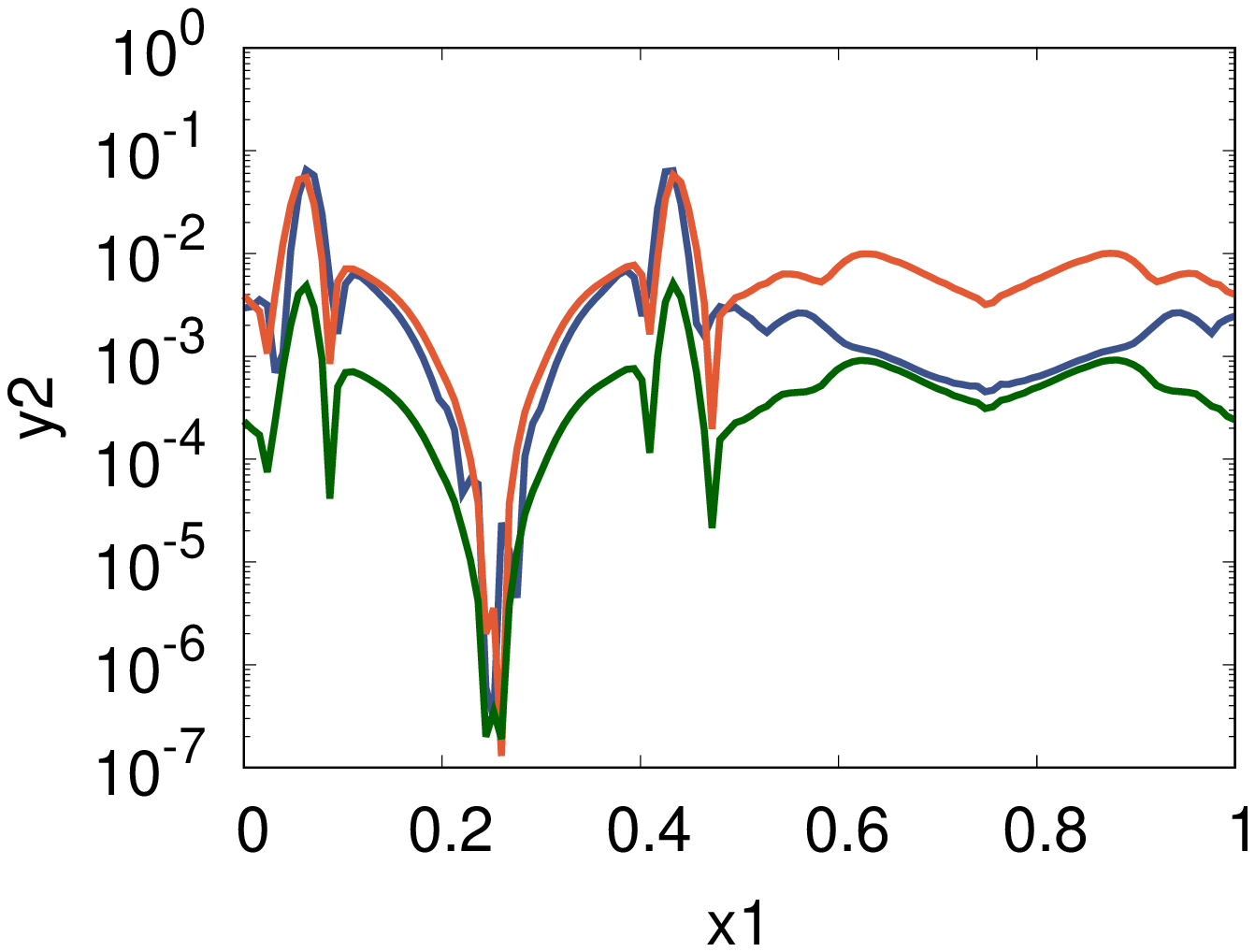} & \psfrag{y3}[b][b][0.9]{$Kn_{\Tp{}}$}
        \hspace{-0.4cm}\includegraphics[scale=0.18,angle=-90]{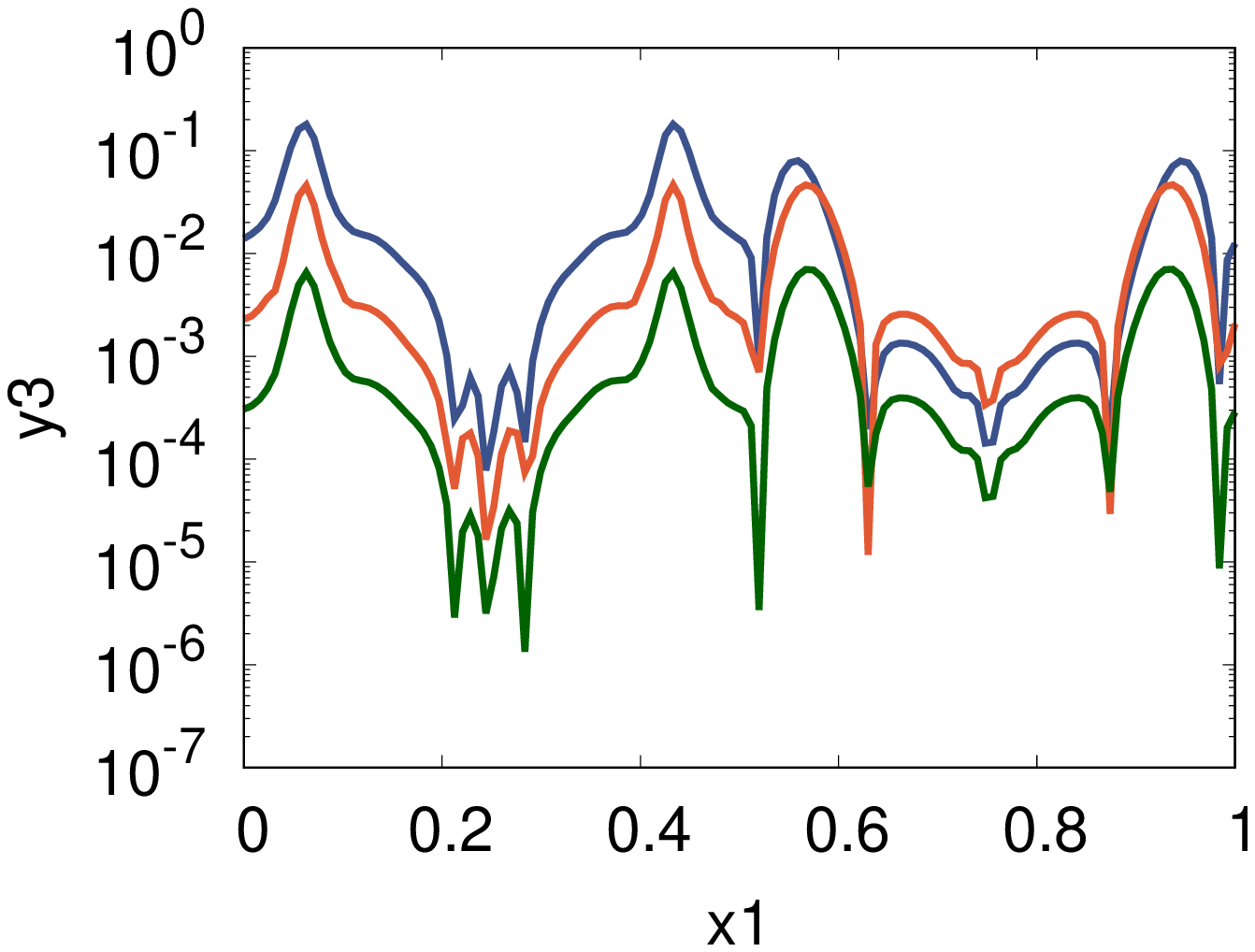}  
    \end{tabular}
    \caption{Profiles of Knudsen number based on (a) phase solid volume fraction, (\ref{knalp}), (b) phase velocity, (\ref{knvel}), and (c) phase granular temperature, (\ref{knthe}), for Case-E at $t^*=25$ (top) and $t^*=50$ (bottom). 
    Blue, red and green lines refer to hydrodynamic variables of phases $i$, $j$, and mean, respectively.\label{knprofile}}
\end{figure}
For the quasi 1-D simulations, instead of computing a $Kn$ value averaged over the computation domain that would be misleading, the Knudsen profiles were computed along $x$-direction at various time instants. An example of the profiles of the three Knudsen numbers defined by \eqref{knalp}, \eqref{knvel} and \eqref{knthe} for each phase ($i$ and $j$) and mixture is shown in figure \ref{knprofile} for Case-E at $t^*=25$ (top) and $t^*=50$ (bottom).  One can see that all three Knudsen numbers are in the range of several orders of magnitude from approximately $10^{-6}$ to $10^{-1}$. So that, the low-Knudsen assumption is valid for our simulation cases. 


\subsection{Wall-bounded segregating bidisperse granular flow}\label{Section:segflow}

\begin{figure}
    \centering
    \includegraphics[scale=0.4,trim=2.cm 1.6cm 2.cm 1cm, clip=true]{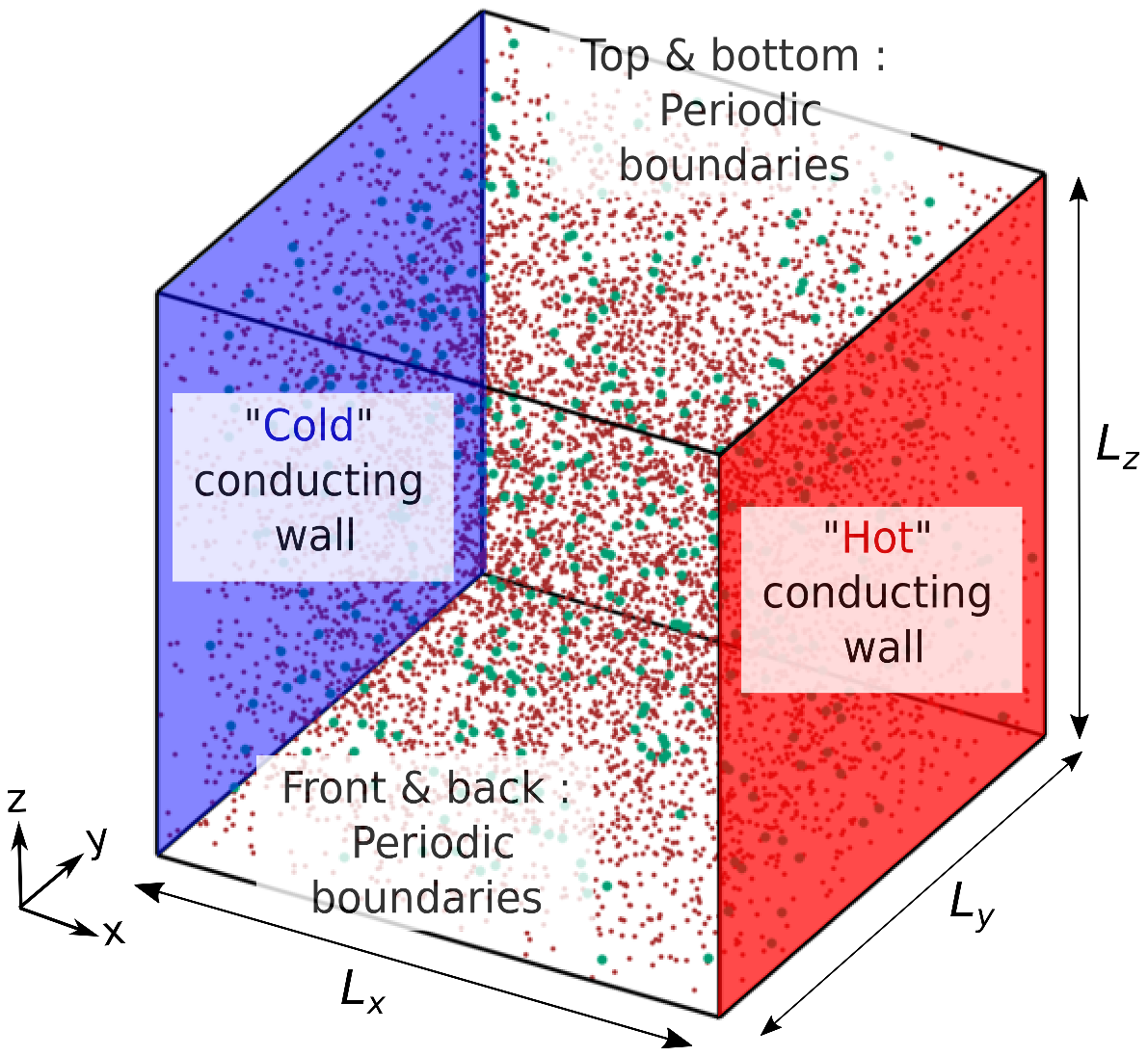}
    \caption{Computational domain of segregating granular flow simulation. Different granular temperatures were imposed to left (cold-blue) and right (hot-red) surfaces. Both surfaces are conducting wall with an effective work function difference between particles and walls of $0.001\,\text{eV}$. Electric potential was equal to zero at walls. Green spheres refer to larger particles while brown particles refer to smaller particles.}
    \label{hotColdWall}
\end{figure}

In the previous validation cases, the granular temperature for each phase evolves to the mixture granular temperature. In this section, we further validate the proposed model with a 3D steady segregating granular flow simulation where the non-equipartition of granular temperature persists. Figure \ref{hotColdWall} shows the computational domain which is a cubic box with a size of $L_x=L_y=L_z=24\,d_{pj}$ ($d_{pj}$ is the larger particle diameter, see table \ref{Table:segregatingFlowCases}). A granular temperature gradient along $x$-direction is imposed by left (cold) and right (hot) stationary conducting walls with different constant granular temperatures. 

By following \cite{galvin2005role}, for the hard-sphere simulation, we imposed post-collision velocities of a particle colliding with a wall as
\begin{eqnarray}
    c_{post,x} & = & \sqrt{-\frac{4}{3}\frac{\Theta_{wall}}{m_{pi}}\ln{z_1}}\cos{(2\pi z_2)}, \label{postCollVelx}\\
    c_{post,y} & = & \sqrt{-\frac{4}{3}\frac{\Theta_{wall}}{m_{pi}}\ln{z_3}}\cos{(2\pi z_4)}, \label{postCollVely} \\
    c_{post,z} & = & \sqrt{-\frac{4}{3}\frac{\Theta_{wall}}{m_{pi}}\ln{z_5}}\cos{(2\pi z_6)}, \label{postCollVelz}
\end{eqnarray}
where $\Theta_{wall}$ is the imposed granular temperature at wall, $z_1-z_6$ are random numbers generated from a uniform distribution within the interval of $[0, 1]$. The post-collision velocity along $x$-direction, $c_{post,x}$, is reversed according to the inward normal vector of the wall. We also imposed the effective work function difference between particles and conducting walls, $\varphi_{w-p}$, to maintain charge in the domain. Periodic boundary conditions are imposed in $y$- and $z$-directions. We do not account for any external force, therefore the time-averaged mean velocity for each phase is equal to zero. 

The particle densities and diameters and the domain-averaged solid volume fraction for each phase are listed for a simulation case (Case-H) in table \ref{Table:segregatingFlowCases}. The granular temperature ratio between cold and hot walls was set to $\Theta_{\text{hot-wall}}/\Theta_{\text{cold-wall}}=6$. The particles were randomly distributed in the domain and velocities were initialised with a Maxwellian distribution with a zero mean velocity for each solid phase. The total number of particles was approximately $\SI{12500}{}$. The particle velocities were updated when they contacted with walls based on \eqref{postCollVelx}, \eqref{postCollVely} and \eqref{postCollVelz} and the kinetic energy dissipated through inelastic collisions between particles (the restitution coefficient, $e_c$, was 0.9). The effective work function difference between particles and walls was set to $\Delta \varphi_{w-p} = \varphi_w - \varphi_{p} = \SI{0.001}{\electronvolt}$. The initial charge on particles was equal to $Q_{pi} = Q_{pj} = \SI{1}{\femto\coulomb}$. 

\begin{table}
    \centering
    \begin{tabular}{cccccccc}
        Case & Phase &$d_{p}$ & $\rho_{p}$  & $\langle \alpha_{p} \rangle$
             &  $\Theta_{\text{hot-wall}}/\Theta_{\text{cold-wall}}$ &  $N_p$ & $n_c/N_p$
        \\
             &      &[\SI{}{\micro\meter}] & [\SI{}{\kilo\gram\per\meter^3}] 
             & [-] & [-] &  [-]  &
        \\
        \hline
        H & $i$ & 150 & 1500 & 0.04  & 6 & $\SI{8450}{}$ & $\approx$ 4100\\
          & $j$ & 300 & 1500 & 0.16  & 6 & $\SI{4028}{}$ & \\ 
        \hline
    \end{tabular}
    \caption{Particle properties and flow parameters for a 3D steady segregating flow configuration. Case-H refers to a bidisperse case with a particle diameter ratio of $R_d$ = 2 and the same particle density (the mass ratio is equal 8). Initial charge for each phase was imposed to $\SI{1}{\femto\coulomb}$. The restitution coefficient was set to $e_c = 0.9$.\label{Table:segregatingFlowCases}}
\end{table}

For the hard-sphere simulation, we monitored the time evolution of domain-averaged kinetic energy and charge to ensure that the flow reached a statistically steady state. We computed the Eulerian variables such as granular temperature, mean charge for each phase and mixture granular temperature for each time-step and the time-averaging of these variables were carried out during a duration while an additional $\SI{4000}{}$ number of collisions per particle occurred. To compute the Eulerian variables, we divided the computational domain into cubic cells with a length of $2\,d_{pj}$ (the mesh configuration is $12\times12\times12$). As we determined the length of the unit cell, we ensured that a further mesh refinement had no effect on the computed variables. The particle Lagrangian quantities such as velocity and charge were mapped by using a simple injection method where any Lagrangian properties were averaged into a cell without using any smoothing function. A further averaging over periodic directions ($y$ and $z$) was undertaken to generate the profiles along $x$-direction. The same mesh configuration was also used to solve the Poisson's equation to compute the electric field at contact point for charge transfer by a finite difference approach during the simulation (see Appendix-B for the Lagrangian hard-sphere method). 

For the charge transfer model, a boundary condition for a conducting wall is necessary. Following our earlier work~\citep{kolehmainen_eulerian_2018}, charge balance for each phase ($h = i,j$) at a conducting wall is computed by:
\begin{equation}
\sigma_{qh,w} \left( \frac{\Delta \varphi_{w-p}}{\delta_c e} - \frac{2Q_{ph}}{\pi \varepsilon_0 d_{ph}^2} \right) + \kappa_{qh} \boldsymbol{n}_w \cdot \boldsymbol{\nabla} Q_{ph} - \left( \sigma_{qh} - \sigma_{qh,w} \right) \boldsymbol{E} \cdot \boldsymbol{n}_w = 0.
\end{equation} 
Here, $\boldsymbol{n}_w$ is the inward normal unit vector of wall and triboelectric conductivity at the wall, $\sigma_{qh,w}$, is given by:
\begin{equation}
    \sigma_{qh,w} = \varepsilon_0 \alpha_{ph} \left( 1 + 2 (1+e_w) \alpha_{ph} g_0 \right)\left( \frac{6}{ \sqrt[10]{2} (1+e_w) d_{ph}^2 } { \left(\frac{15\,m_{ph}(1-\nu_{ph}^2)}{8Y_{ph}\sqrt{d_{ph}}}\right)^{\frac{2}{5}} \Gamma(\frac{9}{10}) } \right) \Theta_{ph}^{9/10},
\end{equation} 
where $e_w$ is the restitution coefficient of wall-particle collisions which is set to the particle-particle restitution coefficient, $e_c$. The symbols $\sigma_{qh}$ and $\kappa_{ph}$ are the triboelectric conductivity and diffusivity defined in (\ref{Eq:sigmaii}) and (\ref{Eq:kappaii}), respectively.

\eqref{Eq:SetSegFlow} shows the complete set of the equations simplified for a 1-D steady wall-bounded flow configuration:

\begin{equation}\label{Eq:SetSegFlow}
\left\{\begin{array}{l}
    \frac{d}{dx}\Big(\frac{1}{\rho_{pi}}\Big[P_{pi}^{kin} + \sum\limits_{l = i,j}\theta_{il}\Big] + \frac{1}{\rho_{pj}}\Big[P_{pj}^{kin} + \sum\limits_{l = i,j}\theta_{jl}\Big] \Big) = 0 \\[2pt]
    \frac{d}{dx}\sum\limits_{l = i,j}q_{hl} = \sum\limits_{l = i,j}\gamma_{hl} \\[2pt]
    \frac{d}{dx}\sum\limits_{l = i,j}\theta_{hl}^{q} = \sum\limits_{l = i,j}\chi_{hl}^q 
    \end{array}  \right.
\end{equation} 

The first line is the mixture momentum balance for which the flux term and the kinetic pressure are defined in \eqref{fluxMom} and \eqref{Eq:kineticPressure}. The energy balance for phase $h$ is given in the second line with the flux and the source terms defined in \eqref{fluxGran} and \eqref{sourceGran}. The last line represents the charge balance for phase $h$ with the flux and source terms defined in \eqref{Eq:FluxCharge} and \eqref{Eq:SourceCharge}. All terms have been simplified for a 1-D flow configuration (see figure \ref{hotColdWall}) where the mean phase velocities are equal to zero.

\begin{figure}
    \centering
    \psfrag{x1}[][][0.8]{$x/L$}
    \begin{tabular}{ccc}
        (a) & (b) & (c) \vspace{-0.25cm}\\
        \psfrag{y1}[][bl][0.9]{$\alpha_{pi}$}
        \hspace{-0.6cm}\includegraphics[scale=0.2,angle=-90]{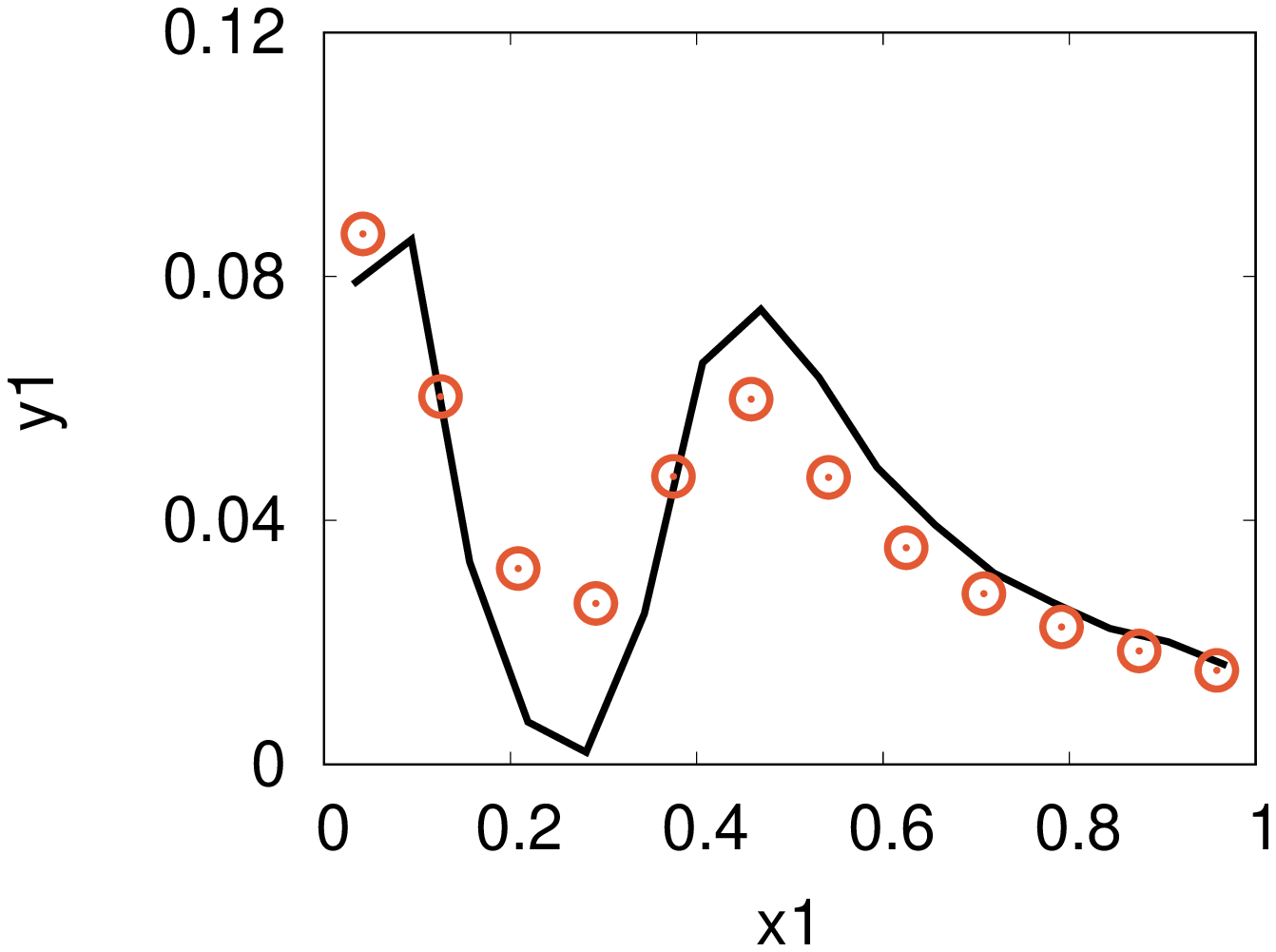} & \psfrag{y1}[][bl][0.9]{$\Theta_{pi}/\Theta_{cold-wall}$}
        \hspace{-0.6cm}\includegraphics[scale=0.2,angle=-90]{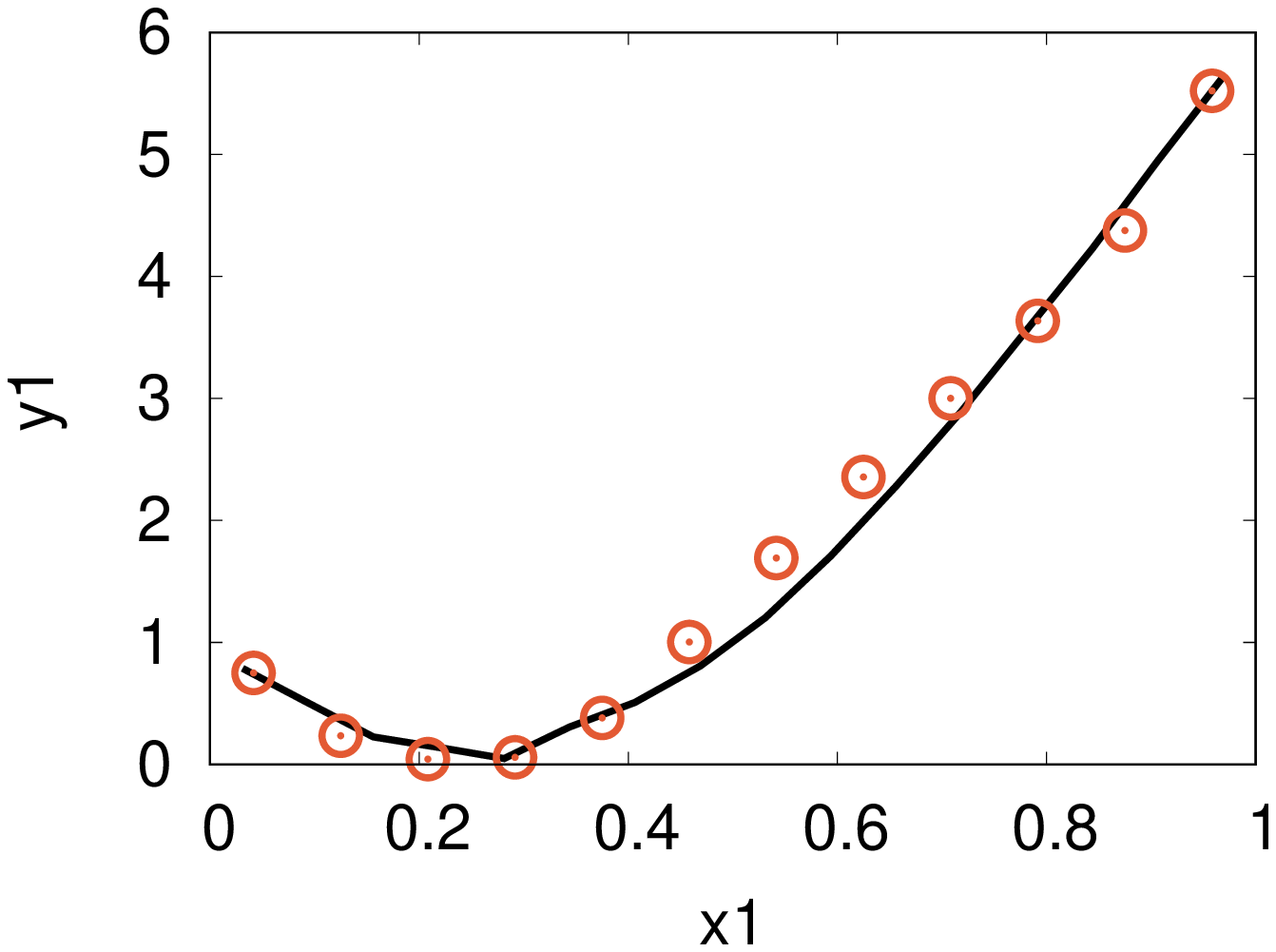} & \psfrag{y1}[][bl][0.9]{$Q_{pi}/Q_{eq,i}$}
        \hspace{-0.6cm}\includegraphics[scale=0.2,angle=-90]{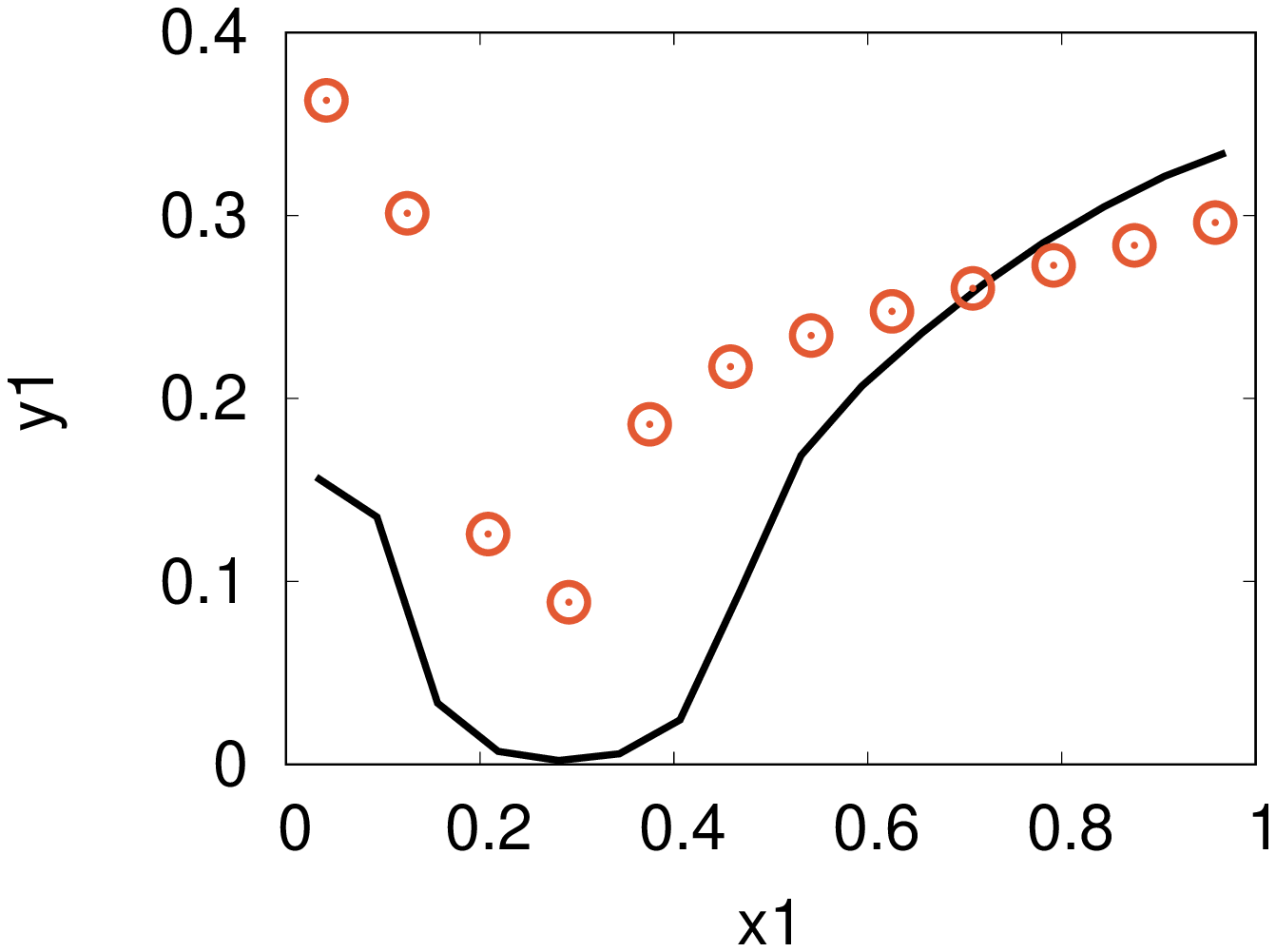}  \\
    \end{tabular}
    \begin{tabular}{ccc}
        (d) & (e) & (f) \vspace{-0.25cm}\\
        \psfrag{y1}[][bl][0.9]{$\alpha_{pj}$}
        \hspace{-0.6cm}\includegraphics[scale=0.2,angle=-90]{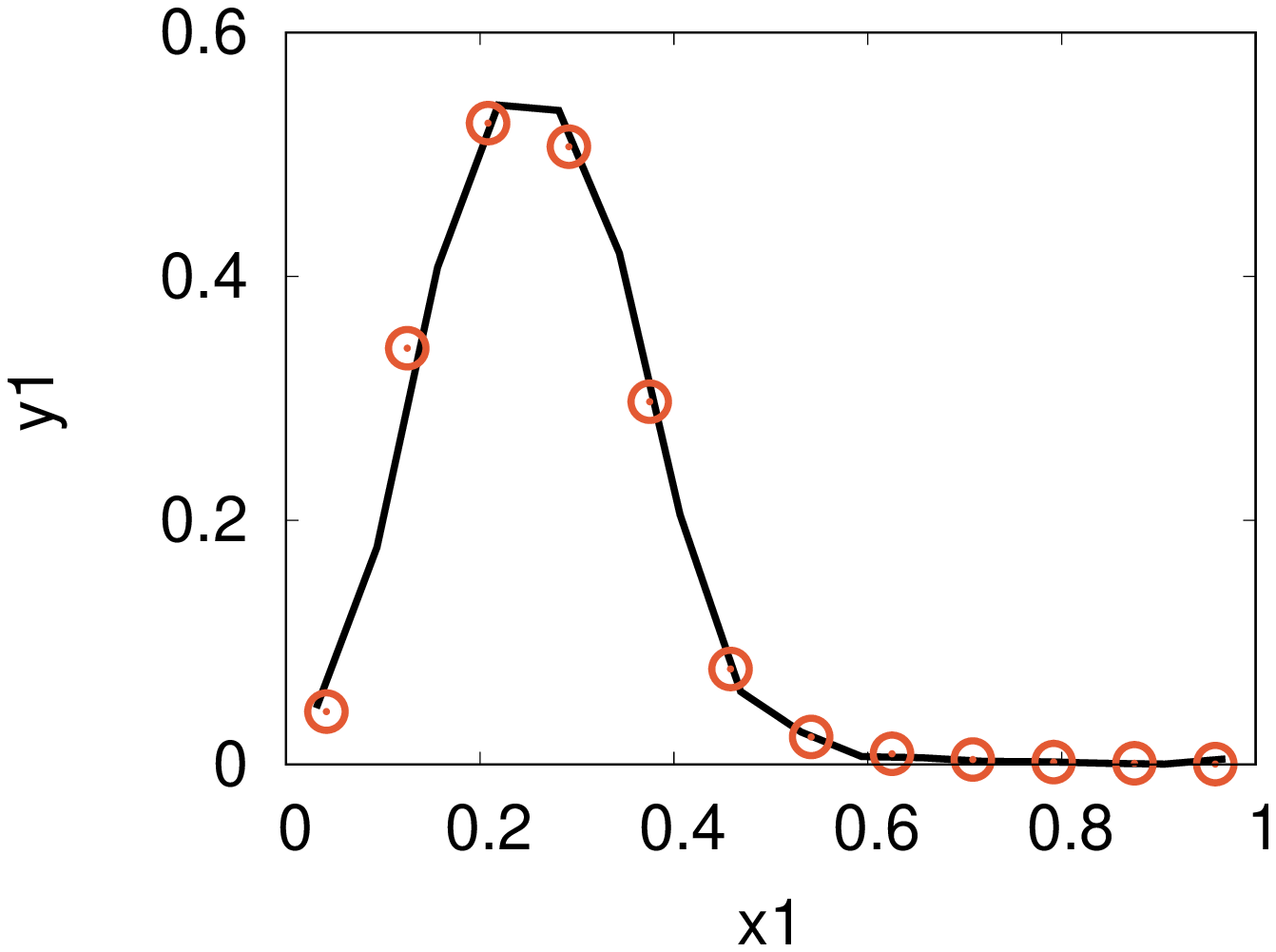} & \psfrag{y1}[][bl][0.9]{$\Theta_{pj}/\Theta_{cold-wall}$}
        \hspace{-0.6cm}\includegraphics[scale=0.2,angle=-90]{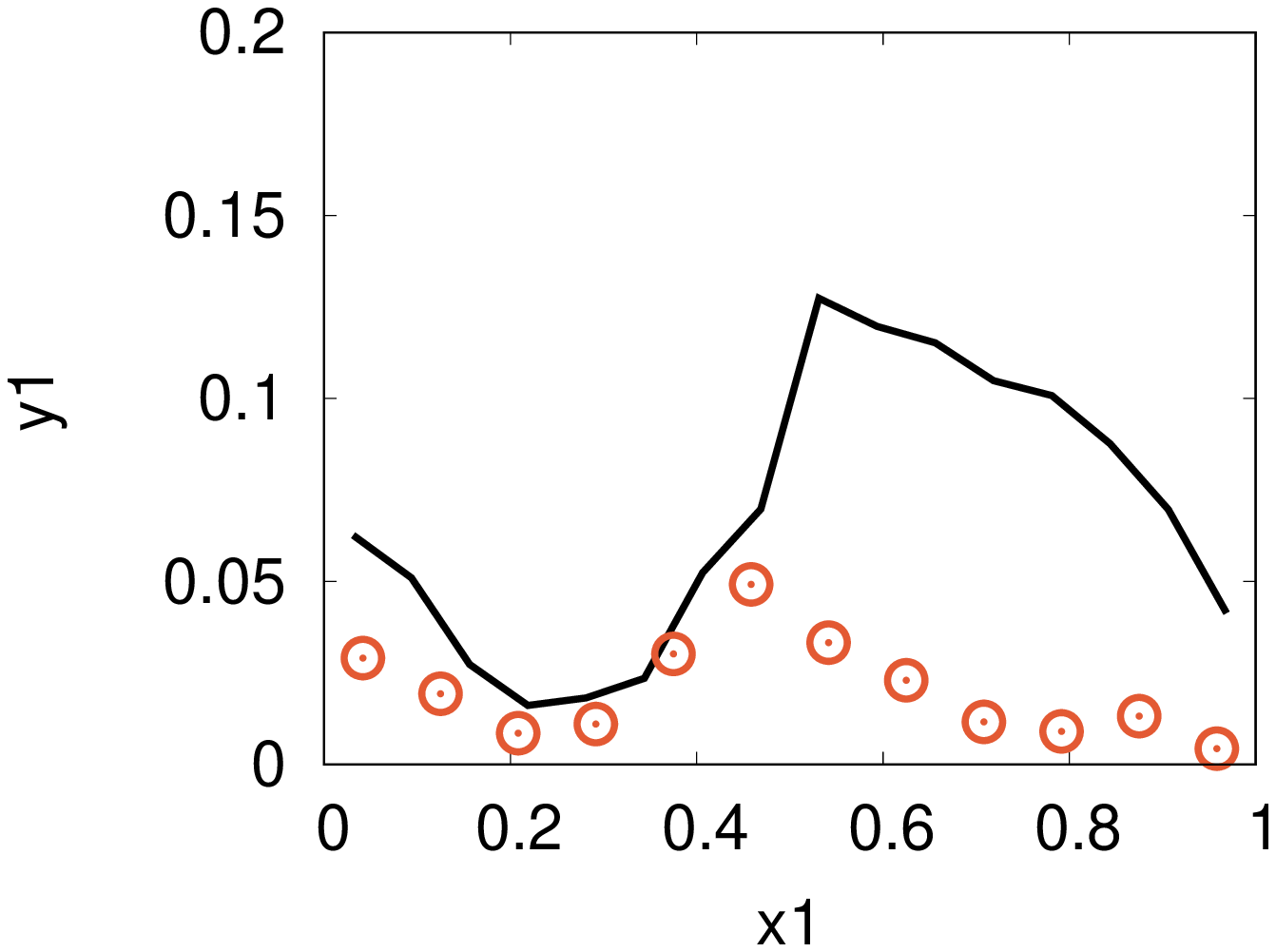} & \psfrag{y1}[][bl][0.9]{$Q_{pj}/Q_{eq,j}$}
        \hspace{-0.6cm}\includegraphics[scale=0.2,angle=-90]{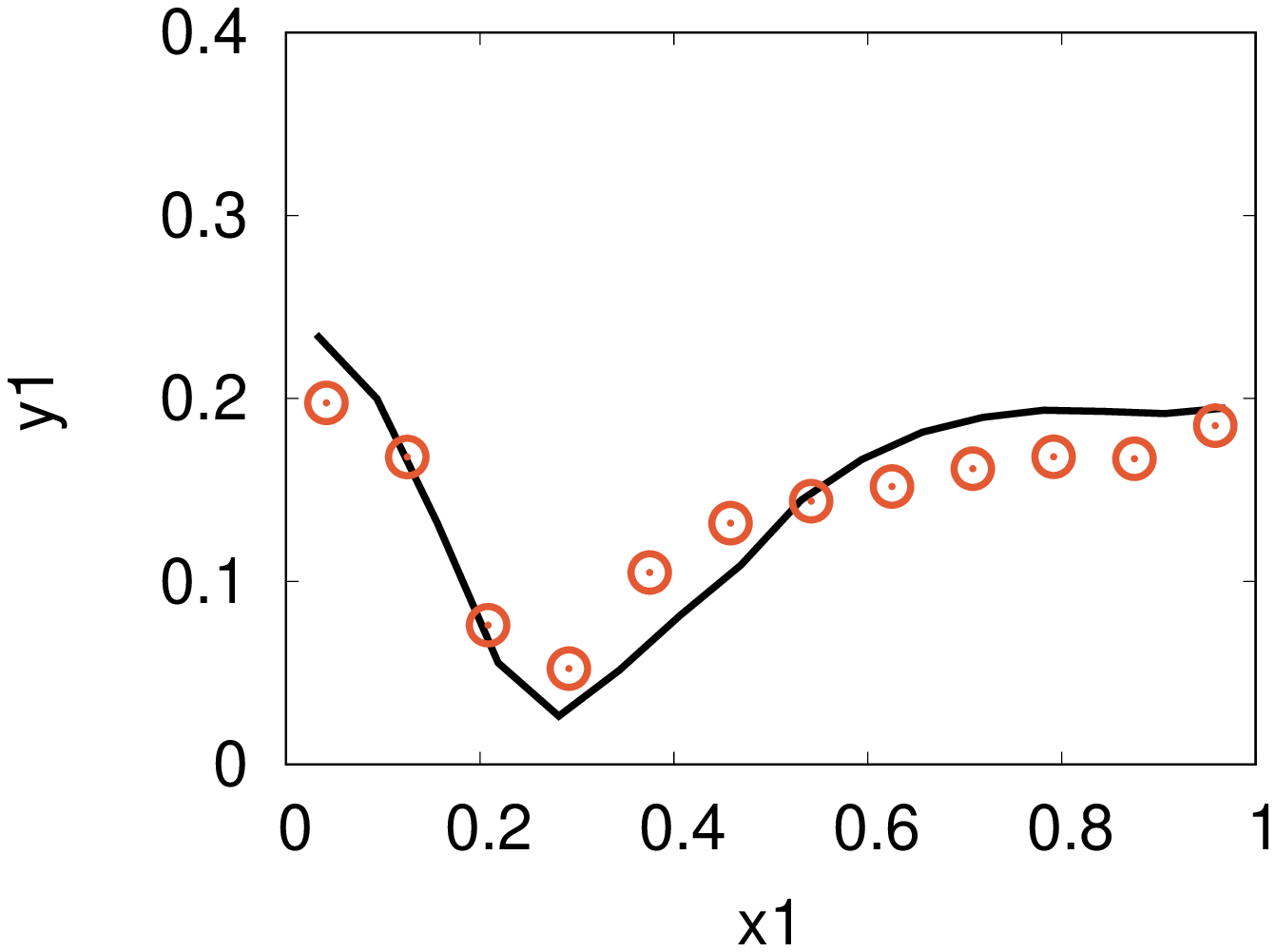}\\
    \end{tabular}
    \caption{Profiles of phase solid volume fractions (a and d), phase scaled granular temperatures (b and e) and phase scaled mean charges (c and f) for Case-H. ${\color{Black}\rule[0.5ex]{10pt}{0.75pt}}$: Eulerian model predictions and ${\color{RedOrange}\boldsymbol{\odot}}$: hard sphere simulation results. Variables were scaled by using $\Theta_{cold-wall}$ and $Q_{eq,h}$ from \eqref{Eq:Qeq}.
    \label{ResultCaseI}}
\end{figure}

Figure \ref{ResultCaseI} shows profiles of the phase solid volume fraction, the scaled phase granular temperature and the scaled phase mean charge for Case-H along $x$-direction. The solid volume fraction distributions (figures \ref{ResultCaseI}(a) and (d)) show a clear segregation of particles. The larger particles (the phase $j$) are mainly located around $x/L\approx 0.3$ and very few of them stay close to the hot wall. The smaller particles (the phase $i$) tend to group around the larger particles ($x/L\approx 0.1$ and $x/L\approx 0.4$) and, similarly to the larger particles, the number of the smaller particles decreases close to the hot wall. The model predictions (${\color{Black}\rule[0.5ex]{10pt}{0.75pt}}$) of both solid volume fraction profiles are good in agreement with hard-sphere simulation results (${\color{RedOrange}\boldsymbol{\odot}}$). The scaled granular temperature (the granular temperature is divided by cold wall temperature) for the small particles (figure \ref{ResultCaseI}(b)) has a v-shape profile with a minimum value at the highest total solid concentration location ($x/L\approx 0.3$). The scaled temperature for the larger particles (figure \ref{ResultCaseI}(e)) has also a v-shape profile up to the half of the domain but it decreases to zero at $x/L>0.5$ due to absence of the larger particles. At $x/L>0.5$, the model overestimated the granular temperature which points out the limitation of the proposed model. It is worth to note that the larger particles are located away from walls and mainly interacted with small particles. Therefore, the imposed wall temperatures have a very limited effect on the fluctuating energy of the larger particles and in the very dilute region ($x/L>0.5$). A small number of particles stay nearly still. 

Figures \ref{ResultCaseI}(c) and (f) present the mean charge profiles scaled by the equilibrium charge  for both phases. The equilibrium charge, $Q_{eq,h}$, is defined as:
\begin{equation}\label{Eq:Qeq}
    Q_{eq,h} = \frac{1}{2}\frac{\pi\,\varepsilon_{0}}{\delta_c\,e}\Delta \varphi_{w-p}\,d_{ph}^2.
\end{equation}

We refer the equilibrium charge as the maximum charge that a particle can acquire by interactions of an isolated particle with a wall (without the electric field effect on the charge transfer) \citep{kolehmainen_triboelectric_2017}. One can see that the maximum value of the scaled charge for both phases is smaller than one and that shows how the electric field hinders the total charge in the domain. As we discussed in previous sections, the proposed model has been assessed for granular flows with a mixture solid volume fraction between $0.2$ and $0.4$. For dilute flows ($\alpha_{pi}<0.1$), the correlation between charge and velocity, or namely the self-diffusion of charge, should be accounted for which is crucial to accurately predict charge distribution at $x/L>0.5$ ($\alpha_{pi}<0.05$ and $\alpha_{pj}<0.005$). By following \cite{kolehmainen_eulerian_2018}, an ad-hoc model for the self-diffusion of charge has been included as:
\begin{equation}
    \kappa_{qh}^{+} = \frac{d_{ph} \sqrt{\Theta_{ph}/m_{ph}}}{9 \sqrt{\pi} g_0 V_{ph}},
\end{equation}
where $V_{ph}$ is the volume of a particle within the solid phase $h$. The proposed model prediction (${\color{Black}\rule[0.5ex]{10pt}{0.75pt}}$) is in very good agreement with the hard-sphere simulation results (${\color{RedOrange}\boldsymbol{\odot}}$) for the larger particle (figure \ref{ResultCaseI}(f)), even for dilute region ($x/L>0.5$). However, the predicted charge profile for the smaller particle is less accurate and the total amount of charge is underestimated (figure \ref{ResultCaseI}(c)). 

\begin{figure}
    \centering
    \psfrag{x1}[][][0.8]{$x/L$}
    \begin{tabular}{ccc}
        (a) & (b) & (c)  \vspace{-0.25cm}\\
        \psfrag{y1}[][bl][0.9]{$\Tp{h}/\Tp{m}$}
        \hspace{-0.6cm}\includegraphics[scale=0.2,angle=-90]{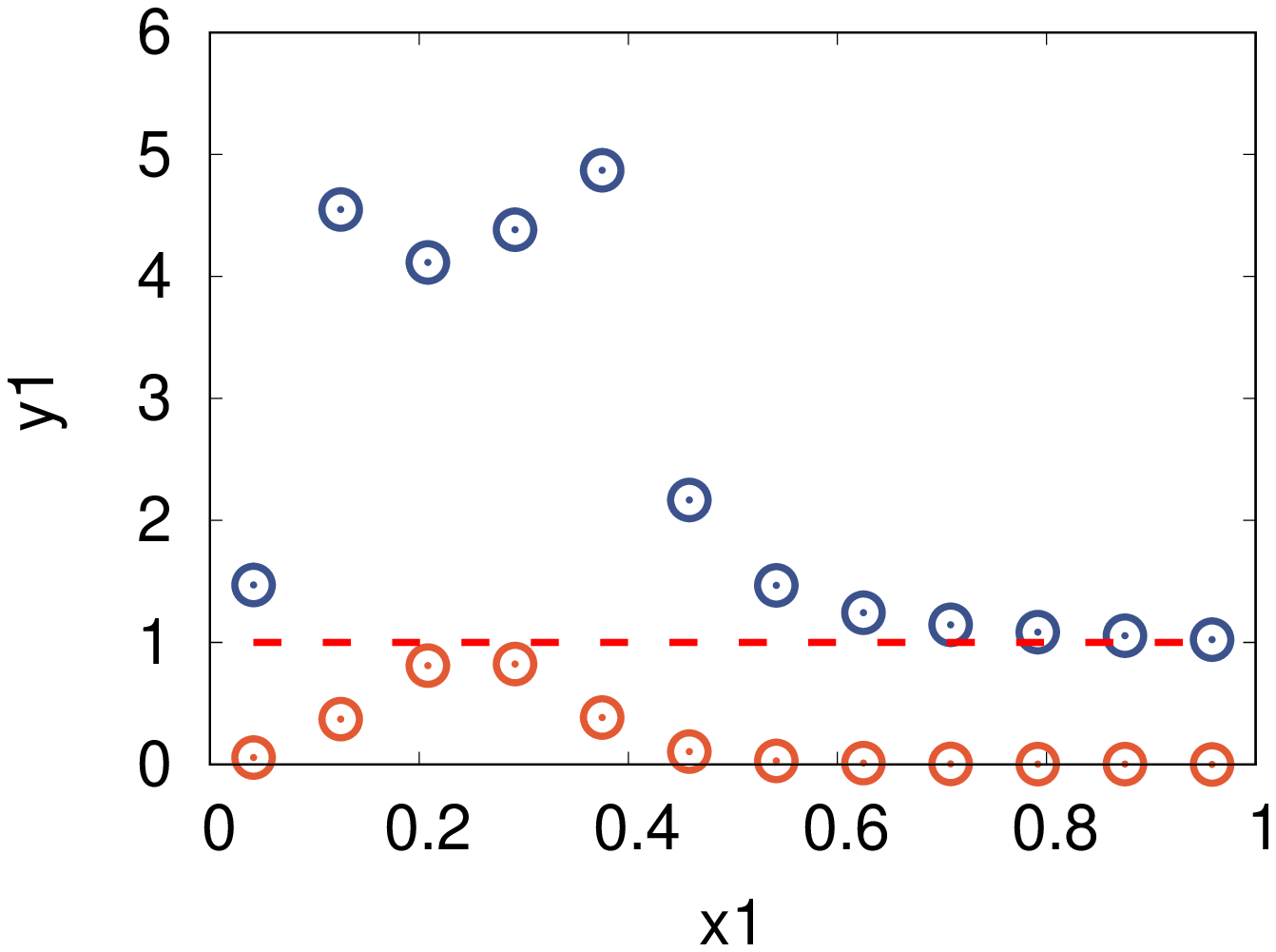} & \psfrag{y1}[][bl][0.9]{$Q_{pi}/Q_{eq,i}$}
        \hspace{-0.6cm}\includegraphics[scale=0.2,angle=-90]{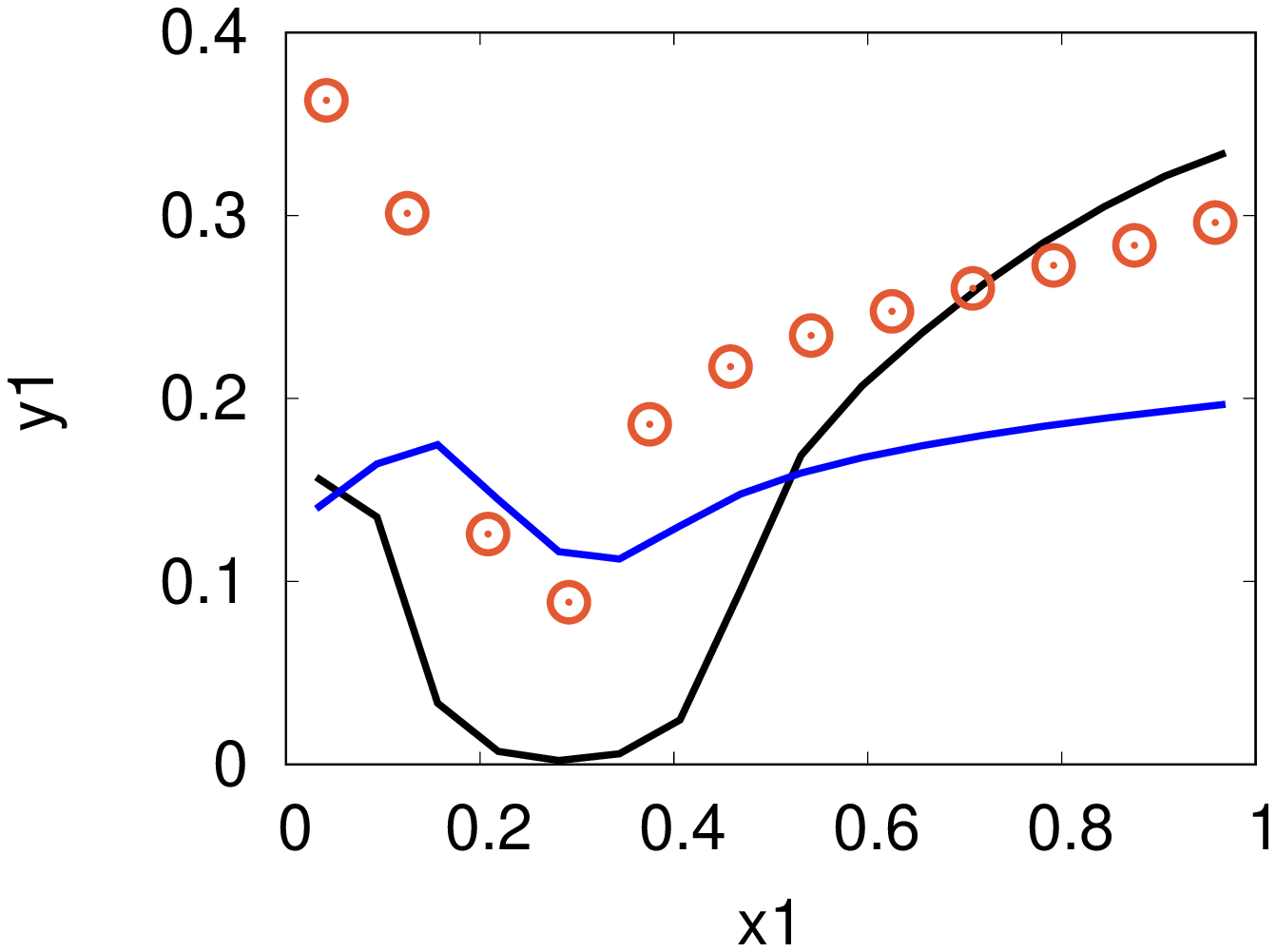} & \psfrag{y1}[][bl][0.9]{$Q_{pj}/Q_{eq,j}$}
        \hspace{-0.6cm}\includegraphics[scale=0.2,angle=-90]{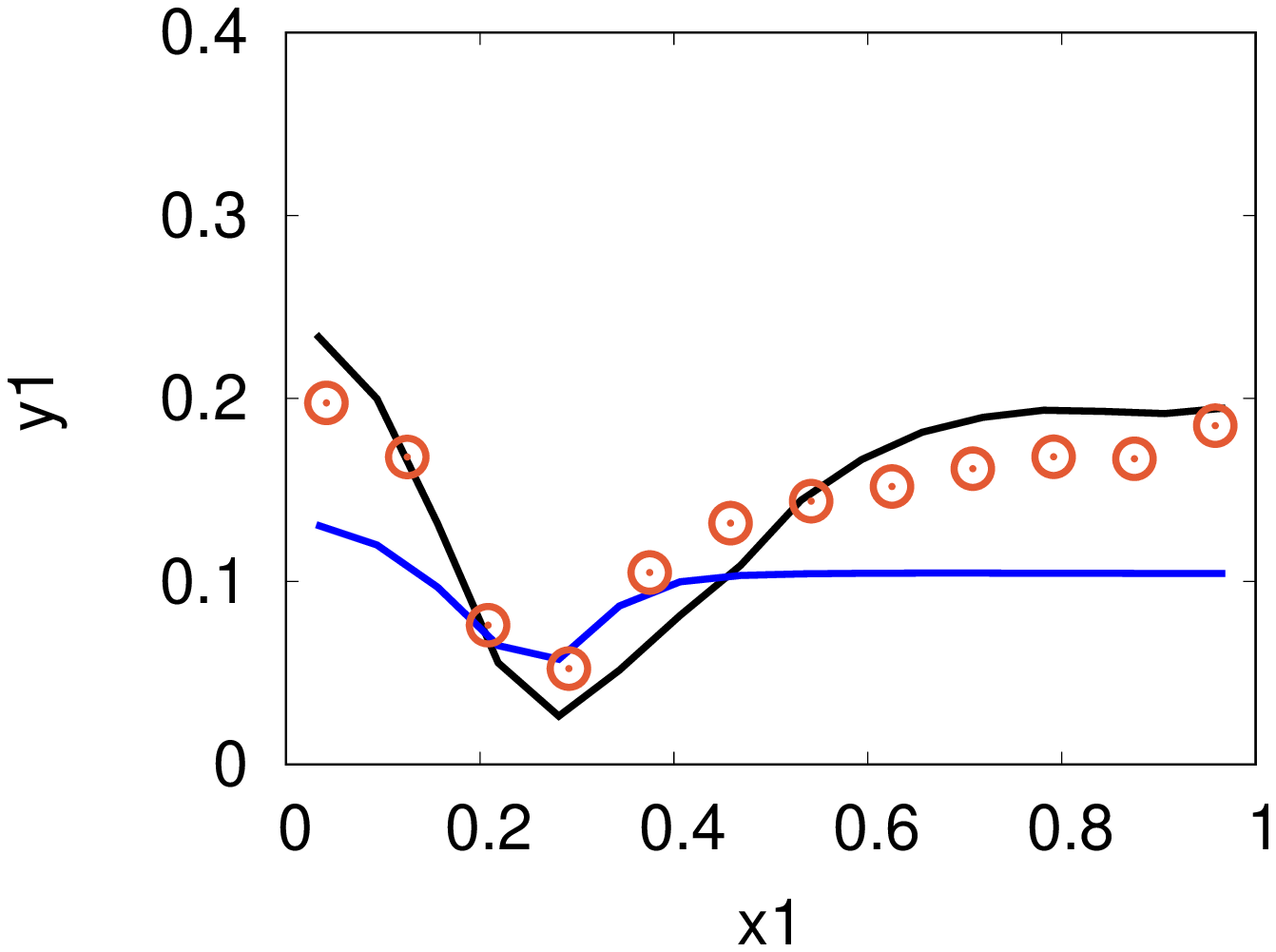} 
        \\
    \end{tabular}
    \caption{(a) Phase granular temperatures scaled by the mixture granular temperature, \eqref{scaledtheta}, with hard-sphere simulations (b) and (c) scaled phase mean charges for Case-H. ${\color{Black}\rule[0.5ex]{10pt}{0.75pt}}$: Eulerian model predictions with the non-equipartition of granular temperature, ${\color{Blue}\rule[0.5ex]{10pt}{0.75pt}}$: Eulerian model predictions with the mixture granular temperature and ${\color{RedOrange}\boldsymbol{\odot}}$: hard sphere simulation results. Phase mean charges are scaled by \eqref{Eq:Qeq}.
    \label{ResultCaseIMix}}
\end{figure}

To discuss why the non-equipartition of granular temperature is crucial for the charge distribution, we computed the mean charge with the proposed model but instead of using different granular temperatures for each phase, we used the mixture granular temperature \eqref{scaledtheta} only for the constitutive equations \eqref{Eq:SetSegFlow} by imposing $\Tp{i}$ = $\Tp{j}$ = $\Tp{m}$ (that leads $B$ = 0). The flux and the source terms in the charge transfer equation are simplified for the equal-partitioning of granular temperature as follows:
\begin{eqnarray}
    & & \theta_{ih}^q =  - n_{pi}n_{ph}\Big(\frac{m_{pi}m_{ph}}{\Tp{i}\Tp{h}}\Big)^{3/2}\mathcal{A}^*\epsilon_0g_0\frac{d_{pih}^3}{8}\frac{5\sqrt{\pi}}{21}N_1 \nonumber \\
    & & \times \bigg[ -E + \frac{d_{pih}}{2} \frac{1}{\pi\epsilon_0}\bigg[ \Big(\frac{Q_{ph}}{d_{ph}^2} - \frac{Q_{pi}}{d_{pi}^2}\Big)\times\frac{d}{dx}\Big(\ln\Big( \frac{n_{ph}}{n_{pi}}\Big)\Big)
     + \frac{d}{dx}\Big(\frac{Q_{ph}}{d_{ph}^2}\Big) + \frac{d}{dx}\Big(\frac{Q_{pi}}{d_{pi}^2}\Big)\bigg]\bigg],\label{Eq:segFluxQ} \\
    & & \chi^{q}_{ih} = n_{pi}n_{ph}\Big(\frac{m_{pi}m_{ph}}{\Tp{i}\Tp{h}}\Big)^{3/2}\mathcal{A}^*\epsilon_0g_0d_{pih}^2\frac{5\sqrt{\pi}}{28}N_1\bigg[ \frac{1}{\pi\epsilon_0}\Big(\frac{Q_{ph}}{d_{ph}^2} - \frac{Q_{pi}}{d_{pi}^2}\Big) - \frac{d_{pih}}{6}E\frac{d}{dx}\Big(\ln\Big( \frac{n_{ph}}{n_{pi}}\Big)\Big)\bigg].\nonumber\\\label{Eq:segSourceQ}
\end{eqnarray}
To emphasise how the non-equipartition of granular temperature takes place, the phase granular temperature with hard-sphere simulation is scaled by the mixture granular temperature and shown in figure \ref{ResultCaseIMix}(a). The larger particles contribute a big portion of the mixture granular temperature at $0.1<x/L<0.3$ whereas the small particles generate the mixture granular temperature at $x/L>0.6$. Figures \ref{ResultCaseIMix}(b) and (c) show profiles of the scaled mean charge for each phase. One can see that using the mixture granular temperature leads to a nearly uniform charge distribution. Particularly, the charge conductivity, $\bm{\upsigma}_{ih}$, is overestimated for the larger particles at $x/L>0.6$ due to overestimation of the granular temperature of the phase $j$ (orange points in figure \ref{ResultCaseIMix}) by using the mixture granular temperature (dash red line in figure \ref{ResultCaseIMix}), therefore it leads to a smoother distribution with an underestimation of mean charge. As shown in \eqref{Eq:segFluxQ}-\eqref{Eq:segSourceQ}, the charge distribution along $x$-direction is driven only by charge difference between solid phases and the gradient of natural logarithm of solid volume fractions for the case with the equipartition of granular temperature. The underestimation of mean charge further reduces charge transfer due to the charge differences between phases. The slight variation on the left side ($0.2<x/L<0.4$) results from the segregation of particles and the formation of band of smaller particles (phase $i$) around the larger particles (phase $j$). These results show that the non-equipartition of fluctuating kinetic energy should be accounted for bidisperse granular flows with charge particles such as wall bounded flows commonly associated with charged particle transport in powder technology.  

\section{Conclusion}\label{section:conclusion}
In this study, we have revisited kinetic-theory based hydrodynamic equations and derived charge transport equation for bidisperse granular flows with tribocharging. Each  solid phase has separate mean velocity, total fluctuating kinetic energy, which is the sum of the granular temperature and the trace of fluctuating kinetic tensor, charge variance and mean charge. To close mass, momentum, total kinetic energy and mean charge balance equations, a Maxwellian distribution for particle velocity and charge without a cross-correlation (an assumption of both velocity and charge are independent variables) has been used for local-averaging of the Boltzmann equation. The constitutive relations of collisional flux and source terms for momentum, granular temperature and charge balance equations, which account for the rate of change of the quantities between phases and within a phase, are then presented. We introduced a finite-volume scheme to discretize and solve the transport equations and assessed the proposed models through various hard-sphere simulations of three-dimensional spatially homogeneous and quasi-one-dimensional spatially inhomogeneous bidisperse granular gases. For these cases, the mixture solid volume fraction were set into a range from 0.2 to 0.4. However, the hard-sphere algorithm cannot handle granular flows with mixture solid volume fraction higher than 0.4 due to time-step limitation. We will improve our hard-sphere time-stepping algorithm to study these flows in a future study. In these simulations, we varied particle diameter and density ratios, initial phase charges and phase granular temperatures. The proposed model predictions were in very good agreement with the hard-sphere simulation results. Finally, a segregating bidisperse granular flow in a wall-bounded domain where the non-equipartition of granular temperature persisted was studied with a steady state solution of the proposed model and hard-sphere simulation. This steady-state solution had an acceptable level of accuracy with the hard-sphere simulation results. This case also showed the importance of accounting for the charge-velocity correlation in the dilute region and the non-equipartition of granular temperature to have an accurate charge distribution.

In this study, we limited hard-sphere simulations and the proposed model predictions to elastic granular flows (except the three-dimensional segregating bidisperse flow). For a further study, we will extend the proposed models by the Chapman-Enskog expansion by following \cite{iddir_modeling_2005} or couple the revisited Enskog theory (e.g. \cite{garzo_enskog_2007}) with the developed charge transfer models to consider a wider range of inelastic bidisperse granular flows.

We will also extend the proposed model with accounting for the charge-velocity correlation which is significant in dilute granular flows~\citep{montilla2020modelling}. In this study, we only focus on the granular flows without interstitial fluid effect but the interstitial fluid has a huge impact on granular material hydrodynamics in particle technology applications such as fluidised bed and pneumatic conveying. Introducing the fluid phase will be also a topic of a future study. Additionally, the charge variance is assumed to be a constant thorough this study but as discussed by \citet{singh2019electrification}, the charge variance plays a role in agglomeration of particles in homogeneous granular gases. It is necessary to develop the transport equation for charge variance and study its effects on the charge transport properties. 

\section*{Acknowledgements}
The authors thank Prof. Sankaran Sundaresan from Princeton University for all of his time, support and fruitful discussions and Josh Williams for proof-reading of the manuscript. 

\section*{Declaration of Interests}
The authors report no conflict of interest.

\clearpage

\appendix

\newpage
\section{Derivations of Flux and Source Terms for Charge Transport Equation}\label{appA}

The total collisional operator for the charge transfer given by \eqref{Eq:CollChargeDef} is decomposed into two parts: same particle-type and different particle-type collisions. In this appendix, the theoretical development for the different particle-type collisions contributions is given. The interested readers are referred to \cite{kolehmainen_eulerian_2018} for same particle-type collisions. With \eqref{Eq:Fluxdefinition} and \eqref{Eq:Sourcedefinition}, the collisional operator between different particle-phase $i$ and $j$ can be recasted as:
\begin{eqnarray}
    \mathcal{C}_{ij}(q_{pi}) & = & g_0\mathcal{A}^*\epsilon_0\bigg(- \nabla\cdot\Big[\frac{\dpp{ij}^3}{2}\bm{\theta}_{ij}^{q,(1)} + \frac{d_{pij}^4}{4}\bm{\theta}_{ij}^{q,(2)}\Big] + d_{pij}^2\chi_{ij}^{q,(1)} + \frac{d_{pij}^3}{2}\chi_{ij}^{q,(2)}\bigg).\label{Eq:collUnlikeCharge}
\end{eqnarray}
The superscript (1) refers to the first part in the joint density function given in \eqref{Eq:JointDensityFunction} and the superscript (2) stands for the natural logarithm of the function extension. Each of the terms given in \eqref{Eq:collUnlikeCharge} is explicitly defined as follows:
\begin{eqnarray}
    \bm{\theta}_{ij}^{q,(1)} & = & \int_{\kdotc > 0} |\kdotc|^{9/5}\kv \bigg[ \Ev\cdot\kv - \bigg(\frac{\varphi_{pi} - \varphi_{pj}}{\delta_c e} + \frac{1}{\upi\epsilon_0}\Big(\frac{q_{pj}}{\dpp{j}^2} - \frac{q_{pi}}{\dpp{i}^2}\Big)\bigg)\bigg]d\Gamma,\label{Eq:flux1Def}\\
    \bm{\theta}_{ij}^{q,(2)} & = & \int_{\kdotc > 0} |\kdotc|^{9/5}(\bfit{k}\otimes\bfit{k})\cdot\nabla\Big(\ln\frac{f_{pj}}{f_{pi}}\Big)\bigg[ \Ev\cdot\kv - \bigg(\frac{\varphi_{pi} - \varphi_{pj}}{\delta_c e} + \frac{1}{\upi\epsilon_0}\Big(\frac{q_{pj}}{\dpp{j}^2} - \frac{q_{pi}}{\dpp{i}^2}\Big)\bigg)\bigg]d\Gamma,\label{Eq:flux2Def} \hspace{0.7cm}\\
    \chi_{ij}^{q,(1)} & = & - \int_{\kdotc > 0} |\kdotc|^{9/5} \bigg[ \Ev\cdot\kv - \bigg(\frac{\varphi_{pi} - \varphi_{pj}}{\delta_c e} + \frac{1}{\upi\epsilon_0}\Big(\frac{q_{pj}}{\dpp{j}^2} - \frac{q_{pi}}{\dpp{i}^2}\Big)\bigg)\bigg]d\Gamma,\label{Eq:source1Def}\\
    \chi_{ij}^{q,(2)} & = & - \int_{\kdotc > 0} |\kdotc|^{9/5}\kv\cdot\nabla\Big(\ln\frac{f_{pj}}{f_{pi}}\Big)\bigg[ \Ev\cdot\kv - \bigg(\frac{\varphi_{pi} - \varphi_{pj}}{\delta_c e} + \frac{1}{\upi\epsilon_0}\Big(\frac{q_{pj}}{\dpp{j}^2} - \frac{q_{pi}}{\dpp{i}^2}\Big)\bigg)\bigg]d\Gamma, \label{Eq:source2Def}
\end{eqnarray}
with $d\Gamma = f_{pi}f_{pj} d\kv d\cv{i}d\cv{j} dq_{pi}dq_{pj}$. Let $\bfit{G}$ be the centre of mass velocity and $\bfit{w}$ be the relative velocity between two colliding particles with masses of $m_{pi}$ and $m_{pj}$:
\begin{eqnarray}
\bfit{G} & = & \frac{m_{pi}(\cv{i} - \Uv{i}) + m_{pj}(\cv{j} - \Uv{j})}{(m_{pi} + m_{pj})},\\
\bfit{w} & = & (\bfit{c}_{pj} - \bfit{U}_{pj}) - (\bfit{c}_{pi} - \bfit{U}_{pi}).
\end{eqnarray}
Then, the infinitesimal phase velocities are
\begin{eqnarray}
d\cv{i}d\cv{j} & = & \mathrm{det}\begin{vmatrix}
                            \frac{\partial \cv{i}}{\partial \bfit{G}} & \frac{\partial \cv{i}}{\partial \bfit{w}} \\
                            \frac{\partial \cv{j}}{\partial \bfit{G}} & \frac{\partial \cv{j}}{\partial \bfit{w}} \end{vmatrix}d\bfit{G}d\bfit{w} = d\bfit{G}d\bfit{w}.
\end{eqnarray}
The Cartesian $z$-coordinate aligns with the relative velocity $\bfit{w}$. The following two rotations and the rotation matrix are used to convert a point from Cartesian coordinates to spherical coordinates
\begin{equation}
        \mathsfbi{R} = \mathsfbi{R}_z(\phi')^T\mathsfbi{R}_y(\theta')^T = \begin{bmatrix}
	\cos(\phi') \cos(\theta') & \sin(\phi') & - \cos(\phi') \sin(\theta')\\
	- \sin(\phi') \cos(\theta') & \cos(\phi') & \sin(\phi') \sin(\theta')\\
	\sin(\theta') & 0 & \cos(\theta')\\\end{bmatrix}.
\end{equation}
The symbol $\kv$ is the unit vector that points from the particle $j$ to the particle $i$ and is defined with the angles $\theta$ and $\phi$ between $\kv$ and $\bfit{w}$:
\begin{eqnarray}
    \kv & = & \begin{bmatrix} \cos(\phi) \sin(\theta)\\
                                \sin(\phi) \sin(\theta)\\
                                \cos(\theta),\end{bmatrix}
\end{eqnarray}
and the solid angle is given as $d\kv = \sin(\theta) d\theta d\phi$. The differentials of the centre of mass velocity and the relative velocity are defined in the spherical coordinates:
\begin{eqnarray}
    d\bfit{G}d\bfit{w} & = & G^2 \sin(\theta^*)d\theta^*d\phi^*dG w^2 \sin(\theta')d\theta'd\phi'dw 
\end{eqnarray}
where $\theta^*$ and $\phi^*$ are the angles between $\bfit{G}$ and $\bfit{w}$. For a probable collision, the constraint is $\kdotc > 0$. The integration upper and lower bounds are then defined for both angles as $\phi=[0:2\upi]$ and $\theta=[0:\upi]$. 

The product of probability density function and the natural logarithm given in \eqref{Eq:flux1Def}-\eqref{Eq:source2Def} are written in the spherical coordinate system. The velocity contribution in the product of the probability density function is defined as:
\begin{equation}
f_{pi,c}f_{pj,c} = \frac{1}{(2\pi)^3}\Big(\frac{m_{pi}m_{pj}}{\Tp{i}\Tp{j}}\Big)^{3/2} \exp\Big(-(AG^2 + D w^2 + 2B G w\cos(\theta^*)) \Big).
\end{equation}
The definitions of coefficient $A$, $D$ and $B$ can be found in table \ref{tab:coefficients}. We use a Taylor expansion for the term $2B G w\cos(\theta^*)$ (the integral is not defined):
\begin{eqnarray}
f_{pi,c}f_{pj,c} & = & \frac{1}{(2\pi)^3}\Big(\frac{m_{pi}m_{pj}}{\Tp{i}\Tp{j}}\Big)^{3/2} \Big(1 - 2B Gw \cos(\theta^*) + 2(BGw)^2(\cos(\theta^*))^2 - \frac{4}{3}(BGw)^3(\cos(\theta^*))^3 \nonumber \\ 
& & + \frac{2}{3}(BGw)^4(\cos(\theta^*))^4 + ...\Big)\times\exp\Big(-(AG^2 + D w^2)\Big).\label{Eq:TaylorExpansionff}
\end{eqnarray}
The natural logarithm terms are then written as:
\begin{eqnarray}
\nabla\ln\Big(\frac{f_{pj}}{f_{pi}}\Big) & = & \nabla\ln\Big(\frac{n_{pj}}{n_{pi}}\Big) + \nabla\ln\Big(\frac{f_{pj,c}}{f_{pi,c}}\Big) + \nabla\ln\Big(\frac{f_{pj,q}}{f_{pi,q}}\Big)\\
\text{with } \nabla\ln\Big(\frac{f_{pj,c}}{f_{pi,c}}\Big) & = & \frac{3}{2}\nabla\ln\Big( \frac{\Tp{i}}{\Tp{j}}\Big) + \Big(m_{pj}\frac{\nabla\Tp{j}}{2\Tp{j}^2} - m_{pi}\frac{\nabla\Tp{i}}{2\Tp{i}^2} \Big) G^2 +  \frac{m_{pi}m_{pj}}{2(m_{pi} + m_{pj})^2}\nonumber\\
& \times & \Big(m_{pi}\frac{\nabla\Tp{j}}{\Tp{j}^2} - m_{pj}\frac{\nabla\Tp{i}}{\Tp{i}^2}\Big) w^2 - \frac{m_{pi}m_{pj}}{(m_{pi} + m_{pj})}\Big(\frac{\nabla\Tp{j}}{\Tp{j}^2} + \frac{\nabla\Tp{i}}{\Tp{i}^2}\Big) Gw\cos(\theta^*) \nonumber\\
& + & \Big(m_{pj}\frac{\nabla\Uv{j}}{\Tp{j}} - m_{pi}\frac{\nabla\Uv{i}}{\Tp{i}}\Big)\cdot\bfit{G} - \frac{m_{pi}m_{pj}}{m_{pi} + m_{pj}}\Big(\frac{\nabla\Uv{j}}{\Tp{j}} + \frac{\nabla\Uv{i}}{\Tp{i}}\Big)\cdot \bfit{w},\nonumber\\\label{Eq:naturalLogff}
\end{eqnarray}
The same procedure is applied for each integration; at first, the integral over $\kv$ is computed (the derivation functions can be found in table \ref{table:kInt}). With the help of the rotation matrix, we transform variables in the spherical coordinates to the Cartesian coordinates. Due to symmetry, the integral over an odd number of matrix turns to be zero. Finally, we compute the integrals over charge and velocity spaces. 

The first term in \eqref{Eq:flux1Def} after the integral over $\kv$ and the rotation is
\begin{eqnarray}
    \bm{\theta}_{ij}^{q,(1)} & = & \frac{20\upi^2}{21} (\Ev\cdot\Ii)\int w^{19/5} G^2 f_{pi}f_{pj}\sin(\theta^*)d\theta^*d\phi^*dG dw dq_{pi}dq_{pj}.
\end{eqnarray}
Using the Taylor expansion given in \eqref{Eq:TaylorExpansionff}, we compute the integrals over the angles $\theta^*$ and $\phi^*$ and the charge $q_{pi}$, $q_{pj}$:
\begin{eqnarray}
    \bm{\theta}_{ij}^{q,(1)} & = & \frac{10}{21}\Big(\frac{m_{pi}m_{pj}}{\Tp{i}\Tp{j}}\Big)^{3/2}  n_{pi}n_{pj}\Ev\int_0^{\infty}\int_0^{\infty} w^{19/5} G^2\Big(1 + \frac{2}{3}(BGw)^2 + \frac{2}{15}(BGw)^4 + ...\Big)\nonumber \\ 
    & \times &\exp\Big(-(AG^2 + D w^2)\Big) dG dw.
\end{eqnarray}
With solving the last integral, we obtain the first contribution for the flux term with the coefficients $N_k$ (k=1,...,5) given in table \ref{tab:coefficients}:
\begin{equation}
    \bm{\theta}_{ij}^{q,(1)} = \frac{5\sqrt{\upi}}{84}n_{pi}n_{pj}\Big(\frac{m_{pi}m_{pj}}{\Tp{i}\Tp{j}}\Big)^{3/2}  N_1\Ev. 
\end{equation}
The derivation of the second contribution of the flux term \eqref{Eq:flux2Def} starts with the integral over $\kv$  
{\small \begin{eqnarray}{\label{thetaq2}}
\bm{\theta}_{ij}^{q,(2)} & = & \frac{50\upi}{551}\Ev\cdot\underbrace{\int w^{19/5}G^2 \mathsfbi{R}\begin{bmatrix}
        \mathsfbi{R}\,[0, ~ 0, ~ 1]^T &  \mathsfbi{R}\,[0, ~ 0, ~ 0]^T &   \mathsfbi{R}\,[1, ~ 0, ~ 0]^T\\
        & \mathsfbi{R}\,[0, ~ 0, ~ 1]^T & \mathsfbi{R}\,[0, ~ 1, ~ 0]^T\\
        &  & \mathsfbi{R}\,[0, ~ 0, ~ \frac{19}{5}]^T\\ \end{bmatrix}\mathsfbi{R}^T\cdot\nabla\Big(\ln\frac{f_{pj}}{f_{pi}}\Big)d\Gamma'}_{I^{\theta}_1}\nonumber\\
& - & \underbrace{\int w^{19/5}G^2 \bigg(\frac{\varphi_{pi} - \varphi_{pj}}{\delta_c e} + \frac{1}{\upi\epsilon_0}\Big(\frac{q_{pj}}{\dpp{j}^2} - \frac{q_{pi}}{\dpp{i}^2}\Big)\bigg) \mathsfbi{R}\begin{bmatrix}
        1 & 0 & 0\\
        0 & 1 & 0\\
        0 & 0 & 14/5\\ \end{bmatrix}\mathsfbi{R}^T\cdot\nabla\Big(\ln\frac{f_{pj}}{f_{pi}}\Big)d\Gamma'}_{I^{\theta}_2}. \hspace{1cm}
\end{eqnarray}}
Here, $d\Gamma'$ is
\begin{equation}
d\Gamma' = f_{pi}f_{pj} \sin(\theta^*)d\theta^*d\phi^*dG \sin(\theta')d\theta'd\phi'dw dq_{pi}dq_{pj}.
\end{equation}
For the sake of the clarity, we decompose \eqref{thetaq2} in two contributions and start with the first integration. When applying rotation for the natural logarithm in \eqref{Eq:naturalLogff}, only the last term depending on the mean velocities and the vector $\bfit{w}$ remains, others are equal to be zero due to the odd number of rotational matrix and symmetry. It remains the integrals over the velocity norms and angles $\theta^*$ and $\phi^*$:
\begin{eqnarray}
I^{\theta}_1 & = & - \frac{3}{50\upi^2}n_{pi}n_{pj}\frac{m_{pi}m_{pj}}{(m_{pi} + m_{pj})}\Big(\frac{m_{pi}m_{pj}}{\Tp{i}\Tp{j}}\Big)^{3/2}\sum_{k = i,j}\bigg[\frac{1}{\Tp{k}}\Big( (\nabla\Uv{k}) + (\nabla\Uv{k})^T + \nabla\cdot\Uv{k}\Ii\Big)\bigg]\nonumber\\
    & \times & \int_0^{2\upi}\int_0^{\upi}\int_0^{\infty}\int_0^{\infty} w^{24/5}G^2\Big(1 - 2B Gw \cos(\theta^*) + 2(BGw)^2(\cos(\theta^*))^2 - \frac{4}{3}(BGw)^3(\cos(\theta^*))^3 \nonumber \\
    & & + \frac{2}{3}(BGw)^4(\cos(\theta^*))^4 + ...\Big)\exp\Big(-(AG^2 + D w^2) \Big)dwdG \sin(\theta^*) d\theta^* d\phi^*\\
    & = & - \frac{3}{100\sqrt{\upi}}n_{pi}n_{pj}\frac{m_{pi}m_{pj}}{(m_{pi} + m_{pj})}\Big(\frac{m_{pi}m_{pj}}{\Tp{i}\Tp{j}}\Big)^{3/2}\sum_{k = i,j}\bigg[\frac{1}{\Tp{k}}\Big( (\nabla\Uv{k}) + (\nabla\Uv{k})^T + \nabla\cdot\Uv{k}\Ii\Big)\bigg]N_5.\nonumber\\
\end{eqnarray}
For the second integral, $I^{\theta}_2$, the rotation cancels the last two terms in \eqref{Eq:naturalLogff}, it remains the gradients of the granular temperature and charge contributions. After that, the integrals over charges from $-\infty$ to $\infty$ are computed and it gives:
\begin{eqnarray}
I^{\theta}_2 & = & \frac{32\upi}{5}n_{pi}n_{pj}\int_0^{2\upi}\int_0^{\upi}\int_0^{\infty}\int_0^{\infty} w^{19/5}G^2\bigg(\bigg[\frac{\varphi_{pi} - \varphi_{pj}}{\delta_c e} + \frac{1}{\upi\epsilon_0}\Big(\frac{q_{pj}}{\dpp{j}^2} - \frac{q_{pi}}{\dpp{i}^2}\Big)\bigg]\times\Big[\nabla\ln\Big( \frac{n_{pj}}{n_{pi}}\Big)\nonumber\\
& + & \frac{3}{2}\nabla\ln\Big( \frac{\Tp{i}}{\Tp{j}}\Big) + \Big(m_{pj}\frac{\nabla\Tp{j}}{2\Tp{j}^2} - m_{pi}\frac{\nabla\Tp{i}}{2\Tp{i}^2} \Big) G^2 +  \frac{m_{pi}m_{pj}}{2(m_{pi} + m_{pj})^2}\nonumber\\
& \times &\Big(m_{pi}\frac{\nabla\Tp{j}}{\Tp{j}^2} - m_{pj}\frac{\nabla\Tp{i}}{\Tp{i}^2}\Big) w^2  - \frac{m_{pi}m_{pj}}{(m_{pi} + m_{pj})}\Big(\frac{\nabla\Tp{j}}{\Tp{j}^2} + \frac{\nabla\Tp{i}}{\Tp{i}^2}\Big) Gw\cos(\theta^*) \Big]\nonumber\\
& + & \frac{\nabla Q_{pj}}{\pi\epsilon_0d_{p_j}^2} + \frac{\nabla Q_{pi}}{\pi\epsilon_0d_{p_i}^2}\bigg) f_{pi,c}f_{pj,c}dGdw\sin(\theta^*)d\theta^*d\phi^*.
\end{eqnarray}
With solving the remaining integrals, we obtain:
\begin{eqnarray}
I^{\theta}_2 & = & \frac{2}{5\sqrt{\upi}}n_{pi}n_{pj}\Big(\frac{m_{pi}m_{pj}}{\Tp{i}\Tp{j}}\Big)^{3/2} \bigg[\bigg(\Big[\frac{\varphi_{pi} - \varphi_{pj}}{\delta_c e} + \frac{1}{\upi\epsilon_0}\Big(\frac{q_{pj}}{\dpp{j}^2} - \frac{q_{pi}}{\dpp{i}^2}\Big)\Big]\times\Big[\nabla\ln\Big( \frac{n_{pj}}{n_{pi}}\Big)\nonumber\\
& + & \frac{3}{2}\nabla\ln\Big( \frac{\Tp{i}}{\Tp{j}}\Big)\Big] + \frac{\nabla Q_{pj}}{\pi\epsilon_0d_{p_j}^2} + \frac{\nabla Q_{pi}}{\pi\epsilon_0d_{p_i}^2} \bigg)N_1 + \frac{3}{4}\Big[\frac{\varphi_{pi} - \varphi_{pj}}{\delta_c e} + \frac{1}{\upi\epsilon_0}\Big(\frac{q_{pj}}{\dpp{j}^2} - \frac{q_{pi}}{\dpp{i}^2}\Big)\Big]\nonumber\\
& \times & \bigg(\Big(m_{pj}\frac{\nabla\Tp{j}}{\Tp{j}^2} - m_{pi}\frac{\nabla\Tp{i}}{\Tp{i}^2} \Big) N_2 +  \frac{m_{pi}m_{pj}}{2(m_{pi} + m_{pj})^2}\nonumber\\
& \times &\Big(m_{pi}\frac{\nabla\Tp{j}}{\Tp{j}^2} - m_{pj}\frac{\nabla\Tp{i}}{\Tp{i}^2}\Big) N_3 + B\frac{m_{pi}m_{pj}}{(m_{pi} + m_{pj})}\Big(\frac{\nabla\Tp{j}}{\Tp{j}^2} + \frac{\nabla\Tp{i}}{\Tp{i}^2}\Big) N_4 \bigg)\bigg].
\end{eqnarray}

The source terms are derived by following the same procedure. We start with the first part of the source term, \eqref{Eq:source1Def}:
\begin{eqnarray}
\chi_{ij}^{q,(1)} & = & \frac{20\upi^2}{7}\int w^{19/5}G^2 \bigg(\frac{\varphi_{pi} - \varphi_{pj}}{\delta_c e} + \frac{1}{\upi\epsilon_0}\Big(\frac{q_{pj}}{\dpp{j}^2} - \frac{q_{pi}}{\dpp{i}^2}\Big)\bigg)f_{pi}f_{pj} \sin(\theta^*)d\theta^*d\phi^*dG dw dq_{pi}dq_{pj}\nonumber\\
& = & \frac{20\upi^2}{7}n_{pi}n_{pj}\int w^{19/5}G^2 \bigg(\frac{\varphi_{pi} - \varphi_{pj}}{\delta_c e} + \frac{1}{\upi\epsilon_0}\Big(\frac{Q_{pj}}{\dpp{j}^2} - \frac{Q_{pi}}{\dpp{i}^2}\Big)\bigg)f_{pi,c}f_{pj,c} \sin(\theta^*)d\theta^*d\phi^*dG dw \nonumber\\
& = & \frac{10}{7}n_{pi}n_{pj}\bigg(\frac{\varphi_{pi} - \varphi_{pj}}{\delta_c e} + \frac{1}{\upi\epsilon_0}\Big(\frac{Q_{pj}}{\dpp{j}^2} - \frac{Q_{pi}}{\dpp{i}^2}\Big)\bigg)\Big(\frac{m_{pi}m_{pj}}{\Tp{i}\Tp{j}}\Big)^{3/2} \int_{0}^{\infty}\int_{0}^{\infty}  w ^{19/5}G^2 \nonumber\\
&\times &\Big(1 + \frac{2}{3}(BGw)^2 + \frac{2}{15}(BGw)^4 + ...\Big)\exp\Big(-(AG^2 + D w^2) \Big)dwdG\nonumber\\
& = & \frac{5\sqrt{\upi}}{28}n_{pi}n_{pj}\bigg(\frac{\varphi_{pi} - \varphi_{pj}}{\delta_c e} + \frac{1}{\upi\epsilon_0}\Big(\frac{Q_{pj}}{\dpp{j}^2} - \frac{Q_{pi}}{\dpp{i}^2}\Big)\bigg)\Big(\frac{m_{pi}m_{pj}}{\Tp{i}\Tp{j}}\Big)^{3/2} N_1.
\end{eqnarray}
For the second part of the source term, \eqref{Eq:source2Def}, we compute the $\kv$ integral:
\begin{eqnarray}
\chi_{ij}^{q,(2)} & = & \frac{10\upi}{19}\underbrace{\int w^{19/5}G^2\mathsfbi{R}\begin{bmatrix}0\\0\\1\\\end{bmatrix}\cdot\nabla\Big(\ln\frac{f_{pj}}{f_{pi}}\Big)\bigg(\frac{\varphi_{pi} - \varphi_{pj}}{\delta_c e} + \frac{1}{\upi\epsilon_0}\Big(\frac{q_{pj}}{\dpp{j}^2} - \frac{q_{pi}}{\dpp{i}^2}\Big)\bigg)d\Gamma'}_{I_1^{\chi}}\nonumber\\
& - & \frac{25\upi}{168} \Ev\cdot\underbrace{\int w^{19/5}G^2\mathsfbi{R}\begin{bmatrix} 1 & 0 & 0\\
0 & 1 & 0\\
0 & 0 & 14/5 \end{bmatrix}\mathsfbi{R}^T\cdot\nabla\Big(\ln\frac{f_{pj}}{f_{pi}}\Big)d\Gamma'}_{I_2^{\chi}}.
\end{eqnarray}
While applying the rotation for the first integral, $I_1^{\chi}$, only the last of \eqref{Eq:naturalLogff} remains, others terms vanish due to the odd number of rotation matrix:
\begin{eqnarray}
I_1^{\chi} & = & - \frac{4\upi}{3}\frac{m_{pi}m_{pj}}{m_{pi} + m_{pj}}\Big(\frac{\nabla\Uv{j}}{\Tp{j}} + \frac{\nabla\Uv{i}}{\Tp{i}}\Big)\nonumber\\
& \times & \int w^{24/5}G^2\bigg(\frac{\varphi_{pi} - \varphi_{pj}}{\delta_c e} + \frac{1}{\upi\epsilon_0}\Big(\frac{q_{pj}}{\dpp{j}^2} - \frac{q_{pi}}{\dpp{i}^2}\Big)\bigg)f_{pi}f_{pj} \sin(\theta^*)d\theta^*d\phi^*dG dw dq_{pi}dq_{pj}\nonumber\\
& = & - \frac{4\upi}{3}n_{pi}n_{pj}\frac{m_{pi}m_{pj}}{m_{pi} + m_{pj}}\Big(\frac{\nabla\Uv{j}}{\Tp{j}} + \frac{\nabla\Uv{i}}{\Tp{i}}\Big)\bigg(\frac{\varphi_{pi} - \varphi_{pj}}{\delta_c e} + \frac{1}{\upi\epsilon_0}\Big(\frac{Q_{pj}}{\dpp{j}^2} - \frac{Q_{pi}}{\dpp{i}^2}\Big)\bigg)\nonumber\\
& \times & \int w^{24/5}G^2f_{pi,c}f_{pj,c} \sin(\theta^*)d\theta^*d\phi^*dG dw \nonumber\\
& = & - \frac{n_{pi}n_{pj}}{12\sqrt{\upi}}\frac{m_{pi}m_{pj}}{m_{pi} + m_{pj}}\Big(\frac{m_{pi}m_{pj}}{\Tp{i}\Tp{j}}\Big)^{3/2}\Big(\frac{\nabla\Uv{j}}{\Tp{j}} + \frac{\nabla\Uv{i}}{\Tp{i}}\Big)\bigg(\frac{\varphi_{pi} - \varphi_{pj}}{\delta_c e} + \frac{1}{\upi\epsilon_0}\Big(\frac{Q_{pj}}{\dpp{j}^2} - \frac{Q_{pi}}{\dpp{i}^2}\Big)\bigg)N_5\nonumber\\
\end{eqnarray}
For $I_2^{\chi}$, the mean velocity terms in \eqref{Eq:naturalLogff} vanishes and it remains all terms with the gradients of granular temperature and charge contributions. After solving the rotation and charge integrals, we obtain:
\begin{eqnarray}
I_2^{\chi} & = & \frac{32\upi}{5}n_{pi}n_{pj}\int w^{19/5}G^2 \bigg[\nabla\ln\Big( \frac{n_{pi}}{n_{pj}}\Big) + \frac{3}{2}\nabla\ln\Big( \frac{\Tp{i}}{\Tp{j}}\Big) + \Big(m_{pj}\frac{\nabla\Tp{j}}{2\Tp{j}^2} - m_{pi}\frac{\nabla\Tp{i}}{2\Tp{i}^2} \Big) G^2\nonumber\\
& & +  \frac{m_{pi}m_{pj}}{2(m_{pi} + m_{pj})^2} \Big(m_{pi}\frac{\nabla\Tp{j}}{\Tp{j}^2} - m_{pj}\frac{\nabla\Tp{i}}{\Tp{i}^2}\Big) w^2 - \frac{m_{pi}m_{pj}}{(m_{pi} + m_{pj})}\Big(\frac{\nabla\Tp{j}}{\Tp{j}^2} + \frac{\nabla\Tp{i}}{\Tp{i}^2}\Big)\nonumber\\
& \times & Gw\cos(\theta^*) \bigg]f_{pi,c}f_{pj,c} \sin(\theta^*)d\theta^*d\phi^*dG dw
\end{eqnarray}
Finally after the last integrals, we obtain:
\begin{eqnarray}
I_2^{\chi} & = & \frac{2}{5\sqrt{\upi}}n_{pi}n_{pj}\Big(\frac{m_{pi}m_{pj}}{\Tp{i}\Tp{j}}\Big)^{3/2}\bigg[\bigg(\nabla\ln\Big( \frac{n_{pi}}{n_{pj}}\Big) + \frac{3}{2}\nabla\ln\Big( \frac{\Tp{i}}{\Tp{j}}\Big)\bigg)N_1 + \frac{3}{4}\Big(m_{pj}\frac{\nabla\Tp{j}}{\Tp{j}^2} - m_{pi}\frac{\nabla\Tp{i}}{\Tp{i}^2} \Big)N_2 \nonumber\\
& & +  \frac{m_{pi}m_{pj}}{2(m_{pi} + m_{pj})^2} \Big(m_{pi}\frac{\nabla\Tp{j}}{\Tp{j}^2} - m_{pj}\frac{\nabla\Tp{i}}{\Tp{i}^2}\Big) N_3 + \frac{m_{pi}m_{pj}}{(m_{pi} + m_{pj})}\Big(\frac{\nabla\Tp{j}}{\Tp{j}^2} + \frac{\nabla\Tp{i}}{\Tp{i}^2}\Big)BN_4\bigg]\nonumber\\
\end{eqnarray}
We arrange all these terms in \eqref{Eq:collUnlikeCharge} and summarise the complete set of equations for the collisional term as
\begin{equation}
    \mathcal{C}(q_{pi}) = \sum_{h = i,j}\Big(- \nabla\cdot\bm{\theta}_{ih}^q(q_{pi}) + \chi_{ih}^q(q_{pi})\Big)
\end{equation}
with the flux term, $\bm{\theta}_{ih}^q$,
\begin{eqnarray}
    & & \bm{\theta}_{ih}^q =  - n_{pi}n_{ph}\Big(\frac{m_{pi}m_{ph}}{\Tp{i}\Tp{h}}\Big)^{3/2}\mathcal{A}^*\epsilon_0g_0\frac{d_{pih}^3}{2}\Bigg[ -\frac{5\sqrt{\upi}}{84}\bfit{E} N_1 + \frac{d_{pih}}{8}\sqrt{\upi}\nonumber\\
    & & \times\Bigg(\frac{5}{21}\bigg[\Big(\frac{\varphi_{pi} - \varphi_{ph}}{\delta_c e} + \frac{1}{\upi\epsilon_0}(\frac{Q_{ph}}{d_{ph}^2} - \frac{Q_{pi}}{d_{pi}^2})\Big)\times\Big(\nabla\ln\Big( \frac{n_{ph}}{n_{pi}}\Big) + \frac{3}{2}\nabla\ln\Big( \frac{\Tp{i}}{\Tp{h}}\Big)\Big) + \frac{\nabla Q_{ph}}{\upi\epsilon_0d_{ph}^2}\nonumber \\
    & & + \frac{\nabla Q_{pi}}{\upi\epsilon_0d_{pi}^2}\bigg] N_1 + \Big(\frac{\varphi_{pi} - \varphi_{ph}}{\delta_c e} + \frac{1}{\upi\epsilon_0}(\frac{Q_{ph}}{d_{ph}^2} - \frac{Q_{pi}}{d_{pi}^2})\Big)\times\bigg[ \frac{5}{28}\Big(m_{ph}\frac{\nabla\Tp{h}}{2\Tp{h}^2} - m_{pi}\frac{\nabla\Tp{i}}{2\Tp{i}^2}\Big)N_2  \nonumber\\
    & & + \frac{5}{42}\frac{m_{pi}m_{ph}}{(m_{pi} + m_{ph})^2} \Big(m_{pi}\frac{\nabla\Tp{h}}{\Tp{h}^2} - m_{ph}\frac{\nabla\Tp{i}}{\Tp{i}^2}\Big)N_3  + \frac{5}{21}B\frac{m_{pi}m_{ph}}{(m_{pi} + m_{ph})}\Big(\frac{\nabla\Tp{h}}{\Tp{h}^2} + \frac{\nabla\Tp{i}}{\Tp{i}^2}\Big)N_4\bigg]\nonumber\\
   & & + \frac{3}{551}\frac{m_{pi}m_{ph}}{(m_{pi} + m_{ph})}N_5\bfit{E}\cdot\sum_{l = i,j}\bigg[\frac{1}{\Tp{l}}\bigg( (\nabla\bfit{U}_{pl}) + (\nabla\bfit{U}_{pl})^T + \nabla\cdot\bfit{U}_{pl}\Ii\bigg)\bigg]\Bigg) \Bigg],\label{Eq:FluxCharge}
\end{eqnarray}
and the source term, $\chi^{q}_{ih}$,    
\begin{eqnarray}  
    & & \chi^{q}_{ih}  =  n_{pi}n_{ph}\Big(\frac{m_{pi}m_{ph}}{\Tp{i}\Tp{h}}\Big)^{3/2}\mathcal{A}^*\epsilon_0g_0d_{pih}^2\frac{5\sqrt{\upi}}{28}\Bigg[\Big(\frac{\varphi_{pi} - \varphi_{ph}}{\delta_c e} + \frac{1}{\upi\epsilon_0}(\frac{Q_{ph}}{d_{ph}^2} - \frac{Q_{pi}}{d_{pi}^2})\Big)N_1 \nonumber\\
    & & - \frac{7d_{pih}}{57}\frac{m_{pi}m_{ph}}{(m_{pi} + m_{ph})}\Big(\frac{\varphi_{pi} - \varphi_{ph}}{\delta_c e} + \frac{1}{\upi\epsilon_0}(\frac{Q_{ph}}{d_{ph}^2} - \frac{Q_{pi}}{d_{pi}^2})\Big) \times\Big(\frac{\nabla\cdot\bfit{U}_{ph}}{\Tp{h}} + \frac{\nabla\cdot\Uv{i}}{\Tp{i}}\Big)N_5\nonumber\\
    & & - \frac{d_{pih}}{6}\bfit{E}\cdot\bigg[\Big[\nabla\ln\Big( \frac{n_{ph}}{n_{pi}}\Big) + \frac{3}{2}\nabla\ln\Big( \frac{\Tp{i}}{\Tp{h}}\Big)\Big]N_1 + \frac{3}{4}\Big(m_{ph}\frac{\nabla\Tp{h}}{\Tp{h}^2} - m_{pi}\frac{\nabla\Tp{i}}{\Tp{i}^2}\Big) N_2\nonumber\\
    & & + \frac{m_{pi}m_{ph}}{2(m_{pi} + m_{ph})^2}\Big(m_{pi}\frac{\nabla\Tp{h}}{\Tp{h}^2} - m_{ph}\frac{\nabla\Tp{i}}{\Tp{i}^2}\Big) N_3 + \frac{m_{pi}m_{ph}}{m_{pi} + m_{ph}}\Big(\frac{\nabla\Tp{h}}{\Tp{h}^2} + \frac{\nabla\Tp{i}}{\Tp{i}^2}\Big)B N_4\bigg]\Bigg].\nonumber\\
    & & \label{Eq:SourceCharge}
\end{eqnarray}
The coefficients $N_k$ ($k=1..5$) and $B$ are listed in table \ref{tab:coefficients}. If we simply these terms for a solid phase with uniform size distribution, we obtain the same equations as \cite{kolehmainen_eulerian_2018}:
\begin{equation}\label{Eq:FluxChargeii}
    \bm{\theta}_{ii}^q =  \sigma_q \bfit{E} - \kappa_q \nabla Q_{pi},
\end{equation}
with the triboelectric conductivity, $\sigma_q$, 
\begin{equation}
    \sigma_q = 2^{14/5}\frac{5\upi\sqrt{\upi}}{21} n_{pi}^2 d_{pi}^3 g_0\epsilon_0 \Gamma\Big(\frac{12}{5}\Big)r_p^*\Big(\frac{15m_p^*}{16Y_p^*\sqrt{r_p^*}}\Big)^{2/5}\Big(\frac{\Tp{i}}{m_{pi}}\Big)^{9/10}\label{Eq:sigmaii}
\end{equation}
and the triboelectric diffusivity, $\kappa_q$,
\begin{equation}
    \kappa_q = 2^{14/5}\frac{5\sqrt{\upi}}{21} n_{pi}^2 d_{pi}^2 g_0 \Gamma\Big(\frac{12}{5}\Big)r_p^*\Big(\frac{15m_p^*}{16Y_p^*\sqrt{r_p^*}}\Big)^{2/5}\Big(\frac{\Tp{i}}{m_{pi}}\Big)^{9/10}.\label{Eq:kappaii}
\end{equation}
The source term is equal to zero:
\begin{equation}\label{Eq:SourceChargeii}
    \chi^{q}_{ii} = 0.
\end{equation} 

\clearpage
\begin{table}
    \centering
    \begin{tabular}{lll}
        {\large$\int_{\kdotc > 0} (\kv\cdot\bfit{w})^{9/5} d\kv$} & = & {\large$w^{9/5}\int_0^{2\upi}\int_0^{\upi/2} (\cos(\theta))^{9/5}\sin(\theta) d\theta d\phi$}\\[1.5ex]
        & = & {\large$w^{9/5} \frac{5\upi}{7}$}\\[1.5ex]
        {\large$\int_{\kdotc > 0} (\kv\cdot\bfit{w})^{9/5}\kv d\kv$} & = & {\large$w^{9/5}\int_0^{2\upi}\int_0^{\upi/2} \kv(\cos(\theta))^{9/5}\sin(\theta) d\theta d\phi$}\\[1.5ex]
        & = & {\large$w^{9/5} \frac{10\upi}{19}(0, ~ 0, ~ 1)^T$}\\[1.5ex]
        {\large$\int_{\kdotc > 0} (\kv\cdot\bfit{w})^{9/5}(\kv\otimes\kv)d\kv$} & = & {\large$w^{9/5}\int_0^{2\upi}\int_0^{\upi/2} (\kv \otimes \kv)(\cos(\theta))^{9/5}\sin(\theta) d\theta d\phi$}\\[1.5ex]
        & = & {\large$w^{9/5} \frac{25\upi}{168}\begin{bmatrix}
	        1 & 0 & 0\\
	        0 & 1 & 0\\
	        0 & 0 & 14/5\\\end{bmatrix}$}\\[1.5ex]
	    {\large$\int_{\kdotc > 0} (\kv\cdot\bfit{w})^{9/5}(\kv\otimes\kv\otimes\kv)d\kv$} & = & {\large $w^{9/5}\int_0^{2\upi}\int_0^{\upi/2} (\kv \otimes \kv\otimes\kv)(\cos(\theta))^{9/5}\sin(\theta) d\theta d\phi$}\\[1.5ex]
        & = &  {\large $\frac{50\upi}{551}w^{9/5}\begin{bmatrix}
        [0, ~ 0, ~ 1]^T &  [0, ~ 0, ~ 0]^T & [1, ~ 0, ~ 0]^T\\
        & [0, ~ 0, ~ 1]^T & [0, ~ 1, ~ 0]^T\\
        &  & [0, ~ 0, ~ \frac{19}{5}]^T\\ \end{bmatrix}$}\\
    \end{tabular}
    \caption{List of integration over the unit vector $\kv$.} 
    \label{table:kInt}
\end{table}

\section{Hard sphere modelling}\label{appB}
The hard sphere model in this study is based on the time-stepped algorithm \citep{kolehmainen_eulerian_2018}. The particles are moved along their trajectories with small steps according to their velocity:
\begin{equation}\label{Eq:predictedPosition}
\boldsymbol{x}_{pi}^{(n+1)*} = \boldsymbol{x}_{pi}^{(n)} + \boldsymbol{v}_{pi}^{(n)} \Delta t,
\end{equation} where $\Delta t$ is the time step; $\boldsymbol{x}_{pi}^{(n)}$ and $\boldsymbol{v}_{pi}^{(n)}$ are the particle position and velocity at time $n \Delta t$; and $\boldsymbol{x}_{pi}^{(n+1)*}$ is the predicted particle location. 
With the predicted locations, we ensure that particles are not overlapping according to their respective radius: $\|\boldsymbol{x}_{pi}^{(n+1)*} - \boldsymbol{x}_{pj}^{(n+1)*}\| > d_{pi} + d_{pj}$. When overlapping occurs, the time is reversed for these two particles to first time of contact. The velocities of particles are updated according to hard sphere model:
\begin{equation} 
\boldsymbol{v}_{pi}^{(n+1)} = \boldsymbol{v}_{pi}^{(n)} + \frac{m_{pi}}{(m_{pi} + m_{pj})}(1 + e_c)\left( (\boldsymbol{v}_{pj}^{(n)}-\boldsymbol{v}_{pi}^{(n)}) \cdot \boldsymbol{k}_{ij} \right) \boldsymbol{k}_{ij},
\end{equation}
where $\boldsymbol{k}_{ij}$ is unit vector pointing from particle $i$ to particle $j$, $e_c$ is the coefficient of restitution and $m_p$ is the mass of particle. The predicted locations in \eqref{Eq:predictedPosition} are computed using the update velocities due to collision. 

During a particle-particle contact, the charge transfer occurs by following \eqref{Eq:qiPlus} and \eqref{Eq:qjPlus}. The electric field at particle positions is computed by mapping all the particle charges into the Eulerian cells to compute charge densities $\rho_q$. For fully periodic simulations, the electric field at cell each is resolved with a spectral method as:

\begin{equation}
\boldsymbol{E}(\boldsymbol{x}) = \mathcal{F}^{-1} \left( \frac{ \mathcal{F}(\rho_q(\boldsymbol{x}) )  }{ \| \lambdabar \|^2 } \lambdabar \right),
\end{equation} 
where $\lambdabar$ is the wave-number vector; and $\mathcal{F}$ and $\mathcal{F}^{-1}$ refer to Fourier transform and inverse Fourier transform respectively. When a collision between a particle and the wall occurs, the particle velocity is updated following \eqref{postCollVelx}-\eqref{postCollVelz}. During a particle-wall contact, the transfer of charge is modelled by \cite{kolehmainen_eulerian_2018} as:
\begin{equation}
    \Delta q = \frac{\beta_q}{\alpha_q}(1 - e^{-\alpha_q\mathcal{A}_{max}})
\end{equation}
where $\alpha_q$ is a geometrical factor depending on the type of collision; $\alpha_q$ = $\frac{2}{\pi d_p^2}$. $\mathcal{A}_{max}$ is the maximum area of contact defined in \eqref{Eq:maxArea} with the following effective parameters; $Y_p^* = \frac{Y_{p}}{1 - \nu_{p}^2}$, $r_p^* = \frac{d_p}{2}$ and $m_p^* = m_p$. $\beta_q$ is defined as:
\begin{equation}
    \beta_q = \epsilon_0\Big(\frac{\Delta\varphi_{w-p}}{\delta_c e} - \frac{2q_p}{\epsilon_0\pi d_p^2} - \Ev\cdot\kv\Big)
\end{equation}
with the first term representing the work function difference between particle and wall, $\Delta\varphi_{w-p}$, the second term is due to the charge carried by particle and the last term is due to the electric field resulting from charge on surrounding particles. For the wall bounded flow configuration, the electric field is solved by finite difference based on \eqref{poisson} and \eqref{Ei}. First, the linear system for the electrical potential is solved using a second-order central difference with the discrete three-dimensional Laplacian matrix $\mathcal{L}$:
\begin{equation}
    \mathcal{L}\,\phi = \frac{\rho_{q,cell}^s}{\epsilon_0}
\end{equation}
where $\rho_{q,cell}^s$ is the interpolated charge density at the surface of the cell. At the walls, we impose the electric potential equal be zero. Finally, the electric field is then computed following \eqref{Ei} with a first-order finite difference method in each direction.

\newpage

\bibliographystyle{jfm}
\bibliography{bibli}

\end{document}